\documentclass[preprint,10pt,3p,twocolumn]{elsarticle}

\journal{Computer Physics Communications}

\usepackage{amsfonts}
\usepackage{amsmath}
\usepackage{amssymb}
\usepackage{eucal}
\usepackage{graphicx}
\usepackage{hyperref}
\usepackage{multirow}
\usepackage{xspace}
\usepackage{xcolor}
\usepackage{soul} 
\usepackage{microtype} 

\renewcommand{\vec}[1]{{\mathbf{#1}}}
\def\akap{\bar\kappa}
\def\br{{\vec{r}}}
\def\bR{{\vec{R}}}
\def\bk{{\vec{k}}}
\def\bq{{\vec{q}}}

\def\PkL{P_{kL}}
\def\tPkL{\tilde{P}_{kL}}
\def\pkl{p_{kl}}
\def\tpkl{\tilde{p}_{kl}}
\def\CikL{C^{(i)}_{kL}}
\def\Ves{V^{\rm es}}
\def\tVes{\tilde{V}^{\rm es}}

\def\d3r{\,d^3r}
\def\lbr{\Bigl\{}
\def\rbr{\Bigr\}}

\def\epxc{\epsilon_{\rm xc}}
\def\Rmt{s_{R}}
\def\dprime{{\prime\prime}}
\def\tprime{{\prime\prime\prime}}
\def\ekn{{\varepsilon_{kn}}}
\def\GLDA{{G^\text{LDA}}}
\def\WLDA{{W^\text{LDA}}}
\def\HLDA{{H^\text{LDA}}}

\newcommand{\myvec}[1]{\vec{#1}}
\newcommand{\ftrns}[1]{\widehat{#1}}
\newcounter{Alist}

\makeatletter
\AtBeginDocument{\@ifpackageloaded{natbib}{
\ifNAT@numbers\if@filesw\immediate\write\@auxout{\string\global\string\NAT@numberstrue}\fi\fi}{}}
\makeatother

\begin{document}

\begin{frontmatter}
\title{Questaal: a package of electronic structure methods based on the linear muffin-tin orbital technique}

\author[kings]{Dimitar Pashov}
\author[kings]{Swagata Acharya}
\author[case]{Walter R. L. Lambrecht}
\author[daresbury]{Jerome Jackson\corref{corr}}
\author[unl]{Kirill D. Belashchenko}
\author[aps]{Athanasios Chantis}
\author[kings]{Francois Jamet}
\author[kings]{Mark van Schilfgaarde}

\cortext[corr]{Corresponding author. E-mail address: jerome.jackson@stfc.ac.uk}

\address[kings]{Department of Physics, King's College London, Strand, London WC2R 2LS, United Kingdom}
\address[case]{Department of Physics, Case Western Reserve University, Cleveland, Ohio 44106, USA}
\address[daresbury]{Scientific Computing Department, STFC Daresbury Laboratory, Warrington WA4 4AD, United Kingdom}
\address[unl]{Department of Physics and Astronomy and Nebraska Center for Materials and Nanoscience, University of
Nebraska-Lincoln, Lincoln, Nebraska 68588, USA}
\address[aps]{American Physical Society, 1 Research Road, Ridge, New York 11961, USA}

\begin{abstract}
This paper summarises the theory and functionality behind \href{https://www.questaal.org}{Questaal}, an open-source
suite of codes for calculating the electronic structure and related properties of materials from first principles.  The
formalism of the linearised muffin-tin orbital (LMTO) method is revisited in detail and developed further by the
introduction of short-ranged tight-binding basis functions for full-potential calculations.  The LMTO method is
presented in both Green's function and wave function formulations for bulk and layered systems.  The suite's
full-potential LMTO code uses a sophisticated basis and augmentation method that allows an efficient and precise
solution to the band problem at different levels of theory, most importantly density functional theory, LDA$+U$,
quasi-particle self-consistent \emph{GW} and combinations of these with dynamical mean field theory.  This paper details
the technical and theoretical bases of these methods, their implementation in Questaal, and provides an overview of the
code's design and capabilities.
\end{abstract}

\begin{keyword} 
Questaal
\sep Linear Muffin Tin Orbital
\sep Screening Transformation
\sep Density Functional Theory
\sep Many-Body Perturbation Theory
\end{keyword}

\end{frontmatter}
{\bf PROGRAM SUMMARY}

\begin{small}
{\em Program Title:} Questaal\\
{\em Licensing provisions:} GNU General Public License, version 3\\
{\em Programming language:} Fortran, C, Python, Shell\\
{\em Nature of problem:} Highly accurate \emph{ab initio} calculation of the electronic structure of periodic solids and
of the resulting physical, spectroscopic and magnetic properties for diverse material classes with different strengths
and kinds of electronic correlation.\\
{\em Solution method:} The many electron problem is considered at different levels of theory: density functional theory,
many body perturbation theory in the \emph{GW} approximation with different degrees of self consistency (notably
quasiparticle self-consistent \emph{GW}) and dynamical mean field theory.  The solution to the single-particle band
problem is achieved in the framework of an extension to the linear muffin-tin orbital (LMTO) technique including a
highly precise and efficient full-potential implementation.  An advanced fully-relativistic, non-collinear
implementation based on the atomic sphere approximation is used for calculating transport and magnetic properties.\\
\end{small}


\section{Introduction}
\label{sec:intro}
\begin{figure}[ht]
\begin{center}
\includegraphics[width=.20\textwidth]{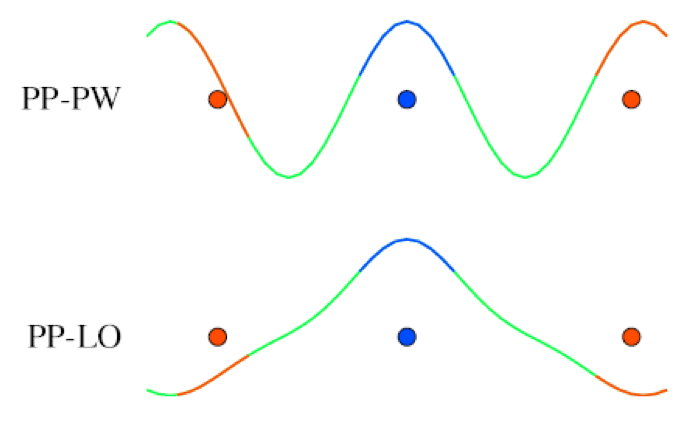}\
\includegraphics[width=.24\textwidth]{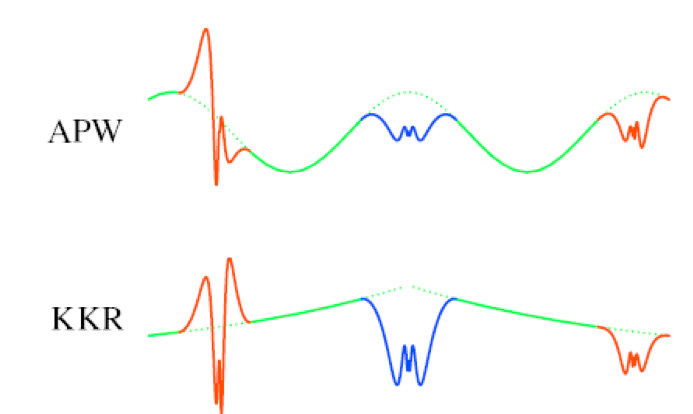}
\end{center}
\caption{$2{\times}2$ rubric for main classifications of basis set.  Nuclei are shown as dots.  The all-electron methods
APW and KKR on the right substitute (augment) the envelope function (green) with numerical solutions of partial waves
inside augmentation spheres (blue and red).  Parts inside augmentation spheres are called ``partial waves.''  The two
figures on the left use a pseudopotential allowing their envelope functions to be smooth, with no augmentation needed.
A pseudopotential's radius corresponds to a characteristic augmentation radius.  The top two figures use plane waves for
envelope functions; the bottom two use atom-centred local basis sets.  We denote localised basis sets as ``KKR'', for
the Korringa-Kohn-Rostocker method~\cite{Korringa94} as it plays a central role in this work; but there are other kinds,
for example the Gaussian orbitals widely favoured among quantum chemists.}
\label{fig:basisrubric}
\end{figure}
\begin{figure}[ht]
\begin{center}
\includegraphics[width=.45\columnwidth]{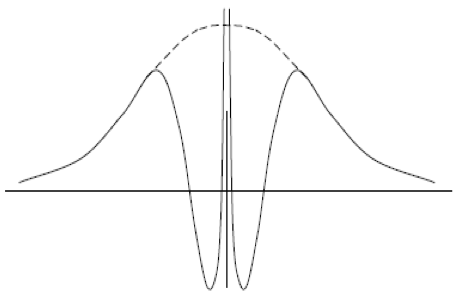}
\raisebox{-2 em}{\includegraphics[width=.45\columnwidth]{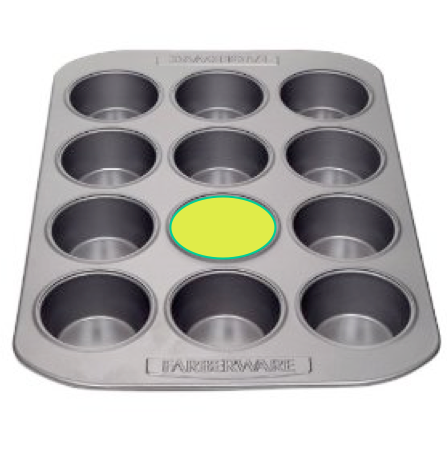}}
\end{center}
\caption{Left: basis function in the vicinity of a nucleus, showing orthogonalisation to core states (solid line), and
corresponding basis function in a pseudopotential (dashed line).  Right: a muffin-tin potential: the
potential is flat in the interstitial region between sites, and spherically symmetric in a volume around each site.
Blue depicts an augmentation radius $\Rmt$ around a sphere centred at some nuclear position $\bR$ where the interstitial
and augmented regions join.}
\label{fig:pseudowf}
\end{figure}

Different implementations of DFT are distinguished mainly by their basis set, which forms the core of any electronic
structure method, and how they orthogonalise themselves to the core levels.  Using these classifications most methods
adopt one of four possible combinations shown in Fig.~\ref{fig:basisrubric}.  In the vast majority of cases, basis sets
consist of either atom-centred spatially localised functions (lower panel of Fig.~\ref{fig:basisrubric}), or plane waves
(PW) (upper).  As for treatment of the core, it is very common to substitute an effective repulsive (pseudo)potential to
simulate its effect, an idea initially formulated by Conyers Herring~\cite{Herring40}.  Pseudopotentials make it
possible to avoid orthogonalisation to the core, which allows the (pseudo)wave functions to be nodeless and smooth.  For
methods applied to condensed matter, the primary alternative method, formulated by Slater in 1937~\cite{SlaterAPW},
keeps all the electrons.  Space is partitioned into non-overlapping spheres centred at each nucleus, with the
interstitial region making up the rest. The basis functions are defined by plane waves in the interstitial, which are
replaced (``augmented'') by numerical solutions of the Schr\"odinger equations (partial waves) inside the augmentation
spheres.  The two solutions must be joined smoothly and differentiably on the augmentation sphere boundary (minimum
conditions for a non-singular potential).  Slater made a simplification: he approximated the potential inside the
augmentation spheres with its spherical average, and also the interstitial potential with a constant. This is called the
Muffin Tin (MT) approximation; see Fig.~\ref{fig:pseudowf}.

Solutions to spherical potentials are separable into radial and angular parts, $\phi_\ell(\varepsilon,r)Y_L(\hat{\br})$.
The $\phi_\ell$ are called partial waves and $Y_L$ are the spherical harmonics.  Here and elsewhere, angular momentum
labelled by an upper case letter refers to both the $\ell$ and $m$ parts.  A lower case symbol refers to the orbital
index only ($\ell$ is the orbital part of $L{=}(\ell,m)$).  The $\phi_\ell$ are readily found by numerical
integration of a one-dimensional wave equation (Sec.\,\ref{sec:MTO}), which can be
efficiently accomplished.

An immense amount of work has followed the original ideas of Herring and Slater.  The Questaal package is an
all-electron implementation in the Slater tradition, so we will not further discuss the vast literature behind the
construction of a pseudopotential, except to note there is a close connection between pseudopotentials and the energy
linearisation procedure to be described below.  Bl\"ochl's immensely popular Projector Augmented-Wave method~\cite{PAW}
makes a construction intermediate between pseudopotentials and APWs.  Questaal uses atom-centred envelope functions
instead of plane waves (Sec.~\ref{sec:fp}), and an augmentation scheme that resembles the PAW method but can be
converged to an exact solution for the reference potential, as Slater's original method did.  The spherical
approximation is still almost universally used to construct the basis set, and thanks to the variational principle,
errors are second order in the nonspherical part of the potential.  The nonspherical part is generally quite small, and
this is widely thought to be a very good approximation, and the Questaal codes adopt it.

For a MT potential ($V_{MT}$ taken to be 0 for simplicity), the Schr\"odinger equation for energy $\varepsilon$ has
locally analytical solutions: in the interstitial the solution can expressed as a plane wave
$e^{i\,\mathbf{k}\cdot\br}$, with $\varepsilon{=}\hbar^2 k^2/2m$.  (We will use atomic Rydberg units throughout,
$\hbar{=}2{m}{=}e^2/2{=}1$).  In spherical coordinates envelope functions can be Hankel functions
$H_{L}(E,\br){=}h_{\ell}(kr)Y_L(\hat{\br})$ or Bessel functions $j_{\ell}(kr)Y_L(\hat{\br})$, except that Bessel
functions are excluded as envelope functions because they are not bounded in space.  Inside the augmentation spheres,
solutions consist of some linear combination of the $\phi_\ell$.

The all-electron basis sets ``APW'' (augmented plane wave) and ``KKR''~\cite{Korringa94} are both
instances of augmented-wave methods: both generate arbitrarily accurate solutions for a muffin-tin potential.  They
differ in their choice of envelope functions (plane waves or Hankel functions), but they are similar in that they join
onto solutions of partial waves in augmentation spheres.  Both basis sets are energy-dependent, which makes them very
complicated and their solution messy and slow.  This difficulty was solved by O.K. Andersen in 1975~\cite{Andersen75}.
His seminal work paved the way for modern ``linear'' replacements for APW and KKR, the LAPW and Linear Muffin Tin
Orbitals (LMTO) method.  By making a linear approximation to the energy dependence of the partial waves inside the
augmentation spheres (Sec.~\ref{sec:linearisation}), the basis set can be made energy-independent and
the eigenfunctions and eigenvalues of the effective one-particle equation obtained with standard diagonalisation
techniques.  LAPW, with local orbitals added to widen the energy window over which the linear approximation is valid
(Sec.~\ref{sec:lo}) is widely viewed to be the industry gold standard for accuracy.  Several well known codes
exist:WIEN2K (\url{http://susi.theochem.tuwien.ac.at}) and its descendants such as the Exciting code
(\url{http://exciting-code.org}) and FLEUR (\url{http://www.flapw.de}).  A recent
study~\citep{Lejaeghere2016a} established that these codes all generate similar results when carefully converged.
Questaal's main DFT code is a generalisation of the LMTO method (Sec.~\ref{sec:fp}), using the more flexible smooth
Hankel functions (Sec.~\ref{sec:defhkl}) instead of standard Hankels for the envelope functions.  Accuracy of the
smooth-Hankel basis is also high (Sec.~\ref{sec:benchmarks}), and though not quite reaching the standard of the LAPW
methods, it is vastly more efficient.  If needed, Questaal can add APW's to the basis to converge it to
the LAPW standard (Sec.~\ref{sec:pmt}).

\subsection{Questaal's History}

Questaal has enjoyed a long and illustrious history, originating in O.K. Andersen's group in the 1980's as the standard
``Stuttgart'' code.  It has undergone many subsequent evolutions, e.g. an early all-electron full-potential
code~\cite{Methfessel92}, which was used in one of the first \emph{ab initio} calculations of the electron-phonon
interaction for superconductivity~\cite{Lichtenstein91}, an efficient molecules code~\cite{Methfessel93} which was
employed in the first \emph{ab initio} description of instanton dynamics~\cite{Katsnelson95}, one of the first
noncollinear magnetic codes and the first \emph{ab initio} description of spin dynamics~\cite{Antropov95}, first
implementation of exact exchange and exact exchange+correlation~\cite{Kotani96}, one of the first all-electron \emph{GW}
implementations~\cite{ferdi94}, and early density-functional implementations of non-equilibrium Green's functions for
Landauer-Buttiker transport~\cite{Faleev05}.  In 2001 Aryasetiawan's \emph{GW} was extended to a full-potential
framework to become the first all-electron \emph{GW} code~\cite{Kotani02}.  Soon after the concept of quasiparticle
self-consistency was developed~\cite{Faleev04}, which has dramatically improved the quality of \emph{GW}.  Its most
recent extension is to combine QS\emph{GW} with DMFT.  It and the code of Kutepov \textit{et al.}~\cite{Choi16} are the
first implementations of QS\emph{GW}+dynamical mean field theory (DMFT); and to the best of our knowledge it has the
only implementation of response functions (spin, charge, superconducting) within QS\emph{GW}+DMFT.

\subsection{Main Features of the Questaal Package}

Ideally a basis set is \emph{complete}, \emph{minimal}, and \emph{short ranged}.  We will use the term \emph{compact} to
mean the extent to which a basis can satisfy all these properties: the faster a basis can yield converged results for a
given rank of Hamiltonian, the more compact it is.  It is very difficult to satisfy all these properties at once.  KKR
is by construction \emph{complete} and \emph{minimal} for a ``muffin-tin'' potential, but it is not short-ranged.  In
1984 it was shown (by Andersen once again!~\cite{Andersen84}) how to take special linear combinations of muffin-tin
orbitals (``screening transformation'') to make them short ranged.  Andersen's screening transformation was derived for
LMTOs, in conjunction with his classic Atomic Spheres Approximation, (ASA, Sec.~\ref{sec:asa}), and screening has
subsequently been adopted in KKR methods also.  The original Questaal codes were designed around the ASA, and we develop
it first in Sec.~\ref{sec:traditionallmto}.  Its main code no longer makes the ASA approximation, and generalises the
LMTOs to more flexible functions; that method is developed in Sec.~\ref{sec:fp}.  These functions are nevertheless
long-ranged and cannot take advantage of very desirable properties of short-ranged basis sets.  Very recently we have
adapted Andersen's screening transformation to the flexible basis of full-potential method (Sec.~\ref{sec:introtb}).
Screening provides a framework for the next-generation basis of ``Jigsaw Puzzle Orbitals'' (JPOs) that will be a nearly
optimal realisations of the three key properties mentioned above.

Most implementations of \emph{GW} are based on plane waves (PWs), in part to ensure completeness, but also because
implementation is greatly facilitated by the fact that the product of two plane waves is analytic.  \emph{GW} has also
been implemented recently in tight-binding forms using e.g., a numerical basis~\cite{Caruso2013}, or a Gaussian
basis~\cite{vanSetten2013}.  None of these basis sets is very compact.  The FHI AIMS code \emph{can} be reasonably well
converged, but only at the expense of a large number of orbitals.  Gaussian basis sets are notorious for becoming
over-complete before they become complete.  Questaal's JPO basis---still under development---should bring into a single
framework the key advantages of PW and localised basis sets.

Questaal implements:
\begin{enumerate}
\item density-functional theory (DFT) based on common LDA and GGA exchange-correlation functionals (other LDA or GGA,
but not meta-GGA, varients are available via \texttt{libxc}\cite{libxc}).  There is a standard full-potential DFT code,
\texttt{lmf} (Sec.~\ref{sec:fp}), and also three codes (\texttt{lm}, \texttt{lmgf}, \texttt{lmpg}) that implement DFT in
the classical Atomic Spheres Approximation~\cite{Andersen75,Andersen84}, presented in Sec.~\ref{sec:asa}.  The latter
use the screened, tight-binding form (Sec.~\ref{sec:asatb}). \texttt{lm}, a descendant of Andersen's standard
\href{https://www2.fkf.mpg.de/andersen/LMTODOC/LMTODOC.html}{ASA package} (Stuttgart code), is an approximate, fast form
of \texttt{lmf}, useful mainly for close-packed magnetic metals; \texttt{lmgf} (Sec.~\ref{sec:fullG}) is a Green's
function implementation closely related to the KKR Green's function, parameterised so that it can be linearised.
Sec.~\ref{sec:lmto} shows how this is accomplished, resulting in an efficient, energy-independent Hamiltonian
\texttt{lm} uses. \texttt{lmgf} has two useful extensions: the coherent potential approximation (CPA,
Sec.~\ref{sec:cpa}) and the ability to compute magnetic exchange interactions.  \texttt{lmpg} (Sec.~\ref{sec:lmpg}) is a
principal layer Green's function technique similar to \texttt{lmgf} but designed for layer geometries (periodic boundary
conditions in two dimensions). \texttt{lmgf} is particularly useful for Landauer-Buttiker
transport~\cite{Chantis07,Chantis07b,spinlossinterface,Belashchenko19}, and it includes a nonequilibrium Keldysh
technique~\cite{Faleev05}.
\item density functional theory with local Hubbard corrections (LDA+U) with various kinds of double-counting correction
(Sec.~\ref{sec:ldau}).
\item the \emph{GW} approximation based on DFT.  Questaal's \emph{GW} package is separate from the DFT code; there is an
interface (\texttt{lmfgwd}) that supplies DFT eigenfunctions and eigenvalues to it.  It was originally formulated by
Aryasetiawan in the ASA~\cite{ferdi92}, derived from the Stuttgart ASA code; and it was the first all-electron \emph{GW}
implementation.  Kotani and van Schilfgaarde extended it to a full-potential framework in 2002~\cite{Kotani02}.  A shell
script \texttt{lmgw} runs the \emph{GW} part and manages links between it and \texttt{lmf};
\item the Quasiparticle Self-Consistent \emph{GW} (QS\emph{GW}) approximation, first formulated by Faleev, Kotani and
van Schilfgaarde~\cite{Faleev04}.  Questaal's QS\emph{GW} is a descendent of Kotani's original code, which with some
modest modifications can be found at \url{https://github.com/tkotani/ecalj/}.  QS\emph{GW} also works synchronously with
\texttt{lmf}, yielding either a static QS\emph{GW} self-energy $\Sigma_0$ which \texttt{lmf} reads as an effective
external potential added to the Hamiltonian, or a dynamical self-energy (diagonal part) which a post-processing code
\texttt{lmfgws}, uses to construct properties of an interacting Green's function, such as the spectral function.  See
Sec.~\ref{sec:gw};
\item extensions to Dynamical Mean Field Theory using a bath based on either DFT or QS\emph{GW}.  Questaal does not have
its own implementation of DMFT, but has interfaces to Haule's CTQMC solver~\citep{KH_ctqmc}, and to the \texttt{TRIQS}
library~\citep{triqs}.  See Sec.~\ref{sec:dmft};
\item an empirical tight-binding model (\texttt{tbe}).  The interested reader is referred to
Refs~\cite{Finnis98,Finnisbook} and the
\href{https://www.questaal.org/docs/code/tbe/#toc-the-empirical-tight-binding-code-tbe}{Questaal web site}; and
\item a large variety of other codes and special-purpose editors to supply input the electronic structure methods, or to
process the output.
\end{enumerate}

\subsection{Outline of the Paper}

The aim of the paper is to provide a consistent presentation of the key expressions and ideas of the LMTO method, and
aspects of electronic structure theory as they are implemented in the different codes in the Questaal suite.
Necessarily this presentation is rather lengthy; the paper is organised as follows.  Section \ref{sec:MTO} describes the
LMTO basis, assuming the muffin-tin and atomic sphere descriptions of the potential.  Together with tail cancellation
and linearisation this comprises therefore a derivation of the traditional (``second generation'') LMTO method.  The
transformation to a short range tight-binding-like basis, and to other representations is described.  The formulation of
the crystal Green's function in terms of the LMTO potential parameters is derived, allowing the use of coherent
potential approximation alloy theory.  Non-collinear magnetism and fully-relativistic LMTO techniques are presented.
Section \ref{sec:fp} describes the full-potential code: its basis, augmentation method, core treatment and other
technical aspects are described in detail.  The LDA$+U$ method, relativistic effects, combined LMTO+APW, and numerical
precision are also discussed.  Section \ref{sec:gw} describes the \emph{GW} method: the importance of self-consistency,
the nature of QS\emph{GW} in particular and its successes and limitations.  Section \ref{sec:dmft} describes Questaal's
interface to different DMFT solvers; section \ref{sec:susceptibility} discusses calculation of spin and charge
susceptibilities.  Our perspective on realising a high fidelity solution to the many-body problem for solids is
described in section \ref{sec:high-fidelity}.  Finally, software aspects of the Questaal project are outlined in section
\ref{sec:software}.  Several appendices are provided with a number of useful and important relations for the LMTO
methodology.

\section{The Muffin-Tin Potential and the Atomic Spheres Approximation}
\label{sec:MTO}

As noted earlier, the KKR method solves the Schr\"odinger equation in a MT potential to arbitrarily high accuracy.  We
develop this method first, and show how the LMTO method is related to it.  In Sec.~\ref{sec:defhkl} we show how the
basis in \texttt{lmf} is related to LMTOs.  The original KKR basis set consists of spherical Hankel (Neumann) functions
$H_{RL}(E,\br)=h_{R\ell}(kr)Y_L(\hat{\br})$ as envelope functions with $k^2=E$, augmented by linear combinations of
partial waves $\phi_{R\ell}(\varepsilon,r)Y_{RL}(\hat{\br})$ inside augmentation spheres (Fig.~\ref{fig:KKRchain}).
(See the Appendix for definitions and the meaning of subscripts $R$ and $L$.)  The envelope must be joined continuously
and differentiably onto augmentation parts, since the kinetic energy cannot be singular.  Note that for large $\ell$,
$\phi_{R\ell}\rightarrow\text{const}{\times}r^\ell$.  This is because the angular part of the kinetic energy becomes
dominant for large $\ell$.
\begin{figure}[ht]
\begin{center}
\includegraphics[width=0.75\columnwidth]{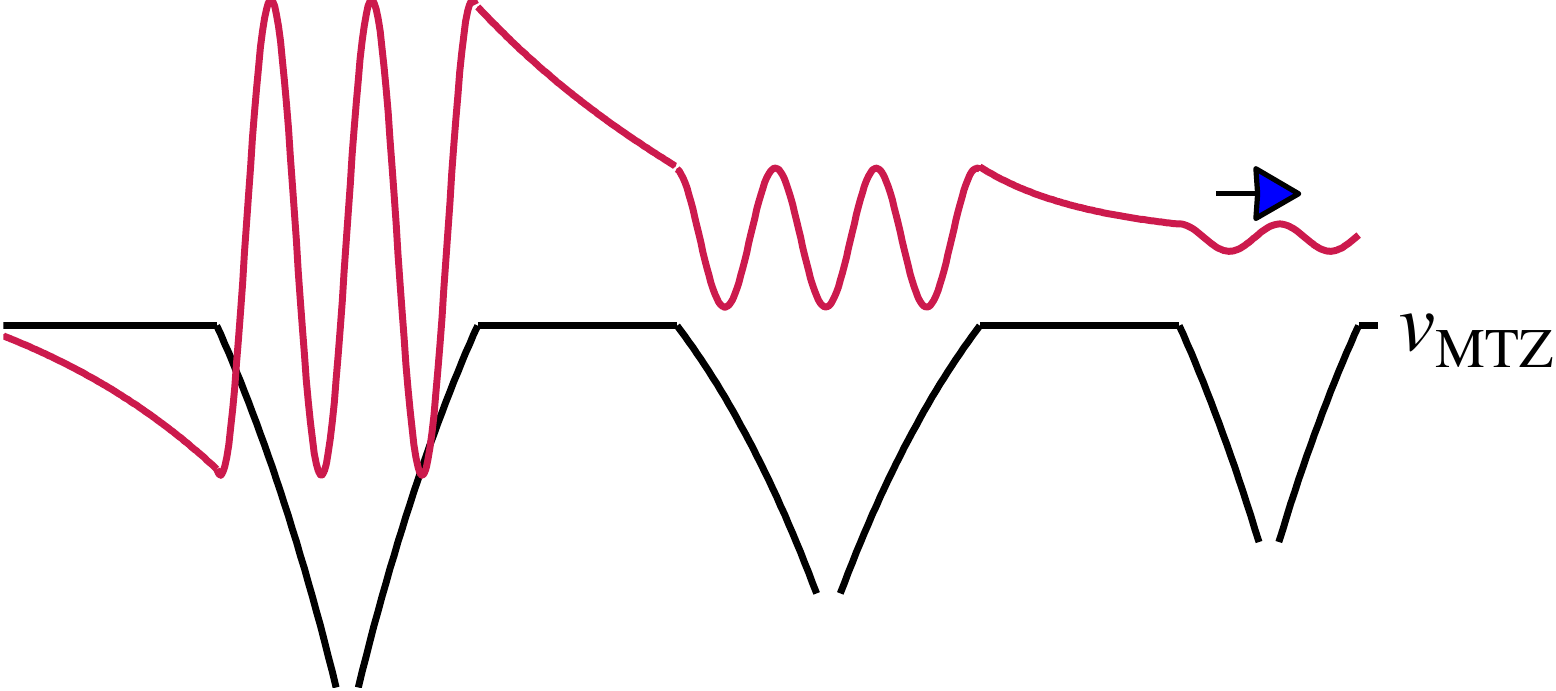}
\end{center}
\caption{Schematic of muffin-tin potential (black) and a solution (red) in a three-atom chain.}
\label{fig:KKRchain}
\end{figure}

\subsection{One-centre Expansion of Hankel Functions}
\label{subsec:onecentre}

An envelope function $H_{L}(E,\br)$ has a ``head'' centred at the origin where it is singular and tails at sites around
the origin.  In addition to the head, tails must be augmented by linear combinations of $\phi_{\ell}(\varepsilon,r)$
centred there, which require $H_{L}$ be expanded in functions centred at another site.  If the remote site is at $\bR$
relative to the head, the one-centre expansion can be expressed as a linear combination of Bessel functions
\begin{align}
H_{L}(E,\br) = \sum_M S_{ML}(E,\bR) J_M(E,\mathbf{r-\bR})
\label{eq:defs0}
\end{align}
This follows from the fact that both sides satisfy the same second order differential equation $(\nabla^2 + E) = 0$.
The two functions centred at the origin satisfying this equation are the Hankel and Bessel functions.  Hankel functions
have a pole at $r=0$, whereas Bessel functions are regular there, so this relation must be true for all $r<|\bR|$.  The
larger the value of $r$, the slower the convergence with $M$.

Expressions for the expansion coefficients $S$ can be found in various textbooks; they of course depend on how $H$ and
$J$ are defined.  Our standard definition is
\begin{align} 
&S_{MK} (E,{\bf R}) = \nonumber\\
&4 \pi \sum_L C_{KLM} (-1)^k (-E)^{(k+m-\ell)/2} H_L(E,{\bf R})
\label{eq:defS}
\end{align}
The $C_{KLM}$ are Gaunt coefficients (Eq.~(\ref{eq:expandsharm})).

To deal with solids with many sites, we write the one-centre expansion using subscript $R$ to denote a nucleus and $\br$
relative to the nuclear coordinate:
\begin{align}
 H_{RL}(E,\br) = \sum_M S_{R^\prime M,RL}(E) J_{R^\prime M}(E,\br)
\label{eq:expandh}
\end{align}
Envelope functions then have two $\ell$ cutoffs: $\ell_b$ for the head at $R$, and $\ell_a$ for the
one-centre expansion at $R^\prime$.  These need not be the same: $\ell_b$ determines the rank of the Hamiltonian, while
$\ell_a$ is a cutoff beyond which $\phi_{R^{\prime}\ell}\times\phi_{R^{\prime}\ell^\prime}$ is well approximated by
$\text{const}{\times}r^{\ell+\ell^\prime}$.  A reasonable rule of thumb for reasonably well converged calculations is to
take $\ell_b$ one number larger than the highest $\ell$ character in the valence bands.  Thus $\ell_b=2$ is reasonable
for $sp$ elements, $\ell_b=3$ for transition metal elements, $\ell_b=4$ for $f$ shell elements.  In the ASA, reasonable
results can be obtained for $\ell_b=2$ for transition metal elements, and $\ell_b=3$ for $f$ shell elements.  As for
$\ell_a$, traditional forms of augmentation usually require $\ell_a=2 \ell_b$ for comparable convergence: \texttt{lmf}
does it in a unique way that converges much faster than the traditional form and it is usually enough to take
$\ell_a=\ell_b+1$~\cite{lmfchap} (Sec.~\ref{sec:augmentation}).

\subsection{Partial Waves in the MT Spheres}
\label{sec:partialwaves}

Partial waves in a sphere of radius $s$ about a nucleus must be solved in the four coordinates, $\br$ and energy
$\varepsilon$.  Provided $\br$ is not too large, $v(\bf{r})$ is approximately spherical, $v(\br){\approx}v(r)$; Even if
the potential is not spherical, it is assumed to be for construction of the basis set, as noted earlier.  Solutions for
a spherical potential are partial waves $\phi_\ell(\varepsilon,r)Y_{L}(\hat{\br})$, where $Y_L$ are the spherical
harmonics.  Usually Questaal uses real harmonics $\text{Y}_{L}(\hat{\br})$ instead (see Appendix for definitions).

$\phi_\ell(\varepsilon,r)$ satisfies Schr\"odinger's equation
\begin{align}
(-{\nabla ^2} + v(r) - {\varepsilon}) \phi_\ell({\varepsilon},r) = 0
\label{eq:sephi}
\end{align}
We are free to choose the normalisation of $\phi$ and use
\begin{align}
\int_0^{s} \phi^2(\varepsilon,r) r^2 dr = 1
\label{eq:phinormalisation}
\end{align}

One way to think about solutions of Schr\"odinger's equation is to imagine each nucleus, with some $v_R(r)$ around it,
as a scatterer to waves incident on it.  Scattering causes shifts in the phase of the outgoing wave.  The condition that
all the sites scatter coherently, which allows them to sustain each other without any incident wave, gives a solution to
Schr\"odinger's equation.  This condition is explicit in the KKR method, and forms the basis for it.  Imagine a ``muffin
-tin'' potential---flat in the interstitial but holes carved out around each nucleus of radius $s$.  The phase shift
from a scatterer at \emph{R} is a property of the shape of $\phi_{R\ell}$ at its MT boundary,
$\phi_{R\ell}(\varepsilon,s)$.  Complete information about the scattering properties of the sphere, if $v(\br){=}v(r)$,
can be parameterised in terms of $\phi_{R\ell}(\varepsilon,s)$ and its slope, as we will see.

For a small change in energy of the incident wave, there can be a strong change in the phase shift.  In the region near
an eigenstate of the free atom the energy dependence is much stronger for the partial wave than it is for the incident
waves striking it, so electronic structure methods focus on the scattering properties of the partial waves $\phi_\ell$.
This information is typically expressed through the logarithmic derivative function
\begin{align}
D_\ell(\varepsilon) \equiv D\{\phi_\ell(\varepsilon)\} = \left( {\frac{r}{\phi_\ell(\varepsilon,r)}}
           {\frac{d\phi_\ell(\varepsilon,r)}{dr}} \right)_{s}
\label{eq:defd}
\end{align}
\begin{figure}[ht]
\begin{center}
\raisebox{1 em}{\includegraphics[width=0.5\columnwidth]{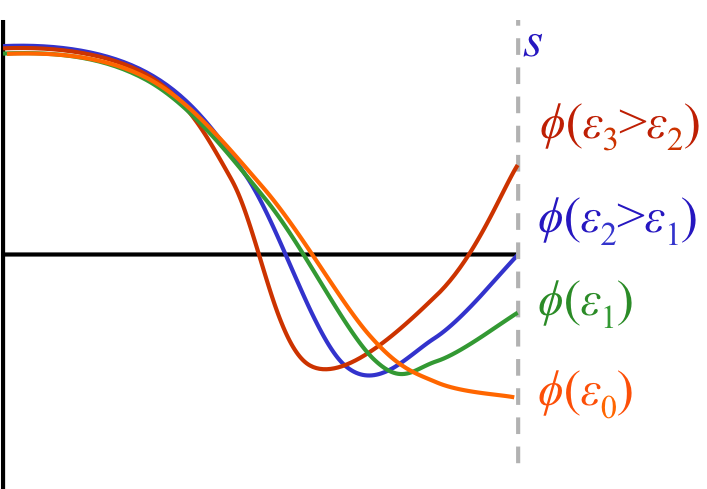}}
\includegraphics[width=0.45\columnwidth]{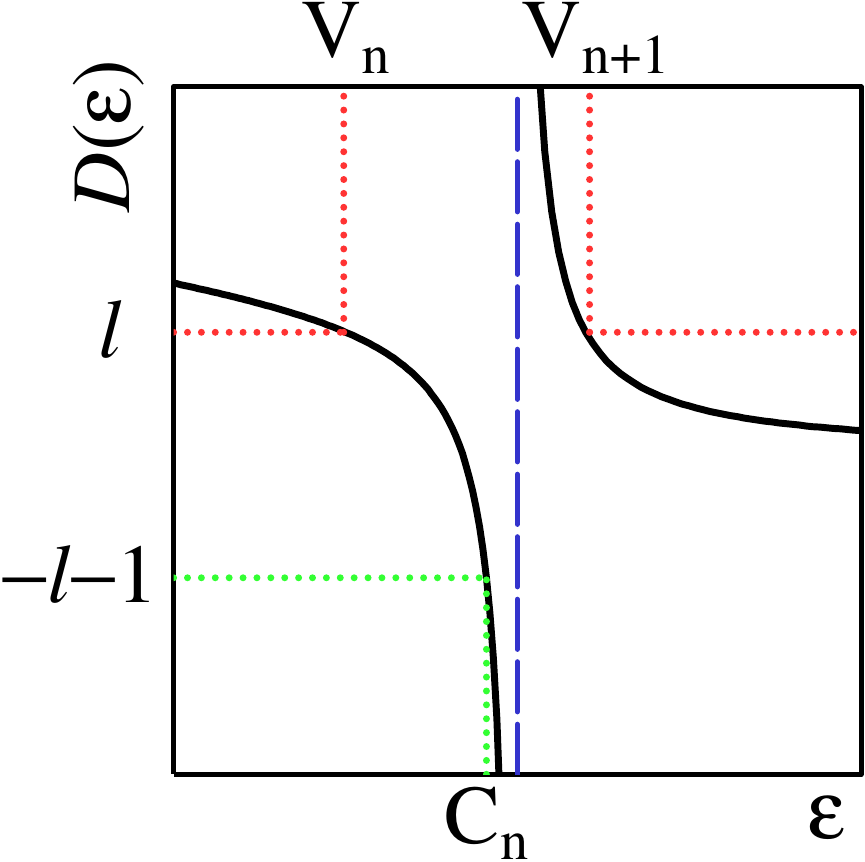}
\end{center}
\caption{Variation of $\phi_{\ell}(\varepsilon,r)$ with $\varepsilon$ for an $s$ orbital.}
\label{fig:figd}
\end{figure}

Consider the change in $\phi_{\ell}(\varepsilon,r)$ with $\varepsilon$ for a given $v(r)$.  As $\varepsilon$ increases
$\phi_{\ell}(\varepsilon,r)$ acquires more curvature (Fig.~\ref{fig:figd}).  In the interval between
$\varepsilon{=}\varepsilon_0$ where $\phi^\prime_{\ell}(\varepsilon_0,s){=}0$ so that $D_\ell=0$, and $\varepsilon_2$
where $\phi_{\ell}(\varepsilon_2,s){=}0$, $\phi^\prime_{\ell}(\varepsilon,s)$ is positive.  $D_\ell$ thus decreases
monotonically, with $D_\ell\to{-}\infty$ as $\varepsilon{\rightarrow}\varepsilon_2$.  At some $\varepsilon_1$ in this
region $D_\ell{=}-\ell-1$, which is the logarithmic derivative of a Hankel function of energy 0. $\varepsilon_1$ is
called the ``band centre'' $C_\ell$ for reasons to be made clear in Sec.~\ref{sec:traditionallmto}: it is close to an
atomic level and in tight-binding theory would correspond to an on-site matrix element.  Increasing from
$\varepsilon_2$, $D_{\ell}$ decreases monotonically from ${+}\infty$ as shown in Fig.~\ref{fig:figd}, reaching 0 once
more, passing through some $\varepsilon_3$ where $D_\ell{=}+\ell$.  This is the logarithmic derivative of a Bessel
function of energy 0, and is traditionally called $V_\ell$.

Thus $D_\ell$ is a monotonically decreasing cotangent-like function of $\varepsilon$, with a series of
poles.  At each pole $\phi_{\ell}(\varepsilon,r)$ acquires an additional node, incrementing the principal quantum number
$n$.  Between poles $n$ is fixed and there are parameters $C_{\ell}$ and $V_{\ell}$ for each $n$.  The linear method
approximates $D_\ell$ with a simple pole structure, which is accurate over a certain energy window.  Similarly
pseudopotentials are constructed by requiring the pesudofunction to match $D_\ell$ of the free atom, in a certain energy
window.

For principal quantum number $n$, $\phi_{\ell}$ has $n{-}\ell{-}1$ nodes and may vary rapidly to be
orthogonal to deeper nodes.  This poses no difficulty: we use a shifted logarithmic radial mesh, with point $i$ given by
\begin{align*}
r_i = b \{ e^{a(i-1)}-1\}
\end{align*}
Typically a few hundred points are needed for accurate integration.  Core and valence waves use the same mesh.

\subsection{Energy Derivative of \texorpdfstring{\emph{D}}{D}}

Consider the matrix element integrated over a sphere of radius $s$:
\begin{align*}
0 = \left\langle {\phi_\ell (\varepsilon_\nu)}
    \right| - \nabla^2  + v- \varepsilon \left| {\phi_\ell (\varepsilon)}\right\rangle
\end{align*}
$\varepsilon_\nu$ is some fixed energy.  Taking into account the boundary condition at $s$, we obtain
\begin{multline*}
\left\langle {\phi_\ell (\varepsilon)} \right|\varepsilon - \varepsilon_\nu\left|
        {\phi_\ell (\varepsilon_{\nu})} \right\rangle =\\
        - [D_\ell(\varepsilon ) - D_\ell\left({\varepsilon_\nu} \right)]s
        \phi_\ell (\varepsilon_\nu,s)\phi_\ell (\varepsilon,s)
\end{multline*}
from Green's second identity.  From this we obtain the energy derivative of $D_\ell$
as~\cite{Andersen75}
\begin{align}
\dot{D}_\ell(\varepsilon) \equiv \mathop{\lim}\limits_{\varepsilon\to\varepsilon_\nu}
        \frac{{D_\ell(\varepsilon)-D_\ell\left({\varepsilon_\nu}\right)}}
        {{\varepsilon-\varepsilon_\nu}} = \frac{{-1}}{{s\phi_\ell^2(\varepsilon,s)}}
\label{eq:derivd}
\end{align}

\subsection{Linearisation of Energy Dependence in the Partial Waves}
\label{sec:linearisation}

An effective way to solve Schr\"odinger's equation is to linearise the energy-dependence of the partial waves
$\phi_\ell(\varepsilon,r)$, as
\begin{align}
\phi_\ell(\varepsilon,r) \approx \phi_\ell(\varepsilon_\nu,r) +
        (\varepsilon-\varepsilon_\nu)\dot\phi_\ell(\varepsilon_\nu,r)
\label{eq:linearisephi}
\end{align}
This was Andersen's most important contribution to electronic structure theory~\cite{Andersen75}: it had a dramatic
impact on the entire field.  We will make extensive use of the linear approximation here.

In the linear approximation, four parameters ($\phi(s)$, $\dot\phi(s)$, and their logarithmic derivatives) completely
characterise the scattering properties of a sphere with $v=v(r)$.  Only three of them turn out to be independent.  To
see this, obtain an equation for $\dot{\phi}$ by differentiating Eq.~(\ref{eq:sephi}) w.r.t. $\varepsilon$:
\begin{align}
( - {\nabla ^2} + v - {\varepsilon}){\dot\phi({\varepsilon},r)} = \phi({\varepsilon},r)
\label{eq:sephidot}
\end{align}
With the normalisation Eq.~(\ref{eq:phinormalisation}), $\phi$ and $\dot{\phi}$ are orthogonal
\begin{align}
\langle\phi(\varepsilon_\nu)\dot\phi(\varepsilon_\nu)\rangle = 0
\label{eq:orthophianddot}
\end{align}

Using the normalisation Eq.~(\ref{eq:phinormalisation}), we can establish
the following relation between $\phi({\varepsilon_\nu},s)$, $\dot{\phi}({\varepsilon_\nu},s)$,
$D\{\phi\}$ and $D\{\dot{\phi}\}$:
\begin{align}
1 &= \langle {{{[\phi ({\varepsilon_\nu })]}^2}} \rangle  =
     \langle{\phi({\varepsilon_\nu})|-{\nabla^2}+v-{\varepsilon_\nu}|\dot\phi({\varepsilon_\nu})}\rangle\nonumber\\
  &= \langle{\dot{\phi}({\varepsilon_\nu})|-{\nabla^2}+v-{\varepsilon_\nu}|\phi({\varepsilon_\nu})}\rangle
     + W\{\phi,\dot{\phi}\}\nonumber\\
  &= W\{\phi(s),\dot{\phi}(s)\}\nonumber\\
  &= [D\{\phi({\varepsilon_\nu})\} - D\{\dot{\phi}({\varepsilon_\nu})\}]
     s\phi({\varepsilon_\nu},s)\dot{\phi}({\varepsilon_\nu},s)
\label{eq:wronskphidot}
\end{align}
The third line follows from Eq.~(\ref{eq:sephi}), and the second from Green's second identity which adds a surface term
when $\phi$ and $\dot{\phi}$ are interchanged.  The Wronskian $W\{a,b\}$ is defined as
\begin{align}
W\{a(s),b(s)\} &\equiv s^2 \left[a(s)b^\prime(s)-a^\prime(s)b(s)\right]\nonumber\\
               &= s a(s)b(s) \left[D\{b\} -  D\{a\}\right]
\label{eq:defwronskian}
\end{align}
for a pair of functions $a(r)$ and $b(r)$ evaluated at point $s$.

\subsection{The Traditional LMTO Method}
\label{sec:traditionallmto}

For historical reasons Anderson constructed the original LMTO formalism with non-standard definitions for the Hankel and
Bessel functions.  We follow those definitions in order to be consistent with the historical literature.  In this paper
they are named $\hat{H}_\ell$ and $\hat{J}_\ell$ and are defined in  Appendix \ref{app:hankel-bessel-defn}.  Here we
follow Andersen's development only in the context of $E{\le}0$, with $\kappa^2{=}-E$.  Note that $E$ can be chosen
freely and need not be connected to the eigenvalue $\varepsilon$.  But for exact solutions in a MT potential, the energy
of $\hat{H}_L(E,\br)$ and $\hat{J}_\ell(E,\br)$ must be chosen so $E{=}\varepsilon{-}v_\text{MTZ}$.

LMTO and KKR basis sets solve Schr\"odinger's equation in a muffin-tin potential, which in the interstitial reduces to
the Helmholtz equation, and have linear combinations of Hankel functions as solutions that satisfy appropriate boundary
conditions.  We defer treatment of the boundary conditions to the next section and continue the analysis of partial
waves for a single scattering centre for now.

As noted, the scattering properties depend much more strongly on the partial waves than the energy dependence of the
envelopes.  In keeping with the traditional LMTO method, we assume that the kinetic energy of the envelopes vanishes
(the energy is taken to be close to the MT potential), then Schr\"odinger's equation reduces further to Laplace's
equation, whose solutions are Hankel and Bessel functions $\hat{H}_\ell$ and $\hat{J}_\ell$ at $E{=}0$, which we denote
as $\hat{H}_\ell(\br)$ and $\hat{J}_\ell(\br)$.  In close-packed solids there is reasonable justification for this
choice: the spacing between spheres is much smaller than the wavelength of a low-energy solution to the wave equation
(one not too far from the Fermi level).  $\hat{H}_\ell(\br)$ and $\hat{J}_\ell(\br)$ are proportional to $r^{-\ell-1}$
and $r^\ell$ respectively (see Appendix) and so $\phi_\ell$ can be continued into the interstitial in the vicinity of
$s$
\begin{multline}
\phi_\ell(\varepsilon,r{\sim}s) = \phi_\ell \left( {\varepsilon,s} \right)\frac{{\ell+1 + D\{ \varepsilon\} }}
        {{2\ell+1}}\left( {\frac{r}{s}} \right)^\ell \\
+ \phi_\ell \left( {\varepsilon,s} \right)\frac{{\ell - D\{ \varepsilon\} }}
        {{2\ell+1}}\left( {\frac{s}{r}} \right)^{\ell+1}
\label{eq:pwoutsiders}
\end{multline}
The first term is proportional to $\hat{J}_\ell(r)$, the second to $\hat{H}_\ell(r)$.  In the remainder of this section
we will develop expressions for the general $\kappa$ case, showing also the $\kappa{\to}0$ limit, and finally focus on
constructing Hamiltonians and Green's functions with $\kappa{=}0$.

\subsection{Energy-Dependent Muffin-Tin Orbitals}

Eq.~(\ref{eq:pwoutsiders}) is not yet a suitable basis because it diverges as $r{\to}{\infty}$ because of the $J_\ell$
term.  However, we can construct a family of ``muffin-tin'' orbitals that are continuous and differentiable
\begin{align}
&\chi_{RL}(\varepsilon,E,{\br}) = {Y_L(\hat{\br}_R)} \times \nonumber\\
&
\begin{cases}
{N_{R\ell}\left({\varepsilon}\right){\phi_{R\ell}}\left( {\varepsilon ,r_R} \right) +
 P_{R\ell}\left({\varepsilon}\right){\hat{J}_\ell}(E ,r_R)}&{{r_R} < {s_R}}\\
{{\hat{H}_\ell}(E,r_R)}&{{r_R} > {s_R}}
\end{cases}
\label{eq:mtovaryeandk}
\end{align}
and that do not diverge for large $r_R$.  $N_{R\ell}$ and $P_{R\ell}$ are coefficients fixed by requiring that the value
and slope are continuous at $s_R$.  Thus for $r{<}s$, $\chi_{RL}(\varepsilon,E,{\br})$ consists of a linear combination
of $\phi_{R\ell}(r)$ and $\hat{J}_{R\ell}(E,r)$ that matches smoothly and differentiably onto $\hat{H}_{R\ell}(E,r)$.
Apart from the ``contaminating'' $\hat{J}_{R\ell}$ term, $\chi_{RL}$ is a solution for a single MT potential at sphere
$R$.  It vanishes at $\varepsilon$ corresponding to the eigenvalue of the MT ``atom,'' which occurs at
$\varepsilon{=}C_\ell$.  (Taking $E{=}0$, $\varepsilon{=}C_\ell$ when $D_{\ell}{=}-\ell-1$; see
Eq.~(\ref{eq:pwoutsiders}).)  In a lattice ${\hat{H}_\ell}$ must also be augmented at all $R^{\prime}{\ne}R$.  To form
an eigenstate, the contaminating term must be cancelled out by tails from $\chi_{R^{\prime}L^{\prime}}$ centred
elsewhere.  Since any $\hat{H}_{R^{\prime}L^{\prime}}(E,r)$ can be expanded as linear combinations of
$\hat{J}_{RL}(E,r)$, Eq.~(\ref{eq:expandh}), it is easy to anticipate how the ``tail cancellation condition,'' which
forms the basis of the KKR method (\ref{sec:tailcancellation}), comes about.

The $P_\ell$ are called ``potential functions'' and play a central role in constructing eigenfunctions.  Expressions for
$P_\ell$ and $N_\ell$ are developed in Sec.~\ref{sec:defpandn}.  Combined with the linearisation of the partial waves,
(Sec.~\ref{sec:linearisation}), information about $P$ can be encapsulated in a small number of parameters; see
Sec.~\ref{sec:lineearizep}.

Note the similarity between the $\chi_{RL}$ and the partial waves.  There is a difference in normalisation, but more
importantly the term proportional to $\hat{J}_{RL}$ for ${r_R{>}s}$ in Eq.~(\ref{eq:pwoutsiders}) must be taken out of
the MTO because it diverges for large $r$, as noted.  Since $\hat{J}_\ell$ is present for $r{<}s$, $\chi_{RL}$ is not a
solution of Schr\"odinger's equation for ${r_R{<}s}$.  However any linear combination of the $\chi_{RL}$
\begin{align}
\Psi(\epsilon,\kappa,\br) = \sum_{{R}{L}} z_{{R}{L}} \chi_{{R}{L}}(\epsilon,\kappa,\br)
\label{eq:trialwf}
\end{align}
can be taken as a trial solution to Schr\"odinger's equation.  The $z_{{R^\prime}{L^\prime}}$ are expansion
coefficients, which become the eigenvector if $\Psi$ is an eigenstate.  For any $z_{{R^\prime}{L^\prime}}$, $\Psi$
solves the interstitial exactly if the potential is flat and $-E$ is chosen to correspond to the kinetic energy in the
interstitial, $E{=}\varepsilon-v_\text{MTZ}$, since each $\chi_{RL}$ individually satisfies Schr\"odinger's equation.

\subsubsection{Tail Cancellation}
\label{sec:tailcancellation}

Inside sphere $R$ there are three contributions to $\Psi(\varepsilon,{\br})$: partial waves from the ``head'' function
$\chi_{RL}$, the Bessel part of that function, and contributions from the tails of $\chi_{R^\prime{}L^\prime}$ centred
at other sites, which are also Bessel functions, Eq.~(\ref{eq:expandh}).  Thus, all the contributions to $\Psi$ inside
some sphere $R$, in addition to the partial wave, consist of some linear combination of Bessel functions.  We can find
exact solutions for the MT potential by finding particular linear combinations $z_{{R}{L}}$ that cause all the
$\hat{J}_{RL}$ inside each augmentation sphere to cancel.

From the definition Eq.~(\ref{eq:mtovaryeandk}), Eq.~(\ref{eq:trialwf}), has a one-centre expansion inside sphere $R$
\begin{multline*}
\phi_{RL}(\varepsilon,{\bf r}) N_{R\ell}(\varepsilon) z_{RL}
        + \hat{J}_{RL}(E,{\bf r}) P_{R\ell}(\varepsilon) z_{RL}\\
- \sum_{L^\prime} S_{RL^\prime,{R^\prime{}L}}(E) \hat{J}_{RL^\prime}(E,{\bf r})z_{RL}
\end{multline*}
The one-centre expansion satisfies Schr\"odinger's equation provided that the second and third terms cancel.  This leads
to the ``tail cancellation'' theorem
\begin{align}
\sum_{RL} \left[ P_{RL}(\varepsilon) \delta_{{R^\prime{}L}^\prime,RL} - S_{{R^\prime{}L}^\prime,RL}(E) \right]z_{RL} = 0
\label{eq:tailcancellation}
\end{align}
and is the fundamental equation of KKR theory.  For non-trivial solutions ($|z_{RL}|{\ne}0$), the determinant of the
matrix $P{-}S$ must be zero.  This will only occur for discrete energies $\varepsilon_i(P)$ for which $P{-}S$ has a zero
eigenvalue.  The corresponding eigenvector $z_{RL}$ yields an eigenfunction, Eq.~(\ref{eq:trialwf}), which exactly
solves Schr\"odinger's equation for the MT potential in the limit $L{\to}\infty$ and if $E$ is taken to be the proper
kinetic energy in the interstitial, $E{=}\varepsilon_i{-}v_{\rm MTZ}$.  In general there will be a spectrum of
eigenvalues $\varepsilon_i$ that satisfy Eq.~(\ref{eq:tailcancellation}).  The quantity
\begin{align}
g_{RL,RL'}(\varepsilon )=[P_{RL}\delta_{RL,RL'}(\varepsilon ) - S_{RL,RL'}]^{-1}
\label{eq:auxiliaryg}
\end{align}
is called the ``auxiliary Green's function'' and is closely related to the true Green's function $G$~\cite{Gunnarsson83}
(poles of $g$ and $G$ coincide).  We will develop the connection in Sec.~\ref{sec:fullG}.  In KKR theory $g$ is called
the ``scattering path operator.''

\subsection{The Atomic Spheres Approximation}
\label{sec:asa}

Andersen realised very early that more accurate solutions could be constructed by overlapping the augmentation spheres
so that they fill space.  There is a trade-off in the error arising from the geometry violation in the region where the
spheres overlap and improvement to the basis set by using partial waves in this region rather than envelope functions.
The Atomic Spheres approximation, or ASA, is a shorthand for three distinct approximations:
\begin{itemize}
\item $v(\br)$ is approximated by a superposition of spherically symmetric $v_R(r)$, with a flat potential in the
interstitial;
\item the MT spheres are enlarged to fill space, so that the interstitial volume is zero.  The resulting geometry
violations are ignored, except that the interstitial can be accounted for assuming a flat potential (the ``combined
correction'' term).  Errors associated with the geometry violation were carefully analysed by Andersen in his NMTO
development~\cite{nmto}; and
\item the envelope functions are Hankel functions with $\kappa{=}0$, augmented by partial waves inside MT spheres.
There is no difficulty in working with $\kappa{\ne}0$, but $\kappa{=}0$ is a good average choice, as noted above.  As
ASA is an approximate method, little is gained by trying to improve it in this way.  Real potentials are not
muffin-tins, and the loss of simplicity does not usually compensate for limited gain in precision.  The full-potential
methods use better envelope functions (Sec.~\ref{sec:fp}).
\end{itemize}

\subsubsection{Tail Cancellation in the ASA}

In the ASA the spheres fill space, making the interstitial volume null.  By normalising the $\chi$ of
Eq.~(\ref{eq:mtovaryeandk}) as defined there, the eigenvectors $z_{RL}$ of $P-S$ ensure that $\Psi$ is properly
normalised if
\begin{align}
\sum\nolimits_{RL} |N_{R\ell} z_{RL}|^2 = 1
\label{eq:pmsnormalisation}
\end{align}
This is because the Bessels all cancel, and the wave function inside sphere $R$ is purely $\phi_{R\ell}N_{R\ell}z_{RL}$.
Normalisation of $z_{RL}$ with the normalisation of $\phi$ (Eq.~(\ref{eq:phinormalisation})) ensures that
$\left<\Psi|\Psi\right>{=}1$.

Making the spheres fill space is a better approximation than choosing spheres with touching radii, even with its
geometry violation.  This is because potentials are not flat and partial waves $\phi_\ell$ are better approximations to
the eigenfunctions than the $\hat{H}_\ell$.  Moreover, it can be shown~\cite{mtpointofview} that the resulting wave
function is equivalent to the exact solution of the Schr{\"o}dinger equation for $v(\br)$ equal to the sum of
overlapping spherical potentials $v(\br){=}\sum\nolimits_{R}v_R(r)$.  This means $v(\br)$ is deeper along lines
connecting atoms with a corresponding reduction of $v(\br)$ along lines pointing into voids, corresponding to the
accumulation (reduction) of density in the bonds (voids).

$\Psi$ varies in a nonlinear way with $\varepsilon$, so the tail cancellation condition entails a nonlinear problem.
Once $P(\varepsilon)$ (more precisely $1/P$) is linearised (Sec.\ref{sec:defpandn}), the cancellation condition
simplifies to a linear algebraic eigenvalue problem.  This provides a framework, through the linear approximation
Eq.~(\ref{eq:linearisephi}), for constructing an efficient, energy-independent basis set that yields solutions from the
variational principle, without relying on tail cancellation.

\subsection{Potential and Normalisation Functions}
\label{sec:defpandn}

The tail cancellation condition Eq.~(\ref{eq:tailcancellation}), is conveniently constructed through the ``potential
function'' $P_\ell(\varepsilon)$, $P_\ell$ is closely related to the logarithmic derivative $D_\ell(\varepsilon)$,
Eq.~(\ref{eq:defd}), which parameterises the partial wave in isolation.  $P_\ell$ depends on both $D_\ell$ and the
boundary conditions, which depend on what we select for the interstitial kinetic energy $E$.  $P_\ell(\varepsilon)$ is
an always increasing tangent like function of energy, and in the language of scattering theory, it is proportional to
the cotangent of the phase shift.

If the potential were not spherical, a more general tail cancellation theorem would still be possible, but $P$ would
need be characterised by additional indices: $P{=}P_{L,L^\prime}(\varepsilon)$ while for a spherical potential $P$
depends on $\ell$ only.  This is the only case we consider here.

$N_{\ell}$ and $P_{\ell}$ are fixed by requiring that $\chi_{\ell}(\varepsilon,\kappa,r)$ be continuous and
differentiable at $s$.  Expressions for $N_{\ell}$ and $P_{\ell}$ are conveniently constructed by recognising that any
function $f(r)$ can be expressed as a linear combination of $a(r)$ and $b(r)$ near $s$ (meaning it connects smoothly and
differentiably at $r{=}s$) through the combination
\begin{align}
f(r) = \left[ W\{f,b\} a(r) - W\{f,a\} b(r) \right] W\{a,b\}^{-1}
\label{eq:matching}
\end{align}
Thus the matching conditions require $N$ and $P$ to be
\begin{align}
P_\ell(\varepsilon) &= \frac{{W\{\hat{H}_\ell,\phi_\ell\} }}{{W\{\hat{J}_\ell,\phi_\ell\} }} \nonumber\\
 & \xrightarrow{\kappa{\to}0}  2(2\ell+1)
 {\left( {\frac{w}{s}} \right)^{2\ell+1}}\frac{{D\{\phi_\ell(\varepsilon)\} + \ell+1}}{D\{\phi_\ell(\varepsilon)\}-\ell}
 \label{eq:defp} \\
N_\ell(\varepsilon) &= \frac{{W\{ {\hat{J}_\ell},{\hat{H}_\ell}\} }}{{W\{ {\hat{J}_\ell},{\phi _\ell}\} }}
 \label{eq:defn}
\end{align}
$w$ is an arbitrary length scale, typically set to the average value of $s$.

With the help of Eq.~(\ref{eq:derivd}) the energy derivative of $P_\ell$ is readily shown to be (Eq. A21,
Ref.~\cite{Gunnarsson83})
\begin{align}
\dot{P}_\ell = \frac{d{P}_\ell}{d\varepsilon} = \frac{w/2}{[W\{\hat{J}_\ell,\phi_\ell\}]^2}
\label{eq:defpdot}
\end{align}
Using the following relation between Hankels and Bessels,
\begin{align*}
W\{\hat{H}_\ell,\hat{J}_\ell\} = w/2
\end{align*}
it is readily seen that
\begin{align*}
\dot{P}_\ell = \frac{2}{w}\frac{[W\{\hat{H}_\ell,\hat{J}_\ell\}]^2}{[W\{\hat{J}_\ell,\phi_\ell\}]^2}
             = \frac{2}{w}N_\ell^2
\end{align*}
and therefore
\begin{align}
N_\ell = \sqrt{\dot{P}_\ell w/2}
\label{eq:relationntop}
\end{align}

\subsubsection{Linearisation of \texorpdfstring{\emph{P}}{P}}
\label{sec:lineearizep}

In this section we consider a single $\ell$ only and drop the subscript.  By linearising the energy dependence of
$\phi$, Eq.~(\ref{eq:linearisephi}), it is possible to parameterise $P(\varepsilon)$ in a simple manner.  First, we
realise \emph{P} is explicitly a function of $D$, and implicitly depends on $\varepsilon$ through
$D{\equiv}D\{\phi_\ell(\varepsilon)\}$.  Writing Eq.~(\ref{eq:defp}) in terms of $D$ we see that
\begin{align}
P[D] = \frac{\hat{H}(s)}{\hat{J}(s)}\cdot\frac{{D - D\{\hat{H}\}}}{{D - D\{\hat{J}\}}}
\label{eq:pvsd}
\end{align}
The linearised $\phi(\varepsilon,r)$ (Eq.~(\ref{eq:linearisephi})) may be re-expressed using $D$ in place of
$\varepsilon$:
\begin{align}
\Phi(D,r) \approx \phi(\varepsilon_\nu,r) + \omega[D] \dot\phi(\varepsilon_\nu,r)
\label{eq:defPhi}
\end{align}
where
\begin{align*}
\omega[D] = -\frac{{\phi({\varepsilon_\nu },s)}}{{\dot{\phi}({\varepsilon_\nu },s)}}\cdot
        \frac{{(D - D\{\phi({\varepsilon_\nu })\} })}{{(D - D\{ \dot{\phi}({\varepsilon_\nu })\} )}}
\end{align*}
Eqs.~(\ref{eq:linearisephi}) and~(\ref{eq:defPhi}) refer to the same object; one is parameterised by $\varepsilon$ while
the other is parameterised by $D$ or $\omega[D]$.  The latter is more convenient because $\omega[D]$ and $P[D]$ both
have a simple pole structure.  That each have this structure imply that their inverses $D[\omega]$ and $D[P]$ also have
a simple pole structure.  This further implies that if $P$ is parameterised not by $D$ but instead by $\omega[D]$,
${P}\{\omega[D]\}$ will also have a simple pole structure in $\omega$, and depends on $\omega[D]$ as:
\begin{align*}
{P}\{\omega[D]\}=P\{\omega[D\{\dot\phi\}]\}\cdot
        \frac{(\omega[D]-\omega[D\{\hat{H}\}])}{(\omega[D]-\omega[D\{\hat{J}\}])}
\end{align*}
This relation follows from the fact that ${P}\{\omega\}$ has a pole structure in $\omega$, that $P$ vanishes when
$D{=}D\{\hat{H}\}$, and $1/{P}$ vanishes when $D{=}D\{\hat{J}\}$.  The prefactor
$P\{\omega[D\{\dot\phi(\varepsilon_\nu\}]\}$ follows from the fact that when $\omega[D]{\to}\infty$,
$D{\to}D\{\dot\phi\}$.

To obtain an explicit form for $P(\varepsilon)$ a relation between $\varepsilon$ and $\omega$ is required.  Matrix
elements and overlap for $\Phi[D]$ are readily obtained from Schr\"odinger's equation, Eq.~(\ref{eq:sephi}) and
(\ref{eq:sephidot}).  With these equations and normalisation relations Eq.~(\ref{eq:phinormalisation}) and
(\ref{eq:orthophianddot}), we can find that
\begin{align} 
\langle{\Phi[D^\prime]\,|\,-{\nabla^2} +v -{\varepsilon_\nu}\,|\,\Phi[D]}\rangle = \omega[D] \label{eq:MEPhi} \\
\langle{\Phi[D^\prime]\,|\,  \Phi[D]}\rangle = 1 + \langle\dot{\phi}^2\rangle\omega[D^\prime]\omega[D] \nonumber
\end{align}
$\varepsilon[D]$ can be obtained from the variational principle
\begin{align}
\varepsilon[D] &= \langle {\Phi [D]\,|\, - {\nabla ^2} + v\,|\,\Phi [D]} \rangle /
        \langle {{{\Phi^2[D]}}} \rangle\nonumber\\
 &= {\varepsilon_\nu } + \omega [D] \{1+\omega^2[D]\langle\dot{\phi}^2\rangle\}^{-1}
\label{eq:epsvsw}
\end{align}

Linearisation of $\phi$ or $\Phi$ has errors of second order in $\varepsilon-\varepsilon_\nu$, which means the
variational estimate for the energy has errors of fourth order.  From inspection of Eq.~(\ref{eq:epsvsw}) we can deduce
that the following linear approximation to $\varepsilon$
\begin{align}
\widetilde{\varepsilon}[D] = {\varepsilon_\nu } + \omega [D]
\label{eq:linomega}
\end{align}
has errors of second order in $\varepsilon-\varepsilon_\nu$. Thus, $P$ can be parameterised to second order in
$\varepsilon$ as
\begin{align*}
P(\varepsilon) \approx \widetilde{P}(\varepsilon)=\frac{1}{\gamma}\frac{\varepsilon-C}{\varepsilon-V}
\end{align*}
where the ``potential parameters'' $\gamma$, $C$ and $V$ are defined as
\begin{align}
\gamma  &= \frac{1}{ P[D\{\dot\phi\}] } = \frac{{W\{\hat{J},\dot{\phi}\}}}{{W\{\hat{H},\dot{\phi}\}}} \nonumber\\
&\xrightarrow{\kappa{\to}0}    \frac{(s/w)^{2\ell+1}}{{2(2\ell+1)}}
        \frac{{D\{\dot{\phi}\}-\ell\}}}{{D\{\dot{\phi}\}+\ell+1}} \label{eq:defgam}\\
C{-}\varepsilon_\nu &= \omega[D\{\hat{H}\}] = - \frac{{W\{\hat{H},\phi\}}}{{W\{\hat{H},\dot{\phi}\}}}\nonumber\\
&\xrightarrow{\kappa{\to}0} - \frac{\phi(s)}{{\dot\phi(s)}}
        \frac{{(D\{\phi\}+\ell+1)}}{{(D\{\dot\phi\}+\ell+1)}}\label{eq:defc}\\
V-\varepsilon_\nu &= \omega[D\{\hat{J}\}] = - \frac{{W\{\hat{J},\phi\}}}{{W\{\hat{J},\dot{\phi}\}}}\nonumber\\
&\xrightarrow{\kappa{\to}0} - \frac{\phi(s)}{{\dot\phi(s)}}\frac{{(D\{\phi\}-\ell)}}{{(D\{\dot\phi\}-\ell)}}
\label{eq:defv}
\end{align}

A simple pole structure can be parameterised in several forms.  A particularly useful one is
\begin{align}
\widetilde{P}(\varepsilon) = \frac{{(\varepsilon  - C)}}{{\gamma(\varepsilon-C)+\Delta}}
= \left( {\frac{\Delta } {{\varepsilon - C}} + \gamma } \right)^{ - 1}
\label{eq:parameterisedP}
\end{align}
where $\Delta{\equiv}(C-V)\gamma$.  $\Delta$ can be expressed in terms of Wronskians as
\begin{align}
\sqrt{\Delta} &= - \left({\frac{2}{w}}\right)^{1/2} W\{\hat{J},\phi\} \nonumber\\
&\xrightarrow{\kappa{\to}0}
-\left({\frac{w}{2}}\right)^{1/2}\frac{{\phi(s)}}{{(2\ell+1)}}\frac{{s^{\ell+1}}}{{w^{\ell+1}}}\left[{D\{\phi\}-\ell}
        \right]
\label{eq:defdel}
\end{align}
This last equation defines the sign of $\Delta^{1/2}$.

Eq.~(\ref{eq:parameterisedP}) is accurate only to second order because of the linear approximation
Eq.~(\ref{eq:linomega}) for $\omega(\varepsilon)$.  From the structure of Eq.~(\ref{eq:epsvsw}), it is clear that
${\omega}(\varepsilon)$, and thus ${P}(\varepsilon)$ can be more accurately parameterised (to third order) by the
substitution
\begin{align}
{P}(\varepsilon) &= \widetilde{P}(\varepsilon^\prime) \\
\varepsilon^\prime &= (\varepsilon + (\varepsilon-\varepsilon_\nu)^3 \langle{\dot{\phi}^2(\varepsilon_\nu)}\rangle)
\label{eq:porder3}
\end{align}
The third-order parameterisation requires another parameter, sometimes called the ``small parameter''
\begin{align}
p = \int_0^{s} \dot{\phi}^2(\varepsilon_\nu,r) r^2 dr = -\frac{\ddot{\phi}(s)}{3\phi(s)}
\label{eq:defsmallp}
\end{align}
Thus $P$ is parameterised to third order by four independent parameters.  It is sometimes convenient in a Green's
function context to use $P$ and $N$ (or equivalently $\dot{P}$; see Eq.~(\ref{eq:relationntop})) in place of $C$ and
$\Delta$.  Green's functions can be constructed without linearising $\phi$; this is the KKR-ASA method.  How
linearisation of $\phi$ resolves $G$ in an energy-independent Hamiltonian is described in Sec.~\ref{sec:introh}.
\begin{table}
\setlength{\tabcolsep}{4.8pt} 
\begin{tabular}{|c|p{17ex}|p{25ex}|}
\hline
         & \emph{Name}         & \emph{Interpretation} \\ \hline
$C$      & band centre         & eigenvalue of MT ``atom'' and resonance in extended system, Eq.~(\ref{eq:defc})\\
$\Delta$ & bandwidth           & bandwidth in the absence of hybridisation with other orbitals, Eq.~(\ref{eq:defdel})\\
$p$      & small parameter     & 3$^\text{rd}$ order correction to second-order potential function
                                $\widetilde{P}$, Eq.~(\ref{eq:parameterisedP})\\
$V$      & ``bottom''          & $\varepsilon$ where $D_{\ell}{=}+\ell$ (free electrons)\\
$\gamma$ & ``transformation to orthogonal basis'' & \multirow{2}{*}{see Sec.~\ref{sec:introtb}}\\
\hline
\end{tabular}
\caption{Potential parameters in the original LMTO-ASA method.  In the tight-binding transformation of this method these
symbols become parameterised by $\alpha$, defining the transformation; see
Sec.~\ref{sec:redefinition}.}
\label{tab:asappars}
\end{table}

\subsubsection{Spin Orbit Coupling as a Perturbation}
\label{sec:soc}

The formalism of the preceding sections can be extended to the Pauli Hamiltonian.  If we include the spin-orbit coupling
perturbatively, and include the mass-velocity and Darwin terms the Hamiltonian becomes
\begin{align*}
-{\nabla ^2} + v(r) - \frac{1}{c^2}(\varepsilon-v(r))^2+\frac{\partial{}v(r)}{\partial{}r}\frac{\partial}{\partial{}r}
        +\xi(r) {\mathbf{\hat{L}}}\cdot{\mathbf{\hat{S}}}
\end{align*}
where
\begin{align*}
\xi(r) = \frac{2}{{c^2}}\frac{{dv(r)}}{{rdr}}
\end{align*}
and
\begin{align*}
{\mathbf{\hat{L}}}\cdot{\mathbf{\hat{S}}} = \frac{1}{2}
\begin{pmatrix} 
        {{L^z}}&{L^{+-}}\\
        {{L^{-+}}}&{-{L^z}}
\end{pmatrix}
\end{align*}
The 2$\times$2 matrix refers to spin space.  In orbital space,
\begin{align*}
L^z    &= \delta_{mm'}m\\
L^{+-} &= {\delta_{m^\prime(m+1)}}\sqrt{(\ell + m + 1)(\ell - m)}\\
L^{-+} &= {{\delta _{m^\prime(m - 1)}}\sqrt {(\ell + m)(\ell - m + 1)}}
\end{align*}
${\mathbf{\hat{L}}}\cdot{\mathbf{\hat{S}}}$ mixes spin components; also matrix elements of Eq.~(\ref{eq:defPhi}) and
${\mathbf{\hat{L}}}\cdot{\mathbf{\hat{S}}}$ depend on both $\ell$ and $m$.   We require matrix elements of the
$\Phi(D)$, analogous to Eq.~(\ref{eq:MEPhi}):
\begin{align*}
\langle {\Phi_{\ell^\prime{}m^\prime}}[{D^\prime }]\,|\,\xi {\bf{\hat L}} \cdot
        {\bf{\hat S}}\,|\,{\Phi _{lm}}[D]\rangle = {\xi _\ell}[D^\prime,D]{\delta _{ll^\prime}}(lm|lm^\prime)
\end{align*}

\subsubsection{How the ASA-Tail Cancellation Reduces to a Linear Algebraic Eigenvalue
Problem}
\label{sec:introh}

The eigenvalue condition is satisfied when $\varepsilon$ is varied so that
\begin{align*}
|P - S| = 0
\end{align*}
$P{-}S$ is a matrix $(P - S)_{RL,R^\prime{}L^\prime}$ with $P$ diagonal in $RL$, and $S$ Hermitian.  If
\emph{P} is parameterised by Eq.~(\ref{eq:parameterisedP})
\begin{align*}
\left| {\left( {\frac{\Delta } {{z - C}} + \gamma } \right)^{ - 1}  - S} \right| = 0
\end{align*}
Multiply on the right by $S^{-1}$ and the left by $P^{-1}$
\begin{align*}
\left| {\frac{\Delta } {{\varepsilon  - C}} + \gamma  - S^{ - 1} } \right| = 0
\end{align*}
Rearrange, keeping in mind that $C$, $\Delta$, and $\gamma$ are real and diagonal in ${RL}$
\begin{align*}
\left| {-\varepsilon + C + \sqrt\Delta \left({S^{-1}-\gamma}\right)^{-1}\sqrt\Delta} \right| = 0
\end{align*}
This has the form of the linear algebraic eigenvalue problem
\begin{align*}
\widetilde{h}\psi = \varepsilon\psi
\end{align*}
with
\begin{align}
\widetilde{h} = C + \sqrt\Delta\left({S^{-1}-\gamma}\right)^{-1}\sqrt\Delta
\label{eq:secondorderh}
\end{align}
$\widetilde{h}$ is a Hermitian matrix, with eigenvalues corresponding to the zeros in $|\widetilde{P}-S|$.  The tilde
indicates that $\widetilde{h}$ is obtained from $\widetilde{P}$ and is thus accurate to second order in
$\varepsilon-\varepsilon_\nu$ ($C$ and $\Delta$ are calculated at $\varepsilon_\nu$,
Eq.~(\ref{eq:defc},\ref{eq:defdel})).  Linear MTO's will be constructed in Sec.~\ref{sec:lmto}, and $\widetilde{h}$ can
be identified with $h^\alpha-\varepsilon_\nu$, Eq.~(\ref{eq:defha}), for $\alpha{=}\gamma$ and if $P$ is parameterised
by $\widetilde{P}$.

\subsection{``Screening'' Transformation to Short-ranged, Tight-binding Basis}
\label{sec:introtb}

Formally, Eq.~(\ref{eq:secondorderh}) is a ``tight-binding'' Hamiltonian in the sense that it is a Hamiltonian for a
linear combination of atom-centred (augmented) envelopes $\chi$, Eq.~(\ref{eq:mtovaryeandk}) taken at some fixed
(linearisation) energy $\varepsilon_{\nu}$.  The Hamiltonian is long-ranged because $\hat{H}_{\ell}(r)$ is long ranged
(see Eq.~(\ref{eq:defokaKJ}) for $\kappa{=}0$) unless $E=-\kappa^2$ is $-$1\,Ry or deeper.  But such a basis is not
accurate: the optimal $E$ falls somewhere in the middle of the occupied part of the bands, and is roughly $+$0.3\,Ry in
close-packed systems; $E{=}0$ is a compromise.

\begin{figure}[ht]
\begin{center}
\includegraphics[width=.75\columnwidth]{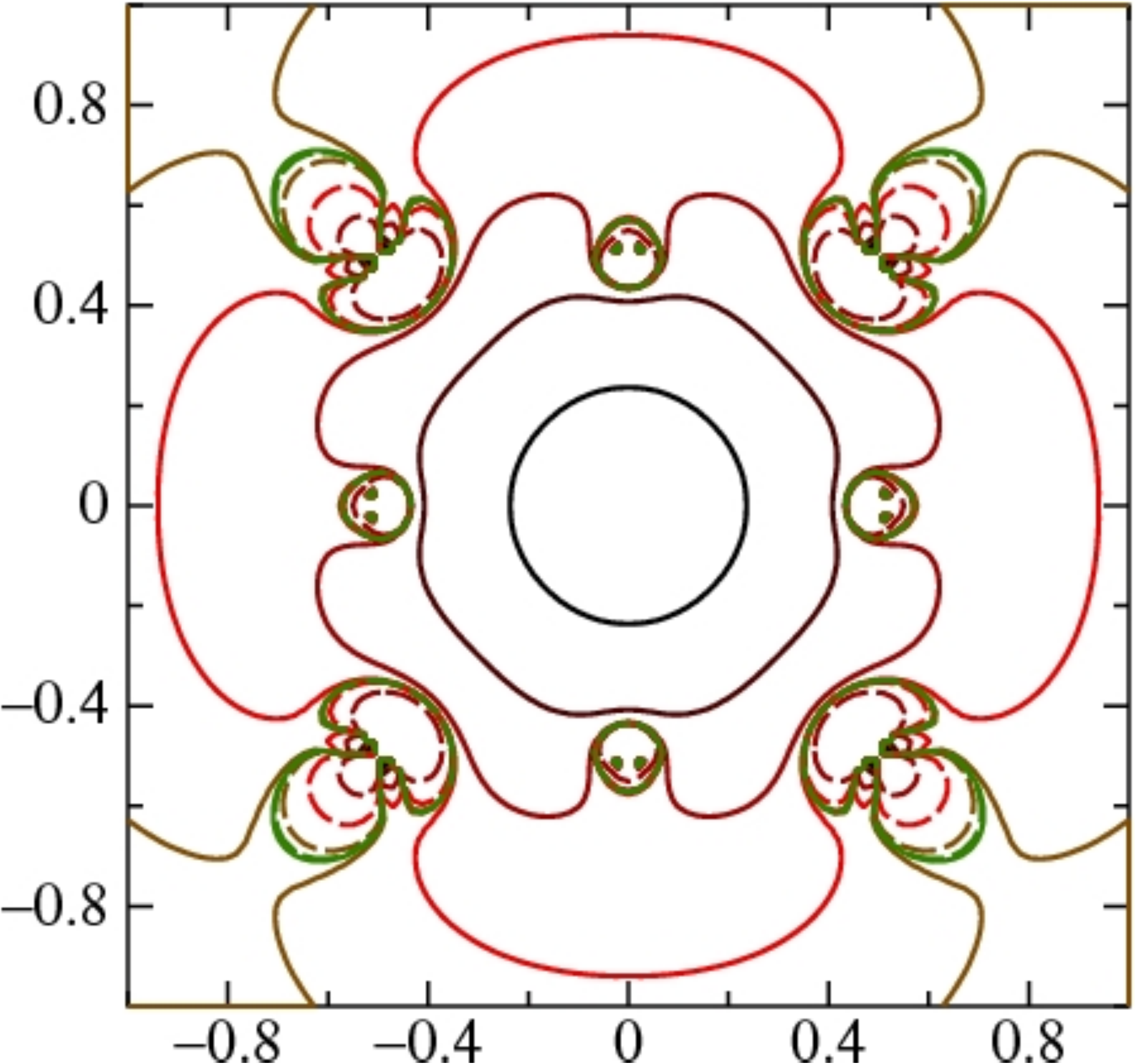}
\end{center}
\caption{Contour of a screened $s$ envelope function in a bcc lattice.  Each contour represents a reduction in amplitude
by a factor of 10.  Dashed lines show contours with negative amplitude.  At the ``hard core'' radius
(Sec.~\ref{sec:hardcoreradius}) on the head site a screened function has pure $\ell$ character; at the ``hard core''
radius on all the tail sites the one-centre expansion of the envelope function vanishes for $\ell<\ell_\text{max}$.}
\label{fig:screenedLMTO}
\end{figure}

The idea behind the ``screening'' or ``tight-binding'' transformation is to keep $E{=}-\kappa^2$ near zero but render
the Hilbert space of the $\chi_{RL}(\varepsilon,\br)$ short range by rotating the basis into an equivalent set
$\{\chi_{RL}(\varepsilon,\br)\}{\to} \{\chi^\alpha_{RL}(\varepsilon,\br)\}$, by particular linear combinations which
render $\chi^\alpha_{RL}(\varepsilon,\br)$ short ranged (or acquire another desirable property, e.g. be orthogonal), see
Fig.~\ref{fig:screenedLMTO}.

Andersen sometimes called the change a ``screening transformation'' because it is analogous to screening in
electrostatics.  Note that $\hat{H}_\ell(E{=}0,r)\propto{}1/r$, has the same form as a charge monopole, with long range
behaviour.  It becomes short ranged if screened by opposite charges in the neighbourhood.  The same applies to higher
order multipoles: they can become short ranged in the presence of multipoles of opposite sign.

The transformation can be accomplished in an elegant manner by admixing to the original (``bare'') envelope (renamed
from $\hat{H}_{RL}$ to $\hat{H}^0_{RL}$ to distinguish it from the ``screened'' one) with amounts of
$\hat{H}^0_{R^\prime{}L^\prime{}}$ in the neighbourhood of $R$:
\begin{align}
\hat{H}_{RL}^\alpha  ({\br}) = \sum\nolimits_{R^\prime{}L^\prime{}}
        {\hat{H}_{R^\prime{}L^\prime{}}^0({\br}) {B^\alpha_{R^\prime{}L^\prime;RL}}}
\label{eq:defkalpha}
\end{align}
Whatever prescription determines $B^\alpha_{R^\prime{}L^\prime{};RL}$, it is evident that the Hilbert space is
unchanged, and that $\hat{H}^\alpha_{RL}{\to}\hat{H}^0_{RL}$ if
${B_{R^\prime{}L^\prime;RL}^\alpha}{\to}{\delta_{R^\prime{}L^\prime;RL}}$

The one-centre expansion of $\hat{H}_{RL}^\alpha$ in any channel ${R^\prime{}L^\prime{}}$ is some linear combination
of Hankel and Bessel functions, because $\hat{H}_{RL}^0$ are Hankels in their head channel $R^\prime{}L^\prime{=}RL$,
and linear combinations of Bessels in other channels $R^\prime{}L^\prime{=}RL$ (see Eq.~(\ref{eq:expandh})).  Thus every
$\hat{H}_{RL}^0$ is expanded in $R^\prime{}L^\prime$ by some particular linear combination of Bessel and Hankel
functions
\begin{align}
J^\alpha_{R^\prime{}L^\prime}(E,\br) =
        J_{R^\prime{}L^\prime}^0(E,\br) - \alpha_{R^\prime{}L^\prime;RL} \hat{H}^0_{R^\prime{}L^\prime}(E,\br)
\label{eq:defjalpha}
\end{align}
$B^\alpha_{R^\prime{}L^\prime;RL}$ determines $\alpha_{R^\prime{}L^\prime;RL}$, and vice-versa.  $\alpha$ used as a
superscript indicates that it determines the screening.

A simple and elegant way to choose the screening transformation is to expand \emph{every} $\hat{H}_{RL}^0$ in channel
$R^\prime{}L^\prime{\ne}RL$, by the \emph{same} function
$J^\alpha_{\ell^\prime}(\kappa,r_{R^\prime})Y_{L^\prime}(\hat{\br}_{R^\prime})$.  Then $\alpha$ need only be specified
by two indices, $\alpha_{R^\prime{}L^\prime;RL}{\to}\alpha_{R^\prime{}\ell^\prime}$.

To sum up, we specify the transformation through $\alpha_{R^\prime{}\ell^\prime}$.  Envelopes $\hat{H}_{RL}^\alpha$ are
expanded at site $R^\prime{}L^\prime{\ne}RL$, as linear combinations
\begin{gather}
\hat{H}_{RL}^\alpha  ({\br}) = -\sum\nolimits_{R^\prime  L^\prime}{\hat{J}_{{R^\prime}L^\prime}^\alpha({\br})
        S^\alpha_{R^\prime{}L^\prime;RL}}\label{eq:kalphaonec}\\
\hat{J}_{{R^\prime}L^\prime}^\alpha({\br}) =
        [\hat{J}^0_{{R^\prime}\ell^\prime}(r) -\alpha_{R^\prime{}\ell^\prime}\hat{H}^0_{{R^\prime}\ell^\prime}(r)]
        Y_{{R^\prime}L^\prime}(\hat{\br})\label{eq:jalph}
\end{gather}
The structure constants $S^\alpha_{R^\prime{}L^\prime;RL}$ are expansion coefficients that will be determined next, but
already it should be evident that $S^\alpha{\xrightarrow{\alpha{\to}0}}S^0$, where $S^0$ are the structure constants of
Eq.~(\ref{eq:expandh}).

In its own ``head'' $\hat{H}^\alpha_{RL}$ must have an additional irregular part.  By expressing the
$B^\alpha_{R^\prime{}L^\prime;RL}$ in Eq.~(\ref{eq:defkalpha}) in terms $S^\alpha_{R^\prime{}L^\prime;RL}$ as
\begin{align}
\hat{H}_{RL}^\alpha({\br}) = \sum_{R^\prime{}L^\prime{}}{\hat{H}_{R^\prime{}L^\prime{}}^0({\br})
\left(\delta _{R^\prime  L^\prime  ;RL}  + \alpha _{R^\prime \ell^\prime  } S_{R^\prime  L^\prime  ;RL}^\alpha  \right)}
\label{eq:newkalpha}
\end{align}
we can see indeed that $\hat{H}_{RL}^\alpha({\br})$ has the required one-centre expansion,
Eqs.~(\ref{eq:kalphaonec},~\ref{eq:jalph}), provided that $S^\alpha$ obey a Dyson-like equation
\begin{align}
S^\alpha &= S^0  + S^0 \alpha S^\alpha  \quad\hbox{or}\nonumber\\
\left. S^\alpha \right.^{-1} &= \left. S^0 \right.^{-1} - \alpha
\label{eq:defsalpha}
\end{align}
In practice $S^\alpha$ is calculated from
\begin{align*}
S^\alpha  = \alpha^{-1}\left(\alpha^{-1} - S^0\right)^{-1}\alpha^{-1} - \alpha^{-1}
\end{align*}
It follows immediately from Eq.~(\ref{eq:defsalpha}) that if there are two screening representations $\alpha$ and
$\beta$, the structure constants connecting them are related by
\begin{align}
\left. S^\alpha\right.^{-1} + \alpha  = \left. S^\beta\right.^{-1} + \beta
\label{eq:salphatobeta}
\end{align}

\subsection{Screened Muffin-Tin Orbitals and Potential Functions}
\label{sec:asatb}

In this section we develop a screened analogue of the MTO's, Eq.~(\ref{eq:mtovaryeandk}) potential functions,
Eq.~(\ref{eq:defp}), normalisation Eq.~(\ref{eq:defn}), and tail cancellation conditions (\ref{eq:tailcancellation}).
Here we mostly concern ourselves with the ASA with $\kappa{=}0$.

\subsubsection{Redefinitions of Symbols}
\label{sec:redefinition}

We have defined a number of quantities in the context of the original MTO basis set, Eq.~(\ref{eq:mtovaryeandk}) that
will have a corresponding definition in a screened basis set.  Several previously defined quantities are now labelled
with a superscript 0 to indicate that their definitions correspond to the unscreened $\alpha{=}0$ representation:
$\hat{H}_{RL}{\equiv}\hat{H}^0_{RL}$, $\hat{J}_{RL}{\equiv}\hat{J}^0_{RL}$, $P_{RL}{\equiv}P^0_{RL}$,
$N_{RL}{\equiv}N^0_{RL}$ and $S_{R^\prime{}L^\prime;RL}{\equiv}S^0_{R^\prime{}L^\prime;RL}$.

The ``potential parameters'' $C$, $\Delta$, and $p$, Eq.~(\ref{eq:defc}-\ref{eq:defsmallp}) and
Table~\ref{tab:asappars}, can also be relabelled with representation-dependent definitions.  It is unfortunately rather
confusing, but the original definitions without superscripts Eq.~(\ref{eq:defc}-\ref{eq:defsmallp}) correspond not to
$C^0$, $\Delta^0$, and $p^0$, but to the particular screening representation $\alpha{=}\gamma$ (dubbed the ``$\gamma$
representation'').  To be consistent with the new superscript convention for $P^0$ and $N^0$, the appropriate
identifications are
\begin{gather} 
\phi(\varepsilon_\nu){\equiv}\phi^\gamma(\varepsilon_\nu),\quad
        \dot{\phi}(\varepsilon_\nu){\equiv}\dot{\phi}^\gamma(\varepsilon_\nu), \quad
        {\text{and}}\nonumber\\
C {\equiv}C^\gamma,\quad
        \Delta {\equiv}\Delta^\gamma,\quad
        p {\equiv}p^\gamma
\label{eq:superscriptp}
\end{gather}
Another unfortunate artefact of the evolution in LMTO formalism is that the meaning of many symbols changed over time.
$\gamma$ is called $Q^{-1}$ in Ref.~\cite{Gunnarsson83}.  In the most recent NMTO formalism, $S$ and $B$ have exchanged
meanings.  Questaal's ASA codes use definitions that most closely resemble the ``second generation'' LMTO formalism
perhaps most clearly expressed in Ref.~\cite{Andersen86}.  Reference 16 of that paper makes correspondences to
definitions laid out in earlier papers.

\subsubsection{Potential and Normalisation Functions for Screened MTO's}

The MTO Eq.~(\ref{eq:mtovaryeandk}) is derived by augmenting the envelope $\hat{H}^0_{R\ell}(\varepsilon,r)$ by matching
it smoothly onto a linear combination of $\phi_\ell(\varepsilon,r)$ and $\hat{J}^0_{R\ell}(r)$.  For the screened case
we match $\hat{H}^0$ to $\phi_\ell(\varepsilon,r)$ and $\hat{J}^\alpha_{R\ell}(r)$, the latter defined by
Eq.~(\ref{eq:defjalpha}): The matching requires
\begin{align}
P^\alpha_\ell(\varepsilon) &= \frac{{W\{\hat{H}^0_\ell,\phi_\ell\} }}{{W\{\hat{J}^\alpha_\ell,\phi_\ell\} }}
        &=& \frac{P^0_\ell(\varepsilon)}{1-{\alpha_{\ell}}P^0_\ell(\varepsilon)}\label{eq:defpa} \\
N^\alpha_\ell(\varepsilon) &=
        \frac{{W\{ {\hat{J}^\alpha_\ell},{\hat{H}^0_\ell}\} }}{{W\{ {\hat{J}^\alpha_\ell},{\phi _\ell}\} }}
        &=& \sqrt{\dot{P}^\alpha_\ell w/2}
\label{eq:defna}
\end{align}
Eq.~(\ref{eq:defpa}) implies
\begin{align}
[P^\alpha_\ell(\varepsilon)]^{-1} &= [P^0_\ell(\varepsilon)]^{-1} - \alpha_\ell
        \to {\frac{\Delta_\ell}{{\varepsilon - C_\ell}} + \gamma_\ell-\alpha_\ell}
\label{eq:defipa}
\end{align}
and is an obvious generalisation of $P^0$, Eq.~(\ref{eq:parameterisedP}).  There is also the analogue of
Eq.~(\ref{eq:salphatobeta}) for $P$:
\begin{align}
[P^\alpha_\ell(\varepsilon)]^{-1} + \alpha_\ell = [P^\beta_\ell(\varepsilon)]^{-1} + \beta_\ell
\label{eq:palphatobeta}
\end{align}
Eq.~(\ref{eq:defipa}) shows that $\partial [P^\alpha(\varepsilon)]^{-1}/\partial\varepsilon$ is independent of the
screening $\alpha$ and
\begin{align}
\left[-\frac{\partial}{\partial\varepsilon}[P^\alpha_\ell(\varepsilon)]^{-1}\right]^{-1/2} &=
        \frac{P^\alpha_\ell(\varepsilon)}{[\dot{P}^\alpha_\ell(\varepsilon)]^{1/2}}\nonumber\\
&= -\sqrt{2/w} {W\{\hat{H}^0_\ell,\phi_\ell\}}\nonumber\\
&{\to} \frac{\varepsilon-C_\ell}{\sqrt{\Delta_\ell}}
\label{eq:pbypdot}
\end{align}
The last forms of Eqs.~(\ref{eq:defipa},\ref{eq:pbypdot}) apply when $P$ is parameterised by $\widetilde{P}$,
Eq.~(\ref{eq:parameterisedP}).

\subsubsection{Screened Muffin-Tin Orbitals}

To define the analogue of the MTO, Eq.~(\ref{eq:mtovaryeandk}), in a screened representation we write
\begin{align}
&N^\alpha_{R\ell}({\varepsilon}) \chi^\alpha_{RL} (\varepsilon,{\br}) = \nonumber\\
&\begin{cases}
N^\alpha_{R\ell}({\varepsilon}){\phi_{RL}}({\varepsilon,\br})\\
        \enskip+\sum\limits_{R^\prime{}L^\prime{}}{\bar{\hat{J}}^\alpha_{R^\prime{}L^\prime{}}({\br})}
        \left[P^\alpha-S^\alpha\right]_{R^\prime{}L^\prime;RL}
        \enskip\text{if }\br\in\{s_R\}\\
\hat{H}^\alpha_{RL} (\kappa,{\br})\hfill\text{if } \br\in\hbox{interstitial}
\end{cases}
\label{eq:tbmtovarye}
\end{align}
The partial wave ${\phi_{RL}}({\varepsilon,\br})$ is understood to vanish outside its own head sphere, and $P$ is a
matrix diagonal in $RL$: $ P^\alpha_{R^\prime{}L^\prime;RL}({\varepsilon}) =
P^\alpha_{R^\prime{}L^\prime}({\varepsilon})\, \delta_{R^\prime{}L^\prime;RL}$.

$\bar{\hat{J}}^\alpha_{R^\prime{}L^\prime{}}$ is the linear combination of $\phi$ and $\dot\phi$ that matches
continuously and differentiably $\hat{J}^\alpha_{R^\prime{}L^\prime}$ defined in Eq.~(\ref{eq:defjalpha}).
Eq.~(\ref{eq:tbmtovarye}) uses it instead of ${\hat{J}}^\alpha$ because when we later construct energy-independent MTO's
we can generate basis sets that accurately solve Schr\"odinger's equation in the augmentation spheres.  Since $P^\alpha$
and $N^\alpha$ depend only on values and slopes at the $\{s_{R}\}$, the substitution has no effect on them.
%
%

\subsubsection{Tail Cancellation in the Tight-binding Representation}
\label{eq:tbtailcancellation}

The energy-dependent $\chi^\alpha_{RL}$, Eq.~(\ref{eq:tbmtovarye}) exactly solve the ASA-MT potential because the trial
function
\begin{align*}
\Psi({\varepsilon,\br}) = \sum\nolimits_{RL} {z^\alpha_{RL} \chi^\alpha_{RL}}({\varepsilon,\br})
\end{align*}
that satisfy the set of linear equations
\begin{align}
\sum\nolimits_{R^\prime{}L^\prime} (P^\alpha - S^\alpha)_{RL;R^\prime{}L^\prime}z^\alpha_{R^\prime{}L^\prime} =0
\label{eq:pms}
\end{align}
and the normalisation
\begin{align}
\sum\nolimits_{RL} |z^\alpha_{RL}|^2 = 1
\label{eq:pmsanormalisation}
\end{align}
simplifies to
\begin{align*}
\Psi({\varepsilon,\br}) = \sum\nolimits_{RL} {N^\alpha_{R\ell} \phi_{RL}}({\varepsilon,\br})
\end{align*}
which is a normalised solution to the SE for $v(\br)=\sum\nolimits_R v_R(r)$.

In practice the solution is inexact because $L$ summations are truncated. The solution is rapidly convergent in the
$L$-cutoff, however; see Ref.~\cite{Andersen86} for an analysis.

\subsubsection{Hard Core Radius}
\label{sec:hardcoreradius}

$\alpha$ can be physically interpreted as equivalent to specifying a ``hard core'' radius where the one-centre expansion
$\hat{H}_{RL}^\alpha$ vanishes in a sphere centred at $R^\prime$.  This is evident from the one-centre expansion
Eq.~(\ref{eq:kalphaonec}) and the form of $\hat{J}_\ell^\alpha$, Eq.~(\ref{eq:jalph}).  $\hat{J}_\ell^\alpha$ vanishes
at the radius where
\begin{align*}
\alpha_{R\ell} = \hat{J}^0_\ell(E,r_R)/\hat{H}^0_\ell(E,r_R)
\end{align*}
In his more recent developments, Andersen defined the screening in terms of the hard core radius $a_\ell$ instead of
$\alpha$, because nearly short-ranged basis functions can be obtained for a fixed $a_{R\ell}{=}0.7s_R$, independent of
$\kappa$ and $\ell$.

It is easy to see how such a transformation can render envelope functions short-ranged. The value of
$\hat{H}_{R\ell}^\alpha$ is forced to be zero in a sea of ${R^\prime{}L^\prime}$ channels surrounding it.  Provided the
$a_{R\ell}$ are suitably adjusted, it quickly drives $\hat{H}_{R\ell}^\alpha(r){\to}0$ everywhere for increasing $r$.
If the $a_{R\ell}{\to}0$, the screening vanishes and $\hat{H}_{R\ell}^\alpha$ returns to the long-ranged
$\hat{H}_{R\ell}^0$; while if the $a_{R\ell}$ becomes comparable to $s_R$ the value on the head must be something like 0
and 1 at the same time (heads and tails meet).  The damping is too large and the $\hat{H}_{R\ell}^\alpha(r)$ ``rings''
with increasing $r$.  For $a_{R\ell}{=}0.7r_{s}$ or thereabouts the ringing is damped and $\hat{H}_{R\ell}^\alpha(r)$
decays exponentially with $r$, even for $\kappa{=}0$.

\subsection{MTO's and Second Order Green's Function}
\label{sec:greena}

Through the eigenvectors $z^\alpha_{RL}$ of Sec.~\ref{eq:tbtailcancellation}, we can construct the Green's function.
This was done in Appendix A of Ref.~\cite{Gunnarsson83}, where the full Green's function, including the irregular parts,
are derived.  Here we will adopt a simpler development along the lines of Ref.~\cite{Andersen75}, after linearising the
$\chi^\alpha_{RL}(\varepsilon,\br)$.

One way to see why $P^\alpha{-}S^\alpha$ have similar eigenvalues for any $\alpha$ is to note that
$[P^\alpha]^{-1}{-}[S^\alpha]^{-1}$ does not depend on $\alpha$, since $[P^\alpha]^{-1}$ and $[S^\alpha]^{-1}$ are
shifted by the same amount (compare Eq.~(\ref{eq:defsalpha}) and (\ref{eq:defipa})).  In scattering theory
$[P^0_\ell(\varepsilon)]^{-1}$ is proportional to tangent of the phase shift, and we realise that the transformation
$(P^0,S^0){\to}(P^\alpha,S^\alpha)$ corresponds merely to a shift of the scattering background.  The pole structures of
$P^\alpha{-}S^\alpha$ can depend on $\alpha$ because of the irregular parts: $P^\alpha{-}S^\alpha$ have the same poles
as $[P^\alpha]^{-1}{-}[S^\alpha]^{-1}$ only where $P^\alpha$ and $S^\alpha$ have no zeros or poles.  Some care must be
taken when generating the Green's function $G$.

The MTOs Eq.~(\ref{eq:tbmtovarye}) form a complete Hilbert space for any $\alpha$, but $\alpha$ can be chosen to satisfy
varying physical requirements.  To make short-ranged Hamiltonians it has been found empirically that the following
universal choice
\begin{align*}
\alpha_s &= 0.34857 &
\alpha_p &= 0.05303\\
\alpha_d &= 0.010714 &
\alpha_\ell &= 0{\text{  for  }}\ell > 2
\end{align*}
yields short-ranged basis functions $\chi^\alpha_{RL}(\br)$ for $\kappa{=}0$, for any reasonably close-packed system.

Another choice is $\alpha{=}\gamma$.  In Sec.~\ref{sec:introh} it was shown how the tail cancellation condition had the
same eigenvalues as a fixed Hamiltonian, Eq.~(\ref{eq:secondorderh}).  We are now equipped make a connection with the
$\chi^\gamma$ basis and Eq.~(\ref{eq:secondorderh}).  Moreover, this connection enables us to construct the second order
Green's function.  First, Eq.~(\ref{eq:defsalpha}) enables us to recognise the quantity
$\left({S^{-1}{-}\gamma}\right)^{-1}$ as $S^\gamma$.  Eq.~(\ref{eq:secondorderh}) then has the simple two-centre form
$\widetilde{h} = C + \Delta^{1/2} \, S^\gamma \, \Delta^{1/2}$

The Green's function corresponding to some fixed $h$ has the simple form
\begin{align*}
\widetilde{G}(z) = [z-\widetilde{h}-i0^+]^{-1}
\end{align*}
for complex energy $z$. It is easy to see that $\widetilde{G}(z)$ can be expressed in the following form:
\begin{align}
\widetilde{G}(z) &= \Delta^{-1/2} \, \widetilde{g}^\gamma \, \Delta^{-1/2}
\end{align}
where
\begin{align}
\widetilde{g}^\gamma &= [\widetilde{P}^\gamma-S^\gamma]^{-1}
\end{align}
This is the analogue of Eq.~(\ref{eq:auxiliaryg}) in the $\gamma$ representation.

$\widetilde{P}^\gamma(z){-}S^\gamma$ has a direct connection with $z-\widetilde{h}$ because of $\widetilde{P}^\gamma(z)$
takes the simple form $(z{-}C)/\Delta$.  $\widetilde{P}^\gamma{-}S^\gamma$ is linear in $z$ since $S^\gamma$ is
independent of it, and $\sqrt{\Delta}(\widetilde{P}^\gamma{-}S^\gamma)\sqrt{\Delta} = z{-}\widetilde{h}$.

\subsubsection{Scattering Path Operator in Other Representations}

To build $G(\varepsilon)$ from general screening representations $\beta$ we need to transform the scattering path
operator $g^\gamma{\to}g^\beta$.  This can be accomplished \cite{Andersen86} using
Eqs.~(\ref{eq:palphatobeta}) (\ref{eq:salphatobeta}).  The result is
\begin{align*}
&{g^\beta } = ({P^\alpha }/{P^\beta }){g^\alpha }({P^\alpha }/{P^\beta }) + (\beta  - \alpha )({P^\alpha }/{P^\beta })
\end{align*}
These transformations require only Eqs.~(\ref{eq:palphatobeta}) and (\ref{eq:salphatobeta}); they do not depend on
parameterisation of $P$.

Since Eq.~(\ref{eq:pbypdot}) is representation-independent, $({P^\alpha}/{P^\beta})$ can equally be written in the
following forms:
\begin{align}
\frac{{{P^\alpha}(\varepsilon)}}{{{P^\beta}(\varepsilon)}} =
 {\left({\frac{{{{\dot{P}}^\alpha}}}{{{{\dot{P}}^\beta}}}}\right)^{1/2}} =
 1+(\alpha-\beta){P^\alpha}(\varepsilon)
\label{eq:formspabypb}
\end{align}

\subsection{The ASA Green's Function, General Representation}
\label{sec:fullG}

As shown in Sec.~\ref{sec:greena} the relation between the Green's function $G(E)$ and the scattering path operator $g$
is particularly simple when potential functions are parameterised to second order.  The relation between the ASA
approximation to $G$ and $g$ can be written more generally as follows.  In the ASA, every point $\br$ belongs to some
sphere $R$ with partial waves $\phi_{RL}(\varepsilon,\br)$ so that
\begin{multline*}
G_{RR'}(\varepsilon,\br,\br') =\\
\sum_{LL'} \phi_{RL}(\varepsilon,\br) \,{G_{RLR'L'}}(\varepsilon)\,\phi_{R'L'}(\varepsilon,\br')
\end{multline*}
where
\begin{multline}
{G_{RLR'L'}}(\varepsilon)=-\frac{1}{2}\frac{{d\ln \dot P_{RL}^\alpha (\varepsilon)}}{{d\varepsilon}}\\
+ \sqrt {\dot P_{RL}^\alpha (\varepsilon)\,} g_{RLR'L'}^\alpha (\varepsilon)\,\sqrt{\dot P_{R'L'}^\alpha (\varepsilon)}
\label{eq:Gfromg}
\end{multline}
$\br$ and $\br'$ (or $R$ and $R'$) are the field and source points, respectively.  It was first shown in
Ref.~\cite{Gunnarsson83}, for the ``bare'' representation $\alpha=0$, and for a screened representation $\alpha$ in
Ref.~\cite{Andersen86}.  The first term cancels a pole appearing in the second term, connected to the irregular part of
$G$ (which we do not consider here).

$P^\alpha(\varepsilon)$ can be computed by integration of the radial Schr\"odinger equation for any $\varepsilon$.  If
this is done, and the structure constants $S$ are taken as energy-dependent, this is the screened KKR method.
Questaal's \texttt{lmgf} and \texttt{lmpg} parameterise $P^\alpha(\varepsilon)$, to second order
(Eq.~(\ref{eq:defipa})), or to third order (Sec.~\ref{sec:thirdorderg}) in $\varepsilon$.

Note that $G$ does not depend on choice of $\alpha$; it can be used to as a stringent test of the correctness of the
implementation.

\subsection{The ASA Hamiltonian: Linearisation of the Muffin-Tin Orbitals}
\label{sec:lmto}

The energy-dependent MTO, Eq.~(\ref{eq:tbmtovarye}), exactly solves the ASA potential for a fixed $\varepsilon$.  To
make a fixed, energy-independent basis set, we constrain the energy-dependent MTO, Eq.~(\ref{eq:tbmtovarye}), to be
independent of $\varepsilon$.  As we saw in Secs.~\ref{sec:introh} and \ref{sec:greena}, there is a simple
energy-independent Hamiltonian Eq.~(\ref{eq:secondorderh}) that has the same eigenvalue spectrum as the second order
$\widetilde{G}$; this is $\chi^\gamma_{RL} (\varepsilon_\nu,{\br})$.

We can construct an energy-independent basis $\chi^\alpha_{RL}({\br}){=}\chi^\alpha_{RL}(\varepsilon_\nu,{\br})$ for any
$\alpha$, by choosing the normalisation $N^{\alpha}_{R\ell}(\varepsilon)$ in such a way that
$\partial{\chi}^\alpha_{RL}(\varepsilon,{\br})/\partial\varepsilon{=}0$ at $\varepsilon=\varepsilon_{\nu}$.  Thus we
require
\begin{gather*}
\frac{\partial}{\partial\varepsilon}
\left[ N^\alpha({\varepsilon}){\phi}({\varepsilon,r}) +
{\bar{\hat{J}}^\alpha({{r}})} P^\alpha({\varepsilon})
 \right]_{\varepsilon=\varepsilon_\nu} = 0\\
{\dot{N}^\alpha }(\varepsilon )\phi (\varepsilon ,{{r}}) + {N^\alpha }(\varepsilon )\dot{\phi}(\varepsilon ,{{r}}) +
{\bar{\hat{J}}^\alpha }({{r}}){\dot{P}^\alpha }(\varepsilon )_{\varepsilon=\varepsilon_\nu} = 0
\end{gather*}
Define
\begin{align}
{\phi^\alpha}(\varepsilon,{{r}})\equiv[{N^\alpha}(\varepsilon)/{N^\alpha}({\varepsilon_\nu})]\phi(\varepsilon,{{r}})
\label{eq:defphia}
\end{align}
Then the condition that $\dot{\chi}^\alpha_{RL}({\br})$ vanish for all $\br$ becomes
\begin{align}
{\bar{\hat{J}}^\alpha}({{r}})= -[{\dot{\phi}^\alpha}(\varepsilon,{{r}})
        {N^\alpha}(\varepsilon)/{\dot{P}^\alpha}(\varepsilon)]_{\varepsilon=\varepsilon_\nu}
\label{eq:defbarj}
\end{align}

Henceforth, when the energy index is suppressed it means that the energy-dependent function or parameter is to be taken
at the linearisation energy $\varepsilon_{\nu}$.  If $\bar{J}$ is replaced by Eq.~(\ref{eq:defbarj}),
Eq.~(\ref{eq:tbmtovarye}) becomes
\begin{align}
&\chi^\alpha_{RL} ({\br}) = \nonumber\\
&\begin{cases}
	{\phi_{RL}}({\br}) + \sum\limits_{R^\prime{}L^\prime{}}{\dot{\phi}^\alpha_{R^\prime{}L^\prime{}}({\br})}
	h^\alpha_{R^\prime{}L^\prime;RL} \quad\text{if }\br\in\{s_R\}\\
	(N^\alpha_{R\ell})^{-1} \hat{H}^\alpha_{RL} (\kappa,{\br}) \hfill\text{if } \br\in\hbox{interstitial}
\end{cases}
\label{eq:tbmto}\\
&h^\alpha = -P^\alpha(\dot{P}^\alpha)^{-1}+[\dot{P}^\alpha]^{-1/2}S^\alpha[\dot{P}^\alpha]^{-1/2}
\label{eq:defha}
\end{align}

The Hilbert space of the $\{\chi^\alpha_{RL} ({\br})\}$ consists of the pair of functions
${\phi}_{R^\prime{}L^\prime{}}$ and $\dot{\phi}_{R^\prime{}L^\prime{}}$ inside all augmentation channels
${R^\prime{}L^\prime{}}$.  Changing $\alpha$ merely rotates the Hilbert space, modifying how much
${\phi}_{R^\prime{}L^\prime{}}$ and $\dot{\phi}_{R^\prime{}L^\prime{}}$ each $\chi^\alpha_{RL}$ contains.

Eq.~(\ref{eq:tbmto}) may be regarded as a Taylor series of $\chi^\alpha_{RL}(\varepsilon,{\br})$,
Eq.~(\ref{eq:tbmtovarye}), to first order in $\varepsilon-\varepsilon_\nu$, with $\dot{\phi}^\alpha(r)\,h^\alpha$
playing the part of $\dot{\chi}^\alpha(r)\,(\varepsilon-\varepsilon_\nu)$.  The eigenvalues of $h^\alpha$ are fact the
eigenvalues of $\chi^\alpha_{RL}(\varepsilon,{\br})$ to first order in $\varepsilon-\varepsilon_\nu$.  If $P$ is
parameterised to second order, $P(\varepsilon){\to}\widetilde{P}(\varepsilon)$, $h^\alpha$ in the $\gamma$
representation becomes the second order ASA Hamiltonian, Eq.(\ref{eq:secondorderh}), when $\alpha{=}\gamma$.

The relation between $N^\alpha(\varepsilon)$ and $P^\alpha(\varepsilon)$ was already established in
Eq.~(\ref{eq:defna}).  To confirm it is consistent with Eq.~(\ref{eq:defbarj}) at $\varepsilon_\nu$, revisit the
definition of $N^\alpha$ using Eq.~(\ref{eq:defbarj})
\begin{align*}
N^\alpha(\varepsilon)&=\frac{{W\{J^\alpha,\hat{H}^0\}}}{{W\{J^\alpha,{\phi}\}}}=
        \frac{{w/2}}{{W\{-{\dot{\phi}^\alpha}{N^\alpha}(\varepsilon)/{\dot{P}^\alpha}(\varepsilon),{\phi}\}}}\\
&=\frac{{{\dot{P}^\alpha}(\varepsilon)w/2}}{{{N^\alpha}(\varepsilon)W\{{\dot{\phi}^\alpha},{\phi}\}}}=
        \frac{{{\dot{P}^\alpha}(\varepsilon)w/2}}{{{N^\alpha}(\varepsilon)}}
\end{align*}
which confirms Eq.~(\ref{eq:defna}).

\subsubsection{Potential Parameters \texorpdfstring{$C^\alpha$, $\Delta^\alpha$, and $o^\alpha$}{C,D and o (alpha
rep.)}}

The LMTO literature suffers from an unfortunate proliferation of symbols, which can be confusing.  Nevertheless we
introduce yet another group because they offer simple interpretations of what is happening as the basis changes with
representation $\alpha$, and also to make a connection with the Green's function.  It is helpful to remember there is a
single potential function $P^\alpha(\varepsilon)$, which determines the normalisation $N^\alpha(\varepsilon)$ though
Eq.~(\ref{eq:defna}).

$P^\alpha(\varepsilon)$ can be parameterised to second order with three independent parameters $C$, $\Delta$, and
$\gamma$ (Eq.~(\ref{eq:defipa})) and to third order with the ``small parameter'' $p$ (Eq.~(\ref{eq:defsmallp})).

The energy derivative of ${\phi}^\alpha(\varepsilon,r)$ at $\varepsilon_\nu$ is
\begin{gather}
{\dot{\phi}_\ell^\alpha}(\varepsilon_\nu,{r}) = {\dot{\phi}_\ell^\alpha}({r}) =
        \dot{\phi}_\ell({r}) + o_\ell\,\phi({r})\label{eq:defdotphia}\\
o^\alpha_\ell \equiv {\dot{N}_\ell^\alpha}/{N_\ell^\alpha} \to
        \frac{\alpha-\gamma}{(\alpha-\gamma)(C-{\varepsilon_\nu})+\Delta}\label{eq:defo}
\end{gather}
The last form applies when $P$ is parameterised to second order.  We have introduced the ``overlap'' potential parameter
$o^\alpha_\ell$.  It vanishes in the $\gamma$ representation and consequently $\dot{\phi}^\gamma{=}\dot{\phi}$.  This
fact provides a simple interpretation of $\chi^\alpha_{RL}({\br})$, Eq.~(\ref{eq:tbmto}) in the $\gamma$ representation.
$\chi^\gamma_{RL}$ acquires pure $\phi$ character for its own head, and pure $\dot{\phi}$ in spheres where
$R^\prime{\ne}R$.  This implies that the $\{\chi^\gamma\}$ basis are orthogonal apart from interstitial contributions
(neglected in the ASA) and small terms proportional to $p$ (Eq.~(\ref{eq:defsmallp})).  $\phi$ and $\dot{\phi}$ combine
in every sphere in the exact proportion Eq.~(\ref{eq:linearisephi}) at each eigenvalue $\varepsilon_i$ of $h^\gamma$.
Thus eigenvalues of $h^\gamma$ are correct to one order in $\varepsilon_i{-}\varepsilon_\nu$ higher than
$h^{\alpha{\ne}\gamma}$.

In the LMTO literature two other parameters are introduced to characterise $h^\alpha$ in a suggestive form:
\begin{multline}
h^\alpha_{RL;R^\prime{}L^\prime} = (C^\alpha-\varepsilon_{\nu})_{R\ell}\,\delta_{RL;R^\prime{}L^\prime}\\
+ \Delta^\alpha_{R\ell}S^\alpha_{RL;R^\prime{}L^\prime}\Delta^\alpha_{R^\prime{}\ell^\prime}
\end{multline}
where
\begin{align}
C^\alpha_{\ell}-\varepsilon_\nu &\equiv -P_\ell^\alpha(\dot{P}_\ell^\alpha)^{-1} \to (C-{\varepsilon_\nu})
        \left[\dots\right]\label{eq:defca}\\
\sqrt{\Delta^\alpha} &\equiv [\dot{P}^\alpha]^{-1/2}\to\sqrt{\Delta}\left[\dots\right]\label{eq:defda}
\end{align}
and
\begin{align*}
\left[\dots\right] = {\left[{1+\frac{{(C-{\varepsilon_\nu})(\alpha-\gamma)}}{\Delta}}\right]}
\end{align*}
$h^\alpha+\varepsilon_\nu$ becomes $\widetilde{h}$ (Eq.~(\ref{eq:secondorderh})) when $\alpha{=}\gamma$.

\subsubsection{How \texorpdfstring{$h^\alpha$}{h alpha} Changes with Representation}

From Eq.~(\ref{eq:defo}) implies that $\dot{\phi}^\alpha$ transforms as
\begin{align*}
\dot{\phi}^\alpha - o^\alpha \phi  = \dot{\phi}^\beta - o^\beta \phi
\end{align*}
Dividing $\chi^\alpha$ in to $\phi$ and $\dot{\phi}$ parts, we realise that $o^\alpha + [h^\alpha]^{-1}$ is independent
of representation and therefore
\begin{align*}
o^\beta + [h^\beta]^{-1} = o^\alpha + [h^\alpha]^{-1}
\end{align*}

\subsubsection{ASA Hamiltonian and Overlap matrix}
\label{sec:asah}

$\chi^\alpha$ (Eq.~(\ref{eq:tbmto})) is an energy-independent basis set and has an eigenvalue spectrum.  Within the ASA
(Sec.~\ref{sec:asa}) matrix elements of the Hamiltonian and overlap are readily obtained.  Using normalisation
Eq.~(\ref{eq:phinormalisation}) and (\ref{eq:defsmallp}), Schr\"odinger's equation for partial waves
Eq.~(\ref{eq:sephi}), the parameterisation of $\dot{\phi}^\alpha$ Eq.~(\ref{eq:defdotphia}), and neglecting the
interstitial parts
\begin{align}
H &= \left< \chi^\alpha | -\nabla^2 + v |\chi^\alpha\right>\nonumber\\
&= h^\alpha(1+o^\alpha{}h^\alpha)\nonumber + (1+h^\alpha{}o^\alpha)\varepsilon_\nu(1+o^\alpha{}h^\alpha)\nonumber\\
&\quad+ h^\alpha\,\varepsilon_\nu\,p\,h^\alpha\label{eq:asah}\\
O &= \left< \chi^\alpha | \chi^\alpha\right> \nonumber\\
  &= (1+h^\alpha{}o^\alpha)(1+o^\alpha{}h^\alpha) + h^\alpha\,p\,h^\alpha\label{eq:asao}
\end{align}

\subsubsection{Third Order Green's Function}
\label{sec:thirdorderg}

$\widetilde{G}(z)$ depends on potential parameters $C$ and $\Delta$.  These can be replaced with the following:
\begin{align}
\Delta=[{\dot{\widetilde{P}}^\gamma(\varepsilon)}]^{-1} \quad\text{and}\quad
\varepsilon-C = \frac{\widetilde{P}^\gamma(\varepsilon)}{{\dot{\widetilde{P}}^\gamma}(\varepsilon)}
\label{eq:relationcanddeltatop}
\end{align}
which follows from Eq.~(\ref{eq:pbypdot}) with $\alpha{=}\gamma$.  Then the substitution $\widetilde{P}{\to}P$ through
the replacement $\varepsilon{\to}\varepsilon^\prime=\varepsilon+(\varepsilon-\varepsilon_\nu)^3p$,
Eqs.~(\ref{eq:porder3}) and (\ref{eq:defsmallp}).  This yields an expression for $G$ to third order---more accurate
than the 2$\mathrm{nd}$ order $\widetilde{G}$~\cite{Andersen86,Gunnarsson83}.  Questaal codes \texttt{lmgf} and
\texttt{lmpg} permit either second or third order parameterisation.

\subsection{Principal Layer Green's Functions}
\label{sec:lmpg}

Questaal has another implementation of ASA-Green's function theory designed mainly for transport.  \texttt{lmpg} is
similar in most respects to the crystal package \texttt{lmgf}, except that is written as a principal-layer technique.
\texttt{lmpg} has a `special direction', which defines the layer geometry, and for which $G$ is generated in real space.
In the other two directions, Bloch sums are taken in the usual way; thus for each $q$ in the parallel directions, the
Hamiltonian becomes one-dimensional and is thus amenable to solution in order-\emph{N} time in the number of layers
\emph{N}.

The first account of this method was presented in Ref.~\cite{pgfcode}, and the formalism is described in detail in
Ref.~\cite{Faleev05}, including its implementation of the non-equilibrium case.  Here we summarise the basic idea and
the main features.

\texttt{lmpg} is similar in many respects to \texttt{lmgf} except for its management of the layer geometry.  The
material consists of an active, or embedded region, which is cladded on the left and right by left and right
semi-infinite leads.
\begin{align*}
 \overbrace{\ldots \,{\rm PL}\ -1\,}^{\rm PLATL}\ \big|\
 \overbrace{{\rm PL}\ 0 \,|\, \ldots \,|\, {\rm PL}\ n{\rm{-}}1}^{\rm PLAT}\ \big|\
 \overbrace{\,{\rm PL}\ n\,\ldots}^{\rm PLATR}
\end{align*}

The end regions are half-crystals with infinitely repeating layers in one direction.  All three regions are partitioned
into slices, or principal layers (PL), along the `special direction'.  The left- and right- end regions consist of a
single PL, denoted $-1$ and $n$, which repeat to $\mp\infty$.  Thus the trilayer geometry is defined by five lattice
vectors: two defining the plane normal to the interface (the potential is periodic in those vectors); one vector
\texttt{PLAT} for the active region and one each (\texttt{PLATL} and \texttt{PLATR}) defining the periodically repeating
end regions.

Far from the interface the potential is periodic and states are Bloch states.  It is assumed that the potential in each
end layer is the same as the bulk crystal (apart from a constant shift) and repeats periodically in lattice vectors
\texttt{PLATL} and (\texttt{PLATR}) to $\mp\infty$.

Partitioning into PL is done because $\varepsilon-H=G^{-1}$ is short-ranged.  It is requirement that a PL is thick
enough so that \emph{H} only connects adjacent PL.  Then \emph{H} is tridiagonal in the PL representation and the work
needed to construct $G$ scales linearly with the number of PL.  Moreover it is possible in this framework to construct
$G$ for the end regions without using Bloch's theorem.

Principal layers are defined by the user; they should be chosen so that each PL is thick enough so that \emph{H}
connects to only nearest-neighbour PL on either side.  (Utility \texttt{lmscell} has a facility to partition the active
region into PL automatically.)

\subsubsection{Green's Function for the Trilayer}

\texttt{lmpg} constructs the auxiliary $g$, and if needed builds $G$ from $g$ by scaling (Sec.~\ref{sec:fullG}).  In
many instances $g$ is sufficient (e.g. to calculate transmission and reflection probabilities~\cite{Faleev05}), although
$G$ is needed to make the charge density.  Note there is a $g$ (or $G$) connecting every layer to every other one; thus
$g$ has two layer indices, $g_{ij}$ ($i$ and $j$ refer to PL here).

$g{=}(P-S)^{-1}$ for the entire trilayer can be constructed in one of two ways.  The first is a difference-equation
method, described in Ref.~\cite{pgfcode}.  The second is simply to invert $(P-S)$ using sparse matrix techniques.  Both
methods require as starting points the diagonal element $g^{s,L}_{-1-1}$ for semi-infinite system (consisting of all
layers between $\infty$ and $-1$, with vacuum for all layers to the right of the L- region) and the corresponding
$g^{s,R}_{nn}$ for the R- region.

The sparse-matrix method is simple to describe.  Supposing the active region is considered in isolation; denote it as
$I$.  Then $g_I^{-1}{=}(P-S)_I$.  The effect of the leads is to modify $g_I^{-1}$ by adding a self-energy to layers 0
and $n$:
\begin{gather*}
g^{-1}    = (P-S)_I + \Sigma_0 + \Sigma_n\\
\Sigma_0  = S_{0,-1}\, g^{s,L}_{-1-1}\, S_{-1,0}\\
\Sigma_n  = S_{n-1,n}\, g^{s,R}_{nn}\, S_{n-1,n}
\end{gather*}
$g^{-1}$ is inverted by a sparse matrix technique.

\texttt{lmpg} implements both the difference-equation and sparse-matrix techniques: both scale linearly with the number
of layers, in memory and in time.  It has been found empirically that they execute at similar speed for small systems,
while the difference-equation method is significantly faster for large systems.

\subsubsection{Green's Functions for the End Regions}
\label{sec:GFends}

To make $g$, the diagonal surface Green's functions $g^{s,L}_{-1-1}$ and $g^{s,R}_{nn}$ are required.  \texttt{lmpg}
implements two schemes to find them: a ``decimation'' technique~\cite{LopezSancho85} and a special-purpose
difference-equation technique applicable for a periodic potential~\cite{Chen89}.

The latter method requires solution of a quadratic algebraic eigenvalue problem, which yields eigenvalues $r$: they
correspond physically to wave numbers as $r = e^{ika}$.  $a$ is the thickness of the PL and $k$ the wave number in the
plane normal to the interface.  $k$ is in general complex since no boundary conditions are imposed; it is real only for
propagating states.  Eigenvalues occur in pairs, $r_{1}$ and $r_{2}$, and in the absence of spin-orbit coupling,
$r_{1}=1/r^*_{2}$.  There is a boundary condition on the end leads for the trilayer, which excludes states that grow
into an end region.  The surface Green's function $g^{s}$ can be constructed from the same eigenvectors that make the
``bulk'' $g$~\cite{Chen89}.

A great advantage of this method is that its solution provides the eigenfunctions of the system; thus the Green's
function can be resolved into normal modes.  A large drawback is the practical problem of finding a solution to the
eigenvalue problem.  It can be converted into a linear algebraic eigenvalue equation of twice the rank; however, the
resulting secular matrix can be nearly singular (especially if $S$ is short-ranged).  Also, the pairs $r_{1}/r_{2}$ can
range over a very large excursion, of unity for propagating states and many orders of magnitude for rapidly decaying
ones.  Capturing them by solving a single eigenvalue problem imposes severe challenges on the eigenvalue solver.

Decimation is recursive and generally efficient; however problems can appear at special values of $k$ energy where the
growing and decaying pair $r_{1}$ and $r_{2}$ become very close to unity.  Unfortunately, those ``hot spots'' are often
the physically interesting ones.

At present Questaal's standard distribution does not have a fully satisfactory, all-purpose method to determine $g^{s}$,
though one has been developed and will be reported in a future work.

\subsection{Contour Integration over Occupied States}

\begin{figure}[ht]
\begin{center}
\includegraphics[width=.75\columnwidth]{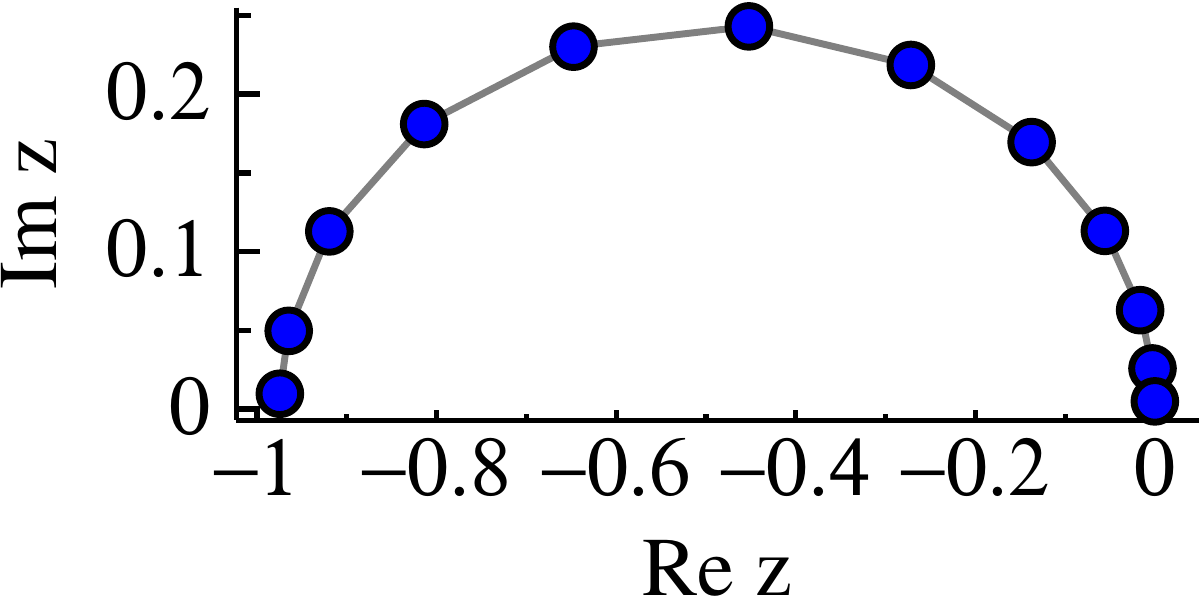}
\end{center}
\caption{Representative contour integration for occupied states.  Twelve points are taken for the interval ($-1,0$).
The contour deformation has eccentricity = 1/2, and a bunching parameter 1/2, giving more weight to the points near
$E_{F}=0$.}
\label{fig:contourintegration}
\end{figure}

Many properties of interest involve integration over the occupied states.  In contrast to band methods which give
eigenfunctions for the entire energy spectrum at once, Green's functions are solved at a particular energy, and must be
numerically integrated on an energy mesh for integrated properties.  The spectral function or density of states is
related to $G$ as
\begin{align}
A(\varepsilon) = \pi^{-1}|\text{Im}\,G(\varepsilon)|
\end{align}
and in the noninteracting case is comprised of a superposition of $\delta$-functions at the energy levels.  Thus
$G(\varepsilon)$ has lots of structure on the real axis, which makes integration along it difficult.  However, since
below the Fermi level $G$ should have no poles in the upper half of the complex plane, the Cauchy theorem can be used to
deform the integral from the real axis to a path in the complex plane (Fig.~\ref{fig:contourintegration}).
\texttt{lmgf} and \texttt{lmpg} use an elliptical path, with upper and lower bounds on the real axis respectively at the
Fermi level and some energy below the bottom of the band.

A Legendre quadrature is used, but the weights can be staggered to bunch points near $E_{F}$ where $G$ has lots of
structure.  Thus five parameters define the mesh: the number of points, the upper and lower bounds, the eccentricity of
the ellipse (between 0 for circle and 1 for a line on the real axis) and bunching parameter which also ranges between 0
and 1.  The integrand on the contour is smooth, except near the endpoint $z{\to}E_{F}$.  Good results can be obtained
with a modest number of points, typically 12-20.

The exact potential function $P^\alpha$ has no poles in this half plane; nor does the second order parameterisation but
spurious poles may appear in the third order parameterisation.  These may be avoided by working in the orthogonal
representation ($\alpha=\gamma$) and/or by choosing fairly large elliptical eccentricities for the contour.

\subsection{Spin-orbit Coupling in the Green's Function}
\label{sec:socgf}

It has been shown in Sec.~\ref{sec:soc} that spin-orbit coupling (SOC) can be added perturbatively to the Hamiltonian,
resulting in the matrix elements containing $\xi_\ell[D',D]$.  These matrix elements are added to the right-hand-side of
the first line in Eq.~(\ref{eq:MEPhi}), while the overlap integrals (second line in Eq.~(\ref{eq:MEPhi})) remain
unchanged.  In order to construct the Green's function with SOC, the resulting modification of the variational energy in
Eq.~(\ref{eq:epsvsw}) needs to be reformulated as a perturbation of the potential parameters.

Because the SOC operator is a matrix (see Sec.~\ref{sec:soc}), the exact solutions $\phi_{\nu lj\kappa}(\mathbf{r})$ of
the radial Pauli equation are linear combinations of spherical waves $|lm\sigma\rangle$ and $|lm'\sigma'\rangle$ with
$m+\sigma=m'+\sigma'=j$. However, our perturbative treatment is still based on basis functions with definite spin that
are calculated without SOC. The energy dependence, however, is modified by allowing the $\omega$ parameter to become a
matrix, so that Eq.~(\ref{eq:defPhi}) is replaced by
\begin{multline}
\Phi_{m\sigma}(D_\uparrow, D_\downarrow,\br) = \phi_{m\sigma}(\epsilon_{\nu\sigma},\br)\\
+\sum_{m'\sigma'}\omega_{m\sigma,m'\sigma'}(D_\uparrow,D_\downarrow)\dot\phi_{m'\sigma'}(\epsilon_{\nu \sigma'},\br)
\label{PhiSOC}
\end{multline}
where we dropped the common $\ell$ superscript because the SOC operator is diagonal in $\ell$.  The summation in
(\ref{PhiSOC}) involves at most two terms with $m+\sigma=m'+\sigma'$.

Instead of the simple variational estimate of $\epsilon(D)$ in Eq.~(\ref{eq:epsvsw}), we now construct a generalised
eigenvalue equation, which leads to
\begin{align}
\hat\omega+\hat V_\text{SO}=
        \epsilon\hat 1 -\hat\epsilon_\nu+\hat\omega^\dagger\hat p (\epsilon\hat 1-\hat\epsilon_\nu)\hat\omega
\label{eq:geveq}
\end{align}
where $\hat\omega$ is the matrix from Eq.~(\ref{PhiSOC}), both $\hat\omega$ and $\hat V_\text{SO}$ are functions of
$D_\uparrow$ and $D_\downarrow$, and $\hat p$ and $\hat\epsilon_\nu$ are diagonal matrices with elements
$\langle\dot\phi^2_{\nu \ell\sigma}\rangle$ and $\epsilon_{\nu\sigma}$, respectively.  The matrices are assumed to
include the full basis set on the given site, i.e., their dimension is $2(2\ell+1)$.

The $\hat\omega$ matrix is found by solving Eq.~(\ref{eq:geveq}).  To first order in $\epsilon-\epsilon_\nu$, it gives
$\hat\omega=\epsilon\hat1-\hat\epsilon_\nu-\hat V_\text{SO}$, so that the matrix elements of $\hat V_\text{SO}$ are
effectively added to $\hat\epsilon_\nu$.  Promoting the potential function $P(\epsilon)$ to a matrix and using the
representation Eq.~(\ref{eq:parameterisedP}) and the definitions Eqs.~(\ref{eq:defgam}-\ref{eq:defv}), we find that the
parameters $\Delta$ and $\gamma$ are unaffected by $\hat V_\text{SO}$ while $C$ is promoted as $C\to \hat C=C\hat 1+\hat
V_\text{SO}$.  However, in order to make the definition of $\hat P(\epsilon)$ unambiguous, we need to fix the correct
order of matrix multiplication.  We also need to ensure that the poles of $G(z)$ have unit residues.  We use the
following definitions:
\begin{align}
G(\epsilon) &= \lambda(\epsilon) +\mu_L(\epsilon) [P(\epsilon)-S]^{-1}\mu_R(\epsilon),\\
P(\epsilon) &= \left[\gamma+\sqrt{\Delta}(\epsilon-\hat C)^{-1}\sqrt{\Delta}\right]^{-1}\label{PV}\\
\mu_L(\epsilon) &=(1-\hat {\dot V}_\text{SO})^{1/2} [\Delta+\gamma(\epsilon-\hat C)]^{-1} \sqrt{\Delta}\label{muL}\\
\mu_R(\epsilon) &=\sqrt{\Delta} [\Delta+(\epsilon-\hat C)\gamma]^{-1} (1-\hat {\dot V}_\text{SO})^{1/2}\label{muR}\\
\lambda &=-\frac12 \mu_R^{-1}\ddot P \mu_L^{-1}
\end{align}
where $\hat {\dot V}_\text{SO}=d\hat {\dot V}_\text{SO}/d\epsilon$ comes from the energy dependence of the SOC
parameters $\xi_\ell[D',D]$.  Note that $\dot P = \mu_R\mu_L$, and the structure of $G(z)$ guarantees that the poles of
$G(z)$ have unit residues.  It is straightforward to check that, with energy-independent $\hat V_\text{SO}$, $G(z)$
becomes the resolvent of the second-order Hamiltonian Eq.~(\ref{eq:secondorderh}), with added $\hat V_\text{SO}$, as
expected.

To third order in $\epsilon-\epsilon_\nu$ and to first order in $\hat V_\text{SO}$, we find from Eq.~(\ref{eq:geveq}):
\begin{align}
\hat\omega^{(3)} &= \epsilon\hat 1 -\hat\epsilon_\nu + \hat p (\epsilon\hat 1-\hat\epsilon_\nu)^3-
        \hat V_\text{SO}^{(3)}\nonumber\\
&= \hat\epsilon'-\hat\epsilon_\nu-\hat V_\text{SO}^{(3)}
\label{omega3}
\end{align}
where $\epsilon'$ is defined in Eq.~(\ref{eq:porder3}), and
\begin{align}
\hat V_\text{SO}^{(3)} = \hat V_\text{SO} + \bigl\{\hat V_\text{SO},\hat p(\epsilon\hat1-\hat\epsilon_\nu)^2\bigr\}
\label{VSO3}
\end{align}
with $\bigl\{\hat A,\hat B\bigr\}=\hat A\hat B+\hat B\hat A$.  Because $\hat V_\text{SO}$ is defined at the fixed values
of the logarithmic derivatives, the spin-orbit coupling parameters $\xi_{\ell,\sigma\sigma'}$ in $\hat V_\text{SO}$ are
calculated with $\epsilon$ replaced by $\epsilon'$:
\begin{align} 
\xi&_{\ell,\sigma\sigma'}(\epsilon)=\langle\phi_{\ell\sigma}|\xi(r)|\phi_{\ell\sigma'}\rangle\nonumber\\
&+(\epsilon'_{\ell\sigma}-\epsilon_{\nu \ell\sigma})
        \langle\dot\phi_{\ell\sigma}|\xi(r)|\phi_{\ell\sigma'}\rangle\nonumber\\
&+(\epsilon'_{\ell\sigma'}-\epsilon_{\nu \ell\sigma'})
        \langle\phi_{\ell\sigma}|\xi(r)|\dot\phi_{\ell\sigma'}\rangle\nonumber\\
&+(\epsilon'_{\ell\sigma}-\epsilon_{\nu \ell\sigma})(\epsilon'_{\ell\sigma'}-
        \epsilon_{\nu \ell\sigma'})\langle\dot\phi_{\ell\sigma}|\xi(r)|\dot\phi_{\ell\sigma'}\rangle
\end{align}
Of course, just as in the non-relativistic case, $\epsilon$ is also replaced by $\epsilon'$ where it appears explicitly
in Eqs.~(\ref{PV}-\ref{muR}). $\hat {\dot V}_\text{SO}$ is always calculated as the \emph{exact} energy derivative of
$\hat {V}_\text{SO}$.

\begin{figure}[ht]
\begin{center}
\includegraphics[height=25ex]{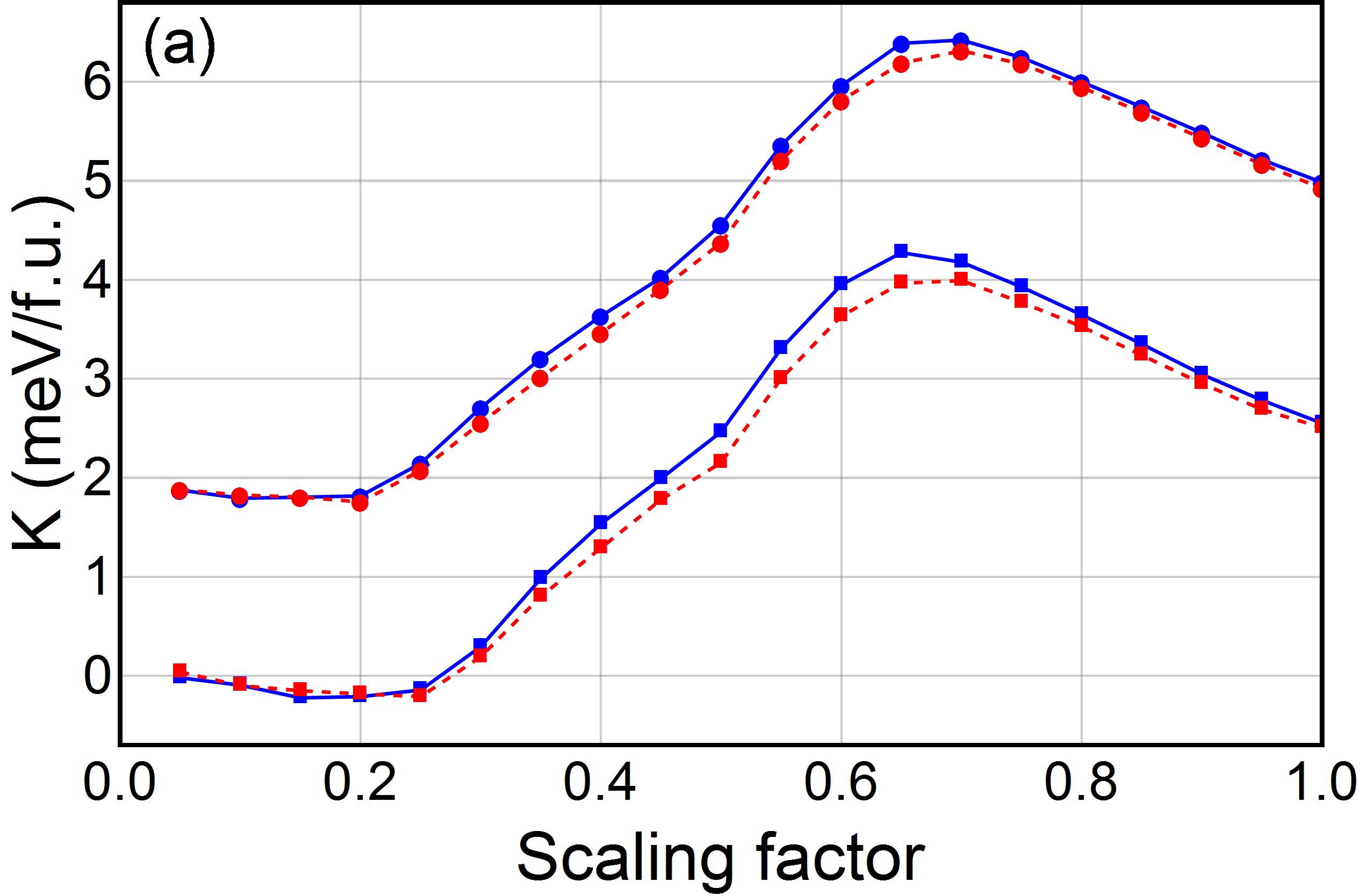} 
\includegraphics[height=25ex]{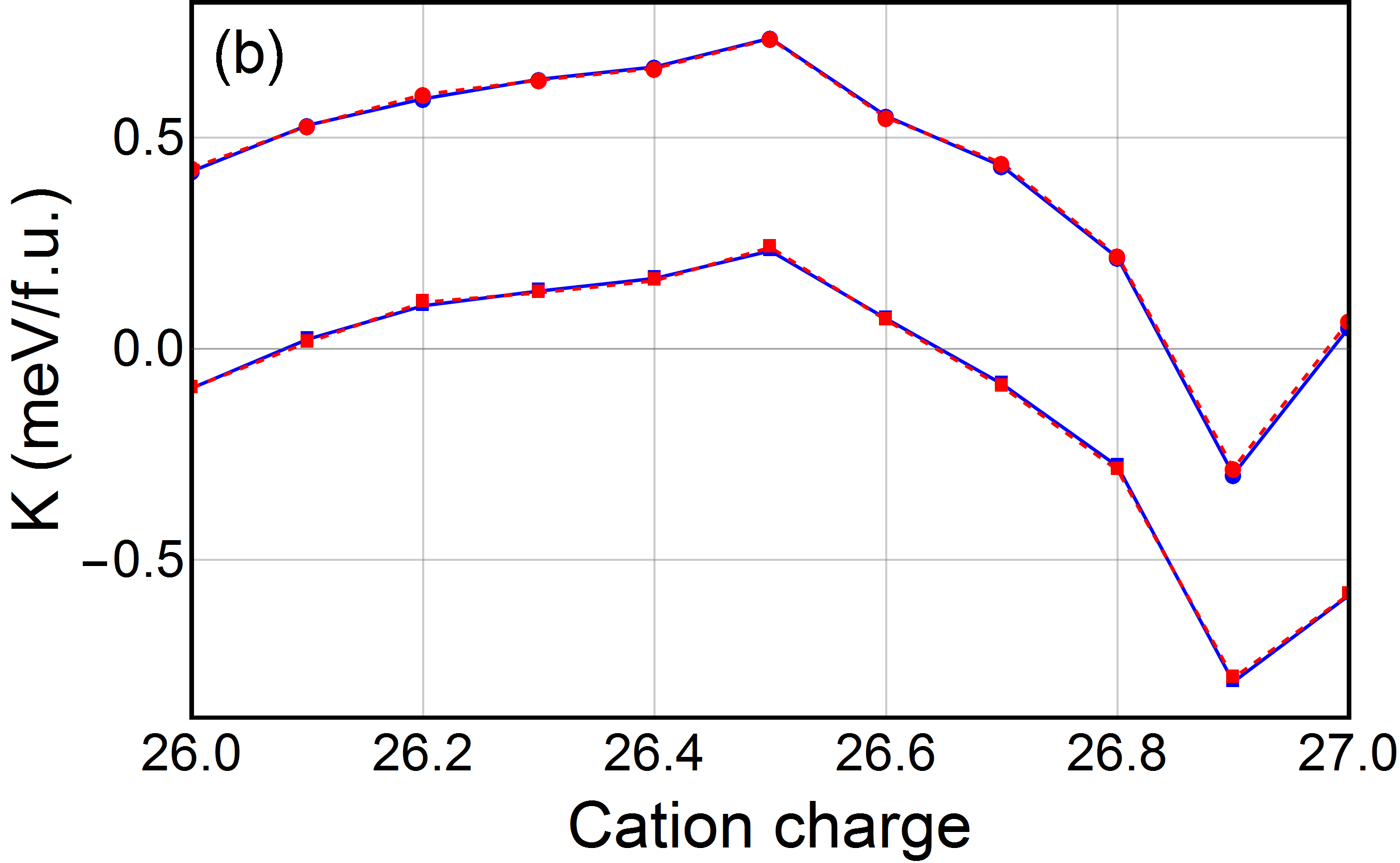}
\end{center}
\caption{Benchmarks for magnetocrystalline anisotropy using \texttt{lm} (blue symbols and solid curves) and
\texttt{lmgf} (red symbols and dashed curves). (a) FePt with the magnetic part of the exchange-correlation field scaled
by a factor between 0 and 1. (b) (Fe$_{1-x}$Co$_x$)$_2$B alloy in the virtual crystal approximation, with the cation
charge varied from 26 to 27.  Lower curves with squares: full three-centre \texttt{lm} and \texttt{lmgf} with
third-order potential functions.  Upper curves (shifted by 2 meV/f.u. in panel (a) and by 0.5 meV/f.u. in panel (b)):
two-centre approximation in \texttt{lm} and second-order potential functions in \texttt{lmgf}.  The charge density in
all calculations for the given material is taken from the full self-consistent \texttt{lm} calculation (without scaling
in the case of FePt).}
\label{fig:socbenchmarks}
\end{figure}

Figure \ref{fig:socbenchmarks} shows the comparison of the magnetocrystalline anisotropy calculated using \texttt{lm}
and \texttt{lmgf} for two benchmark systems.  Two cases are displayed: \texttt{lmgf} with second-order potential
functions compared with the corresponding two-centre approximation in \texttt{lm}, and \texttt{lmgf} with third-order
potential functions compared with the full three-centre \texttt{lm} calculation. The agreement in both cases for FePt
[panel (a)] is very good, while for the (Fe$_{1-x}$Co$_x$)$_2$B alloy in the virtual crystal approximation it is
essentially perfect.

\subsection{Fully Relativistic LMTO-ASA}
\label{sec:frasa}

We have developed a fully relativistic extension of the LMTO-ASA code within a relativistic generalisation of the
density functional formalism~\cite{Raja73,Vignale87,Vignale88,Ebert89,Turek97}. In the most general case, one needs to
solve the Kohn-Sham Dirac equation
\begin{align}
H\Psi(\varepsilon,\vec{r})=\varepsilon\Psi(\varepsilon,\vec{r})
\label{KS-Dirac-1}
\end{align}
with
\begin{align}
H=c \vec{\alpha} \vec{p} + (\beta - I_4) mc^{2}+V(\vec{r})I_4+\mu_{B} \beta\vec{B}_{\text{eff}}(\vec{r}) \vec{\Sigma}
\label{H-Dirac-1}
\end{align}
where
\begin{align}
\alpha=
\begin{pmatrix}
        0&\vec{\sigma}\\
        \vec{\sigma}&0
\end{pmatrix},\;
\beta=
\begin{pmatrix}
        I_2&0\\
        0&-I_2
\end{pmatrix},\;
\Sigma=
\begin{pmatrix}
        \vec{\sigma}&0\\
        0&\vec{\sigma}
\end{pmatrix}\;
\end{align}
Here, $\vec{\sigma}$ is the vector of Pauli matrices, $\vec{p}$ is the momentum operator,
$\vec{B}_{\text{eff}}(\vec{r})$ is an effective spin-dependent potential acting on electrons.  It should be noted that
this is a simplified form of the relativistic Kohn-Sham equation in which the orbital contribution to the 4-component
relativistic current is neglected.  This simplification is necessary in order to avoid the significant formulaic and
computational complications that arise in the relativistic \emph{current density} formulation of the density functional
theory~\cite{Turek97}.

For a spherically symmetric potential $V(r)=1/2[V_{\uparrow}(r)+V_{\downarrow}(r)]$ inside a single MT sphere, the
direction of the magnetic field can be assumed to point along the $z$ direction, $\vec{B}_{\text{eff}}=B(r)\hat{z}$,
where $B(r)=1/2[V_{\uparrow}(r)-V_{\downarrow}(r)]$.  Then, Eq.~(\ref{H-Dirac-1}) can be written as
\begin{align}
H=[c \vec{\alpha} \vec{p} + (\beta - I_4) mc^{2}+V(r)I_4+\mu_{B}\beta B(r)\Sigma_z]
\label{KS-Dirac-2}
\end{align}
The solutions of the Kohn-Sham Dirac equation (\ref{KS-Dirac-2}) are linear combinations of bispinors:
\begin{gather}
\Psi_{\mu}(\vec{r},\varepsilon) = \sum_{\kappa}\Psi_{\kappa\mu}(\vec{r},\varepsilon)\\
\Psi_{\kappa\mu}(\vec{r},\varepsilon) =
\begin{pmatrix}
        g_{\kappa\mu}(\varepsilon,r)\Omega_{\kappa\mu}(\hat{r})\\
        if_{\kappa\mu}(\varepsilon,r)\Omega_{-\kappa\mu}(\hat{r})
\end{pmatrix}
\end{gather}
Here, $\Omega_{\kappa\mu}(\hat{r})$ are the spin spherical harmonics, $\mu$ is the projection of the total angular
momentum, and $\kappa$ is the relativistic quantum number, $\kappa^{2}=J(J+1)+1/4$.  The radial amplitudes
$g_{\kappa\mu}(\varepsilon,r)$ and $f_{\kappa\mu}(\varepsilon,r)$ satisfy the following set of coupled differential
equations:
\begin{multline*}
\left[\frac{d}{dr}+\frac{1+\kappa_{1}}{r}\right]           g_{\kappa_{1}\mu}(\varepsilon,r)=\\
\left[1+\frac{\varepsilon - V(r) + u{'} B(r)}{c^2}\right] cf_{\kappa_{1}\mu}(\varepsilon,r)
\end{multline*}
\vspace*{-\belowdisplayskip}\vspace*{-\abovedisplayskip} 
\begin{multline*}
\left[\frac{d}{dr}+\frac{1-\kappa_{1}}{r}\right]          cf_{\kappa_{1}\mu}(\varepsilon,r)=\\
\left[-\left(\varepsilon - V(r)\right) - u B(r)\right]    cg_{\kappa_{1}\mu}(\varepsilon,r)\\
-\sqrt{1-u^{2}}B(r)                                        g_{\kappa_{2}\mu}(\varepsilon,r)
\end{multline*}
\vspace*{-\belowdisplayskip}\vspace*{-\abovedisplayskip} 
\begin{multline*}
\left[\frac{d}{dr}+\frac{1+\kappa_{2}}{r}\right]           g_{\kappa_{2}\mu}(\varepsilon,r)=\\
\left[1+\frac{\varepsilon - V(r) + u{''} B(r)}{c^2}\right]cf_{\kappa_{2}\mu}(\varepsilon,r)
\end{multline*}
\vspace*{-\belowdisplayskip}\vspace*{-\abovedisplayskip} 
\begin{multline}
\left[\frac{d}{dr}+\frac{1-\kappa_{2}}{r}\right]          cf_{\kappa_{2}\mu}(\varepsilon,r)=\\
\left[-\left(\varepsilon - V(r)\right) - u B(r)\right]    cg_{\kappa_{2}\mu}(\varepsilon,r)\\
-\sqrt{1-u^{2}}B(r)                                        g_{\kappa_{1}\mu}(\varepsilon,r)
\label{radial_Dirac}
\end{multline}
where
\begin{align}
u=\frac{\mu}{\ell+1/2},\quad u{'}=\frac{\mu}{\ell-1/2},\quad u{''}=\frac{\mu}{\ell+3/2}
\end{align}

In the general case, in the presence of a magnetic field, one has to solve a system of two infinite sets of mutually
coupled differential equations because the magnetic field couples radial amplitudes with different relativistic quantum
numbers $\kappa$.  Specifically, states with $\kappa_1$ to those with $\kappa_2=\kappa_1$ and $\kappa_2=-\kappa_1-1$,
i.e., states of the same $\ell$, but also states with $\kappa_1$ to those with $\kappa_2=1-\kappa_1$, i.e., states with
different $\ell$'s, $\Delta \ell=\pm2$.  To avoid this complication, this coupling is neglected.  In this case, the set
simplifies into coupled equations for each pair $\ell\mu$.  For $|\mu|=\ell+1/2$, there is no coupling, so similar to
the non-relativistic case, there is only regular solution with quantum numbers $\kappa\mu$, while for $|\mu|<\ell+1/2 $,
we need to solve the set of four coupled equations (\ref{radial_Dirac}) for the four unknown radial functions
$g_{\kappa_{1}\mu}$, $f_{\kappa_{1}\mu}$, $g_{\kappa_{2}\mu}$, and $f_{\kappa_{1}\mu}$.  The coefficients $u$, $u^{'}$,
and $u{''}$ in Eq.~(\ref{radial_Dirac}) result from matrix elements of the type
$\langle\kappa\mu|\sigma_z|\kappa^{'}\mu\rangle$~\cite{Turek97}.

Physically, the neglected coupling corresponds to a magnetic spin-orbit interaction given by a term $2 c^{-2} r^{-1}
dB/dr \vec{L}\cdot\vec{S}$ in the weak relativistic domain~\cite{feder83}.  Since this is proportional to the product of
two small quantities ($c^{-2}$ and $dB/dr$) its omission is justified in most cases.

The construction of the MT orbitals and the corresponding boundary conditions proceeds along the same principles as in
the non-relativistic case described in the previous sections. The tail cancellation condition is conveniently formulated
in terms of relativistic extensions of the potential and normalisation functions:
\begin{multline}
N_{\mathcal{L}}(\varepsilon)=(2\ell+1)\left(\frac{w}{s}\right)^{\ell+1}g_{\mathcal{L}}^{-1}(\varepsilon,s)\\
\times\left(D_{\mathcal{L}}(\varepsilon,s)-I \ell\right)^{-1}
\end{multline}
\begin{multline}
P_{\mathcal{L}}(\varepsilon)=2(2\ell+1)\left(\frac{w}{s}\right)^{2\ell+1}\\
\times\left(D_{\mathcal{L}}(\varepsilon,s)+I_\ell+I\right)\left(D_{\mathcal{L}}(\varepsilon,s)-I \ell\right)^{-1}
\label{frP}
\end{multline}
The logarithmic derivative matrix being given in terms of the small and large components of the radial amplitude:
\begin{gather}
D_{\mathcal{L}}(\varepsilon,r)=scf_{\mathcal{L}}(\varepsilon,s)g_{\mathcal{L}}^{-1}(\varepsilon,s)-\kappa - I\\
\nonumber
\kappa=
\begin{pmatrix}
\kappa_{1} & 0 \\
0          & \kappa_{2}
\end{pmatrix}
\end{gather}
For states with $|\mu|<\ell+1/2$, $N_{\mathcal{L}}(\varepsilon)$, $P_{\mathcal{L}}(\varepsilon)$,
$D_{\mathcal{L}}(\varepsilon)$, $g_{\mathcal{L}}(\varepsilon)$, and $f_{\mathcal{L}}(\varepsilon)$ are 2$\times$2
matrices for each subblock $\mathcal{L}=(\ell,\mu)$ with the general form
\begin{align}
A_{\mathcal{L}}(\varepsilon)=
\begin{pmatrix}
        A_{\alpha_{1}\kappa{1}}(\varepsilon)& A_{\alpha_{1}\kappa_{2}}(\varepsilon)\\
        A_{\alpha_{2}\kappa{1}}(\varepsilon)& A_{\alpha_{2}\kappa_{2}}(\varepsilon)
\end{pmatrix}
\label{2x2}
\end{align}
Note that the indices in this matrix have different physical meaning. Although the values of index $\alpha$ are
numerically equal to those of $\kappa$, index $\alpha$ stands for different behaviour of the radial function at the
origin~\cite{Ebert89,Turek97} while $\kappa$ stands for solutions with different quantum states.  Therefore, these
matrices are not symmetric and do not commute with each other.

The linearisation can be formulated in a matrix form too [in the following we will leave out the subblock index
$\mathcal{L}$; all presented matrices have the form of Eq.~(\ref{2x2})]. Within second-order approximation, the radial
amplitudes are expanded around a linearisation energy:
\begin{gather}
g(\varepsilon,r)\approx g_{\nu}(r)+\left(\varepsilon-\varepsilon_{\nu}\right) I \dot{g_{\nu}}(r)\\
f(\varepsilon,r)\approx f_{\nu}(r)+\left(\varepsilon-\varepsilon_{\nu}\right) I \dot{f_{\nu}}(r)
\end{gather}
Then, a symmetric matrix form of the linearisation of the logarithmic derivative can be written as
\begin{align}
\left(D(\varepsilon)-D_{\nu}\right)^{-1} = -\frac{s}{\epsilon}g_{\nu}g_{\nu}^{T}+A
\label{linear_frD}
\end{align}
where
\begin{align}
A&=-\frac{1}{2}\left[\left(D_{\nu}-D_{\dot{\nu}}\right)^{-1}+\left(D_{\nu}-D_{\dot{\nu}}^{T}\right)^{-1}\right]
        \nonumber\\
&= -\frac{1}{2}s\left(\dot{g}_{\nu}g_{\nu}^{T}+g_{\nu}\dot{g}_{\nu}^{T}\right)
\end{align}
By direct substitution of (\ref{linear_frD}) into (\ref{frP}) we obtain the parameterisation of the potential function:
\begin{align}
P(\varepsilon)=R\left(V-\varepsilon\right)^{-1}R^{T}+Q
\label{rel_P_param}
\end{align}
or equivalently
\begin{align}
P^{-1}(\varepsilon)=W\left(C-\varepsilon\right)^{-1}W^{T}+\gamma
\end{align}
where
\begin{gather}
V=\varepsilon_{\nu} + s g_{\nu}^{T} \left[A+\left(D_{\nu}-Il\right)^{-1}\right]^{-1}g_{\nu}\\
R=\sqrt{2s}\left(\frac{w}{s}\right)^{\ell+\frac{1}{2}}\left(2\ell+1\right)\left[A\left(D_{\nu}-Il\right)+I\right]^{-1}
        g_{\nu}
\end{gather}
\vspace*{-\belowdisplayskip}\vspace*{-\abovedisplayskip} 
\begin{multline}
Q=2\left(2\ell+1\right)\left(\frac{w}{s}\right)^{2\ell+1}\\
\times\left[I+\left(2\ell+I\right)\left[\left(D_{\nu}-Il\right)+A^{-1}\right]^{-1}\right]
\end{multline}
and
\begin{align}
\gamma=Q^{-1},\quad W=\gamma R,\quad C=V+R^{T}\gamma R
\end{align}
For an arbitrary representation $\alpha$:
\begin{align}
{P^{\alpha}}^{-1}(\varepsilon)=W\left(C-\varepsilon\right)^{-1}W^{T}+\gamma-\alpha
\end{align}
which is the relativistic analogue of Eq.~(\ref{eq:defipa}).  In general, the matrices $V$, $R$, $Q$, $C$, $\gamma$ and
$W$ are non-diagonal, some are symmetric and some are not, so they do not all commute with each other. This is related
to the different physical origin of the matrix indices discussed above.  The parameter $W$ is the relativistic analogue
of $\sqrt{\Delta}$ introduced earlier [see Eq.~(\ref{eq:defdel})], and Eq.~(\ref{rel_P_param}) is analogous to the
non-relativistic Eq.~(\ref{eq:parameterisedP}).  We should note that in the code, we have also included a third-order
parameterisation of the potential function similar to the non-relativistic case, Eq.~(\ref{eq:porder3}).

The physical Green function and Hamiltonian are written in terms of the potential function and scalar relativistic
structure constants matrices after a transformation of the former from $\kappa\mu$ to $\ell m m_{s}$ basis:
\begin{align}
G^{\alpha}=-\frac{1}{2}\ddot{P}^{\alpha}\dot{P}^{\alpha}+\frac{2}{w}{N^{\alpha}}^{T}g^{\alpha}N^{\alpha}
\end{align}
where
\begin{align}
g^{\alpha}=\left(P^{\alpha}-S^{\alpha}\right)^{-1}
\end{align}
these are the relativistic analogues of Eq~(\ref{eq:Gfromg}) and the scattering path operator.

Within the framework of the TB-LMTO and principal layer approach, the fully relativistic version of the green function
for layered geometry is constructed straightforwardly from the site diagonal fully relativistic potential
function~\cite{Turek97}.

\subsection{Coherent Potential Approximation}
\label{sec:cpa}

The coherent potential approximation (CPA) is a Green's function-based method used to describe the electronic structure
of disordered substitutional alloys.  Questaal's \texttt{lmgf} code implements the CPA in the Atomic Spheres
Approximation, following the formulation of Refs.~\cite{kudrnovsky.prb1990,Turek97}.  Any lattice site $i$ can be
occupied by any number of components $a$ with probabilities (concentrations) $c^a_i$, which must be supplied by the
user.  These components can have different atomic sphere radii, and each has its own charge density, atomic potential,
and diagonal matrix of potential functions $P^a_i(\epsilon)$.  Each site with substitutional disorder (``CPA site'') is
also assigned a coherent potential matrix $\EuScript{P}_{i}(\epsilon)$ that has the same orbital structure as
$P^a_i(\epsilon)$ but is off-diagonal, with the restriction that it must be invariant under its site's point group.  The
elements of the $\EuScript{P}_{i}(\epsilon)$ matrix are complex even if $\epsilon$ is real, and it is fixed by the CPA
self-consistency condition.

The configurational average of the scattering path operator is given by
\begin{align}
\bar g(\epsilon,\mathbf{k})=[\EuScript{P(\epsilon)}-S(\mathbf{k})]^{-1}
\label{barg}
\end{align}
where the site-diagonal matrix $\EuScript{P}(\epsilon)$ absorbs the coherent potential matrices for the CPA sites and
the diagonal matrices of the conventional potential functions for the sites that are occupied deterministically
(``non-CPA sites'').  The matrix in Eq.~(\ref{barg}) is integrated over the Brillouin zone in the usual way, and its
site-diagonal blocks $\bar g_{ii}$ are extracted.

For non-CPA sites, the full Green's function, density matrix, and the charge density are obtained from $\bar g_{ii}$ in
the usual way.  For each CPA site $i$, we define the scattering path operator $\bar g^a_i$ (separately for each
component $a$), which is the statistical average under the restriction that site $i$ is occupied deterministically by
component $a$, while all other sites in the infinite crystal are occupied statistically, according to their average
concentrations.  Such quantities are called ``conditionally averaged.''  The charge density for the component $a$ on
site $i$ is obtained from the site-diagonal block $\bar g^a_{ii}$ on site $i$ of the conditionally averaged $\bar
g^a_i$.  This site-diagonal block can be found from the matrix equation on that site:
\begin{align}
(\bar g^a_{ii})^{-1} = \bar g_{ii}^{-1}+P^a_i-\EuScript{P}_i
\end{align}
and the CPA self-consistency condition reads
\begin{align}
\bar g_{ii} = \sum_i c^a_i \bar g^a_{ii}
\label{CPAsc}
\end{align}

Iteration to self-consistency is facilitated~\cite{kudrnovsky.prb1990,Turek97} by introducing the \emph{coherent
interactor} matrix $\Omega_{i}$, for each CPA site, defined through $\bar g_{ii}=(\EuScript{P}_{i}-\Omega_{i})^{-1}$.
The conditionally averaged $\bar g^a_{ii}$ is then $\bar g^a_{ii}=(P^a_i-\Omega_i)^{-1}$.  The latter two equations can
be used to re-express the self-consistency condition (\ref{CPAsc}) in terms of $\EuScript{P}_i$, $\Omega_i$, and $P^a_i$
without an explicit reference to the scattering path operator:
\begin{align}
(\EuScript{P}_{i}-\Omega_{i})^{-1} = \sum_i c^a_i (P^a_i-\Omega_i)^{-1}
\label{CPAscO}
\end{align}

The $\Omega_i$ matrices (for each required complex energy point) are converged to self-consistency using the following
procedure.  At the beginning of the CPA iteration, Eq.~(\ref{CPAscO}) is used to obtain $\EuScript{P}_{i}$ for each CPA
site from $\Omega_i$, and then $\EuScript{P}_{i}$ is inserted in Eq.~(\ref{barg}).  After integration over $\mathbf{k}$,
the site-diagonal block $\bar g_{ii}$ is extracted for each CPA sites and used to obtain the next approximation for
$\Omega_i=\EuScript{P}_{i}-\bar g_{ii}^{-1}$, closing the self-consistency loop.  The output $\Omega_i$ matrices are
linearly mixed with their input values.  The mixing parameter can usually be set to 1 for energy points that are not too
close to the real axis.

The CPA loop is repeated until the $\Omega_i$ matrices converge to the desired tolerance, after which a charge iteration
is performed.  At the beginning of the calculation, the $\Omega_{i}$ matrices are initialised to zero unless they have
already been stored on disk.  The CPA loop can sometimes converge to an unphysical symmetry-breaking solution; this can
usually be avoided by symmetrising the coherent potentials using the full space group of the crystal.

\begin{figure}[ht]
\begin{center}
\includegraphics[width=0.49\columnwidth]{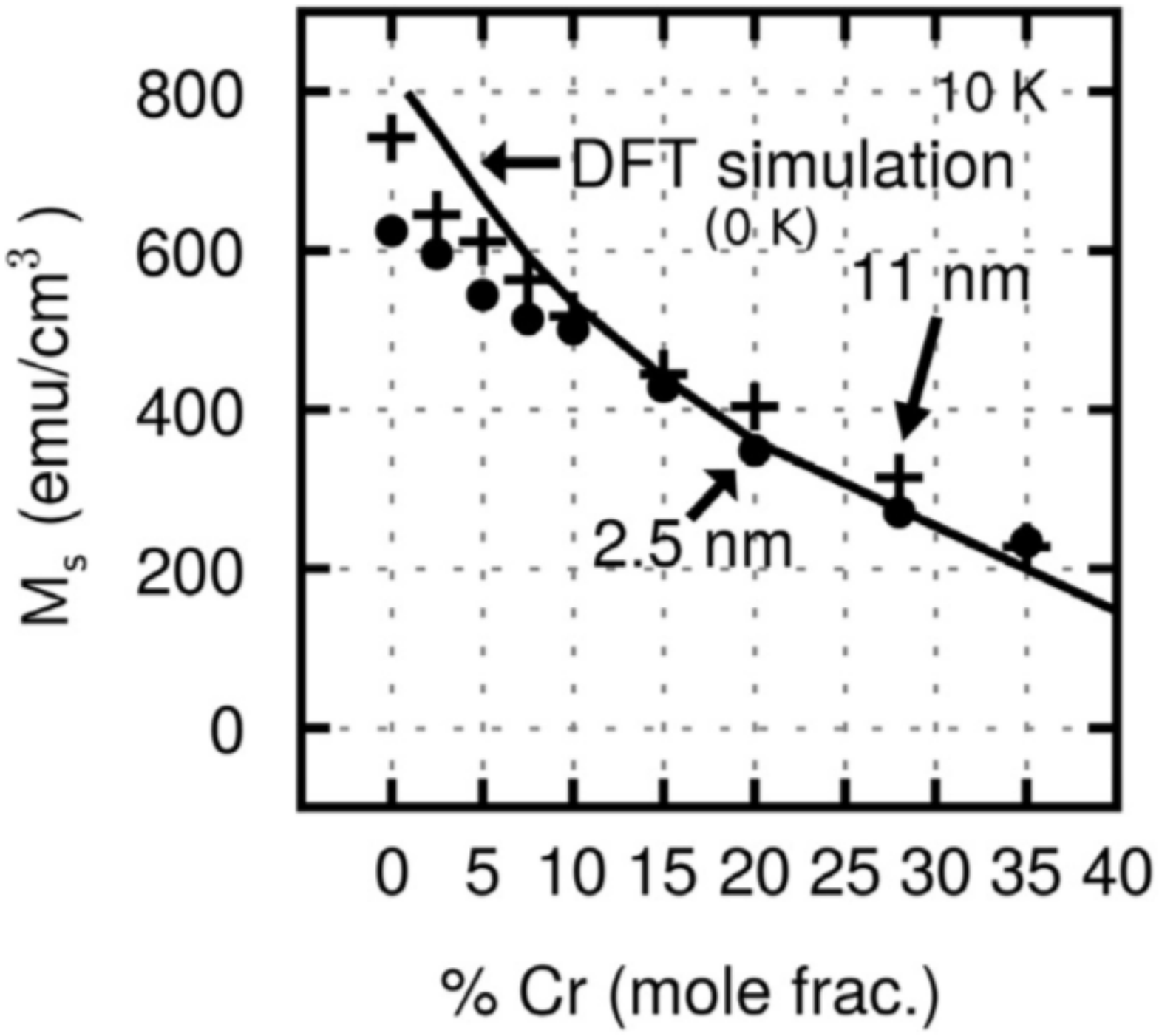}
\includegraphics[width=0.49\columnwidth]{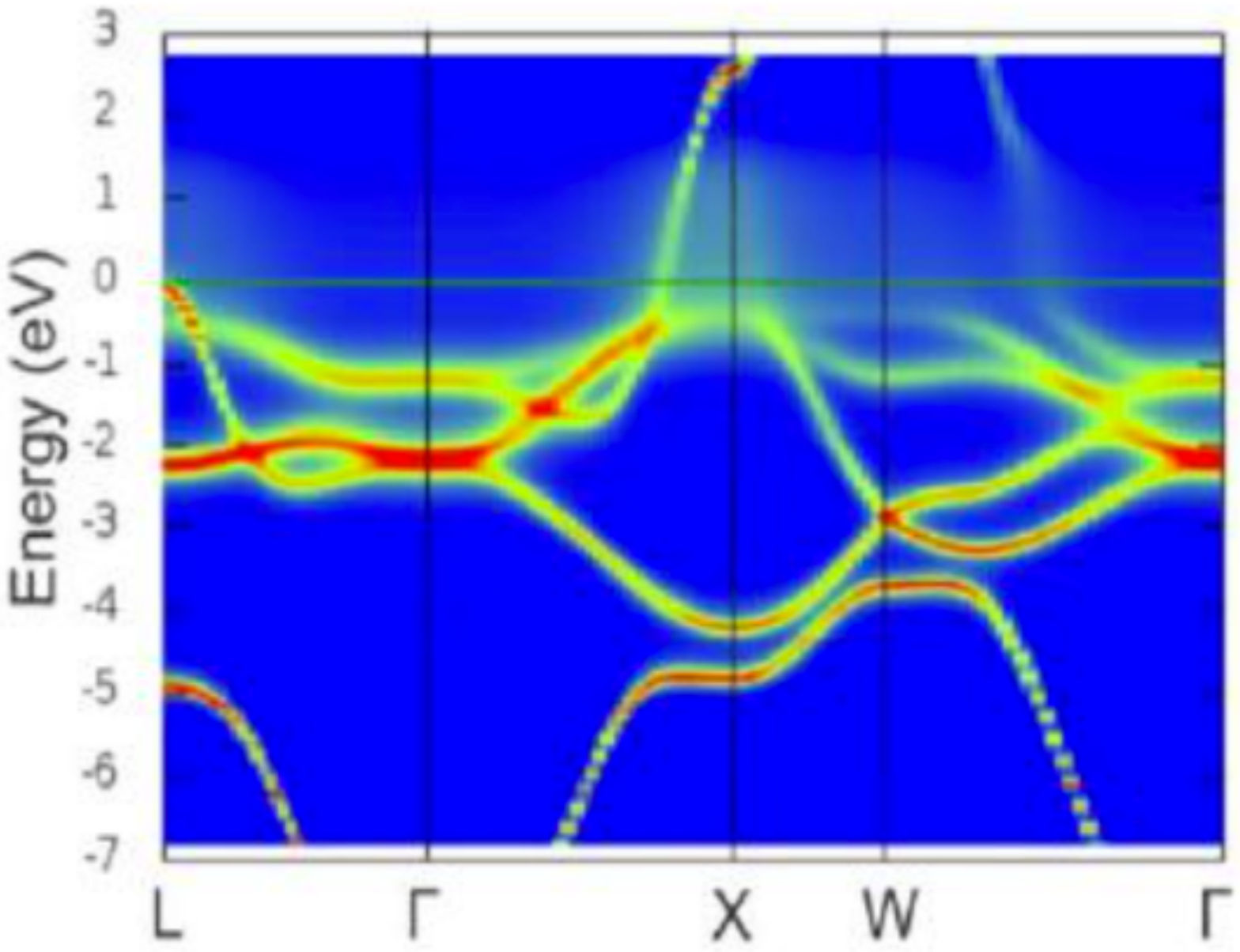}\\
\vspace*{-3 em}
\caption{Left: DFT magnetic moment in (Ni$_{80}$Fe$_{20}$)$_{1-x}$Cr$_{x}$, compared to experiment
(films with 11 nm (crosses) and 2.5 nm (dots) thickness) at 10K.  Right: Majority-spin CPA spectral function at
$x$=20\%.  Adapted from Ref.~\cite{Devonport18} and Vishina's PhD thesis~\citep{alena_thesis}.}
\label{fig:permalloy} 
\end{center}
\end{figure}

Fig.~\ref{fig:permalloy} illustrates one recent application of Questaal's implementation of the CPA.  A number of new,
potentially high-impact technologies, Josephson MRAM (JMRAM) in particular, performs write operations by rotating a
patterned magnetic bit.  These operations consume a significant portion of a system's power.  One way to reduce the
power consumed by write operations is to substitute materials with smaller saturation magnetisation $M_{s}$.  The
minimum usable $M_{s}$ is limited by the need to maintain large enough energy barrier to prevent data loss due to
thermal fluctuations.  Low $M_{s}$ can greatly reduce power particularly for JMRAM~\cite{Qader14}, which operates at
around 4~K.  One way to reduce $M_{s}$ in permalloy (Ni$_{80}$Fe$_{20}$ alloys, the most commonly used JMRAM material),
is to admix Cr into it, as Cr aligns antiferromagnetically to Fe and Ni.  Fig.~\ref{fig:permalloy} shows some results
adapted from a recent joint experimental and theoretical study of Py$_{1-x}$Cr$_{x}$,~\citep{Devonport18}.  The CPA
calculations of $M_{s}$ track measured values fairly well.  Also shown is the CPA energy band structure for the majority
spin.  Alloy scattering causes bands to broaden out, and scattering lifetimes can be extracted.  A detailed account can
be found in Vishina's PhD thesis~\cite{alena_thesis}.

\subsection{Noncollinear Magnetism}
\label{sec:noncollinear}

The ASA codes \texttt{lm}, \texttt{lmgf} and \texttt{lmpg} are fully noncollinear in a rigid-spin framework, meaning
that the spin quantisation axis is fixed within a sphere, but each sphere can have its own axis.  The formulation is a
straightforward generalisation of the nonmagnetic case.  Potential and Hamiltonian-like objects become $2{\times}2$
matrices in spin space.  The structure matrix $S$ is diagonal matrix in this space and independent of spin, while
potential functions $P$ and potential parameters of Sec.~\ref{sec:lineearizep}, become $2{\times}2$ matrices that are
diagonal in $\ell$ but off-diagonal in spin.  (Alternatively, a local spin quantisation axis can be defined which makes
$P$ diagonal in spin; then $S$ is no longer diagonal.)  How Questaal constructs the Hamiltonian in the general
noncollinear case, and also for spin spirals, is discussed in Ref.~\cite{Antropov96}.

\begin{figure}[ht]
\begin{center}
\includegraphics[width=1.00\columnwidth]{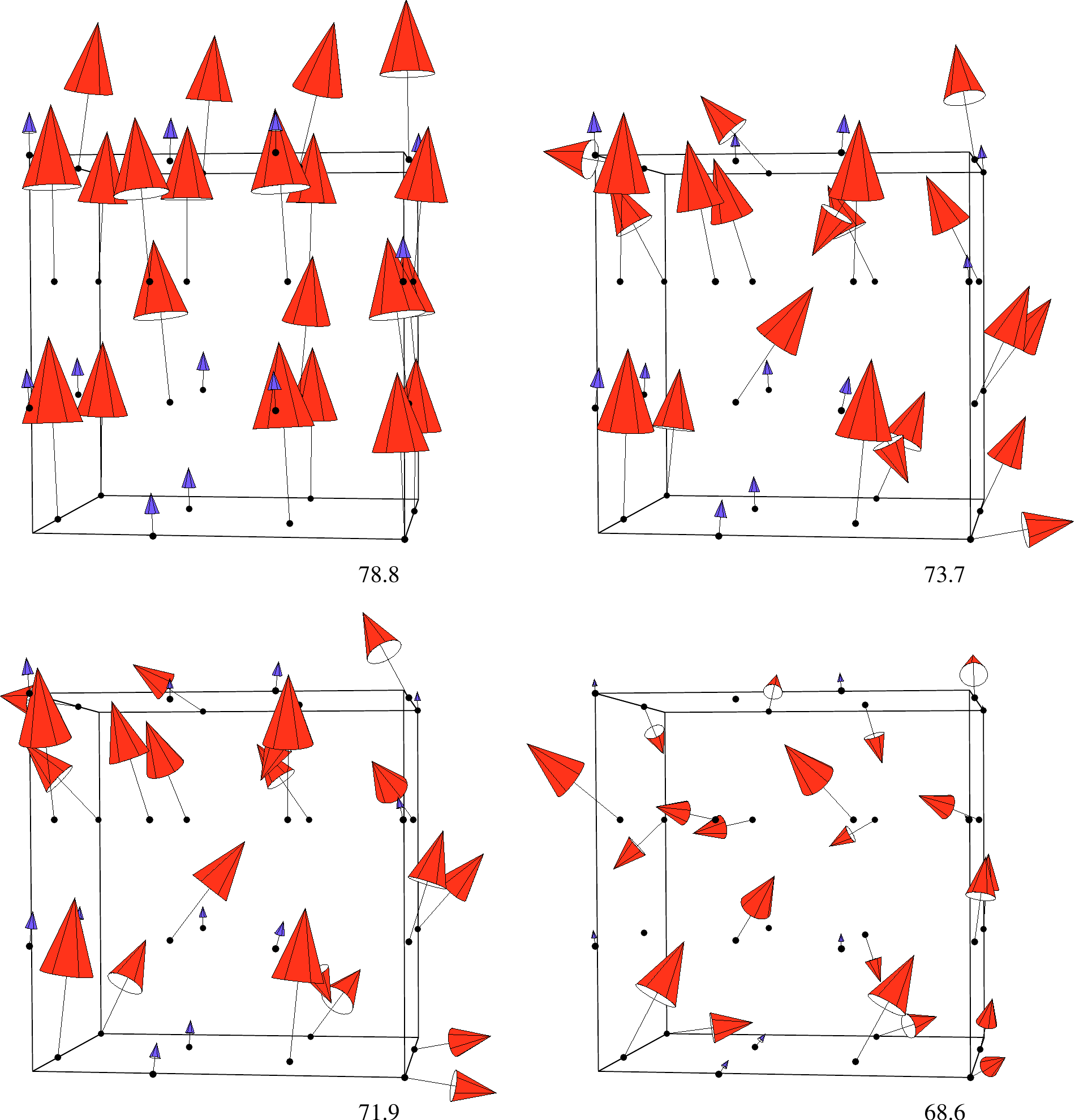}
\caption{Self-consistent magnetic spin configurations of the fcc Fe-Ni alloy at the four volumes 78.8, 73.7, 71.9, and
68.6 a.u..  Red and blue arrows show magnetic moments on Fe and Ni atoms, respectively. Taken from Ref.~\cite{INVAR99}.}
\label{fig:invar}
\end{center}
\end{figure}

These codes have implemented spin dynamics~\cite{Antropov95}, integrating the Landau-Lifshitz equation using a solver
from A. Bulgac and D. Kusnezov \cite{Bulgak90}, the spin analogue of molecular dynamics and molecular statics.  The
formulation is described in detail in Ref.~\cite{Antropov96}.  Codes also implement spin statics.  Torques needed for
both are obtained in DFT from the off-diagonal parts of the spin-density matrix.  A classic application is the study of
INVAR.  Fe-Ni alloys with a Ni concentration around 35 atomic \% (INVAR) exhibit anomalously low, almost zero, thermal
expansion over a considerable temperature range. In Ref.~\cite{INVAR99} it was shown that at 35\% composition, Fe-Ni
alloys are on the cusp of a collinear-noncollinear transition, and this adds a negative contribution to the
Gr{\"u}neisen parameter.  Fig.~\ref{fig:invar} is reproduced from that paper.

\section{Full Potential Implementation}
\label{sec:fp}

Questaal's primary code \texttt{lmf} is an augmented-wave implementation of DFT without shape approximations.  It also
handles the one-body part of the quasiparticle self-consistent \emph{GW} approximation, by adding the (quasiparticlised)
\emph{GW} self-energy to the DFT part.  Closely related codes are \texttt{lmfgwd}, a driver supplying input for the
\emph{GW}, \texttt{lmfdmft}, a driver supplying input for a DMFT solver, and \texttt{lmfgws}, a post-processing tool
that generates quantities using the one-body part from \texttt{lmf} and the dynamical self-energy from \emph{GW} or
DMFT.  This code uses different definitions for classical functions (see Appendix B), and we shift to those definitions
in what follows.

\begin{figure}[ht]
\begin{center}
\includegraphics[width=.75\columnwidth]{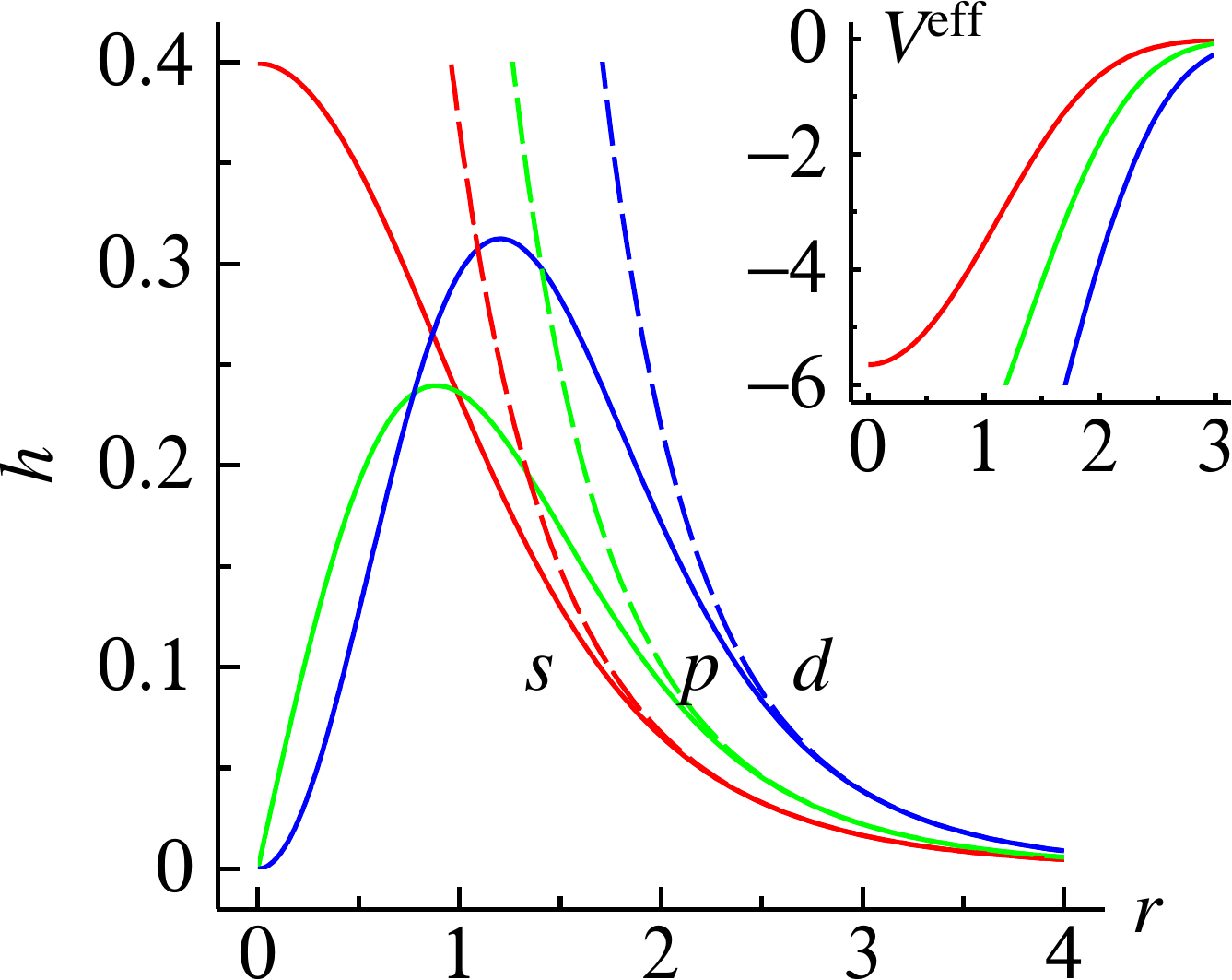}
\end{center}
\caption{Radial part of smooth Hankel functions $\mathcal{H}_L$ for $s$, $p$, $d$ orbitals (solid lines) and
corresponding ordinary Hankel functions $H_L$ (dashed lines).  Smoothing radius and energy were chosen to be $r_s=+1$
and $E=-1$, respectively.  For $r{\gg}r_s$, $\mathcal{H}_L\,{\to}\,H_L$, while for $r{\ll}r_s$,
$\mathcal{H}_L\,{\propto}\,r^\ell$ and ${H}_L\,{\propto}\,r^{-\ell-1}$.  The ${H}_L$ satisfy the Helmholtz wave equation
(Schr\"odinger equation for constant potential), while the $\mathcal{H}_L$ satisfy a Schr\"odinger equation
corresponding to a potential $V^\text{eff}=-4\pi G_L/\mathcal{H}_L$, as shown in the inset.}
\label{fig:smhankel}
\end{figure}

As for the DFT part, Questaal's unique features are:
\begin{itemize}
\item it uses a three-component augmentation, following most of the original method of Ref.~\cite{lmfchap}.  The
augmentation is reviewed in Sec.~\ref{sec:augmentation}.  Ref.~\cite{Kotani15} offers a slightly different presentation
and shows how it is connected with the PAW method~\cite{PAW}.
\item it has a more general basis set.  Ordinary Hankel functions solve Schr\"odinger's equation for a MT potential, but
real potentials vary smoothly into the interstitial (Fig.~\ref{fig:smhankel}).  The envelope function of a minimal basis
set must adapt to this potential.  The traditional Questaal basis uses smooth Hankel functions~\cite{Bott98}, which may
be thought of as a convolution of a Gaussian function and a traditional Hankel function; they are developed in
Sec.~\ref{sec:defhkl}.  This traditional basis works very well for most systems; however when the system is very open,
it is slightly incomplete (Sec.~\ref{sec:benchmarks}).  One way to surmount the incompleteness is to combine smooth
Hankel functions with plane waves; this is the ``Plane-wave Muffin Tin'' (PMT) basis~\cite{KotaniPMT10} (see also
Ref.~\cite{Kotani15}).  While it would seem appealing, PMT suffer from two serious drawbacks: first, it tends to become
over-complete even with a relatively small number of plane waves, and second, it is not \emph{compact} (minimal, short
ranged and as complete as possible for the relevant energy window).
\item Questaal's most recent development is the ``Jigsaw Puzzle Orbital'' (JPO) basis, which uses information from the
augmentation to construct an optimal shape for the envelope functions.  To the best of our knowledge, JPO's are the
closest practical realisation of compactness.  This is particularly important in many-body treatments where the efficacy
of a theory hinges critically on compactness.  This new basis will be presented more fully elsewhere.  In
Sec.~\ref{sec:tblmf} we show how a transformation to a tight-binding representation can be carried out in a
full-potential framework.  In the present version the basis set is merely a unitary transformation of the original one.
\end{itemize}

\subsection{Smooth Hankel Functions}
\label{sec:defhkl}

In his PhD dissertation Michael Methfessel introduced a class of functions $\mathcal{H}_{kL}(E,r_s,\br)$.  As limiting
cases they encompass both ordinary Hankel functions and Gaussian functions (see Eq.~(\ref{eq:defh0l}) and
(\ref{eq:defgl}) below).  They are explained in detail in Ref.~\cite{Bott98}; here we present enough information for
development of the Questaal basis set.  They are all connected to the smooth Hankel function for ${\ell{=}0}$.  Its
Fourier transform is
\begin{align}
\ftrns{h}_{0}(E,r_s;{q}) = -\frac{4\pi}{E-q^2}e^{r_s^2(E-q^2)/4}
\label{eq:ftrnsh0}
\end{align}
which is a product of Fourier transforms of ordinary Hankel function for ${\ell{=}0}$ with a Gaussian function of width
$2/r_{s}$.

We use standard definitions of Fourier transforms
\begin{align}
\ftrns{f}({\bq}) &= \int {{e^{ - i{\bq} \cdot {\br}}}} f({\br})\d3r\nonumber\\
{f}({\br}) &= \frac{1}{(2\pi)^3} \int {{e^{ + i{\bq} \cdot {\br}}}} \ftrns{f}({\bq})\,{d^3}q
\label{def:ft}
\end{align}
Thus in real space, ${h}_{0}(E,r_s;r)$ is a convolution of a ordinary Hankel function and a Gaussian function.  It has
an analytic representation
\begin{gather}
h_0(E,r_s;r) = \frac{1}{2r}\left(u_+ - u_{-}\right)\label{eq:defh0}\\
\dot{h}_0(E,r_s;r) = \frac{1}{4\akap}\left(u_+ + u_{-}\right)\label{eq:defh0dot}\\
u_{\pm} = e^{\mp\akap{}r}\left[1-{\rm{erf}}\left(\frac{r_s\akap}{2}\mp{}\frac{r}{r_s}\right)\right] \\
E = -\akap^2 {\text{  if  }} E<0
\label{eq:defkap}
\end{gather}
For small $r$, $h_0$ behaves as a Gaussian and it evolves smoothly into an ordinary Hankel when $r{\gg}r_s$.
$\dot{h}_0$ is the energy derivative of ${h}_{0}$.

Questaal's envelope functions are generalised Hankel functions ${\mathcal{H}}_{L}(E,r_s;{\br})$.  Their Fourier
representation has a closed form
\begin{align}
\ftrns{\mathcal{H}}_{L}(E,r_s;{\bq}) &=
        \mathcal{Y}_L(-i{\bq})\frac{-4\pi\;}{E-q^2}e^{r_s^2(E-q^2)/4} \nonumber\\
&= \mathcal{Y}_L(-i{\bq}) \ftrns{h}_{0}(E,r_s;{q})
\label{eq:defhlq}
\end{align}
$\mathcal{Y}_L({\br})$ is a spherical harmonic
polynomial (Appendix~\ref{app:sharm}).  Following the usual rules of Fourier transforms this function has a real-space
representation
\begin{align}
\mathcal{H}_{L}(E,r_s;{\br}) = \mathcal{Y}_L(-\nabla) h_0(E,r_s;r)
\label{eq:defhl}
\end{align}
$\mathcal{Y}_L({\br}),\ {\br}=(x,y,z)$ is a polynomial in $(x,y,z)$, so is meaningful to talk about
$\mathcal{Y}_L(-\nabla)$.
The extended family $\mathcal{H}_{kL}({\br})$ is defined through powers of the Laplacian operator:
\begin{align}
\ftrns{\mathcal{H}}_{kL}(E,r_s;{\bq}) &= (-iq)^{2k}\,{\ftrns{\mathcal{H}}_{L}(E,r_s;{\bq})}\nonumber\\
\mathcal{H}_{kL}(E,r_s;{\br}) &= \nabla^{2k} \mathcal{H}_{L}(E,r_s;{\br})
\label{eq:defhkl}
\end{align}
A recursion relation for $\mathcal{H}_{L}(E,r_s;{\br})$ for $\ell{>}0$ can be derived from properties of Fourier
transforms in spherical coordinates.  Any function whose Fourier transform factors as
\begin{align}
\hat{F}({\bq}) = {{\hat{f}(q)}}{Y_L}({\mathbf{\hat{q}}})
\label{eq:fq}
\end{align}
has a real-space form
\begin{align}
F({\br}) = {{f(r)}}{Y_L}({\mathbf{\hat{r}}})
\label{eq:fr}
\end{align}
where $f$ and $\hat{f}$ are related by~\cite{Bott98}
\begin{align*}
f(r) &= \frac{{4\pi {i^\ell }}}{{{{(2\pi )}^3}}}\int_0^\infty  {q\hat{f}(q)\,{j_\ell }(qr)}\, qr\,dq \\
\hat{f}(q) &= 4\pi {( - i)^\ell }\int_0^\infty  {rf(r)\,{j_\ell }(qr)}\, qr\,dr
\end{align*}
and $j_\ell$ is the spherical Bessel function.  We can therefore express $\mathcal{H}_{L}$ in a Slater-Koster form
\begin{align}
\mathcal{H}_{L}({\br}) &= h_{\ell}(r)\,{Y}_L({\hat{\br}}) \equiv \chi_{\ell}(r)\,\mathcal{Y}_L({\br}) \\
\chi_{\ell}(r) &= (1/r^\ell)\,h_{\ell}(r)
\label{eq:hlslaterkoster}
\end{align}

For envelope functions used in basis sets, e.g. $\mathcal{H}_{L}(E,r_s;{\br})$, the radial portion of the Fourier
transform does not depend on $\ell$; therefore corresponding real-space part depends on $\ell$ only through the
$j_\ell$.  This is a very useful fact.  It is possible to derive a recurrence relation to obtain $\chi_\ell$ for
$\ell{>}1$; this provides an efficient scheme for calculating them.  The recurrence is derived in Ref.~\cite{Bott98}
(see Eq.~6.21):
\begin{multline*}
\chi_{\ell+1} = \frac{{2\ell+1}}{{{r^2}}}{\chi_\ell} - \frac{E}{{{r^2}}}{\chi_{\ell-1}}\\
-\frac{{4\pi}}{r^2(\pi r_s^2)^{3/2}}\left(\frac{2}{r_s^2}\right)^{\ell-1}{e^{r_s^2E/4-{{(r/{r_s})}^2}}}
\end{multline*}
$\chi_{0}$ and $\chi_{-1}$ are needed to start the recursion.  $\chi_{0}$ is written in
Eq.~(\ref{eq:defh0}) and $\chi_{-1}$ can be obtained by combining Eqs.~(\ref{eq:defh0dot}),
(\ref{eq:rhl}) and (\ref{eq:hlslaterkoster}).

By taking limiting cases we can see the connection with familiar functions, and also the significance of parameters $E$
and $r_s$.
\setcounter{Alist}{0}
\begin{list}{({\rm\roman{Alist}})\,}{\leftmargin 18pt \itemindent 0pt \usecounter{Alist}\addtocounter{Alist}{0}}
\item $k{=}r_s{=}0$:
$\ftrns{H}_{0}(E,0;{\bq}){=}-{4\pi/(E-q^2)}\mathcal{Y}_{00}({\bq})$.  This is the Fourier transform of
$H_{0}(E,0;r)=\mathcal{Y}_{00}({\br})\exp(-\akap{}r)/r$, and is proportional to the $\ell{=}0$ spherical Hankel function
of the first kind, $h_\ell^{(1)}(z)$.  For general $L$ the relation defines Questaal's standard definition of ordinary
Hankel functions
\begin{align}
H_{L}(E,0;{\br})&=H_{0L}(E,0;{\br})\nonumber\\
&=-i^\ell\akap^{\ell+1}h_\ell^{(1)}(i\akap{}r)\text{Y}_L(\hat{\br})
\label{eq:defh0l}
\end{align}
\item $k{=}1$ and $E{=}0$:
$\ftrns{\mathcal{H}}_{10}(0,r_s;{\bq})=-{4\pi} e^{-r_s^2q^2/4} $.\\
This is the Fourier transform of a Gaussian function of width $r_s$.  For general $L$ we can define the family as
\begin{gather}
G_{L}(E,r_s;{\br}) = \mathcal{Y}_L(-\nabla) g(E,r_s;r)\\
g(E,r_s;r) = \left({\pi{}r_s^2}\right)^{-3/2} e^{E{}r_s^2/4}e^{-r^2/r_s^2}
\label{eq:defgl}
\end{gather}
\end{list}
Evidently $\ftrns{\mathcal{H}}_L({\bq})$ is proportional to the product of the Fourier transforms of a conventional
spherical Hankel function of the first kind, and a Gaussian.  By the convolution theorem, ${\mathcal{H}_L}({\br})$ is a
convolution of a Hankel function and a Gaussian.  For $r{\gg}r_s$, ${\mathcal{H}_L}({\br})$ behaves as a Hankel function
and asymptotically tends to $\mathcal{H}_L({\br})\to r^{-\ell-1}\exp(-\sqrt{-E}r)\text{Y}_L(\hat{\br})$.  For
$r{\ll}r_s$ it has structure of a Gaussian; it is therefore analytic and regular at the origin, varying as
$r^\ell\text{Y}_L(\hat{\br})$.  Thus, the $r^{-\ell-1}$ singularity of the Hankel function is smoothed out, with $r_s$
determining the radius for transition from Gaussian-like to Hankel-like behaviour.  Thus, the smoothing radius $r_s$
determines the smoothness of $\mathcal{H}_L$, and also the width of $G_{L}$.

By analogy with Eq.~(\ref{eq:defhkl}) we can extend the $G_{L}$ family with the Laplacian operator:
\begin{align}
G_{kL}(E,r_s;{\br}) &= \nabla^{2k}\, G_{L}(E,r_s;{\br})\label{eq:defgkl}\\
&= \mathcal{Y}_L(-\nabla) \nabla^{2k} g(E,r_s;r)\label{eq:defgklb}\\
&= \mathcal{Y}_L(-\nabla)\left(\frac{1}{r}\frac{\partial^2}{\partial r^2}r\cdot\right)^k g(E,r_s;r)\label{eq:defgklc}\\
\ftrns{G}_{kL}(E,r_s;{\bq}) &= \mathcal{Y}_L(-i{\bq})(-q^2)^k\,e^{r_s^2(E-q^2)/4}\label{eq:defgklq}
\end{align}
Eq.~(\ref{eq:defgklc}) shows that $G_{kL}$ has the structure (polynomial of order $k$ in $r^2$)$\times G_L$.
Specifically~\cite{Bott98}
\begin{align*}
{G_{kL}}(E,{\br}) &= \frac{{{2^{k + \ell }}(2k + 2\ell  + 1)!!}}{{r_s^{2k + \ell }(2\ell  + 1)!!}}
        {p_{k\ell }}(r_s;r){G_L}(E,{\br})
\end{align*}
where
\begin{align}
{p_{k\ell }}(r_s;r) = \frac{{{{( - 1)}^k}(2\ell  + 1)!!{2^k}k!}}{{r_s^\ell (2k + 2\ell  + 1)!!}}
        L_k^{(\ell  + 1/2)}({r^2/r_s^2})
\label{eq:defpkl}
\end{align}
and
\begin{align}
{G_{L}} = \left({\pi}r_s^2\right)^{-3/2} e^{E{}r_s^2/4} \left(2/r_s^{2}\right)^\ell e^{-r^2/r_s^2}
\end{align}
$L_k^{(\ell+1/2)}(u)$ are generalised Laguerre polynomials~\cite{Bott98}, which have the following orthogonality
relation
\begin{multline*}
\int {du} \,L_k^{(\ell  + 1/2)}(u)L_{k'}^{(\ell  + 1/2)}(u)\,{e^{ - u}}{u^{(\ell  + 1/2)}} =\\
\Gamma (k + \ell  + 3/2){\delta _{kk'}}
\end{multline*}
and as a consequence the ${G_{kL}}$ are orthogonal in the following sense:
\begin{multline*}
\int {{G_{kL}}} {G_{k'L'}}\,{e^{{a^2}{r^2}}}\d3r =\\
        {\frac{2^{3k+\ell}k!(2k+2\ell+1)!!}{{4\pi}{\left({\pi}r_s^2\right)^{3/2}}{r_s^{4k + 2\ell}}}}
        {\delta_{k{k^\prime }}}{\delta _{\ell {\ell ^\prime }}}
\end{multline*}
This can be also written as
\begin{align}
\int {{G_{kL}}} {P_{k'L'}}\d3r = \frac{2^{2k}k!(2\ell+1)!!}{{4\pi r_s^{2k-\ell}}}
        {\delta_{k{k^\prime }}}{\delta _{\ell {\ell ^\prime }}}\label{eq:biorthogonality}
\end{align}
where
\begin{align}
P_{kL}(r_s;\br) = p_{k\ell}(r_s;r) \mathcal{Y}_L({\br})\label{eq:defcapPkL}
\end{align}
We will need this relation for one-centre expansion of the $\mathcal{H}_{L}$ around remote sites
(Sec.~\ref{sec:augmentation}) since the simple expansion theorem Eq.~(\ref{eq:defs0}) does not apply to them.

Comparing the last form Eq.~(\ref{eq:defgklq}) to Eq.~(\ref{eq:defhkl}) and the definition of $\mathcal{H}_{kL}$
Eq.~(\ref{eq:defhkl}), we obtain the useful relations
\begin{align}
\mathcal{H}_{k+1,L}(E,r_s;{\br})+ & E \mathcal{H}_{kL}(E,r_s;{\br})\nonumber\\
&=\left(\nabla^2+E\right)\mathcal{H}_{kL}(E,r_s;{\br})\nonumber\\
&=-4\pi G_{kL}(E,r_s;{\br})
\label{eq:diffhkl}
\end{align}
This shows that $\mathcal{H}_{kL}$ is the solution to the Helmholtz operator $-\nabla^2+E$ in response to a source term
smeared out in the form of a Gaussian.  A conventional Hankel function is the potential from a point multipole at the
origin (see Eq.~6.14 in Ref.~\cite{Bott98}): smearing out the singularity makes $\mathcal{H}_{L}$ regular at the origin,
varying as $r^\ell$ for small $r$ instead of $r^{-\ell-1}$.  $\mathcal{H}_{kL}$ is also the solution to the
Schr\"odinger equation for a potential that has an approximately Gaussian dependence on $r$ (Ref.~\cite{Bott98},
Eq.~6.30).

\subsection{Gradients of Smooth Hankel Functions}
\label{sec:gradh}

Gradients of the $\mathcal{H}_{L}$ are needed in several contexts, e.g. for forces and for matrix elements of the
momentum operator, which enters into the dielectric function and velocity operator.  Also, the energy derivative is
needed in several contexts, e.g. for integration of some matrix elements (Sec.~\ref{sec:appendixintegrals}).

The form of Eq.~(\ref{eq:defhlq}) suggests that gradient and position operators acting on $\ftrns{\mathcal{H}}$ return
functions of the same family.  This turns out to be the case, and it makes possible matrix elements of these operators
with two such functions.  Consider the energy derivative first. $\ftrns{\mathcal{H}}_{L}(E,r_s;{\bq})$ is readily
differentiated
\begin{align}
\ftrns{\dot{\mathcal{H}}}_{L}(E,r_s;{\bq}) =
\ftrns{{\mathcal{H}}}_{L}(E,r_s;{\bq})
\left(\frac{1}{E-q^2} - {r_s^2/4} \right)\quad
\label{eq:defhld}
\end{align}
In real space, the energy derivative is most easily derived from an integral representation of $h_\ell$
(Eq.~A5 in Ref.~\cite{Bott98})
\begin{align}
{h_\ell(r)} = \frac{2^{\ell+1}r^\ell}{{\sqrt\pi}}\int_0^{1/{r_s}}{{\xi^{2\ell}}{e^{E/(4{\xi^2})}}
        {e^{-{r^2}{\xi^2}}}} d\xi
\label{eq:integralrep}
\end{align}
It is readily seen that
\begin{gather}
r{h_\ell}(r) = 2{{\dot h}_{\ell+1}}(r)\nonumber\\
(\frac{d}{dr}-\frac{\ell}{r})\,h_\ell(r) = {h_{\ell+1}}(r)
\label{eq:rhl}
\end{gather}
Noting that the three components of the real harmonic polynomials $\mathcal{Y}_{1m}(\br)$, Eq.~(\ref{eq:defylpoly}) for
$\ell$=1, are ($y$,$z$,$x$), the gradient and position operators can be obtained from
$-\sqrt{4\pi/3}\,\mathcal{Y}_{1p}(-\nabla)$ and
$\sqrt{4\pi/3}\,\mathcal{Y}_{1p}(\br)$ respectively,
with the appropriate permutation of $p$.  Using Eq.~(\ref{eq:defhl}),
\begin{align}
\mathcal{Y}_{1p}(\nabla) \mathcal{H}_{L} = -\mathcal{Y}_{1p}(-\nabla) \mathcal{Y}_L(-\nabla) h_0(E,r_s;r)
\label{eq:defhgrad}
\end{align}
The operator product $\mathcal{Y}_L({-\nabla}) \mathcal{Y}_K({-\nabla})$ can be expanded in the same manner as the
product of polynomials $\mathcal{Y}_L({\br}) \mathcal{Y}_K({\br})$ (Appendix~\ref{app:sharm}).

More generally, for a function of the form Eq.~(\ref{eq:fr}), it is possible to show that
\begin{multline}
\nabla_p f_\ell(r) \mathrm{Y}_L = f_\ell^{(+)} \sum_{i = 1,2} {\overline{C}^{(+)}_{i;\ell,m_\ell;p}}
        \mathrm{Y}_{\ell+1m_i'}\\
+ f_\ell^{(-)} \sum_{i=1,2} {\overline{C}^{(-)}_{i;\ell,m_\ell;p}} \mathrm{Y}_{\ell-1m_i'}
\label{eq:skgrad}
\end{multline}
where
\begin{align}
f_\ell^{(+)} &= [df_\ell/dr - \frac{\ell}{r}f_\ell]\nonumber\\
f_\ell^{(-)} &= [df_\ell/dr + \frac{\ell+1}{r}f_\ell]
\label{eq:flplusminus}
\end{align}
${\overline{C}^{(\pm)}_{i;\ell,m_{\ell};p}}$ is a sparse representation of a linear transformation that maps
$\mathrm{Y}_{\ell,m}(\br)$ to a linear combination of two $\mathrm{Y}_{\ell+1,m'}(\br)$ and two
$\mathrm{Y}_{\ell-1,m'}(\br)$.  It is more compact to write each sum in a standard matrix form, as $\sum_K
{{C}^{(\pm)}_{K{L};p}}\mathrm{Y}_K$.  ${{C}^{(\pm)}_{K{L};p}}$ is a family of three rectangular matrices ($p$=1,2,3),
which for a particular $L$, has two nonzero elements corresponding to $m_{i'}$  (Table~\ref{tab:coffmprime}).  An
alternative definition of ${{C}^{(\pm)}_{K{L};p}}$ is given in Appendix~\ref{app:sharm}.
\begin{table}[ht]
\begin{center}
\begin{tabular}{|c|c|c|c|}
\hline
$i$ &  $p{=}x$   &    $p{=}y$    &  $p{=}z$\\
\hline
1   &  $m_\ell{-}1$ &   $-m_\ell{-}1$  &    $m_\ell$ \\
2   &  $m_\ell{+}1$ &   $-m_\ell{+}1$  &    $-$ \\
\hline
\end{tabular}
\label{tab:coffmprime}
\end{center}
\vspace*{-1.0\baselineskip}
\caption{Index $m'_i$ corresponding to coefficient $\overline{C}^{(\pm)}_{i{=}1..2;\ell,m_\ell;p}$ in
Eq.~(\ref{eq:skgrad}).  Index $K$ in ${{C}^{(\pm)}_{K{L};p}}$ corresponding to $m'_i$ is
$K=(k{\pm}1)^2+(k{\pm}1)+1+m'_i$.}
\end{table}

The position operator $\br_p$ applied to $f_\ell(r) \mathrm{Y}_K$ results in the same form Eq.~(\ref{eq:skgrad}) but
with radial functions $f^{(\pm)}{\to}rf(r)$.  Note the close similarity with the gradient operator and in particular
${rf_\ell = {r^2}(f_\ell^{(+)} -f_\ell^{(-)})/(2\ell+1)}.$  This operator appears for optical matrix elements of
Hamiltonians with nonlocal potentials.  It is more convenient to evaluate it in reciprocal space
(Appendix~\ref{sec:appendixintegrals}).

Functions of type Eq.~(\ref{eq:fr}) have the Slater-Koster form with another special property, namely that the Fourier
transform $\hat{f}(q)$, Eq.~(\ref{eq:fq}), is independent of $\ell$, from which it follows that $f_\ell(r)$ satisfies
the following differential equations (Ref.~\cite{Bott98}, Eq.~4.7 and 4.20)
\begin{align*}
-{f_{\ell+1}}(r) &= \frac{{\partial {f_{\ell}}(r)}}{{\partial r}} - \frac{\ell}{r}{{f_{\ell}}(r)}\\
-{\nabla^2f_{\ell-1}}(r) &= \frac{{\partial {f_\ell}(r)}}{{\partial r}} + \frac{(\ell+1)}{r}{{f_\ell}(r)}
\end{align*}
Thus $f^{(+)}(r){=}-f_{\ell+1}(r)$ and $f^{(-)}(r){=}-\nabla^2 f_{\ell-1}(r)$.

The $\mathcal{H}_{kL}(\br)$ family is of this type, and moreover, $\nabla^2 \mathcal{H}_{kL}(\br)$ merely maps
$\mathcal{H}_{kL}$ to another member of the family (Eq.~(\ref{eq:defhkl})).  Another useful instance is the real
harmonic polynomials $\mathcal{Y}_L$ (Eq.~(\ref{eq:defylpoly})).  Equations in this section, e.g.  Eq.~(\ref{eq:rhl}),
yield explicit forms for $f^{(+)}$ and $f^{(-)}$ for a number of functions of interest here, summarised in the table
below.
\begin{table}[ht]
\setlength{\tabcolsep}{1.8pt} 
\begin{tabular}{|c|c|c|c|c|}
\hline
$F(\br)$                    & $f(r)$        & $f^{(+)}$                 & $f^{(-)}$               & $rf(r)$\\\hline
$\mathcal{Y}_L$             & $r^\ell$      & 0                         & $(2\ell{+}1)r^{\ell-1}$ & $r^{\ell+1}$\\
$r^{-2\ell-1}\mathcal{Y}_L$ & $r^{-\ell-1}$ & $(-2\ell{-}1)r^{-\ell-2}$ & 0                       & $r^{-\ell}$\\
$\mathcal{H}_{kL}$          & $h_{kl}$      & $-h_{k,\ell+1}$           & $-h_{k+1,\ell-1}$       &
        $2{{\dot h}_{\ell+1}}$\\
$H_L(E{=}0)$                & \multicolumn{4}{c|}{Same as $\mathcal{Y}_L/r^{2\ell+1}$}\\
$J_L(E{=}0)$                & \multicolumn{4}{c|}{Same as $\mathcal{Y}_L$}\\
\hline
\end{tabular}
\label{eq:posgradop}
\caption{Mapping gradient and position operators of smoothed Hankel functions to functions in the same family.}
\end{table}

In general, e.g. for partial waves $\phi_\ell$ inside augmentation spheres, $f^{(+)}$ and $f^{(-)}$ must be determined
by numerical differentiation.  For the pair of functions $\{r^\ell,\,r^{-\ell-1}\}$, $f_\ell^{(+)}$ ($f_\ell^{(-)}$) for
the first (second) function vanishes.  Any function can thus be matched at some $r$ to this pair
(Eq.~(\ref{eq:defwronskian})), which determines the projection of its gradient onto $Y_{\ell{\mp}1m'}$.

\subsection{Two-centre Integrals of Smoothed Hankels}
\label{sec:inthkl}

One extremely useful property of the $\mathcal{H}_{kL}$ is that the product of two of them, centred at different sites
${\bR}_1$ and ${\bR}_2$, can be integrated in closed form.  The result is a sum of other $\mathcal{H}_{kL}$, evaluated
at the connecting vector ${\bR}_1-{\bR}_2$.  This follows from Parseval's (Plancharel's) identity
\begin{multline}
\int \mathcal{H}^*_1({\br}-{\bR}_1) \mathcal{H}_2({\br}-{\bR}_2)\d3r =\\
(2\pi)^{-3}\int \ftrns{\mathcal{H}}^*_1({\bq}) \ftrns{\mathcal{H}}_2({\bq}) e^{i{\bq}\cdot({\bR}_1-{\bR}_2)}\,d^3q
\label{eq:powertheorem}
\end{multline}
and the fact that $\ftrns{\mathcal{H}}^*_{k_1L_1}({\bq})\ftrns{\mathcal{H}}_{k_2L_2}({\bq})$ can be expressed as a
linear combination of other $\ftrns{\mathcal{H}}_{kL}({\bq})$, or their energy derivatives.  Derivations of these and
related integrals are taken up in Appendix~\ref{sec:appendixintegrals}.

\subsection{Smoothed Hankels of Positive Energy}
\label{sec:smooth-hank-posit}

The smooth Hankel functions defined in (Ref.~\cite{Bott98}) for negative energy also apply for positive energy.  We
demonstrate that here, and show that the difference between the conventional and smooth Hankel functions are real
functions.

(Ref.~\cite{Bott98}) defines $\akap$ in contradistinction to usual convention for $\kappa$
\begin{align*}
\akap^2 = -\varepsilon \hbox{\quad with }\akap>0
\end{align*}
and restricts $\varepsilon < 0$.  According to usual conventions $\kappa$ is defined as
\begin{align*}
\kappa = \sqrt\varepsilon \hbox{, \quad Im}(\kappa) >= 0
\end{align*}
We can define for any energy $\akap = -i \kappa$ and therefore
\begin{align*}
\begin{cases}
        \akap  \text{ real and positive, } \varepsilon<0\\
        \kappa \text{ real and positive, } \varepsilon>0
\end{cases}
\end{align*}
Then
\begin{align*}
u_{\pm}(\varepsilon,r_s;r) &= e^{\mp\akap{}r}    {\rm{erfc}}\left({\akap{r_s}}/{2}\mp{}{r}/{r_s}\right)\\
&= e^{\pm{}i\kappa{}r}{\rm{erfc}}\left({-{i}\kappa{}r_s}/{2}\mp{}{r}/{r_s}\right)
\end{align*}
The smoothed Hankels for $\ell=0,-1$ are real
\begin{align*}
h^s_0 (r) &= (u_+ - u_-) / 2r \\
h^s_{-1}(r) &= (u_+ + u_-) / 2\akap
\end{align*}
as defined in Ref.~\cite{Bott98}.

To extend the definition to any energy we define $U_{\pm}$ as:
\begin{align*}
U_{\pm} = e^{\pm{}i\kappa{}r}\, {\rm{erfc}}\left({r}/{r_s}\pm{{i}\kappa{r_s}}/{2}\right)
\end{align*}
The following relations are useful:
\begin{align*}
\hbox{erfc}(-x^*) &= 2-\hbox{erfc}^*(x)\\
\hbox{erfc}( x^*) &=   \hbox{erfc}^*(x)
\end{align*}

Then for $\varepsilon<0$, $i\kappa$ is real and
\begin{align*}
U_{+} &= 2e^{i\kappa{}r} - u_{+}\\
U_{-} &= u_{-}
\end{align*}
are also real.  The difference in unsmoothed and smoothed Hankels is for $\ell=0,-1$
\begin{align*}
h_0 - h^s_0 &= [e^{i \kappa r} - u_+/2 + u_-/2] /r\\
&= [U_+/2 + U_-/2] /r\\
h_{-1} - h^s_{-1} &= [e^{i \kappa r} - u_+/2 - u_-/2] /\akap\\
&= [U_+/2 - U_-/2] /(-i \kappa)
\end{align*}

For $\varepsilon>0$, $\kappa$ is real and $U_+$ = $U_-^*$.  Then the difference in unsmoothed and smoothed Hankels for
$\ell=0,-1$:
\begin{align*}
h_0  - h^s_0      &= [U_+/2 + U_-/2] /r = \hbox{Re}(U_+) /r\\
h_{-1} - h^s_{-1} &= [U_+/2 - U_-/2] / (-i \kappa) = -\hbox{Im} (U_+) / \kappa
\end{align*}
are real, although $h_0$ and $h_{-1}$ are complex.

\subsection{One-Centre Expansion}
\label{sec:smooth-hank-onec}

Ordinary Hankel functions have a one-centre expansion in terms of Bessel functions, Eq.~(\ref{eq:defs0}): the shape of
the radial function does not change as the vector $\bR$ connecting source and field point changes, only the expansion
coefficients $S$.  This is also true for a plane wave.  The $\mathcal{H}$ are more complicated: the $r$-dependence
depend on $\bR$, tending to Bessel functions only when $|\bR|{\gg}r_s$.  This complicates the one-centre expansion of
$\mathcal{H}$.  They can, however, be expanded in the polynomials $P_{kL}$ (Eq.~(\ref{eq:defcapPkL})) using the
bi-orthogonality relation Eq.~(\ref{eq:biorthogonality}).  For any smooth function \emph{F}, its one-centre expansion
can be written
\begin{align}
F(\br) = \sum_{kL} {C_{kL} p_{k\ell}(r) \mathcal{Y}_L(\br)}
\label{eq:expandF}
\end{align}
where
\begin{align*}
C_{kL} &= \frac{{4\pi r_s^{2k + \ell }{2^{2k}}}}{{{2^{2k}}k!(2\ell  + 1)!}}\int {\d3r\,{G_{kL}}({r_s};{\br})} F({\br})
\end{align*}
Expressions for integrals $G_{kL}$ with the $\mathcal{H}$ are written in Sec.~\ref{sec:inthkl}.

Eq.~(\ref{eq:expandF}) expands $\mathcal{H}_L$, which generalises Eq.~(\ref{eq:defs0}) which expands ${H}_L$ (plane
waves can similarly be expanded in Bessel functions).  Eq.~(\ref{eq:expandF}) is more cumbersome, and it introduces
another cutoff $k_\text{max}$.  These are drawbacks of this method, but it also brings with it a significant advantage:
envelope functions of very different types can be constructed.  This adds a considerable amount of flexibility to the
method, as smooth Hankels $\mathcal{H}_L$ and plane waves can be combined in a uniform manner; this is the PMT
basis~\cite{KotaniPMT10}.  Perhaps more important, basis functions can be tailored to the potential.  This enables them
to be very accurate whilst staying minimal.  This is the JPO basis (Sec.~\ref{sec:tblmf}).

If $r_{s}$ becomes small $G_{kL}$ becomes sharply peaked and the polynomial expansion becomes a Taylor expansion about
the origin.  But by allowing $r_{s}$ to be a significant fraction of the MT radius $\Rmt$, the error gets distributed
over the entire sphere and $k_\text{max}$ can be set relatively low.  Note that $r_{s}$ used for this expansion need not
be (and in practice is not) the same as $r_{s}$ used to make the envelope functions.

\subsection{Three-component Augmentation}
\label{sec:augmentation}

As noted in the introduction, the pseudopotential method shares many features in common with augmentation.  Bl\"ochl's
PAW method~\cite{PAW} makes the connection explicit.  Questaal starts from an envelope function $F^{(0)}_{RLi}(\br)$
that extends everywhere in space, augmented as follows:
\begin{align}
F_{RLi}(\br) = F^{(0)}_{RLi}(\br) + F^{(1)}_{RLi}(\br) - F^{(2)}_{RLi}(\br)
\label{eq:augm}
\end{align}
Index ${RLi}$ labels the envelope function: $R$ and $L$ mark the channel where functions are centred, and the angular
momentum, while $i$ marks the kind of envelope function at $R$.  General envelope functions (e.g. plane waves) need not
be atom-centred, for such functions we adopt the labelling convention $R{=}L{=}0$.  Unlike the ASA, there is typically
more than radial envelope function per $R$ and $L$.  (We will sometimes label $F$ with a single Roman index, e.g. using
$i$ to implicitly refer to the entire label $RLi$.), $F^{(1)}$ is some linear combination of partial waves matching
value and slope of $F^{(0)}(\br)$, and $F^{(2)}(\br)$ is the one-centre expansion of $F^{(0)}(\br)$. $F^{(0)}$, and
$F^{(2)}$ are truncated at some $\ell_\text{max}$.

Conventional augmented methods construct matrix elements of a local or semilocal operator $X$ straightforwardly as
\begin{align*}
\langle F_i^{(0)} + F_i^{(1)} - F_i^{(2)} | X | F_j^{(0)} + F_j^{(1)} - F_j^{(2)}\rangle
\end{align*}
Evaluating the cross terms is unwieldy and for that reason conventional augmentation methods require $F_i$ be expanded
to very high $L$.  However, it has been observed independently by several authors, the first being Soler and
Williams~\cite{Soler89} that the integrand can be approximated in the following three-component form:
\begin{align}
F_i^{(0)} X F_j^{(0)} + F_i^{(1)} X F_j^{(1)} - F_i^{(2)} X F_j^{(2)}
\label{eq:component3}
\end{align}
Note the minus sign in the last term.  The missing terms may be written as
\begin{align*}
(F_i^{(0)}-F_i^{(2)}) X (F_j^{(1)} - F_j^{(2)})\\ + (F_i^{(1)}-F_i^{(2)}) X (F_j^{(0)} - F_j^{(2)})
\end{align*}
Write the $L$ projection of $F$ as $\mathcal{P}_L F$.  By construction $\mathcal{P}_L F^{(0)} = \mathcal{P}_L F^{(2)}$
so if the augmentation is carried out to $L{\to}\infty$, $F^{(0)}$ and $F^{(2)}$ are identical and the missing terms
vanishes identically.  This form makes it clear that not only does Eq.~(\ref{eq:component3}) converge to the exact
result in the limit $\ell_\text{max}{\to}\infty$, but also that the error converges much more rapidly in
$\ell_\text{max}$ than does conventional augmentation.

\begin{figure}[ht]
\includegraphics[width=\columnwidth]{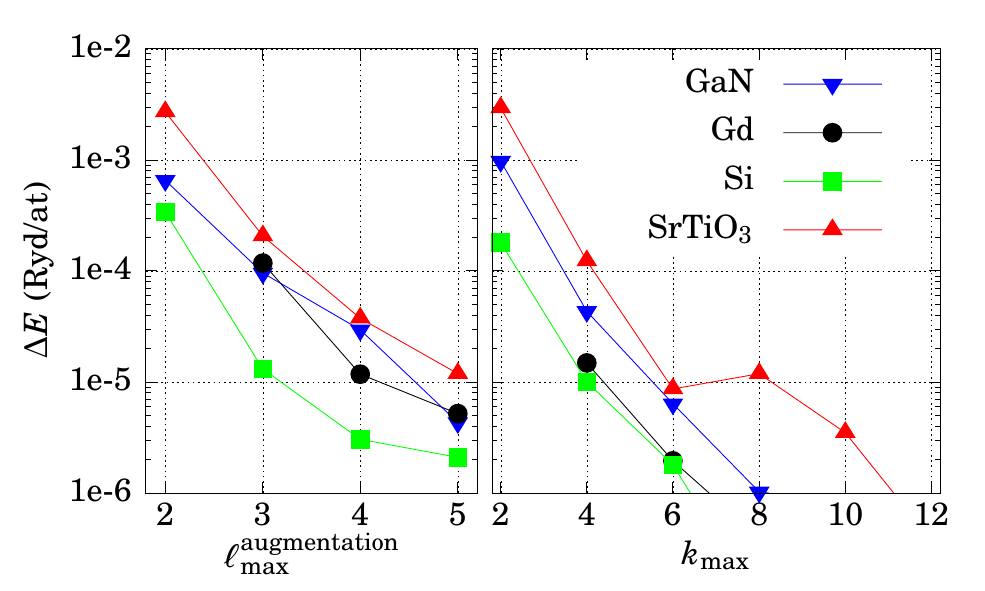}
\caption{Convergence of the total energy with respect to augmentation $\ell$-cutoff and polynomial degree,
Eq.~(\ref{eq:Faug}).}
\label{fig:lmf_lmxa}
\end{figure}

This occurs for two reasons that work synergistically.  The operator $X$ is in practice nearly spherical, coupling parts
of $F$ with unequal $L$ only weakly.  Consider the one-centre expansion of a particular $L$ of one of these components,
and consider for simplicity the unit overlap operator.  Near the augmentation boundary all components have the same
value and slope by construction, and the ($\ell$-projected) $F^{(0)}{-}F^{(2)}$ or $F^{(1)}{-}F^{(2)}$ vanishes to
second order in $r{-}s$.  Second, the ($\ell$-projected) $F$ varies as $r^{\ell}$ for large $\ell$ for all three
components of $F$, because the angular momentum becomes the dominant contribution to the differential operator.  Thus
for large $\ell$ the product of two projected functions varies as $r^{2\ell}$ for all components, which is heavily
weighted to the outer parts of the sphere, in the region where the projections of differences $F^{(0)}-F^{(2)}$ and
$F^{(1)}-F^{(2)}$ are small.  For such $L$ projections of $F_i^{(0)} X F_j^{(0)}$ are then a good approximation to the
(exact) $F_i^{(1)} X F_j^{(1)}$.  By truncating both local components, only the (0) component remains, which as we have
seen accurately represents the (exact) basis function, especially near $\Rmt$ where it is dominant.  In other words, the
projections of the envelope functions contain projections from $\ell_\text{max}{+}1$ to $\infty$, and do so with high
accuracy starting at a relatively low $\ell_\text{max}+1$; indeed convergence is very rapid as shown in
Fig.~\ref{fig:lmf_lmxa} (see also Fig.~3 of Ref.~\cite{lmfchap}).
This explains why the pseudopotential and PAW approximations can be truncated at much lower $\ell$ than the conventional
augmented wave methods.  Indeed, the present approach has much in common with PAW, but does not make approximations or
involve pseudo-partial waves or projector functions.

\subsection{Secular Matrix}

To construct the secular matrix, we need matrix elements of Hamiltonian and overlap:
\begin{gather*}
H_{ij}=\int \chi_i^*(\br) [-\nabla^2+V(\br)]\chi_j(\br)\d3r\\
S_{ij}=\int \chi_i^*(\br) \chi_j(\br)\d3r
\end{gather*}
At this stage we need not specify what $\chi_i$ is, except to note that it has an envelope part $\chi_i^{(0)}$ that
extends everywhere in space, augmented in spheres taking the form Eq.~(\ref{eq:augm}).  Specifically
\begin{align}
{\chi}_i(\br) = \chi^{0}_i(\br) + \sum_{kL} \CikL \lbr \tPkL(\br) - \PkL(\br) \rbr
\label{eq:Faug}
\end{align}
where $\CikL$ are the coefficients (Eq.~(\ref{eq:expandF})) that expand the smooth envelope in $\chi$ as polynomials
$\PkL(\br)$ (Eq.~(\ref{eq:defcapPkL})) and
\begin{align}
\tPkL(\br) &= \Bigl[A_{kl}\phi_\ell(r)+B_{kl}\dot{\phi}_\ell(r)\Bigr] Y_L(\hat r)\nonumber\\
&\equiv \tpkl(r) Y_L(\hat r)
\end{align}
$\phi_\ell(r)$ and $\dot{\phi}_\ell(r)$ are the linearised partial waves, Sec.\ref{sec:linearisation}.  Coefficients
$A_{kl}$ and $B_{kl}$ are chosen so that $\tpkl$ and $\pkl$ have the same value and slope at $\Rmt$.

\subsubsection{Overlap and Kinetic Energy, and Output Density}

We use Eq.~(\ref{eq:component3}) to construct matrix elements of augmented basis functions.  Matrix elements of overlap
and kinetic energy may be written
\begin{multline}
\int \chi^*_i [-\nabla^2]\chi_j \d3r =\int {\chi_i^{(0)}}^* [-\nabla^2] \chi_j^{(0)} \d3r\\
+\sum_{kk'L} C^{(i)*}_{kL} \tau_{kk'l} C^{(j)}_{k'L} \label{eq:tmat}
\end{multline}
\vspace*{-\belowdisplayskip}\vspace*{-\abovedisplayskip} 
\begin{multline}
\int \chi^*_i \chi_j \d3r = \int {\chi_i^{(0)}}^* \chi_j^{(0)} \d3r\\
+\sum_{kk'L} C^{(i)*}_{kL} \sigma_{kk'\ell} C^{(j)}_{k'L}
\label{eq:smat}
\end{multline}
where
\begin{gather}
\tau_{kk'\ell} = \int_S \lbr \tilde{P}_{kL}[-\nabla^2]\tilde{P}_{k'L}- P_{kL}[-\nabla^2]P_{k'L} \rbr\d3r\label{eq:tau}\\
\sigma_{kk'\ell} = \int_{\Rmt} \lbr \tilde{P}_{kL}\tilde{P}_{k'L} - P_{kL} P_{k'L} \rbr \d3r
\label{eq:sigma}
\end{gather}

The local kinetic energy matrix $\tau_{kk'l}$ is symmetric, even while the individual terms in Eq.~(\ref{eq:tau}) are
not, because the integral is confined to a sphere.  However the surface terms from the true and smooth parts cancel
because $\tPkL$ and $\PkL$ match in value and slope there.

Note also that if $\phi_\ell$ and $\dot{\phi}_\ell$ are kept frozen, e.g. computed from a free atom as we might do to
mimic a PAW or a pseudopotential method, $\tau_{kk'\ell}$ and $\sigma_{kk'\ell}$ are independent of environment.  Though
we won't pursue it here, it suggests a path to constructing a unique and transferable pseudopotential.

Matrix elements of the potential are more complicated and will be discussed in the next section.  For now we take them
as given, along with the overlap and kinetic energy.  We can diagonalise the secular matrix and obtain eigenfunctions
and the output density as a bi-linear combination of basis functions
\begin{align}
\psi_n(\br) &= \sum_i Z_{in} {\chi}_{i}(\br) \nonumber\\
n^\text{out}(\br) &= \sum_n w_n \left| \psi_n(\br) \right|^2 \nonumber\\
& = \sum_{ij} \lbr \sum_n w_n Z^*_{in} Z_{jn} \rbr  \chi^*_i(\br) \chi_j(\br)
\end{align}
where the sum runs over the occupied eigenstates with $w_n$ the occupation probability.  The bi-linear product may be
written in a three-component form
\begin{align} 
&{\chi_i}^* {\chi}_j  = {\chi_i^0}^* \chi^0_j\nonumber\\
&+\hspace{-0.3em}\sum_{Rkk'LL'}\hspace{-0.3em} C^{(i)*}_{RkL}\lbr\tilde{P}_{RkL}\tilde{P}_{Rk'L'}
        - P_{RkL} P_{Rk'L'}\rbr C^{(j)}_{Rk'L'}
\label{eq:bilinearchi}
\end{align}
The first term yields a smooth $n_0$ which extends everywhere, and then two local terms, true and one-centre expansion
of $n_0$ to some $L$ cutoff
\begin{align}
n^\text{val}(\br) = n^\text{val}_0(\br) + \sum_{RL} \lbr n^\text{val}_{1RL}(\br)-n^\text{val}_{2RL}(\br) \rbr
\label{eq:density}
\end{align}
As pointed out before, we expect the higher $L$ components be carried accurately by $n^\text{val}_0$ even for low
$\ell_\text{max}$, which provides a simple justification for the pseudopotential and PAW approximations.

\subsubsection{Core}
\label{sec:core}

The core levels and core density can be computed either scalar-relativistically or fully relativistically from the Dirac
equation (Sec.~\ref{sec:frasa}).  The core density is constructed in one of two ways: \emph{Self-consistent core:} a
core partial wave $\phi^c_{\ell}$ is integrated in the true potential, subject to the boundary condition $\phi_c(\Rmt) =
\phi'_c(\Rmt) = 0$.  It was once thought that the advantage of determining the core in the true potential outweighed the
inaccuracy originating from the artificial boundary condition at $\Rmt$.  \emph{Frozen overlapping core approximation:}
$\phi^c_{\ell}$ is computed in the free atom and kept frozen---experience shows this to be a better approximation.  We
develop this case here.

The core density has three components, which add to the valence density:
\begin{align}
\tilde{n}_0(\br) &= \tilde{n}^\text{val}_0(\br) + \sum_R g^\text{cn}_R(\br) \nonumber\\
n_{1R}(\br) &= n^\text{val}_{1R}(\br) + n^\text{core+nuc}_{1R}(\br) \nonumber\\
\tilde{n}_{2R}(\br) &= \tilde{n}^\text{val}_{2R}(\br) + g^\text{cn}_R(\br)
\end{align}
where
\begin{align}
g^\text{cn}_R(\br) = C^G\,G_{L{=}0}(r_g;\br) + C^H\mathcal{H}_{L{=}0}(E_c;r_h;\br)
\label{eq:gcore}
\end{align}
The second term is a smooth Hankel function with $C^H$ fixed to fit the spill-out of the core density into the
interstitial.  It is important that this be taken into account when constructing the total energy and potential.  The
alternative, to renormalise the core and confine it to the sphere, is a much cruder approximation.  This makes it
possible for shallow cores to be treated accurately, without needing to include them in the secular matrix
(unfortunately at the loss of strict orthogonality between these and the valence states).

The pseudocores $g^\text{cn}_R$ in $\tilde{n}_0$ and $\tilde{n}_{2R}$ nearly cancel, but they need to be included to
construct $V^\text{eff}$ and the total energy described below.  Also the two kinds of $g^\text{cn}_R$ do not exactly
cancel because it extends into the interstitial in the former case, but is truncated in the latter.

Smoothing radius $r_g$ must be small, so the core density is almost completely confined to the sphere.  In practice we
use the same Gaussians as for $g^\text{val}_R$ entering into the valence multipole moments (see Eq.~(\ref{eq:n0tilde})
below).  In that way $g^\text{cn}_R$ and $g^\text{val}_R$ can be merged together.  The Hankel smoothing radius $r_{h}$
is chosen independently, but it is also typically small, of order $\Rmt/2$.  The result should depend minimally on the
choice of either.

\subsubsection{Local Orbitals}
\label{sec:lo}

Inside augmentation spheres, the usual Hilbert space consists of partial waves $\{\phi_\ell,\,\dot{\phi}_\ell\}$.
Questaal has the ability to extend this space by one local orbital (LO), to
$\{\phi_\ell(\varepsilon_\nu,r),\,\dot{\phi}_\ell(\varepsilon_\nu,r),\,\phi^z_\ell(\varepsilon_z,r)\}$.  $\phi^z_\ell$
consists of a partial wave evaluated at some energy $\varepsilon_z$ far above or far below the linearisation energy
$\varepsilon_\nu$ for $\phi_\ell$.  Addition of a LO greatly extends the energy window over which the band structure is
valid; see for example Fig.~1, of Ref.~\cite{mark06adeq}, where the band structure for Si was compared to LAPW, and to a
full (non-linearised) APW. Properties of Questaal's LO are explained in that reference in a \emph{GW} context; here we
outline the main features.

LO have one each of the following attributes:
\begin{itemize}
\item either core-like or high in energy, with principal quantum number $\mp$ 1 that of valence
$\phi_\ell$, respectively;
\item either a conventional LO, which consists $\phi_z$ with some $\phi_\ell$ and $\dot{\phi}_\ell$ admixed to make the
value and slope vanish at $\Rmt$ \cite{Singhbook}, or an extended LO, which consists of $\phi_z$ with a smooth Hankel
function tail that extends continuously and differentiably into the interstitial.  The extended LO can be applied only
to the core-like case; and
\item either a scalar relativistic LO, or a Dirac LO with $\mu{=}1/2$.  This is not usually important, but it can have
consequences for heavy elements; see how it affects LDA and \emph{GW} gaps in Sec.~\ref{sec:relativistic}.  This method
was first introduced by Kune\u{s} \emph{et al.}\cite{Kunes01}.
\end{itemize}

High-lying states are not usually important for DFT, because states above $E_{F}$ do not contribute to the potential.
Core states of order $E_{F}{-}2$\,Ry and deeper can usually be treated quite adequately as core levels integrated
independently of the secular matrix (Sec.~\ref{sec:core}).  There are mild exceptions: for high resolution such as
needed for the Delta Codes project (Sec.~\ref{sec:benchmarks}), both deep and high-lying LO were used (in the latter
case, particularly adding $4d$ LO to the $3d$ transition metals).

\emph{GW} is more sensitive to states far from $E_{F}$ than DFT.  Including low-lying cores at $E_{F}{-}2$\,Ry or even
somewhat deeper with LO can modify valence QP levels by $\sim$0.1\,eV.  High-lying states are also more important in the
\emph{GW} context.  ZnO is an extreme case where they were found to shift the gap by several tenths of an
eV~\cite{Friedrich11}, and many high-lying LOs are required to reach convergence\cite{Jiang16}.

\subsubsection{Effective Potential}
\label{sec:lmfpot}

We seek a three-component form of the total energy functional.  Its variational derivative with respect to the density
yields an effective potential $V^\text{eff}$, which should give the first-order variation of the electrostatic and
exchange-correlation energy with respect to the trial density.  The density has three components, the first interacts
with a smooth potential $\tilde{V}_0$, the second with the local true potential $V_{1R}$, and the third with the local
projection of the potential $V_{2R}$. Correspondingly, the accumulated sums over the first, second, and third terms
produce $n^{\rm out}_0$, $n^{\rm out}_{1R}$, and $n^{\rm out}_{2R}$, respectively.

Construction of $V^\text{eff}$ is complicated by the fact that the electrostatic potential at some point depends on the
density everywhere.  Therefore it is not possible to construct the potential in a sphere solely from the density inside
this sphere, or the smooth potential from $n^\text{val}_0$ alone.  We can solve this difficulty by defining a
pseudodensity $\tilde{n^\text{val}}_0$ which has the same multipole moments inside augmentation sphere $R$ as the true
density ${n^\text{val}}_{1R}$.  Suppose the local sphere density $n^\text{val}_{1R}-n^\text{val}_{2R}$ has multipole
moments $q_M$, where $M$ = angular momentum.  Then define
\begin{gather}
\tilde{n}^\text{val}_0(\br) = n^\text{val}_0(\br) + \sum_{R} g^\text{val}_{R}(\br) \nonumber\\
g^\text{val}_{R}(\br) = \sum_{M} q_{RM} G_{RM}(\br)
\label{eq:n0tilde}
\end{gather}
where $G_{RM}$ is a Gaussian of angular momentum $M$ centred at $R$, with unit moment.  $G_M$ must be sufficiently
localised inside $\Rmt$ that the potential from the second term at $\Rmt$ is negligibly different from the potential of
a unit point multipole at $R$ (typically $\Rmt/4$).  Results should not depend on $r_g$, but in practice if it is made
too small, numerical integration of the $G$ become inaccurate, while if too large charge is not confined to $\Rmt$.
Provided it is confined, the compensating Gaussians ensure that the electrostatic potential
$V^\text{es}[\tilde{n}^\text{val}_0]$ is exact in the interstitial.

The local representation $n^\text{val}_{2R}$ of the smoothed density must be equally compensated
\begin{align}
\tilde{n}^\text{val}_{2R} &= n^\text{val}_{2R}+ g^\text{val}_{R}
\end{align}
The $q_{RM}$ can be decomposed into a sum of partial moments
\begin{align}
q_{RM} = \sum_{kk'LL'} Q_{Rkk'LL'M} \nonumber
\end{align}
\vspace*{-\belowdisplayskip}\vspace*{-\abovedisplayskip} 
\begin{multline} 
Q_{Rkk'LL'M} = \int_{\Rmt} \lbr\tilde{P}_{kL}\tilde{P}_{k'L'}- P_{kL} P_{k'L'}\rbr\\
        \times r^m Y_M(\hat r) \d3r
\end{multline}
By decomposing $q_{RM}$ this way, we can make any possible variation in the local density in the Hilbert space spanned
by Eq.~(\ref{eq:bilinearchi}), and obtain the potential from the electrostatic corresponding to the variation.

Matrix elements of the local potential are determined from variation of the electrostatic energy (Sec.~\ref{sec:totale})
\begin{align}
\pi_{1Rkk'LL'} =& \int_{\Rmt} \tilde{P}_{RkL}V_{1R}\tilde{P}_{Rk'L'} \d3r \nonumber\\
\pi_{2Rkk'LL'} =& \int_{\Rmt}  P_{RkL} \tilde{V}_{2R} P_{Rk'L'}  \d3r \nonumber\\
&+ \sum_{RM} Q_{Rkk'LL'M}  \int_{\Rmt} \tilde{V}_{2R} G_{RM} \d3r
\label{eq:piloc}
\end{align}
The true potential $V_{1R}$ is the response to true partial density $\tilde{P}_{RkL}\tilde{P}_{Rk'L'}$, and
$\tilde{V}_{2R}$ the response to the smooth partial density $P_{RkL} P_{Rk'L'} + \sum Q_{Rkk'LL'M} G_{RM}$.  These two
partial densities have the same multipole moments by construction.

We solve the Poisson equation for each of the three components
\begin{align}
\nabla^2 \tVes_{0} &= -8\pi \tilde{n}_{0} \nonumber\\
\nabla^2 \Ves_{1R} &= -8\pi n_{1R} \nonumber\\
\nabla^2 \tVes_{2R} &= -8\pi \lbr n_{2R} + \sum\nolimits_{M} q_{RM} G_{RM}\rbr
\label{eq:poisson}
\end{align}
All density components must include core (or pseudocore) contributions (Sec.~\ref{sec:core}).  Poisson's equation for
$\tVes_0$ is most easily solved by fast Fourier transforming $\tilde{n}_0$ to reciprocal space, scaling each Fourier
component $n_G$ by $-8\pi/G^2$ and transforming back.  In principle the electrostatic potential can be calculated this
way but in practice the compensation Gaussians in $\tilde{n}_0$ can be fairly sharply peaked.  Some parts are separated
out and evaluated analytically to keep the plane wave cutoff low, as discussed in the next section.
$\tilde{V}^\text{es}_0(\br)$ is exact in the interstitial region and extends smoothly through the spheres.  $\Ves_{1R}$
and $\tVes_{2R}$ can be solved up to an arbitrary homogeneous solution which is determined by the boundary condition at
$\Rmt$.  This adds a term proportional to $r^m Y_M(\br)$, which is determined by resolving $\tVes_{0}$ into a one-centre
expansion at $\Rmt$.

Pseudopotentials $\tVes_{0}$ and $\tVes_{2R}$ approximately cancel each other, but not exactly.  This is in part because
the latter has a finite $L$ cutoff, and in part because $\tVes_{0}$ carries the tails of the core densities spilling
into the interstitial.  These incomplete cancellations are the key to rapid $\ell$ convergence of this method, and also
allow relatively shallow cores not to be included in the secular matrix, with minimal loss in accuracy.

\subsection*{Three Component Total Energy in DFT}
\label{sec:totale}

The total energy in DFT is comprised of the kinetic energy, electrostatic energy, and exchange-correlation energy
\begin{align}
E_\text{tot} = \left<T\right> + U + E_\text{xc}
\label{eq:etot}
\end{align}
Questaal implements two functionals for $E_\text{tot}$.  The traditional Kohn-Sham functional is evaluated at the output
density: $U$ and $E_\text{xc}$ are evaluated from $n_\text{out}$ generated by the secular matrix, and
$\left<T\right>_\text{HKS} = \sum_i^\text{occ} \left<\psi_i|-\nabla^2|\psi_i\right> + T_\text{core}$.
${T}_\text{core}=\sum\epsilon_i^{\rm core}-\int n^{\rm core}V^\text{eff}$ is the kinetic energy of the core.  If the
core is kept frozen, it can be evaluated from the free atom.

Harris~\cite{Harris85} and independently Foulkes and Haydock~\cite{Foulkes89} showed that the kinetic energy can be
expressed in terms of $n_\text{in}$: $U{=}U[n_\text{in}]$ and $E_\text{xc}{=}E_\text{xc}[n_\text{in}]$ and
\begin{align}
\left<T\right>_\text{HF} = \sum_{i}^\text{occ}\epsilon_i^\text{val}-\int n_\text{in}^\text{val}V_\text{in}^\text{eff}
        + T_\text{core}
\label{eq:EHF}
\end{align}
Both $E_\text{HKS}$ and $E_\text{HF}$ are calculated, though usually the latter is preferred because it converges more
rapidly with deviations from self-consistency.  It is nevertheless useful to have both energies, because they are
calculated differently in Questaal.  $U$ and $E_\text{xc}$ are computed from different densities, but more,
$\left<T\right>_\text{HKS}$ and $\left<T\right>_\text{HF}$ are calculated in different ways.  It is not trivial that at
(nominal) self-consistency $E_\text{HKS}{=}E_\text{HF}$.  The two always differ slightly; when they differ more than a
small amount, it is an indication that some parameter (e.g. plane-wave cutoff) is set too low.  Thus,
$E_\text{HKS}{-}E_\text{HF}$ is one measure of the convergence of the basis representing the charge density.

In Questaal's three-component framework, $nV^\text{eff}$ a sum of three terms bilinear in the three components
$\tilde{n}_0\tVes_0$, $n_{1R}\Ves_{1R}$ and $\tilde{n}_{2R}\tVes_{2R}$.  $\left<T\right>$, $U$ and $E_\text{xc}$ each
have independent contributions from three components, for example the contribution to the electrostatic energy from the
valence electrons is
\begin{multline}
E^\text{es} =  \frac{1}{2} \int \tilde n_0^\text{val}(\br) \tilde{V}_0^\text{es}(\br)\\
+\sum_R \int_{S_R} \lbr n^\text{val}_{1R} \Ves_{1R}\d3r - \tilde{n}^\text{val}_{2R} \tVes_{2R}\rbr\d3r
\label{eq:estat}
\end{multline}
The last term cancels most of the first inside the augmentation spheres, except the first retains $L$ components above
$\ell_\text{max}$ to all orders.  This is because $\tilde{V}_{2R}$ should be a one-centre expansion of $\tilde{V}_{0}$
up to $\ell_\text{max}$, and $\tilde{n}^\text{val}_{2R}$ should be a one-centre expansion of $\tilde{n}^\text{val}_{0}$.
Because $V^\text{es}[\tilde{n}^\text{val}_0]$ is exact in the interstitial, including at $\Rmt$, and it continued inside
$\Rmt$ by adding by $\Ves_{1R}\d3r - \tVes_{2R}$, the true potential is correct everywhere.  Note that the derivative of
$E^\text{es}$ wrt coefficients $C^{(i)}_{Rk'L'}$ in Eq.~(\ref{eq:bilinearchi}) yield the local potentials
Eq.~(\ref{eq:piloc}), so that the effective one-body Hamiltonian corresponds to the functional derivative of the energy.

The total electrostatic energy must include the core.  Write it analogously to Eq.~(\ref{eq:estat}) but add the (pseudo)
core and nuclear density to $n^\text{val}$ (Sec.~\ref{sec:core})
\begin{align}
U = \tilde{U}_0 + \sum_R \{U_R - \tilde{U}_R\}
\end{align}
The first term is the electrostatic energy of the smooth density, evaluated on a uniform mesh.  It involves a valence
and pseudo-(core+nucleus) term
\begin{gather*}
\tilde{U}_0            = \tilde{U}_0^\text{val} + \tilde{U}_0^\text{c+n}\\
\tilde{U}_0^\text{val} = \frac{1}{2} \int \d3r\,\tilde{n}_0^\text{val}(\br) \, \tilde{V}^\text{es}(\br)\\
\tilde{U}_0^\text{c+n} = \sum_R (C^G_{R} - Z_R)V^G_{R0} + C^H_RV^H_R
\end{gather*}
where
\begin{align*}
V^G_{RL} &= \int \d3r\, G_{RL} \tilde{V}^\text{es}[\tilde{n}_0] \\
V^H_{R}  &= \int \d3r\, \mathcal{H}_{R0} \tilde{V}^\text{es}[\tilde{n}_0]
\end{align*}
Besides being needed for ${U}_0^\text{c+n}$, $V^G_{RL}$ is used in assembling $\tilde{U}_0^\text{val}$.
$\tilde{U}_0^\text{c+n}$ and $\tilde{U}_0^\text{val}$ must be evaluated with some care to avoid numerically integrating
sharply peaked functions (core pseudodensity and the multipole contribution to $\tilde{n}_0$ (Eq.~(\ref{eq:n0tilde})).
$V^G_{RL}$ and $V^H_{R}$ are evaluated by splitting the potential into two terms $ \tilde{V}^\text{es}[\tilde{n}_0] =
\tilde{V}^\text{es}[{n}_0] + \{\tilde{V}^\text{es}[\tilde{n}_0] - \tilde{V}^\text{es}[{n}_0]\}$.  Integrals of the
second term can be performed analytically (the interested reader is referred to Section XI of Ref.~\cite{lmfchap}), thus
avoiding integrals of sharply peaked functions.  All of the other integrals are much smoother and are evaluated on the
grid of $G$ vectors.

$U_R$ and $\tilde{U}_R$ are readily integrated in the sphere together with the valence parts in Eq.~(\ref{eq:estat}).

The exchange-correlation potential is made analogously with the electrostatic potential:
\begin{align}
E&_{xc}  = \int n_0 \epxc[n_0] \d3r\nonumber\\
&+\sum_{R} \int_{\Rmt} \lbr n_{1R} \epxc[n_{1R}] - n_{2R} \epxc[n_{2R}] \rbr \d3r
\end{align}
It is also divided into valence + core contributions.

We reiterate that differences $\int n_{1R} \Ves_{1R}\d3r - \tilde{n}_{2R} \tVes_{2R}$ and $\int n_{1R} \epxc(n_{1R}) -
n_{2R} \epxc(n_{2R})$ converge much more rapidly with $\ell$-cutoff than either term separately, and as a consequence
the three-component augmentation is much more efficient than the standard one.

\subsubsection{Forces}

Our original formulation~\cite{lmfchap} derived a simple expression for the derivative of $H_\text{HF}$ when a nucleus
changes from $\bR$ to $\bR{+}\delta\bR$, assuming that the partial waves $\phi_{R\ell}$ shift rigidly with $\delta\bR$.
(\texttt{lmf} has a ``frozen $\phi$'' switch, which if imposed, guarantees $\phi_{R\ell}$ is rigid.  In practice we have
found that the differences are small.)

The original expression yielded forces that converge only linearly with deviations
${\Delta}n{=}n_\text{out}{-}n_\text{in}$ from self-consistency.  It is possible to achieve faster convergence in
${\Delta}n$, by adding a correction ${\delta}V{\Delta}n$, where ${\delta}V$ is determined from a change
${\delta{n}_\text{in}}$ in $n_\text{in}$, but there is no unambiguous way to determine ${\delta{n}_\text{in}}$.  If the
static susceptibility (Sec.~\ref{sec:susceptibility}) were calculated, a good estimate could be made.  But
$\chi(\br,\br',0)$ is a second derivative of $H_\text{HF}$: it is far more cumbersome to make than $H_\text{HF}$ itself,
and approximation of $\chi$ by models such as the Lindhard function worked less well than hoped.  A practicable, and
simple alternate ansatz is to assume that ${\delta{n}_\text{in}}$ is given by a change in the Mattheis density
construction (superposition of free-atomic densities), arising ${\delta}\bR$.  This has worked very well in practice: it
dramatically improves the convergence of the forces with iterations to self-consistency.  Here we omit the derivation
but refer to the original paper~\cite{lmfchap}, and also to Appendix B of Ref.~\cite{Kotani15} for an alternate
derivation including the correction term.  The final expression for shift of nucleus $R$ from $\bR{\to}\bR{+}\delta\bR$
is
\begin{align}
\delta E_{\rm H} = \int  \tilde{V}_0\,\delta{g^\text{cn}_R} + \delta^{\rm R} \sum_{i}^\text{occ}\epsilon_i^\text{val}
        + {\delta}V_\text{in} (n_\text{out}{-}n_\text{in})
\label{eq:hfforce}
\end{align}
$\tilde{V}_0$ is the sum of the smooth electrostatic and exchange-correlation potential.  This term describes the force
of the smooth density on the Gaussian lumps which represent the core and the nucleus at each site, and is the analogue
of the Hellmann-Feynman theorem.

The eigenvalue shifts account for a change in the atom-centred basis set in the shift in $\bR$.  This the price for
taking a tailored basis, but one whose Hilbert space changes when $\bR$ shifts.  Here the smooth mesh potential is fixed
and augmentation matrices $\tau$, $\sigma$, and $\pi$ (Eqs.~(\ref{eq:tau}),~(\ref{eq:sigma}) and ~(\ref{eq:piloc}))
shift rigidly with the nucleus.  Gradients only include redistribution of plane waves owing to the shifts in envelope
functions, and in the expansion coefficients $C^{(j)}_{Rk'L'}$ (Eq.~(\ref{eq:Faug})).

\begin{figure}[ht]
\begin{center}
\includegraphics[width=.7\columnwidth]{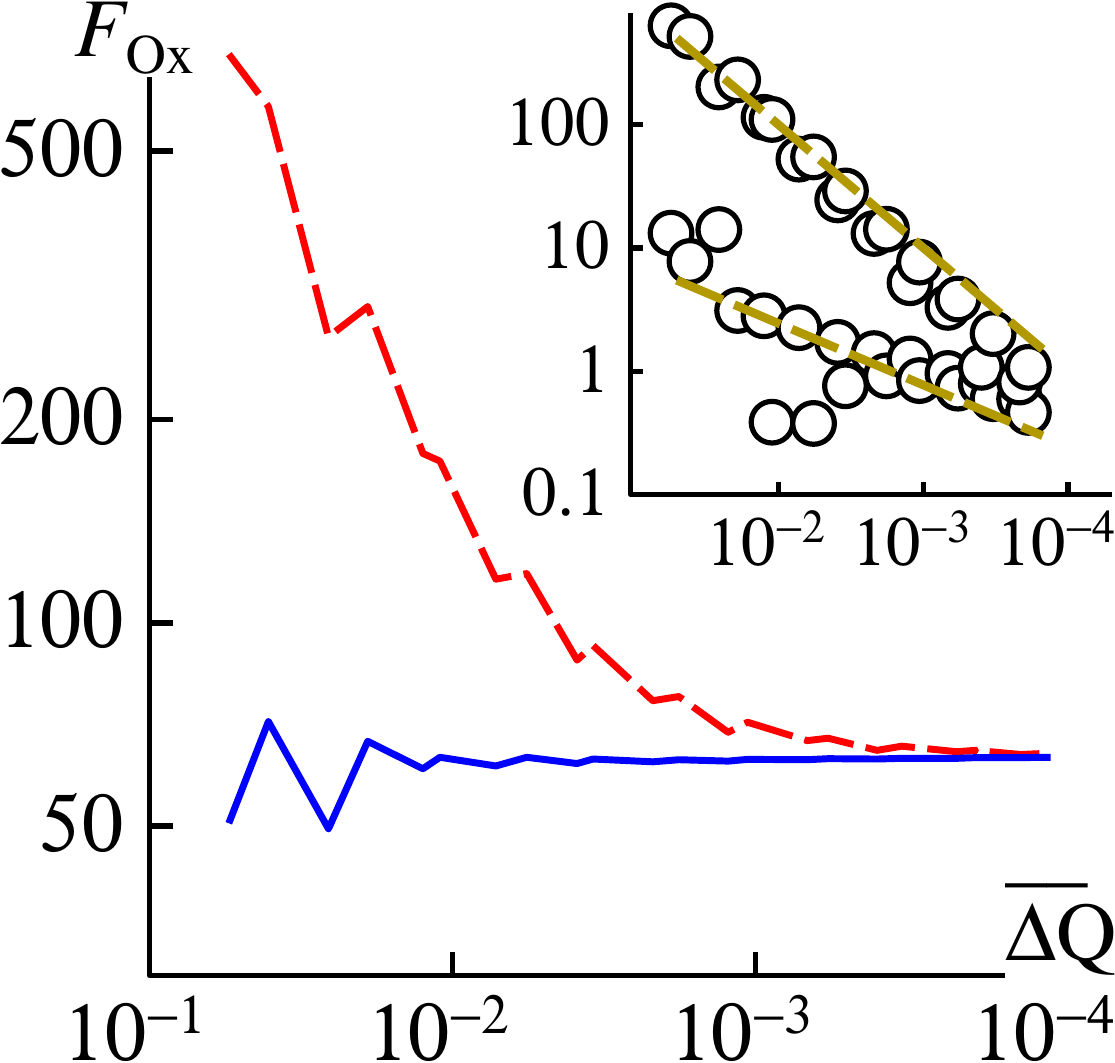}
\end{center}
\caption{Harris force $F_\text{Ox}$ (mRy/au) on O neighbouring Co in Co-doped TiO$_{2}$.  $\overline{\Delta\mathrm{Q}}$
is a measure of the RMS change $n_\text{out}{-}n_\text{in}$ integrated over the unit cell.  Solid and dashed lines are
Eq.~(\ref{eq:hfforce}) with and without the third term. Inset shows deviation in $F_\text{Ox}$ from the converged value.
Yellow lines show $\overline{\Delta\mathrm{Q}}$ and $(\overline{\Delta\mathrm{Q}})^2$ as guides to the eye.}
\label{fig:harrisforce}
\end{figure}

The last term is the correction noted above.  In future, it will probably be replaced by an efficient calculation of
$\chi$; meanwhile the Mattheis-shift ansatz works rather well.  Fig.~\ref{fig:harrisforce} shows the force $F_\text{Ox}$
on an O neighbouring Co substituted for Ti in a 12-atom unit cell of TiO$_{2}$, as \texttt{lmf} iterates to
self-consistency.  The deviation ${\Delta}F$ from the self-consistent force is vastly smaller with the correction term
(compare solid and dashed lines), and $F_\text{Ox}$ is reasonable already for the Mattheis construction (first
iteration).  The inset compares the two kinds of ${\Delta}F$ with iteration to self-consistency to show that without and
with the correction, ${\Delta}F$ scales as
${\Delta}F^\mathrm{uncorr}_{\mathrm{Ox}}{\sim}\overline{\Delta\mathrm{Q}}$ and
${\Delta}F^\mathrm{corr}_{\mathrm{Ox}}{\sim}(\overline{\Delta\mathrm{Q}})^2$, the latter being the same rate of
convergence as the energy itself.

\subsection{LDA+U}
\label{sec:ldau}

The LDA+U method, introduced originally by Anisimov \textit{et al.}~\cite{Anisimov91,Anisimov93,Anisimov97} for open
shell $d$- or $f$-shell materials including on-site Coulomb and exchange interactions by means of the Hubbard model is
implemented in both the \texttt{lm} and the \texttt{lmf} codes following the rotationally invariant approach described
in~\cite{Liechtenstein95}.  Rotational invariance means that the occupations of the different $d_m$ orbitals are
specified by a density matrix ${\bf n}^\sigma=n_{mm^\prime}^\sigma$ independent of the specific choice of Cartesian axes
defining the spherical harmonics.  For brevity we refer to the orbitals on which $U$ is applied as the $d$ orbitals
although the code is written sufficiently general that $U$ terms can be applied to any $nl$ of choice and even on
multiple sets of orbitals.

The on-site Coulomb interactions are added to the LSDA total energy and a double-counting term is subtracted.
\begin{multline}
E^{LDA+U}[\rho^\sigma({\br}),{\bf n}^\sigma]= E^{LSDA}[\rho^\sigma({\bf r})]+E^U[{\bf n}^\sigma]\\
        - E_{dc}[{\bf n}^\sigma]
\end{multline}
Several schemes for the double counting have been
proposed~\cite{Anisimov91,Anisimov93,Liechtenstein95,Czyzyk,Petukhov03} and are implemented in \texttt{lmf} but the
prevailing approach is the fully localised limit (FLL) in which the double-counting term and $U$-terms cancel for the
fully localised (atomic) limit in which the density matrix becomes a diagonal matrix with integer occupations.  This
means that in principle the total energy is already well described in the LSDA in the atomic limit but the orbital
energies are not.  The task of the LSDA+U approach is to optimise the density matrix when the orbitals on which $U$ is
applied are allowed to hybridise with the other orbitals in the system.  It hence describes strictly speaking the
orbital polarisation in the system.  Note that the total energy is then separately a functional of the spin-dependent
electron density and the local density matrix on the orbitals for which $U$ terms are added.  The density matrix or
orbital occupations also influence the spin-charge densities and so the two contributions are not really independent.
The $U$ terms in LSDA+U are treated in the static mean-field Hartree-Fock approximation in contrast to the DMFT method
where they are treated dynamically.  The expressions for the $E^U[{\bf n}^\sigma]$ part of the total energy  can be
found in \cite{Liechtenstein95} and are given in terms of density matrices $n^\sigma_{mm^\prime}$ and matrix elements of
the Coulomb interaction $\langle m,m^\dprime|V_{ee}|m^\prime,m^\tprime\rangle$.  They fully take into account the
rotational aspects of the $Y_{lm}$ of the $d$-states included.
The double counting term, on the other hand, is written in terms of average Coulomb $U$ and exchange $J$ parameters and
involves only the the trace of the density matrix $n^\sigma=Tr({\bf n}^\sigma)$.\cite{Liechtenstein95}
$U$ and $J$ are the screened Coulomb and exchange parameters, which can for example, be estimated by a separate
self-consistent supercell calculation treating the atoms in which the $d$-orbital occupation is changed as an impurity
problem~\cite{Gunnarsson91}. or by means of the linear response approach~\cite{Cococcioni05}.  Most often they are
treated empirically and $U$ is adjusted to spectroscopic splittings, for example.

The bare Coulomb interaction matrix elements can be evaluated exactly in terms of the Slater $F^k(ll)$ integrals and
combinations of Gaunt coefficients as worked out for instance in Condon and Shortley~\cite{Condon}.
The Slater $F^k(ll)$-integrals,
\begin{align}
F^k(ll)=\int\int r_1^2dr_1r_2^2dr_2 R_\ell(r_1)^2R_\ell(r_2)^2(r_<^k/r_>^{k+1})
\end{align}
where $r_<$, $r_>$ are the smaller and larger or $r_1$ and $r_2$ could easily be calculated directly  in terms of the
partial waves inside the spheres.  However, this would then not take into account the screening of the Coulomb
interaction by the other orbitals in the system.  Therefore it is customary to use the relation between the average $U$
and $J$ to the $U_{mm^\prime}$ and
$J_{mm'}$
\begin{gather}
U=\frac{1}{(2\ell+1)^2}\sum_{mm'}U_{mm'}=F^0 \nonumber\\
U-J=\frac{1}{2\ell(2\ell+1)}\sum_{mm^\prime}(U_{mm^\prime}-J_{mm'})
\end{gather}
given by Anisimov \textit{et al.}~\cite{Anisimov93} and where $U_{mm'}=\langle mm^\prime|V_{ee}|mm^\prime\rangle$,
$J_{mm'}=\langle mm^\prime|V_{ee}|m^\prime m\rangle$.  Using the tables of Condon and Shortley~\cite{Condon} one finds
for $d$ electrons
\begin{align}
14J=F^2(dd)+F^4(dd)
\end{align}
and for $f$ electrons
\begin{align}
3J=\frac{2}{15}F^2(ff)+\frac{1}{11}F^4(ff)+\frac{1}{858}F^6(ff)
\end{align}
For $p$-orbitals $J=F^2(pp)/5$ and for $s$-orbitals $J=0$.  Examining the tabulated values of the Slater $F^k$ integrals
from Hartree-Fock calculations, one finds that the ratios $F^4(dd)/F^2(dd)\approx0.625$ and
$F^6(ff)/F^2(ff)\approx0.494$, $F^4(ff)/F^2(ff)\approx0.668$ are approximately constant.  Using these fixed ratios and
the relation to the average exchange integral $J$ one then can fix the $F^k$ values.  The advantage of doing it this way
is that the screening of the $J$ and $U$ can then be taken into account and there are only two empirical parameters.  On
the other hand one finds in practice that while $U$ is strongly reduced from the atomic bare $F^0$, the $J$ is
approximately unscreened, so one might as well use the unscreened directly calculated $F^k$ for $k\ge2$.  From LDA
calculated atomic wave functions, we find $F^4(dd)/F^2(dd)=0.658\pm0.004$ for $3d$ atoms,
$F^4(ff)/F^2(Ff)=0.685\pm0.001$, $F^6(ff)/F^2(ff)=0.518\pm0.001$ for rare-earth RE$^{3+}$.  (Until recently, Questaal
used a different convention; see note~\cite{ldauerr}.)

By minimising the LDA+U total energy with respect to the density matrix elements, one obtains a non-local potential
$V^\sigma_{mm^\prime}$ which can also be found in \cite{Liechtenstein95}.
In the ASA \texttt{lm}-version it is straightforward to add this nonlocal potential directly to the Hamiltonian inside
the spheres because the orbitals defining the matrix $V^\sigma_{mm^\prime}$ are just the single $lm$ channel partial
waves inside the sphere.  It is a little more complex in \texttt{lmf}.  Essentially, we now add the operator
$|\phi^\sigma_{lm}\rangle V^\sigma_{mm^\prime}\langle \phi^\sigma_{lm^\prime}|$.  We add the corresponding augmentation
matrix elements both for the $\phi$ and $\dot\phi$ and for local orbitals if these are present for this $\ell$ channel.

The calculation within LDA+U starts from a set of initial occupation numbers of the $m$ orbitals for each channel $\ell$
for which $U$ and $J$ parameters are provided.  These define the initial density matrix.  The calculation then proceeds
by making this density matrix self-consistent simultaneously with making the standard spin density $\rho^\sigma({\bf
r})$ self-consistent.  We note that the configuration one converges to may depend on the starting density matrix and
hence one should in principle consider different starting points and find the one which minimises the energy or use
guidance from physical insight.  In rare-earth (RE) based systems one finds for example that the lowest energy
corresponds to the configuration which obeys Hund's rules~\cite{Larson07}.

We have here described only the FLL.  The around-mean-field approach~\cite{Czyzyk,Anisimov91} and a mixture of the
two~\cite{Petukhov03} are also implemented and make use of the same ingredients.  Recently, Keshavarz  \textit{et
al.}~\cite{Keshavarz18} argued that adding $U$ to spin-unpolarised LDA, thus LDA+U may have advantages over adding them
to LSDA, for example to extract inter-atomic exchange interactions independent of the reference system used (AFM or FM).
This is presently not supported in the lm and lmf codes but might be useful to add in the future.  Note that in
Ref.~\cite{Anisimov93} also a LDA+U rather LSDA+U approach was used.

A simpler version of the LDA+U is described by Dudarev \textit{et al.}~\cite{Dudarev98}.  In that case, only a single
parameter, namely $\tilde U=U-J$ comes into play.  One can use this scheme in the codes by setting $J=0$ and adjusting
the provided $\tilde U$ parameter.  This means essentially that we neglect the orbital polarisation due to the
anisotropic Coulomb interaction exchange terms but keep the Hartree-Fock like feature that empty states feel a different
potential from occupied states.  In fact, in this case the additional potential is
\begin{align}
V^\sigma_{mm^\prime}=\tilde U[{\scriptstyle\frac{1}{2}}\delta_{mm^\prime}-n^\sigma_{mm^\prime}]
\end{align}
Or if the density matrix is actually diagonal $V^\sigma_m=\tilde U[{\scriptstyle\frac{1}{2}}-n^\sigma_m]$.  This means
that when a spin-orbital $m\sigma$ on a given site is empty it is shifted up by $\tilde U/2$ and when it is filled, it
is shifted down by $\tilde U/2$.  This corresponds to the simplest form of the LDA+U formalism.  It allows one for
example to include a shift of a fully occupied $d$ band.  This is useful for example in Zn-containing systems.  In LDA,
the binding energy of these orbitals is underestimated.  The deeper a band lies below the Fermi level, the higher its
downward shift by the self-energy and this can be simply mimicked by the LDA+U method in this form.  It has the effect
for example of reducing the hybridisation of the Zn-$3d$ with the valence band maximum and slightly increases the gap,
affects things like band-offsets and the valence band maximum crystal field and spin-orbit induced fine splitting.

One can even use this approach for $s$ or $p$ electrons and thereby shift up the mostly cation-$s$ like conduction band
minimum in a semiconductor to adjust the gap.  While this is of course purely empirical and shifting the gaps for the
wrong reason, and hence far less accurate than the QS\emph{GW} approach, it is sometimes useful in defect calculations
because a defect level that would otherwise lie as a resonance in the conduction band can now become a defect level in
the gap and allow one to properly control its occupation for different charge states of the defect.  Using a $U$ on
$p$-orbitals of anions like O or N, this can also have the effect that the empty defect levels pushed out the valence
band and localising on a single $p$ orbital are pushed deeper into the gap. The essential function of the LDA+U terms in
these applications is to introduce a Hartree-Fock like orbital dependence of the potential. This is important to reduce
the self-interaction error of LDA or GGA and allows one in a simpler and much less expensive way to simulate what a
hybrid functional would do.  If one picks these shifts carefully to adjust the bulk band structure to the QS\emph{GW}
bands, it provides a rather efficient way to study defects with a corrected band gap and specific orbital
self-interaction, two of the main errors of LDA plaguing defect calculations.  An example of this can be found in
Boonchun \textit{et al.}~\cite{Boonchun}.

\subsection{Relativistic Effects}
\label{sec:relativistic}

The radial solvers all solve by default a scalar Dirac equation, which incorporates the dominant relativistic effects.
All the DFT codes (\texttt{lmf}, \texttt{lm}, \texttt{lmgf} and \texttt{lmpg}) have the ability to incorporate
spin-orbit coupling perturbatively.  In the band codes the perturbation is straightforward (See Sec.~\ref{sec:soc}); the
Green's function scheme is more subtle, particularly with respect to third order parameterisation, but a formulation is
possible (Sec.~\ref{sec:socgf}).  For \texttt{lmf}, the $\myvec{L}\cdot\myvec{S}$ term is added to the true local
potential $V_{1R}$ (see Sec.~\ref{sec:lmfpot}).  As of this writing, only the spin diagonal part of the output density
is kept in the self-consistency cycle.

Additionally the ASA codes \texttt{lm}, \texttt{lmgf} and \texttt{lmpg} have a fully relativistic implementation
(Sec.~\ref{sec:frasa}).  The full-potential code \texttt{lmf} does not as yet; however it does have a facility to
compute the core levels fully relativistically.  It follows a method similar to that developed by Ebert~\cite{Ebert89}.

As of this writing, the QS\emph{GW} code allows spin-orbit coupling only in a very restricted manner.  Typically SOC has
only a minor effect on the self-energy $\Sigma$: this is because SOC is a modification to the kinetic energy and only
affects the potential in a higher order perturbation.  We have found very good results by generating QS\emph{GW}
self-energies leaving out SOC all together, and then as a post-processing step include it in the band structure.

Additionally, \texttt{lmf} has a special mode where the only diagonal parts of $\myvec{L}\cdot\myvec{S}$ (see
Sec.~\ref{sec:soc}) are added to the one-particle Hamiltonian.  After diagonalisation the spin off-diagonal parts are
used to modify the one-particle levels in a special kind of perturbation theory, as described in the Appendix of
Ref.~\cite{Brivio14}.  This has the advantage that the eigenvectors are kept spin-diagonal, which significantly reduces
the cost of the \emph{GW} code.  Tests show that at the LDA level the modification to the band structure is very similar
to that of the full $\myvec{L}\cdot\myvec{S}$, and for purposes of determining the effect on $\Sigma$ it should be quite
reliable except in very special cases.  Once a modified $\Sigma$ has been found (as a one-shot correction to QS\emph{GW}
computed without SOC), \texttt{lmf} can be run with the full $\myvec{L}\cdot\myvec{S}$, including the modification of
$\Sigma$ approximately in the manner just described.  We have found the difference to be negligible except when elements
are very heavy: CH$_{3}$NH$_{3}$PbI$_{3}$ is the most extreme case we have found so far: the effect of SO on the band
structure through changes in $\Sigma$ reduced the gap by 0.1 eV~\cite{Brivio14}

Finally \texttt{lmf} has an ability to fold in approximately the effects of the proper Dirac Hamiltonian on the valence
states.  The difference between the full Dirac and scalar Dirac partial waves can be important for very small $r$, where
the SOC contributions are the largest (see $\xi(r)$ in Sec.~\ref{sec:soc}).  For small $\br$, the true Dirac wave
function varies as $r^{\gamma},\ \gamma^2={\kappa^2-(2Z/c)^2}$, whereas in the scalar Dirac case it varies as $r^\ell$.
As a result, matrix elements $\xi(r)$ are underestimated.  The effect is typically very small, but it becomes
non-negligible in heavy semiconductors such as CdTe.  To ameliorate this error, \texttt{lmf} has the ability to use a
fully relativistic partial wave in place of a standard local orbital, choosing $\mu=-1/2$.  Most important is the
$p_{1/2}$ state with $\kappa=1$.
\begin{table}[ht!]
\begin{center}
\setlength{\tabcolsep}{4.8pt} 
\begin{tabular}{|lcccccc|}
\hline
                        &     &        &\multicolumn{2}{c}{LDA}&\multicolumn{2}{c|}{QS\emph{GW}} \\
\emph{Material}         &\raisebox{2pt}{${\substack{n_c\\n_a}}$}&\raisebox{2pt}{${\substack{\gamma_c\\\gamma_a}}$} &
        $\Delta_0$ & $E_G$ & $\Delta_0$ & $E_G$ \\
\hline
GaAs (SR)               & 5   & 0.97   & 0.34       & 0.24   & 0.33      & 1.77  \\
GaAs ($p_{1/2}$)        & 5   & 0.97   & 0.35       & 0.24   & 0.34      & 1.75  \\
GaAs (expt)             &     &        & 0.34       & 1.52   & 0.34      & 1.52  \\[8pt]
ZnSe (SR)               & 5   & 0.98   & 0.39       & 0.95   & 0.39      & 3.05  \\
ZnSe ($p_{1/2}$)        & 5   & 0.97   & 0.40       & 0.95   & 0.40      & 3.00  \\
ZnSe (expt)             &     &        & 0.40       & 2.82   & 0.40      & 2.82  \\[8pt]
CdTe (SR)               & 6   & 0.94   & 0.84       & 0.26   & 0.83      & 1.83  \\
CdTe ($p_{1/2}$)        & 6   & 0.93   & 0.92       & 0.26   & 0.88      & 1.77  \\
CdTe (expt)             &   &          & 0.90       & 1.61   & 0.90      & 1.61  \\[8pt]
PbTe (SR)               & 7   & 0.80   & 1.12       & -.16   & 1.00      & 0.40  \\
PbTe ($p_{1/2}$)        & 6   & 0.93   & 1.30       & -.57   & 1.06      & 0.22  \\
PbTe (expt)             &     &        &            & 0.19   &           & 0.19  \\
\hline
\end{tabular}
\end{center}
\caption{Spin orbit splitting $\Delta_0$ of the valence band at $\Gamma$, and energy bandgaps computed in the LDA and by
QS\emph{GW}, with a scalar relativistic (SR) local orbital on the cation and anion $p$ states, and a Dirac $p_{1/2}$
local orbital.  In the zincblende semiconductors the two band edges are at $\Gamma$, but in PbTe case they are both at L
(Fig.~\ref{fig:pbte}).  $n$ are the principal quantum numbers of the cation and anion local $p$ orbital;
$\gamma=[1-(Z/c)^2]^{1/2}.$}
\label{tab:diraclo}
\end{table}

\begin{figure}[ht]
\begin{center}
\includegraphics[width=.48\columnwidth]{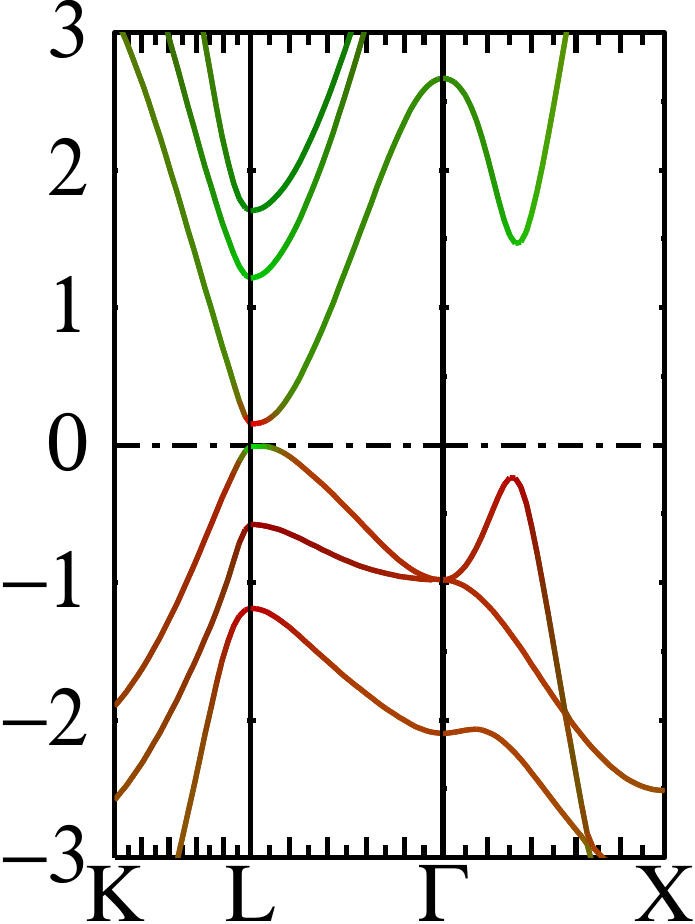}
\includegraphics[width=.48\columnwidth]{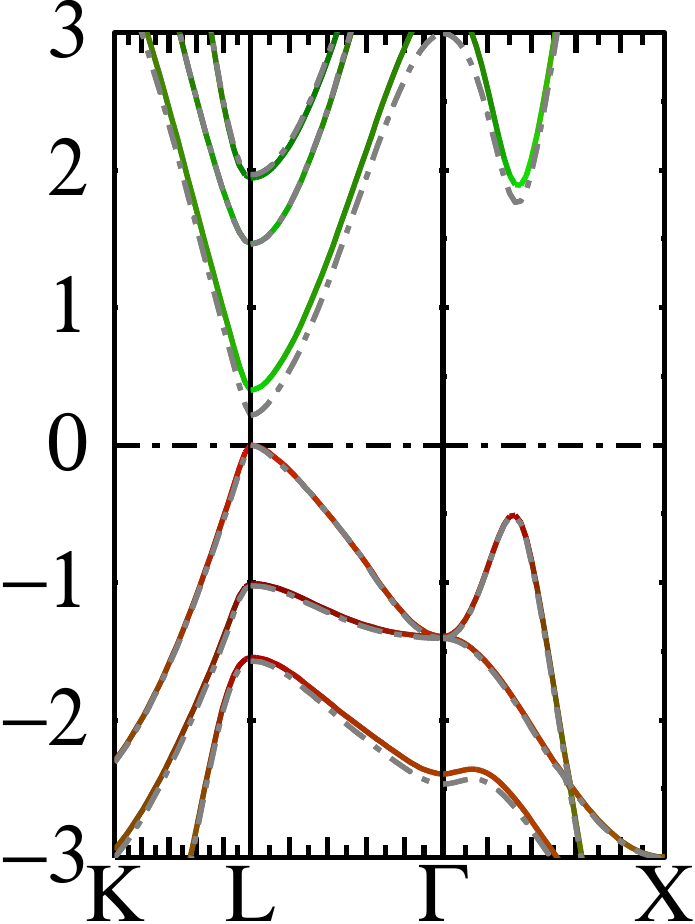}
\end{center}
\caption{Energy band structure of PbTe in eV for (left) the LDA and (right) QS\emph{GW} approximations.  Colours
correspond to projections onto L$_{6}^{\pm}$ symmetry.  L$_{6}^{+}$ is comprised of Pb $s$ and the $t_{2g}$ part of the
$d$ orbital, and Te $p$.  L$_{6}^{-}$ is comprised of Te $s$ and the $t_{2g}$ part of the $d$ orbital, and Pb $p$.  The
dashed lines show the effect of using a Dirac $p_{1/2}$ local orbital instead of the usual scalar relativistic one.
Except for the addition of the Dirac $p_{1/2}$ orbital, the basis is similar to that used in Ref.~\cite{Svane10b}.  It
is a slight enlargement over the default basis, including $g$ orbitals on Pb and Te and a $6s$ local orbital on Te.
These extra orbitals widen the gap relative to the default basis by $\sim$0.1\,eV.}
\label{fig:pbte}
\end{figure}

SO coupling is well described for light semiconductors such as GaAs and ZnSe (see Table~\ref{tab:diraclo}), as $Z$
increases $\gamma$ deviates progressively from unity and the discrepancy with experiment increases.  It is already
significant for CdTe, and in PbTe the effect exceeds 0.1\,eV.  Replacing the scalar Dirac LO with the Dirac $p_{1/2}$ LO
significantly decreases this error in CdTe.  In PbTe, the valence band edges are both at L: the two band edges have
symmetry L$_{6}^{+}$ and L$_{6}^{-}$, respectively (Fig.~\ref{fig:pbte}).  The LDA gap is nominally positive, but
L$_{6}^{+}$ and L$_{6}^{-}$ are inverted very near L as can be seen by the change in colours at the band edges; thus the
gap is actually inverted.  Anticipating the discussion in Sec.~\ref{sec:gw}, if a diagonal-only \emph{GW} were carried
out on top of this, the L$_{6}^{+}$ and L$_{6}^{-}$ would order properly, but the wrong topology would manifest itself
as a band crossing near L, as shown for Ge in Fig. 6 of Ref.~\cite{mark06adeq}.  QS\emph{GW} orders L$_{6}^{+}$ and
L$_{6}^{-}$ correctly, somewhat overestimating the gap~\cite{Svane10b}.  Adding a Dirac $p_{1/2}$ local orbital reduces
the gap by nearly 0.2\,eV.  In the LDA the gap at L \emph{widens} as a consequence of the fully relativistic partial
wave (Table~\ref{tab:diraclo}), another reflection of the inversion of L$_{6}^{+}$ and L$_{6}^{-}$.

\subsection{The PMT Method}
\label{sec:pmt}

\begin{figure}[ht]
\begin{center}
\includegraphics[width=.75\columnwidth]{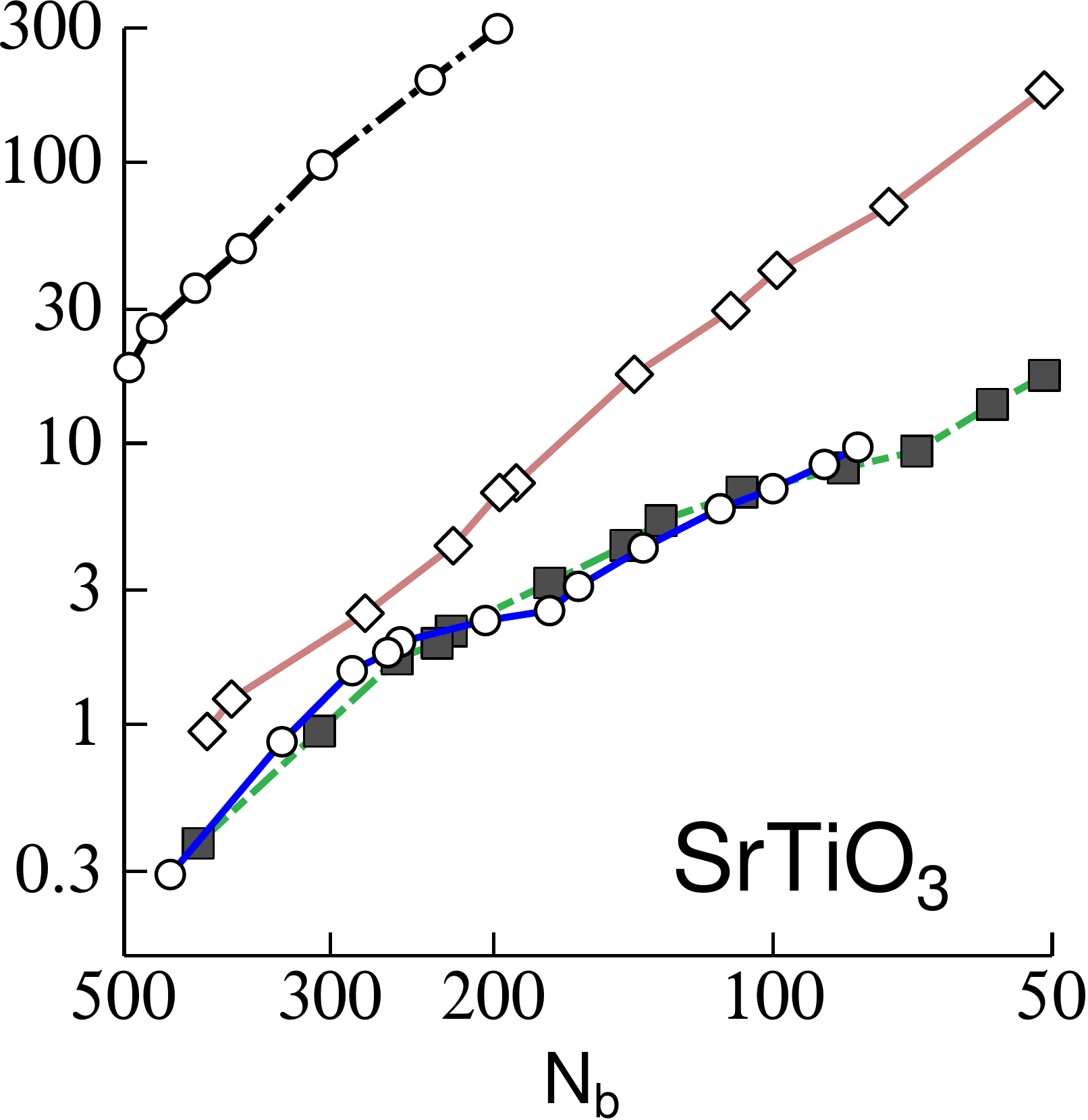}
\end{center}
\caption{Deviation $\Delta{}E$ from fully converged total energy/atom, in mRy, \emph{vs} total number of basis functions
(PW + $\mathcal{H}_L$ + LO) $N^{\Gamma}_{\rm b}$ at $\Gamma$.  Black-dashed, peach-solid, green-dashed, and blue-solid
correspond respectively to the null, hyper, small, and large basis sets.  $\Delta{}E$ is drawn on a log-log scale as a
function plane-wave cutoff, denoted as $1/\text{N}_\text{b}$.  $\text{N}_\text{b}$ is the total number of basis
functions PW+$\mathcal{H}_L$+LO at the $\Gamma$ point.}
\label{fig:scaling}
\end{figure}

In the introduction it was noted that the LAPW and Questaal's generalised LMTO method are essentially the same except
for the choice of envelope functions.  LMTO's are much more compact, and require a much smaller Hilbert space, but they
suffer from problems with basis set completeness.  An obvious alternative is to combine the two basis sets; this is the
``PMT basis''~\citep{KotaniPMT10}.  This can be accomplished in practice quite neatly, because Questaal's standard
$\mathcal{H}_L$ require a one-centre expansion, Eq.~(\ref{eq:expandF}), more general than simple LMTO's
(Eq.~(\ref{eq:defs0})).  Plane waves can be expanded in polynomials $P_{kL}$ in a similar manner, as can other kinds of
envelope functions.  Thus the method provides a unified framework to seamlessly mix different kinds of envelope
functions.  Expressions for total energy, forces, assembling output density, etc, remain unchanged.  The procedure is
described in more detail in Ref.~\cite{KotaniPMT10}; here we focus on the primary strengths and weaknesses of the
method, using the total energy calculated for SrTiO$_{3}$ shown in Fig.~\ref{fig:scaling} for discussion. It was redrawn
from Ref.~\cite{KotaniPMT10}.  Four basis sets were chosen, ranging from pure LAPW ('null') to a moderately large basis
of 75 smooth Hankels $N(\mathcal{H}_L)$ + 6 local orbitals.
\begin{table}
\setlength{\tabcolsep}{4.8pt} 
\begin{tabular}{|c|c|c|}
\hline
\emph{basis} & \emph{character of} $\mathcal{H}_L$          & $N(\mathcal{H}_L)+N(\text{LO})$\\\hline
null         & no $\mathcal{H}_L$                           &  6\\
hyper        & \emph{sp} on O only                          & 18\\
small        & \emph{spd} on all atoms                      & 51\\
large        & +2$^\text{nd}$ \emph{spd} (\emph{spd} on Ti) & 81\\
\hline
\end{tabular}
\caption{Nature and size of different basis configurations for the case of SrTiO$_3$.} 
\end{table}

\paragraph{Advantages of PMT} 

As expected, Questaal's purely atom-centred basis is vastly more compact than a pure LAPW basis
(Fig.~\ref{fig:scaling}).  Even a tiny basis of just O \emph{sp} states dramatically improves on the convergence of the
LAPW basis.  PMT offers a marked advantaged from LAPW perspective: by augmenting that basis by a few functions.  From
the atom-centred perspective Questaal's standard moderately large basis misses about 5\,mRy/atom in total energy,
showing that it is not complete.  Finally the figure shows that adding $\mathcal{H}_L$ continues to dramatically improve
the convergence until the standard minimal basis of $\mathcal{H}_L$ is reached (single \emph{spd} function on each
atom).  Beyond that initial rapid gain, it doesn't seem to matter much whether the basis is increased by adding PW or
more $\mathcal{H}_L$ (compare blue-solid and green-dashed lines).

\paragraph{Drawbacks of PMT} 

While they are far more efficient than ordinary LAPWs, PMTs show less promise than was initially hoped for.
\begin{itemize}
\item Compactness: LAPWs are not ``intelligent'', i.e. tailored to the potential, and therefore convergence is slow in
the basis size.  Also their extended range precludes the advantages of short-ranged functions.
\item Over-completeness: PMT has difficulties when $\mathcal{H}_L$ and PWs are both sizeable.  This is because they span
much the same Hilbert space.  Workarounds (reducing the Hilbert space by projecting out parts that contribute to a small
overlap) have been implemented, but they are inefficient and somewhat finicky---e.g. energy may not vary smoothly with
the change in a nuclear coordinates.
\item Interpolating $\Sigma_0$: in QS\emph{GW} context, the ability to interpolate the QS\emph{GW} self-energy
$\Sigma_0$ is degraded.  A workaround has been made for this problem also (Ref.~\cite{Kotani14}) by projecting
$\Sigma_0$ onto the MTO part of the eigenfunctions, interpolating it, and re-embedding.  Information is lost in the
projection/re-embedding step, resulting in ambiguities.  A much better way to address difficulties with interpolation is
to construct very short-ranged, compact functions.  This is shown clearly in Sec.~\ref{sec:tblmf}.
\end{itemize}

In summary, PMT does best when treated as a slight augmentation of either the LAPW limit, or the MTO limit.  Adding a
few PWs with a low cutoff (2\,Ry) was useful in our participation in the Delta Codes project (Sec.~\ref{sec:benchmarks})
particularly for the molecular solids such as N$_{2}$ which consist of 90\% interstitial, or even more.  Questaal's pure
MTO basis does not handle those compounds that accurately.

\subsection{Floating Orbitals}
\label{sec:floating}

While the PMT method can make the basis nearly complete, it causes severe difficulties when interpolating the
QS\emph{GW} $\Sigma_0$ to an arbitrary $k$.  An alternative is to add ``empty'' sites --- points outside augmentation
spheres where extra $\mathcal{H}_{L}$ can be added to the basis.  These sites have no augmentation spheres; they only
enhance completeness in the interstitial.  They are more \emph{ad hoc} than plane waves (Sec.~\ref{sec:pmt})
particularly because there is no systematic path to convergence.  Nevertheless they can be implemented efficiently; and
Questaal has an automatic procedure to locate points to fill voids in the interstitial.  In practice we use floating
orbitals to converge QS\emph{GW} self-energies with respect to the single-particle basis, in open systems.  Their effect
at the LDA level is usually not large, unless systems are very open or very accurate total energies are required.  But
for QS\emph{GW}, where the potential depends on unoccupied states as well as occupied ones, it makes more of a
difference, e.g correcting the fundamental gap in Si by of order 0.1\,eV.  We expect that the need for floating orbitals
in such systems will be obviated by the new JPO basis.  A precursor to it is discussed next.

\subsection{Screened, Short-ranged Orbitals in the Full-Potential Framework}
\label{sec:tblmf}

\begin{figure}[ht]
\begin{center}
\includegraphics[width=\columnwidth]{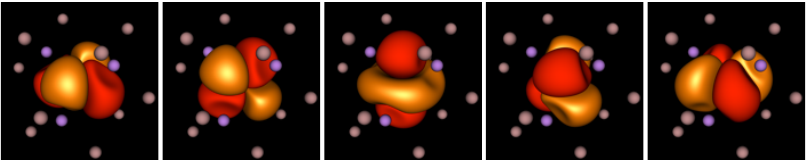}
\end{center}
\caption{Screened $d$ envelope functions, $xy$, $yz$, $3z^2{-}1$, $xz$, and $x^2{-}y^2$, for a zincblende lattice.  See
also Fig.~\ref{fig:screenedLMTO}.}
\label{fig:screened-lmto-d}
\end{figure}
Building on the success of the ``screening'' of LMTOs in the ASA (Sec.\ref{sec:introtb}), we have developed an analogue
within the full-potential LMTO framework.  As in the ASA case, ``screening'' does not alter the Hilbert space but
renders basis functions short ranged.  Screening is a precursor to the next-generation JPO basis, which will alter the
Hilbert space, rendering it significantly more accurate for the same rank of Hamiltonian.  JPOs are still in
development; the method will be presented in a future work.  They have the following properties:
\begin{itemize}
\item they solve the SE with a (nearly) optimal number of basis functions for a given accuracy in the
four dimensions $(\br,E)$;
\item they are very short ranged (see Figs.~\ref{fig:screenedLMTO} and \ref{fig:screened-lmto-d}); and
\item they are atom centred with a definite $L$ character on their own MT sphere.
\end{itemize}
This provides a framework for exploiting the many advantages inherent in a short ranged basis sets.  They can be used as
projectors, replacing Wannier functions; they can be used to construct minimal Hamiltonians; near-sightedness can be
exploited to make very efficient $O(N)$ solvers in DFT and $O(N^3)$ solvers in \emph{GW} and \emph{GW}+BSE.  That the
basis functions are associated with a definite $L$ character is important, for example, when singling out a correlated
subspace.

Screening of the $\mathcal{H}_L$ works in a manner similar to screening traditional $H_L$ because $\mathcal{H}_L$
asymptotically approaches a $H_L$ for large $r$ (Fig.~\ref{fig:smhankel}).  $\mathcal{H}_L - H_L$ decays approximately
as a Gaussian of radius $r_{s}$: for traditional values of $r_{s}$ ($r_{s}\lesssim2s/3$) it becomes negligible beyond
first-neighbour distances.  The screened $\mathcal{H}^\alpha_L$ employ the expansion coefficients $S^\alpha$ in the same
manner as in the ASA~(Sec.~\ref{sec:introtb}), so they have essentially the same range.  What is sacrificed is the
reinterpretation of screening in terms of hard core radius (Sec.~\ref{sec:hardcoreradius}), but this has no effect
because the Hilbert space is unchanged.  JPOs improve on the screened $\mathcal{H}^\alpha$ by restoring this condition
and exploiting it to make the kinetic energy continuous everywhere.

Screening makes it possible to avoid Ewald summation otherwise necessary in the unscreened FP case (Appendix C).  Also,
one-centre expansions are made more efficient.  By writing
$\mathcal{H}^\alpha=\left\{\mathcal{H}^\alpha-{H}^\alpha\right\}+{H}^\alpha$, the first term which involves
$\mathcal{H}$, must be expanded in the more cumbersome polynomials $P_{kL}$ (Eq.~(\ref{eq:Faug})).  In the screened case
the difference becomes short-ranged and can be dealt with efficiently in real space.  The latter can use the simpler
expansion Eq.~(\ref{eq:newkalpha}).

A restriction in transformation is that all functions of a given $\kappa$ must share the same Hankel energy, while the
traditional basis allows arbitrary choice for each orbital.  However, this restriction is fairly mild because of the
extra flexibility $\mathcal{H}_L$ have through the choice of $r_s$, which the ${H}_L$ do not.  The $r_s$ strongly affect
the kinetic energy near the MT boundary (Fig.~\ref{fig:smhankel}), and the results are more sensitive to that choice
than to the Hankel energy.

\begin{figure}[ht]
\begin{center}
\includegraphics[width=\columnwidth]{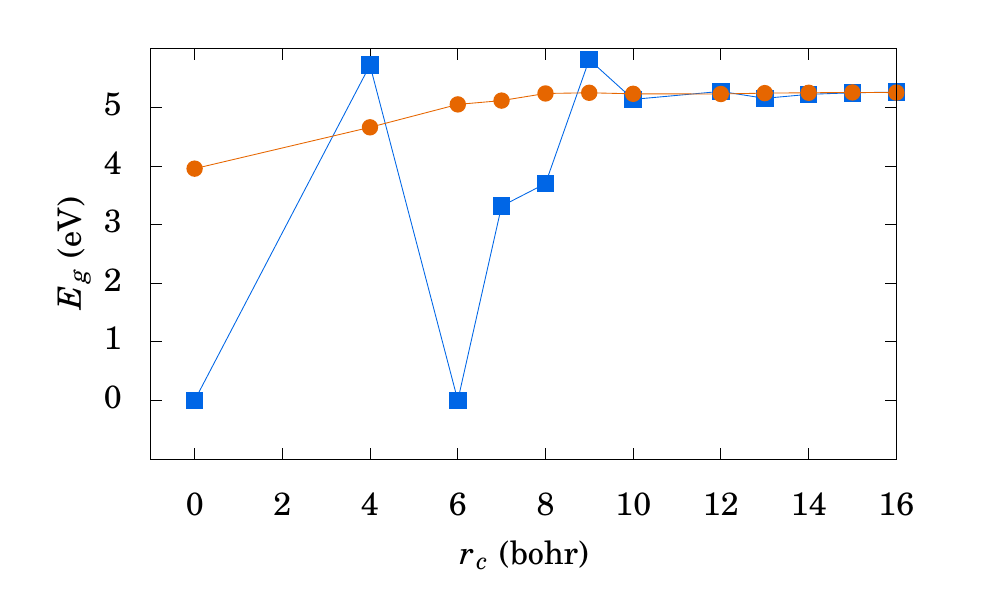}
\end{center}
\caption{NiO bandgap in eV as a function of the range truncated self-energy represented in the conventional and screened
basis sets.  The first point contains onsite terms only, the second adds first neighbours etc.  The conventional basis
shows erratic behaviour until at least the fourth neighbours are present.}
\label{fig:niogap}
\end{figure}

One large advantage of the transformation is that $\mathcal{H}^\alpha$ becomes short enough ranged so that matrix
elements
\begin{align*}
\Sigma_{0;RL,R'L'}^\alpha = \left< \mathcal{H}_{RL}^\alpha | \Sigma_{0} | \mathcal{H}_{R'L'}^\alpha\right>
\end{align*}
are limited by the range of the physical $\Sigma_{0}(\br,\br')$, and not by the basis set.  $\Sigma_{0;RL,R'L'}^\alpha$
turns out not to be very long ranged, as suggested some time ago by Zein \textit{et al.}~\cite{Zein06}.
Fig.~\ref{fig:niogap} shows how the bandgap in NiO evolves with range cutoff $R_\text{cut}$, defined as follows.
$\Sigma_{0;RL,R'L'}^\alpha$ is initially computed without range truncation.  Then matrix elements of
$\Sigma_{0;RL,R'L'}^\alpha$ are set to zero for $|\bR-\bR'|>R_\text{cut}$, and the band structure is
calculated.  As Fig.~\ref{fig:niogap} shows, the bandgap $E_{G}$ converges slowly with
$R_\text{cut}$ in the traditional $\Sigma_{0;RL,R'L'}^{\alpha{=}0}$ case, but for $\Sigma_{0;RL,R'L'}^\alpha$, $E_{G}$
is already reasonably converged including first and second neighbours.  Note that for $R_\text{cut}{\to}\infty$, the two
methods yield the same result, as the screening transformation does not change the Hilbert space.

One powerful feature of \texttt{lmf} is its ability to interpolate $\Sigma_{0}$.  As noted in Sec.~\ref{sec:questaalgw},
some compromises must be made because the high-energy parts of $\Sigma^{nn'}_{0}$ do not interpolate well, and elements
above an energy threshold (of order 2\,Ry) must be set to zero.  The screening mitigates this difficulty; it appears
that the energy threshold can be taken to $\infty$.  The difference is not large, but in small bandgap cases where the
precision of the method is very high, agreement with experiment may be limited by this difference.

The short range of $\Sigma_{0;RL,R'L'}^\alpha$ can be used to much advantage: we realise from the properties of Fourier
transforms that the number of $k$ points directly translates into the number of neighbours in real space.  Because
$\Sigma_{0;RL,R'L'}^\alpha$ is short ranged, it converges faster in the $k$ mesh than $\Sigma_{0;RL,R'L'}^{\alpha{=}0}$,
which saves significantly on computational effort.  Also explicitly semilocal algorithms can be constructed, greatly
reducing the cost to make $\Sigma_{0}$.

At present the short ranged screened smooth Hankel functions shine the most when used to represent the QS\emph{GW}
self-energy $\Sigma(\omega, q)$ and interpolate it at intermediate $q$ points not directly calculated in the heavy
\emph{GW} step.  The interpolation process is significantly simplified and ambiguity is removed because the high energy
$\Sigma$ no longer needs diagonal approximation, there is no need to define energy threshold and there is no loss of
information due to discarding the off-diagonal parts.  This does add to the computing effort for directly computed
QS\emph{GW} $\Sigma(\omega, q)$ because no states are excluded and matrix sizes increase respectively.  To counter that,
significantly fewer $\Sigma(q)$ need to be computed exactly, in our experience an equivalent result can be achieved by
using only a half or in cases even a quarter of the original BZ sampling in each direction, for cells with few symmetry
operations the savings add up quickly considering that the \emph{GW} walks over pairs of $q$ points (quadrupling the
total number of points for low symmetry cells).  Performance can be further improved by still utilising the diagonal
approximation to $\Sigma(\omega, q)$ because the interpolation is excellent and the high energy part only weakly affects
states near $E_{F}$.  To show the quality of the QS\emph{GW} $\Sigma(\omega, q)$ expansion in the new, screened basis
and the conventional unscreened, we plot the dependence of the NiO bandgap as a function of the truncated self energy in
Fig. \ref{fig:niogap}.
%

At the LDA level, the better spatial localisation of screened basis function paves the way to efficient, real-space
assembly of matrices (1 and 2 particle) while maintaining, and in cases improving, accuracy.  The screened functions are
also significantly less linearly dependent.  The implementation is heavily vectorised and at present can handle a couple
hundred atoms on a single node, with the scalability properly utilised the numbers can be an order of magnitude larger.
For very large systems, it will be possible to construct $O(N)$ methods, but empirically it appears so far that systems
of a few hundred atoms are probably still too small for such methods to be advantageous.

To summarise, our current screening transformation offers advantages of short-ranged basis sets.  It is not yet
optimally compact.  However, the JPO basis, to be reported on soon, we believe will be nearly the most compact (optimal
convergence for a given rank of Hamiltonian, and short range). We believe these advantages will be very significant, and
obviate the need for enhancements such as plane waves (Sec.~\ref{sec:pmt}) or floating orbitals
(Sec.~\ref{sec:floating}) to obtain high accuracy.

\subsection{Delta Codes Validation Exercise}
\label{sec:benchmarks}

The Delta Codes project~\cite{Lejaeghere2016a} is a continuing effort to mutually validate the different electronic
structure codes used in condensed matter and materials modelling.  The first comparison has focused on the equations of
state calculated for a selection of elemental solids.  The relative agreement of different codes, including
\texttt{lmf}, can be examined on the project site \url{https://molmod.ugent.be/deltacodesdft}.  Agreement is quantified
using an average ``Delta value''---the integrated difference between equation of state curves over a specified volume
range.

The calculated equations of states (scalar relativistic and PBE-GGA exchange-correlation are specified) agree extremely
closely, particularly among the all-electron codes.  The pseudopotential projects have also demonstrated precision very
similar to the all-electron methods.

The Delta Codes test set has been carried out for \texttt{lmf} with an automatic setup, avoiding any system-specific
modifications.  The main settings defining the calculations are:

\paragraph{Sphere radii:} $\Rmt$ corresponding to touching spheres are used, evaluated at the smallest volume tested.

\paragraph{$\ell_\text{max}$:} The maximum $\ell$ for the basis functions is 1 for H and He, 2 for $Z<18$ and 3
otherwise.  In each case, one higher $\ell$ is included in the partial wave expansion.

\paragraph{Semi-core states:} Local orbitals are effective for including semicore states in the valence.  We recommend
treating states as semicore whenever the leaked charge exceeds $0.002e$, or the eigenvalue is higher than $\sim -2$\,Ry.
For transition metals and f-electron systems, the inclusion of high-lying $\ell=2$ or $3$ conventional local orbitals
can significantly improve accuracy and these are added by default for these groups of elements.

\paragraph{Molecular cases:} For cases where the touching muffin-tin spheres fill less than 30\% of the cell volume, an
additional LAPW basis has been included.  The LMTO basis was not designed for molecular systems --- but the shortcomings
of the LMTO basis can be easily rectified by the addition of small number of plane-waves.

\paragraph{Basis setup:}  Basis parameters determined automatically as described below.

\subsubsection{Choice of Basis Parameters}
Questaal's full-potential code allows two LMTO basis functions to be used per $\ell$.  Each of these is defined by two
parameters which must be chosen: the Hankel energy and the smoothing radius.  Because the size of the LMTO basis is
small, the choice of basis parameters can have a significant impact on accuracy.  Different schemes for choosing basis
parameters have evolved, of which the simplest is to fit the basis functions to the free atom wave functions.  This
provides one set of basis parameters for each $\ell$.  The second set is obtained from the first by shifting the Hankel
energy (typically by $-0.5$ to $-0.8$\,Ry).  This is a heuristic approach but it is automatic and effective and is the
default.

An alternative method chooses smoothing radii directly from the potential. The gradient of the smoothed Hankel functions
at typical muffin-tin sizes is sensitive to $r_s$ and variational freedom is obtained by differentiating the two sets of
basis functions by $r_s$ instead of the Hankel energy.  The smoothing radius describes the point where the basis
function begins to deviate from the exponential, Hankel-like tail: this is similar to the behaviour of atomic
eigenfunctions at their classical turning points.

Associating the smoothing radius with the classical turning point allows basis functions to be constructed that resemble
atomic-like states at energies different to the atomic eigenvalues and the question of choosing suitable $r_s$ is
translated into a determining a pair of reference energies---one each for the two basis sets per $L$.  Energies of the
valence states are typically in the range $0$ to $-2$\,Ry, and choices for the reference energies of $E_1=-0.5$ and
$E_2=-2.0$ result in accurate basis sets for most materials.  In this scheme, the Hankel energy remains a free parameter
but similar results are found for Hankel energies in the range $-0.1$ to $-1$\,Ry.

The choice of reference energies can be automated by expressing them in terms of the atomic potential at the muffin-tin
radius.  One basis set is chosen with a higher reference energy, giving a more diffuse basis function, the other is
setup at a smaller, more negative, energy which yields a smaller $r_s$ and a more tightly bound basis function.  Because
$v(\Rmt)$ varies significantly across elements and materials, a proportional scheme is more suitable: e.g.,
$v(r_s^1)=2v(\Rmt)$ and $v(r_s^2)=\frac{1}{2}v(\Rmt)$; this algorithm is used for automating the DeltaCodes tests.

\subsubsection{Comparison with LAPW Results}

\begin{figure}[ht]
\includegraphics[width=\columnwidth]{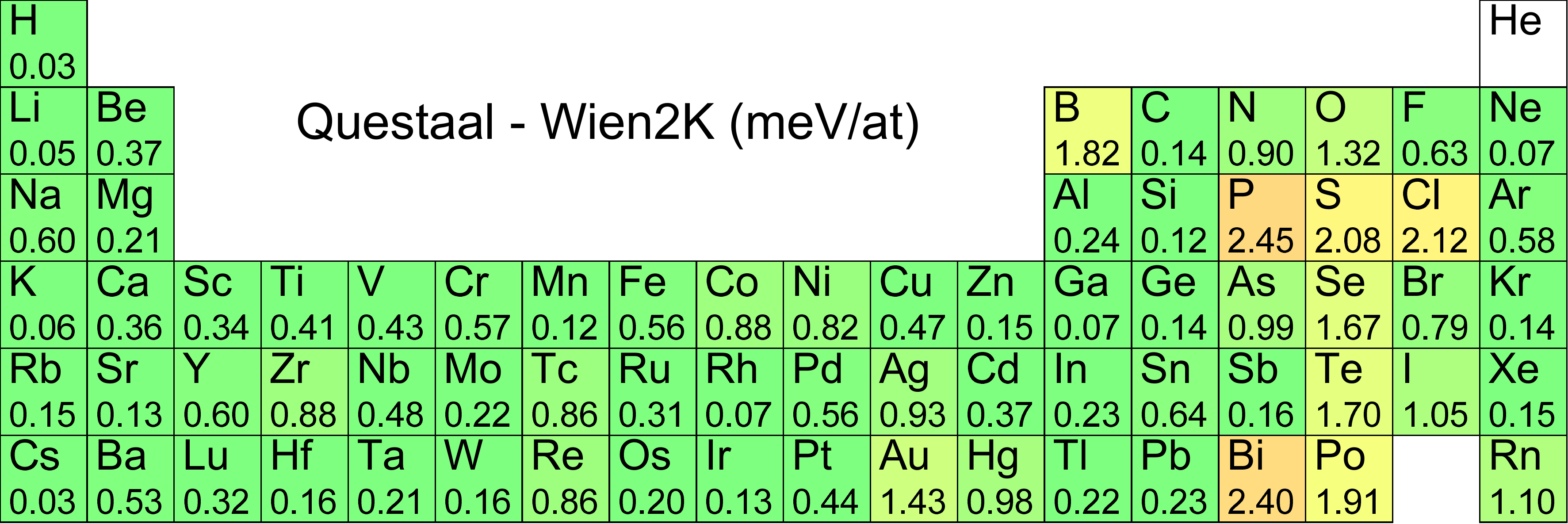}
\caption{``Delta Codes'' tests: integrated energy difference between the energy-volume curves calculated by
\texttt{lmf} and \texttt{WIEN2K}~\cite{wien2k}, expressed in meV/atom; the average value is
0.62meV/atom.}
\label{fig:deltaWien2k}
\end{figure}

Figure \ref{fig:deltaWien2k} shows the ``Delta values'' for Questaal with respect to the LAPW code WIEN2K.  The level of
agreement is typical (or better) than most all electron methods involved in the validation exercise.  In particular, the
transition elements are extremely well represented by the smooth Hankel basis.  Some of the first period and group VII
cases involve molecular problems for which the muffin-tins fill a small fraction of the unit cell: the performance of
the LMTO basis in these cases becomes sensitive to $r_s$ and $E$ choices.  We do not attempt to choose optimal
parameters, instead we include an additional LAPW (see Sec.~\ref{sec:pmt}), with cut-off 2\,Ry or 4\,Ry depending on the
packing fraction.  Au and group VI are outliers: in both cases the automatic basis setup generates basis functions that
are rather too smooth to describe the tightly bound states in these elements.  If necessary, improved performance can be
obtained, thereby converging $V_0$ with respect to the basis, by adding plane waves.

In addition to confirming that the smooth Hankel basis is capable of high precision, the testing procedure also
demonstrates that the automatic procedure for setting up the basis and setting various numerical parameters is highly
reliable.

\section{\texorpdfstring{\emph{GW} and QS\emph{GW}}{GW and QSGW}}
\label{sec:gw}

The \emph{GW} approximation (\emph{G}=Green's function, \emph{W}=screened Coulomb interaction) is the first-order term
in the formally exact diagrammatic expansion around some one-particle Hamiltonian $H_{0}$~\cite{hedin65}.

Questaal's \emph{GW} formalism is explained in some detail in Ref.~\cite{mark06qsgw}, and its implementation of
quasi-particle self-consistent \emph{GW} (QS\emph{GW}) is explained in Ref.~\cite{Kotani07}.  Here we recapitulate only
the main points.  From an implementation point of view, \emph{GW} requires two-point functions, such as the
susceptibility from which $W$ is made (Sec.~\ref{sec:susceptibility}).  This entails an auxiliary \emph{product basis}
of wave function products.  Most commonly plane waves are used to implement \emph{GW}.  In such a case the same basis
for two-particle objects can be used, because a product of plane waves is another plane wave, but for all-electron
methods this is not possible.  Representation of these objects requires a basis spanning the Hilbert space of wave
function products.  Four-centre integrals (e.g. Eq.~(\ref{generalchi01q})) can be evaluated as integrals of pairs of
product basis functions.  This was first accomplished by Aryasetiawan~\cite{prodbasis94} in an ASA framework, and
extended by us~\cite{Kotani02} into a full-potential scheme.

In the \emph{ab initio} context, \emph{GW} has traditionally referred to a perturbation around the LDA: i.e.
$H_{0}=\HLDA$ so that $GW=\GLDA\WLDA$, though in recent years hybrid functionals or LDA+U have become popular.  It has
become increasingly recognised that results depend rather strongly on the choice of $H_{0}$, or equivalently the
$V_\text{xc}$ entering into it, by which we mean the beyond-Hartree parts of the effective potential.  Thus the
self-energy of $H_{0}$ is $\Sigma=V_\text{xc}$ in the language of many-body perturbation theory.  The implications of
starting point dependence are profound: the quality of \emph{GW} depends on the quality of the reference.  Moreover it
is increasingly accepted that $\HLDA$ is only a good choice for $H_{0}$ for a very restricted materials class.

$GW$ is a perturbation theory, and usually only the lowest order perturbation (diagonal part of $\Sigma$ in the basis of
the reference Hamiltonian) is kept.  As a result eigenfunctions are not perturbed and the change in QP level $\ekn$ is
\begin{align}
\delta\ekn = Z_{kn}[\langle\Psi_{kn}|\Sigma(\br,\br',\ekn)-V_{\rm xc}^{\rm{LDA}}(\br)|\Psi_{kn} \rangle]
\label{eq:e1shot}
\end{align}
$Z_{kn}$ is the quasi-particle (QP) renormalisation factor
\begin{align}
Z_{kn}=\left[ 1-\langle\Psi_{kn}|\frac{\partial}{\partial\varepsilon}
        \Sigma(\br,\br',\ekn)|\Psi_{kn} \rangle \right]^{-1}
\label{eq:defzfac}
\end{align}
and accounts for the fact that $\Sigma$ is evaluated at $\ekn$ rather than at the QP energy $\ekn{+}\delta\ekn$.
Eq.~(\ref{eq:e1shot}) is the customary way QP energies are evaluated in $GW$ calculations.  In Sec.~VI of
Ref.~\cite{mark06adeq}, we show that omitting the $Z$ factor is an approximate way to incorporate self-consistency, and
thus should be a better choice than including it.  As a practical matter it is also the case that results improve using
$Z$=1.

Ref.~\cite{mark06adeq} notes a serious drawback in the diagonal-only approximation: when levels in the reference
Hamiltonian are wrongly ordered as they are, e.g. at the $\Gamma$ point in Ge, InN, and CuInSe$_{2}$, and the L point in
PbTe~\cite{Svane10b} the wrong starting topology results in an unphysical band crossing in the \emph{GW} QP levels.
(See, for example an illustration for Ge in Fig. 6 of Ref.~\cite{mark06adeq}).  Something similar will happen for PbTe
(Fig.~\ref{fig:pbte}).  Moreover, the off-diagonal parts of $\Sigma$ can significantly modify the eigenfunctions and
resulting charge density.  This reflects itself in many contexts, e.g. strong renormalisation of the bandgaps in polar
insulators such as TiSe$_{2}$ and CeO$_{2}$ ~\cite{mark06adeq}.  It modifies orbital character of states near the Fermi
energy in La$_{2}$CuO$_{4}$, and significantly affects how the metal-insulator transition comes about, when \emph{GW} is
combined with DMFT~\cite{Acharya18}.

\subsection{Need for Self-consistency}
\label{sec:qsgw}

The arbitrariness in the starting point means that there is no unique definition of the \emph{GW} approximation.  Errors
in the theory (estimated by deviations from experiment) can be fairly scattered for a particular choice of reference,
e.g. LDA, making it unclear what the shortcomings of $GW$ actually are.  Arbitrariness in the starting point can be
surmounted by iterating $G$ to self-consistency, that is, by finding a $G$ generated by \emph{GW} that is the same as
the $G$ that generates it ($G^\text{out}$=$G^\text{in}$).  But it has long been known that full self-consistency can be
quite poor in solids~\cite{Shirley96,holm98}.  A recent re-examination of some semiconductors~\cite{Grumet18} confirms
that the dielectric function (and concomitant QP levels) indeed worsen when $G$ is self-consistent, for reasons
explained in Appendix A in Ref.~\cite{Kotani07}.  Fully sc$GW$ becomes more problematic in transition
metals~\cite{BelashchenkoLocalGW}.  Finally, even while sc\emph{GW} is a conserving approximation in the Green's
function $G$, in $W$ it is not: it violates the sum rule~\cite{Tamme99} and loses its usual physical meaning as a
response function.  As a result it washes out spectral functions in transition metals~\cite{BelashchenkoLocalGW}, often
yielding worse results than the LDA.

A simple kind of self-consistency is to update eigenvalues but retain LDA eigenfunctions, as was first done by
Aryasetiawan and Gunnarsson~\cite{AryasetiawanNiO}.  This greatly improves \emph{GW}, especially in systems such as NiO
where the LDA starting point is very poor.  But it is only adequate in limited circumstances.  In TiSe$_{2}$ it strongly
overestimates the bandgap (see below), as it does in NiO~\cite{AryasetiawanNiO}.  An extreme case is CeO$_{2}$ where the
position of the Ce $4f$ levels is severely overestimated; see Fig. 3 in Ref.~\cite{mark06qsgw}.  The off-diagonal
elements in $\Sigma$, which are needed to modify the eigenfunctions, can be very important.

Questaal employs \emph{quasiparticle} self-consistency~\cite{Faleev04,mark06qsgw,Kotani07}: by construction, the
reference $H_{0}$ is determined within \emph{GW} as in the fully self-consistent case.  But the proper $\Sigma{=}iGW$
cannot be used because it is energy dependent and non-Hermitian, thus falling outside of an independent particle
picture.  This causes \emph{GW} to degrade, and higher-order diagrams are needed~\cite{Shirley96} to restore the quality
of \emph{GW}.  But by quasiparticlising $\Sigma$ we can stay within a framework of perturbation around $H_{0}$ but
choose $H_{0}$ in an optimal manner.  QS\emph{GW} quasiparticlises $\Sigma(\varepsilon)$ by approximating it by a static
Hermitian potential as
\begin{align}
\Sigma_0 = \frac{1}{2}\sum_{ij}
        |\psi_i\rangle
        \left\{ {{\rm Re}[\Sigma({\varepsilon_i})]_{ij}+{\rm Re}[\Sigma({\varepsilon_j})]_{ij}} \right\}
        \langle\psi_j|
\label{eq:sigma0}
\end{align}
which is the QS\emph{GW} form for a static $V_\text{xc}$.  This process is carried out to self-consistency, until
$\Sigma_0^\text{\,out}=\Sigma_0^\text{\,in}$.  Given $\Sigma_0$, a new effective Hamiltonian can be made, which
generates a new $\Sigma_0$:
\begin{align}
{\Sigma_0} \to {G_0} \to \Sigma  = i{G_0}W[{G_0}] \to {\Sigma_0} \dots
\end{align}
$\Sigma_0$ is nonlocal with off-diagonal components; both of these features are important.

TiSe$_{2}$ is an interesting case where self-consistency becomes critical.  It is a layered diselenide with space group
$P\bar{3}m1$.  Below $T_{c}$=200K, it undergoes a phase transition to a charge density wave, forming a commensurate
$2{\times}2{\times}2$ superlattice ($P\bar{3}c1$) of the original structure.  It has attracted a great deal of interest
in part because the CDW may to be connected to superconductivity.
\begin{figure}[ht!]
\centering
\includegraphics[width=0.480\columnwidth]{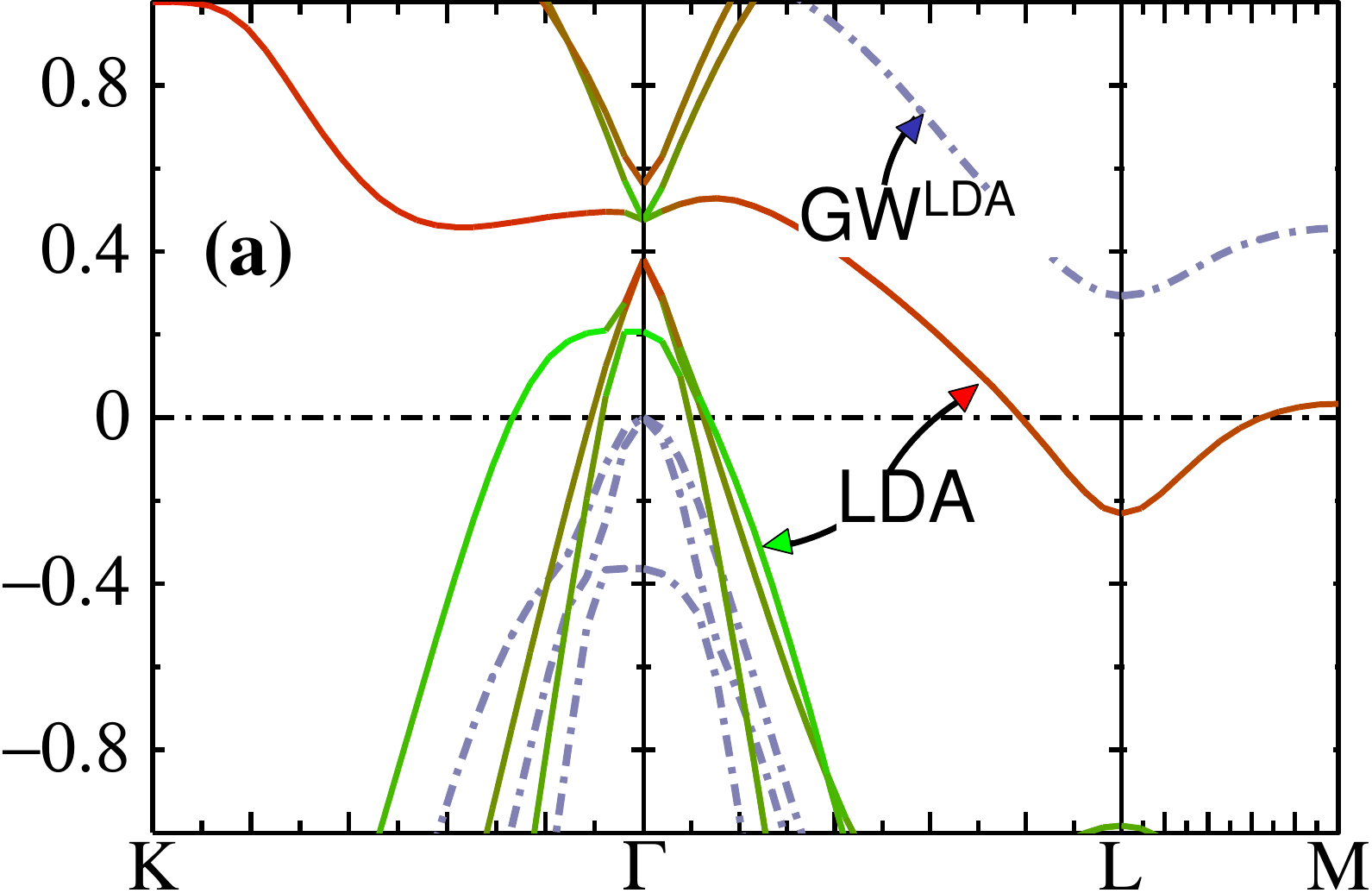} 
\includegraphics[width=0.435\columnwidth]{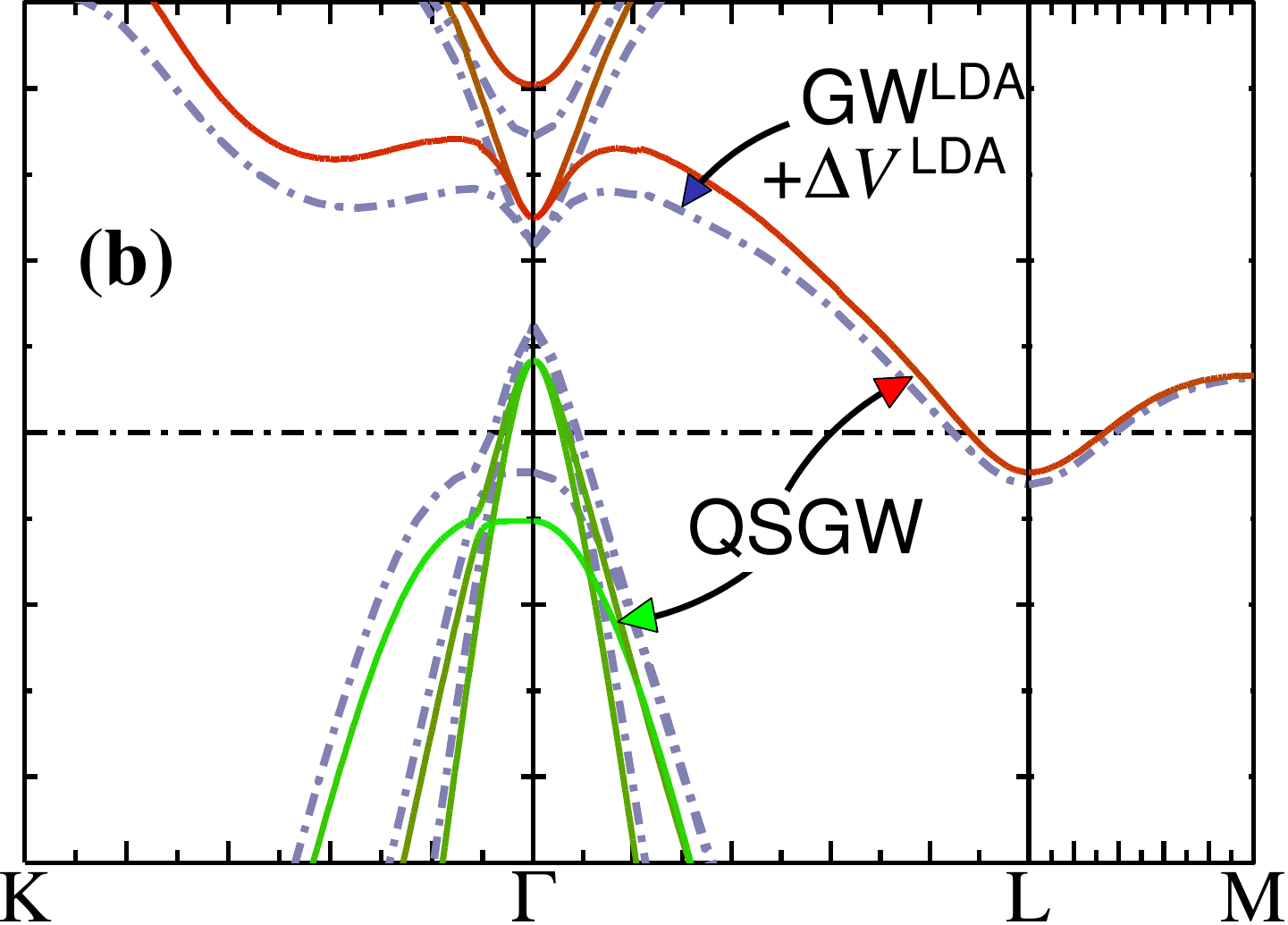}
\caption{TiSe$_2$ energy bands in eV for the undistorted $P\bar{3}c1$ structure.  $(a)$: solid lines are LDA results,
with red and green depicting a projection onto Ti and Se orbital character, respectively.  Blue dashed line shows shifts
calculated in the $GW$ approximation based on the LDA. $(b)$: blue dashed line shows results from $GW$ based on the LDA
(same self-energy as in panel $(a)$), with an extra potential $\Delta V^\text{LDA}$ deriving from a charge density shift
computed from the rotation of the LDA eigenvectors.  Solid lines are QS\emph{GW} results, with the same colour scheme as
in panel $(a)$.}
\label{fig:tise2}
\end{figure}

How the CDW affects the energy band structure is a source of great controversy.  It is accepted experimentally that in
the $P\bar{3}c1$ phase, TiSe$_{2}$ is semiconductor with a gap of $\sim$0.15\,eV.  Above $T_{c}$ whether intrinsic
TiSe$_{2}$ has a gap is not settled.  On the theoretical side, the CDW makes TiSe$_{2}$ an unusual materials system: DFT
predicts a metal in both $P\bar{3}m1$ and $P\bar{3}c1$, as expected, but a recent $\GLDA\WLDA$
calculation~\cite{Cazzaniga12} predicts $P\bar{3}m1$ to have a small positive gap.  We have revisited this problem and
confirmed the findings of Ref.~\cite{Cazzaniga12} at the $\GLDA\WLDA$ level.  However, at the QS\emph{GW} level,
TiSe$_{2}$ is a \emph{metal} in the ideal $P\bar{3}m1$ phase.  This is atypical for insulators: usually self-consistency
\emph{widens} the gap, as has long been known (see Fig.~1 in Ref.~\cite{mark06qsgw}).  The origin can be traced to the
modification of the LDA eigenfunctions by off-diagonal elements $\Sigma_{n \ne n'}$, which modify the density $n(r)$.
To see approximately the effect of the density change, assume LDA adequately describes ${\delta}V^\text{eff}/{\delta}n$
(this is the inverse susceptibility, the charge analogue of Eq.~(\ref{eq:eqmxb})).  For a fixed $\Sigma^{0}$, we can
estimate the effect of the change ${\delta}n$ renormalising the $V^\text{eff}$ through the change
\begin{align*}
\delta{\Sigma} \approx \left\{{\Sigma_0} - V_{\text{xc}}^{\text{LDA}}[{n^{\text{LDA}}}]{\text{ }}\right\}
        + V_{\text{xc}}^{\text{LDA}}[n^{GW}]
\end{align*}
This is accomplished in a natural way with the Questaal package.  The quantity in curly brackets is generated by
\emph{GW} in the first QS\emph{GW} cycle, and \texttt{lmf} treats this term as an external perturbation.  Running
\texttt{lmf} to self-consistency allows the system to respond to the potential and screen it, yielding $n^{GW}$.

The result of this process is shown as a dashed line in Fig.~\ref{fig:tise2}(b).  It bears a close resemblance to the
full QS\emph{GW} result, including the negative gap.  We will show elsewhere that TiSe$_{2}$ is an insulator in the
$P\bar{3}c1$ phase only as a consequence of lattice displacements relative to the symmetric $P\bar{3}m1$ phase.

\begin{figure}[ht]
\begin{center}
\includegraphics[width=0.24\columnwidth]{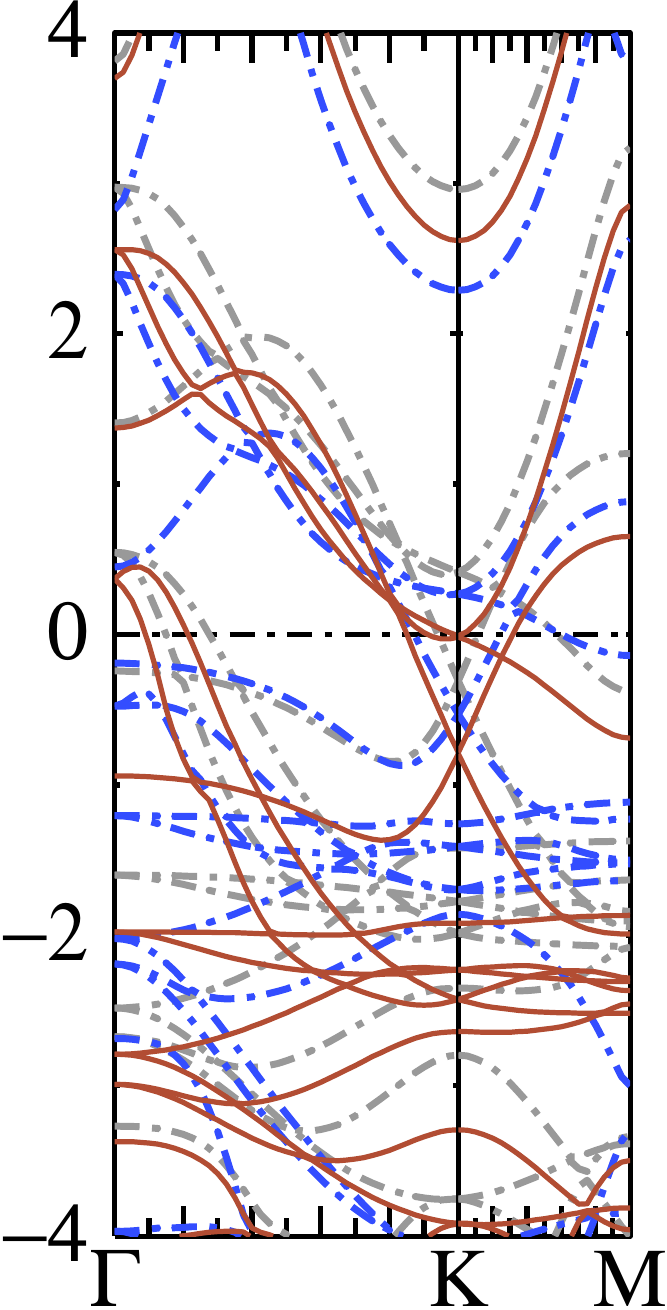}
\includegraphics[width=0.24\columnwidth]{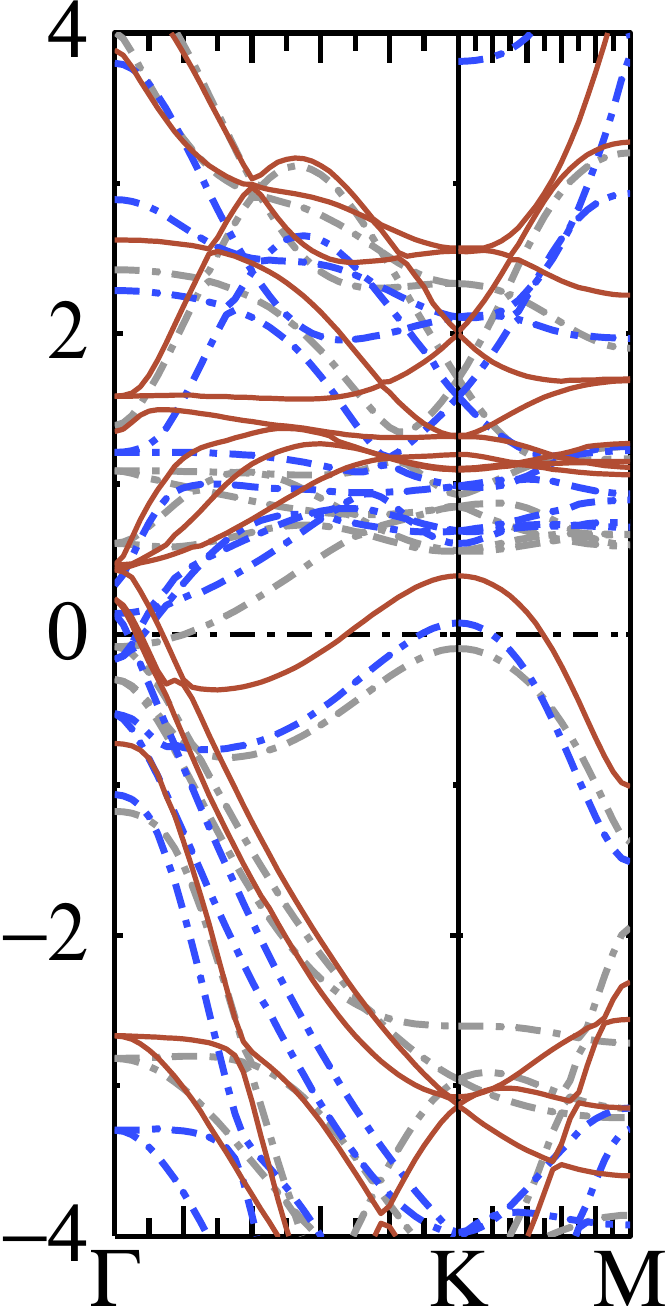}
\includegraphics[width=0.49\columnwidth]{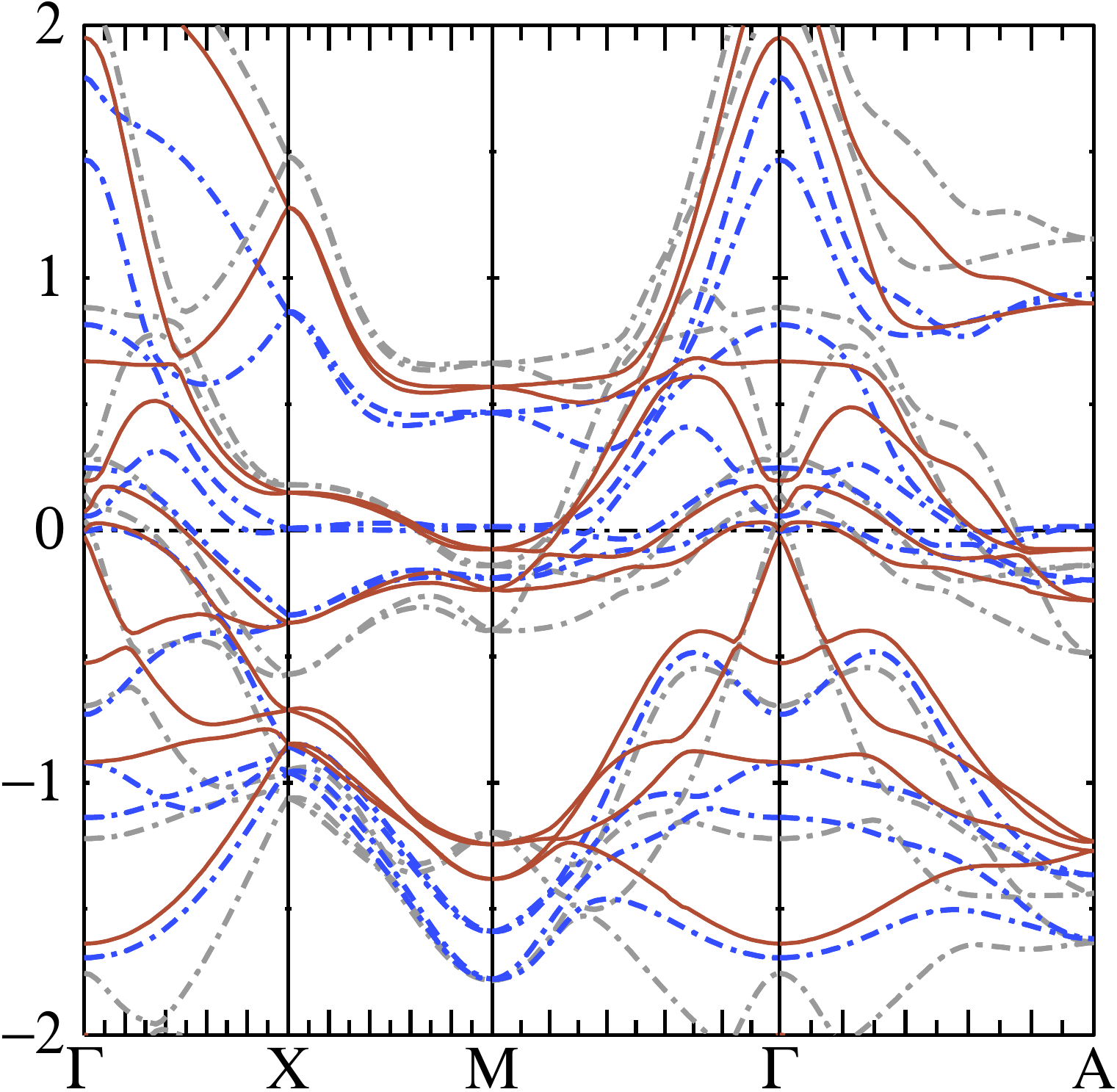}
\end{center}
\caption{Energy band structure, in eV, of majority-spin MnAs (left), minority-spin MnAs (middle) and nonmagnetic FeTe
(right) in (a) the QS\emph{GW} approximation (orange solid lines), $\GLDA\WLDA$ (dashed blue lines), and the LDA (light
dashed grey lines).  $\GLDA\WLDA$ was calculated from the LDA, but including the off-diagonal components of $\Sigma$
with $Z$=1.  It is the 0$^\mathrm{th}$ iteration of the QS\emph{GW} self-consistency cycle.}
\label{fig:fete}
\end{figure}

\emph{GW} is used less often in metals.  As in the insulating case, \emph{GW} based on DFT can sometimes work well, as
it apparently does for SrVO$_{3}$ \cite{Sakuma13}.  But the range of applicability is limited, and indeed $\GLDA\WLDA$
can yield catastrophically bad results.  Two cases in point are MnAs and FeTe.  Fig.~\ref{fig:fete} shows LDA,
$\GLDA\WLDA$, and QS\emph{GW} energy bands near $E_{F}$; they are mostly of Mn or Fe $3d$ character.  In the
$\GLDA\WLDA$ case, $E_{F}$ must be adjusted to conserve charge.  (The shift is large, of order 1\,eV.)  Consider MnAs
first.  The majority and minority $3d$ bands lie at about $-$1.5\,eV and +1\,eV in the LDA.  This exchange splitting is
underestimated: QS\emph{GW} increases the splitting, putting bands at about ($-$2.5,1.5)\,eV while $\GLDA\WLDA$ does the
opposite, reducing the splitting relative to LDA.  Also, the Fermi surface topology is poor: the hole pocket at $\Gamma$
disappears, and the one at K is similar to the LDA.  FeTe fares no better.  $\GLDA\WLDA$ is somewhere intermediate
between LDA and QS\emph{GW}.  Most important is the unphysical, dispersionless band at $E_{F}$ on the X-M line, which
yields a nonsensical Fermi surface.  In both of these materials, the LDA describes the electronic structure better than
$\GLDA\WLDA$.

\begin{figure}[ht]
\begin{center}
\includegraphics[width=0.75\columnwidth]{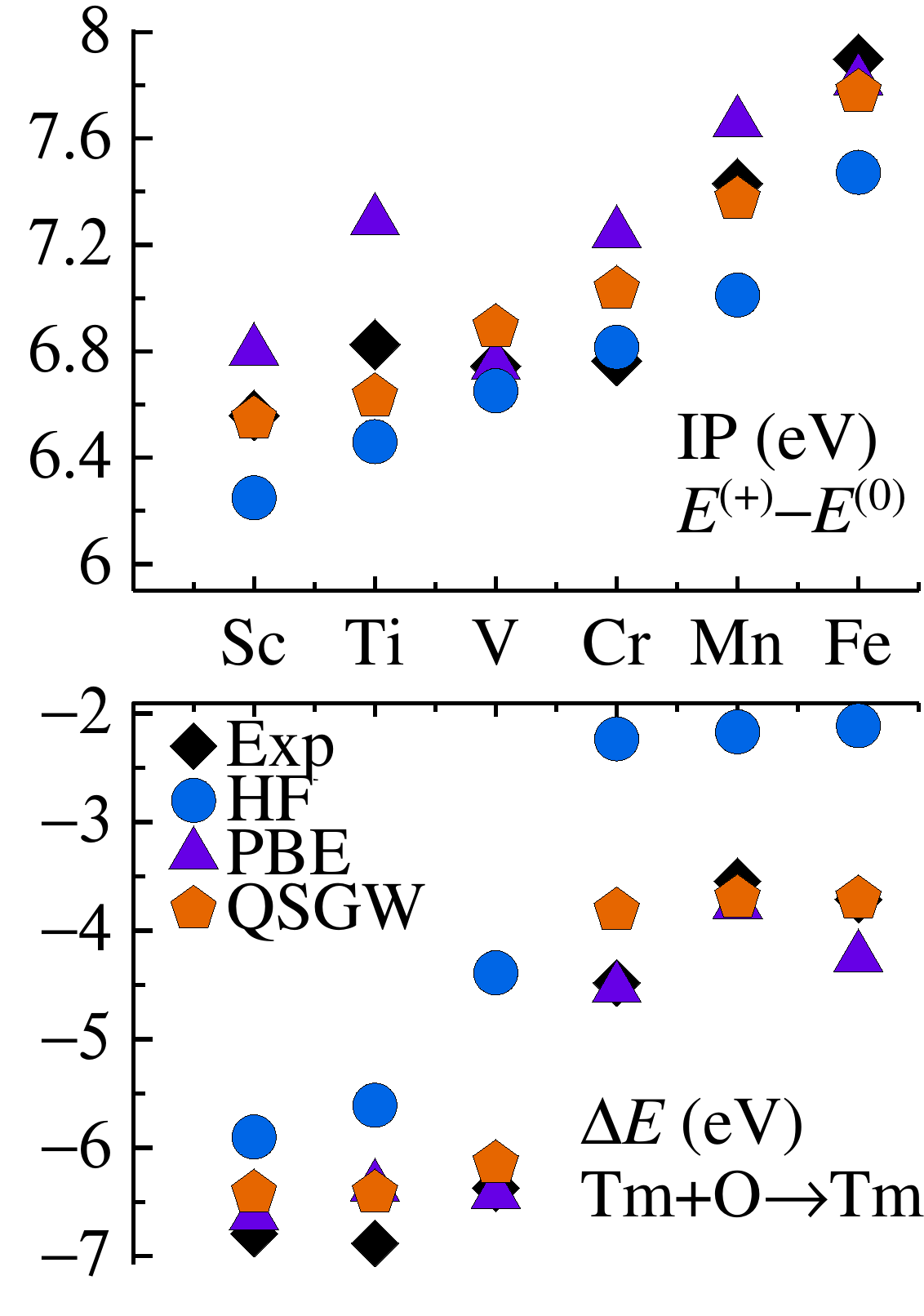} 
\caption{Ionisation potential, IP, of TM atom and heat of formation ${\Delta}E$ of transition-metal dimers computed
within the RPA, using the \texttt{molgw} code selecting various choices of starting point.  Note the dramatic difference
between Hartree-Fock and the PBE starting points.  Of the common semi-empirical functionals, HSE06 (not shown) is the
closest to QS\emph{GW}.  Data were compared to reference CCSD(T) for the ionisation potential and experiment for
${\Delta}E$.}
\label{fig:tmmolbenchmark}
\end{center}
\end{figure}

Ambiguities in the starting point make it difficult to know what errors are intrinsic to the theory and which are
accidental.  The literature is rife with manifestations of this problem; see e.g.
Ref.~\cite{PhysRevMaterials.2.075003}.  Self-consistency (mostly) removes the starting-point dependence so that the
errors intrinsic to the $GW$ approximation can be best elucidated.  Moreover, QS\emph{GW} should yield better RPA total
energies on average than those calculated from other starting points, because the path of adiabatic connection is
optimally described through QS\emph{GW}~\cite{Kotani07}.  To better illustrate these points
Fig.~\ref{fig:tmmolbenchmark} shows the ionisation potential of 3$d$ transition metal atoms and the heat of formation of
3$d$-O dimers, computed by the \texttt{molgw} code~\cite{Bruneval16}.  We focus on these properties because it is known,
e.g. from Ref.~\cite{OlsenRPA+ALDAEnergy}, that the RPA tends to systematically overbind, and the error is connected
with short-ranged correlations.  The ionisation potential and the dimer formation energy, both of which benefit from
partial cancellation of such errors, are much better described.

Note the dramatic difference between choice of a Hartree-Fock or the PBE functional.  As for the superiority of
QS\emph{GW}, Fig.~\ref{fig:tmmolbenchmark} generally bears this argument out, though there are some surprises, e.g.  the
heat of formation from RPA@PBE deviates less from CCSD(T) results than QS\emph{GW}.  QS\emph{GW} does not describe Cr
well, probably because spin fluctuations are important there (Sec.~\ref{sec:qsgwlimits}), which skews the statistics
with this small sample.  The discrepancies are small enough that incomplete convergence in the basis set (results in
Fig.~\ref{fig:tmmolbenchmark} were obtained with a triple-$\zeta$ basis) are of the same order and could account for
some of the results.  The ionisation potential of simple $sp$ atoms was analysed in Ref.~\cite{bruneval_jcp2012}, where
it was computed from RPA total energy differences, compared to the eigenvalue of the \emph{GW} one-particle Hamiltonian.
There also it was shown QS\emph{GW} significantly improves on RPA@HF, and also that QS\emph{GW} is slightly biased
towards Hartree-Fock, reflecting a slight tendency to underestimate screening.  This is consistent with the tendency for
QS\emph{GW} to overestimate bandgaps.  Koval \textit{et al.}\cite{koval_prb2014} found somewhat different results for
small, second-row molecules.

QS\emph{GW}-RPA is not accurate enough to reach quantum chemical accuracy, $\sim$1 kcal/mol. The addition of ladders
should considerably improve on the RPA's known inadequacy in describing short-ranged correlations.  It has long been
known that ladders dramatically improve the dielectric function $\epsilon(\omega)$ ($\epsilon(\omega)$ and \emph{G}
enter in the coupling constant integration for total energy) and it would be interesting to explore if such low-order
diagrams in a QS\emph{GW} framework are sufficient to capture total energy to this accuracy.

While self-consistency does largely surmount starting-point dependence (it was recently shown to be the case for some
insulators using Hartree-Fock and LDA as starting points~\citep{nora_thesis}), but it is not strictly true that it does.
The QS\emph{GW} $H_{0}$ or $G_{0}$ has a Hartree Fock structure, and it can get stuck in metastable valleys in the same
way as Hartree Fock; an example are the dual low-spin and high-spin states found for FeSe (Sec.~\ref{sec:qsgwlimits}).
Also, in strongly correlated systems $\Sigma(\varepsilon)$ can vary rapidly with energy.  There can be more than one
$\epsilon_i$ that forms a stationary point.  This happens, for example, in CuO.  It is usually an indication that the
ground state is not well described by a single Slater determinant. QS\emph{GW} restricts $G_0$ to a single determinant,
and there is no longer an unambiguously optimal choice when there are two or more equally strongly competing ones.

There is no unique prescription for quasiparticlisation, but Eq.~(\ref{eq:sigma0}) minimises the difference between the
full $G$ and $G_0$ according a particular choice of norm.  Recently it was shown~\cite{Beigi17} that
Eq.~(\ref{eq:sigma0}) minimises the absolute value of the gradient of the Klein functional over all possible spaces of
non-interacting Green's functions.  There are formal reasons ($Z$ factor cancellation~\cite{Kotani07} and conserving sum
rule~\cite{Tamme99}) why quasiparticlisation should be better; there is also abundant empirical evidence, that
quasiparticle self-consistency performs better than full self-consistency~\cite{BelashchenkoLocalGW,Kutepov17,Grumet18},
and the particular construction Eq.~(\ref{eq:sigma0}) is an optimum one.

QS\emph{GW} has been implemented in numerous electronic structure packages, e.g. the \texttt{spex}
code~\cite{Aguilera15}, the \texttt{Abinit} code~\cite{Shaltaf08}, \texttt{VASP}~\cite{Shishkin07}, recently in the
Exciting code~\cite{nora_thesis}, and also in molecules codes such as \texttt{molgw}~\cite{Bruneval16} and
\texttt{NWChem}~\cite{koval_prb2014}.  For solids perhaps the most rigorous implementation is the \texttt{spex} code,
which can use a local Sternheimer method to rapidly converge the calculation of the dielectric
function~\citep{Freidrich-dynamical-sternheimer}.  As attempts to make efficient schemes that are also well converged,
e.g. to calculate the RPA total energy, the local Sternheimer method will likely emerge as being very significant over
time.

\subsection{Questaal's QSGW Implementation}
\label{sec:questaalgw}

\begin{figure}[ht]
\begin{center}
\includegraphics[width=.3\textwidth]{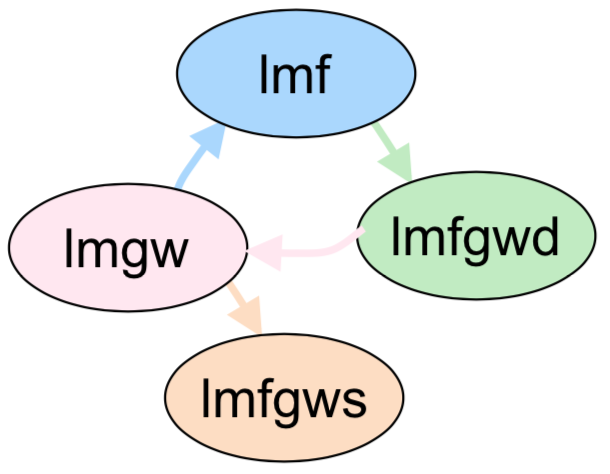}
\end{center}
\caption{Questaal's QS\emph{GW} cycle. \texttt{lmf} generates a new self-consistent density and noninteracting $H_{0}$
for a fixed $\Sigma_0$; \texttt{lmfgwd} generates information about the eigenfunctions $\Psi$ corresponding to $H_{0}$,
and \texttt{lmgw} receives the output of \texttt{lmfgwd} and uses it to make a new $\Sigma_0$.
\texttt{lmgw} can also generate response functions and the fully dynamical self-energy $\Sigma^{nn}(\omega)$.
\texttt{lmfgws} is a post-processing code that yields observables computed from the interacting $G$, where
$G^{-1}={G_0}^{-1}+\Sigma^{nn}(\omega)-{\Sigma_{0}}^{nn}$.}
\label{fig:qsgwcycle}
\end{figure}

Questaal's implementation of QS\emph{GW} evolved from the \texttt{ecalj} package, developed by Kotani and coworkers in
the early 2000's out of Aryasetiawan's \emph{GW}-ASA code~\cite{ferdi94}, which in turn was developed from the
``Stuttgart'' LMTO code.  The original \texttt{ecalj} package can now be found at
\url{https://github.com/tkotani/ecalj/}.  As of this writing we maintain the \emph{GW} code as a separate branch
(Fig.~\ref{fig:qsgwcycle}).  In its present form the \emph{GW} part of Questaal's QS\emph{GW} implementation operates
independently, receiving information about eigenfunctions and eigenvalues, and returning response functions or a
self-energy.  As for the QS\emph{GW} cycle, one cycle occurs in three parts.

Rather than storing $\Sigma_0$ on disk, $\Sigma_0-V_\text{xc}$ is stored ($V_\text{xc}$ usually being the LDA
potential).  Thus \texttt{lmf} generates its customary LDA Hamiltonian, and adds $\Sigma_0-V_\text{xc}$ to it.  In
practice this is accomplished by reading $\Sigma_0-V_\text{xc}$ on the mesh of $k$ points where it was generated, and
inverse Bloch-summing to make $\Sigma_0-V_\text{xc}$ in real space.  Then $\Sigma_0-V_\text{xc}$ can be interpolated to
any $k$ by a forward Bloch sum.  This is a unique and powerful feature, as it makes it possible to generate QS\emph{GW}
eigenfunctions and eigenvalues at any $k$.  There are some subtleties in this step: the interpolation does not work well
if all the $\langle n|\Sigma_0^{nn'}|n'\rangle$ are included.  It is solved by rotating to the LDA eigenfunctions, and
zeroing out $\Sigma_0^{n{\ne}n'}$ for $n$ or $n'$ whose energy exceeds a threshold, $E_\text{cut}$; i.e. diagonal-only
approximation for high-energy subblocks of $\Sigma_0$.  $E_\text{cut}$ is typically ${\sim}2$\,Ry; convergence can be
checked by varying $E_\text{cut}$.  The prescription is explained in detail in Sec.~IIG of Ref.~\cite{Kotani07}.  The
error is connected to the relatively long range of the smooth Hankel envelopes.  Preliminary tests using short-ranged,
screened Hankels indicate that this truncation is no longer needed.

Typically $\Sigma_0$ is a smoother function of $k$ than the kinetic energy.  Thus the $k$ mesh on which $\Sigma_0$ is
made can usually be coarser than the $k$ mesh.  Because of its ability to interpolate, \texttt{lmf} and \texttt{lmgw}
operate with independent meshes.

\subsection{Successes of QSGW}

QS\emph{GW} has the ability to calculate properties for a wide variety of materials classes in a manner that no other
single theory can equal.  It consistently shows dramatic improvement relative to DFT and extensions such as hybrid
functionals or LDA-based \emph{GW}.  Atomic ionisation potentials~\cite{bruneval_jcp2012}, quasiparticle (QP) levels,
Dresselhaus coefficients in semiconductors~\cite{Chantis06a,Chantis08a,Luo10}; band offsets at the Si/SiO$_{2}$
interface~\cite{Shaltaf08}, magnetic moments and spin wave excitations~\cite{KotaniSW08,Ke11}; tunnelling
magnetoresistance~\cite{Faleev12}; impact ionisation~\cite{KotaniII}; electric field gradients and deformation
potentials~\cite{Christensen10}, spectral functions~\cite{Sponza17}, and dielectric response~\cite{Cunningham18}.  Its
superior ability to yield QP levels and DOS made it possible to accurately determine valence maximum in
SrTiO$_{3}$~\cite{Chambers04}.  In contrast to {{${G^\text{LDA}}{W^\text{LDA}}$}}, QS\emph{GW} is uniformly reliable
with systematic errors (see Sec.~\ref{sec:qsgwlimits}).  Fundamental gaps are usually in very good agreement with
experiments, though they are systematically overestimated (see Fig.~1 in Ref.~\cite{mark06qsgw}), even in the strongly
correlated M1 phase of VO$_{2}$~\cite{Gatti07,Weber19,vo2}, where spin fluctuations are not important.  It properly
narrows the bandwidth of localised $d$ bands and widens it in wide-gap semiconductors~\cite{Chambers04} and
graphene~\cite{MvSgraphene11}.

\begin{figure}[ht]
\begin{center}
\includegraphics[width=.45\columnwidth]{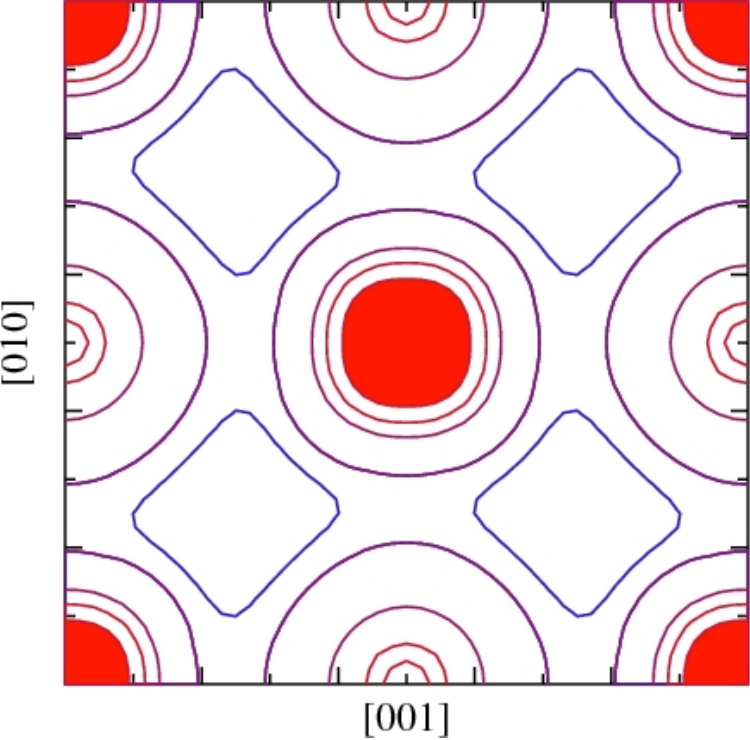}\quad 
\includegraphics[width=.45\columnwidth]{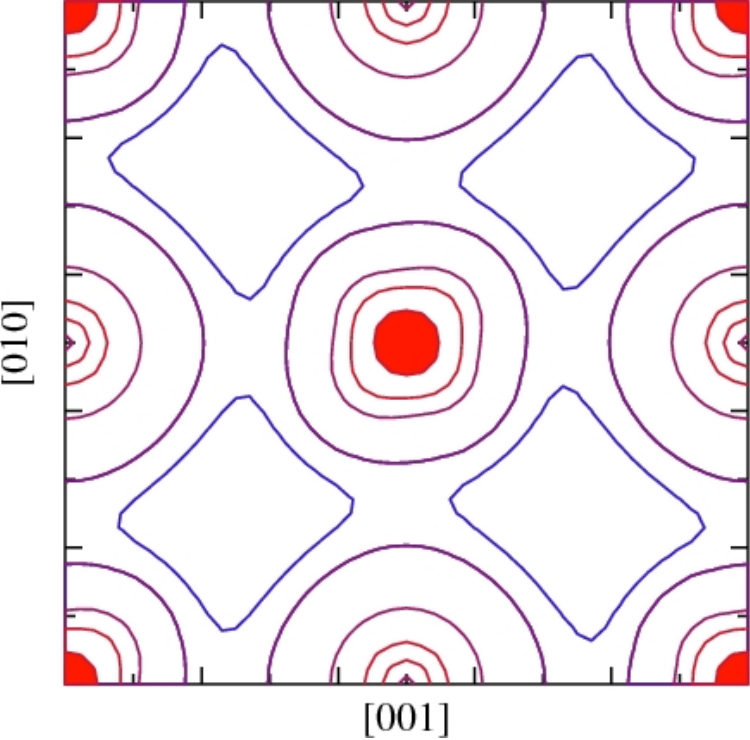}
\end{center}
\caption{Self-consistent charge density in antiferromagnetic NiO (left) and CoO (right), in the plane normal to [001],
computed by QS\emph{GW}.  The Ni (Co) nucleus lies at the centre of the Figure.}
\label{fig:niocoodensity}
\end{figure}

QS\emph{GW} can predict properties inaccessible to {{${G^\text{LDA}}{W^\text{LDA}}$}}, such as magnetic ground
state~\cite{Sponza17} and charge density.  Fig.~\ref{fig:niocoodensity} shows the density calculated by QS\emph{GW} in
the plane normal to [001] for antiferromagnetic NiO and CoO.  Two prominent features are seen: (1) NiO is much more
spherical than CoO, and the density contours for CoO are elongated along the [110] line.  Both of these findings are
consistent with $\gamma$ ray measurements~\cite{Jauch02,Jauch04}.

\begin{figure}[ht]
\begin{center}
\includegraphics[width=0.65\columnwidth]{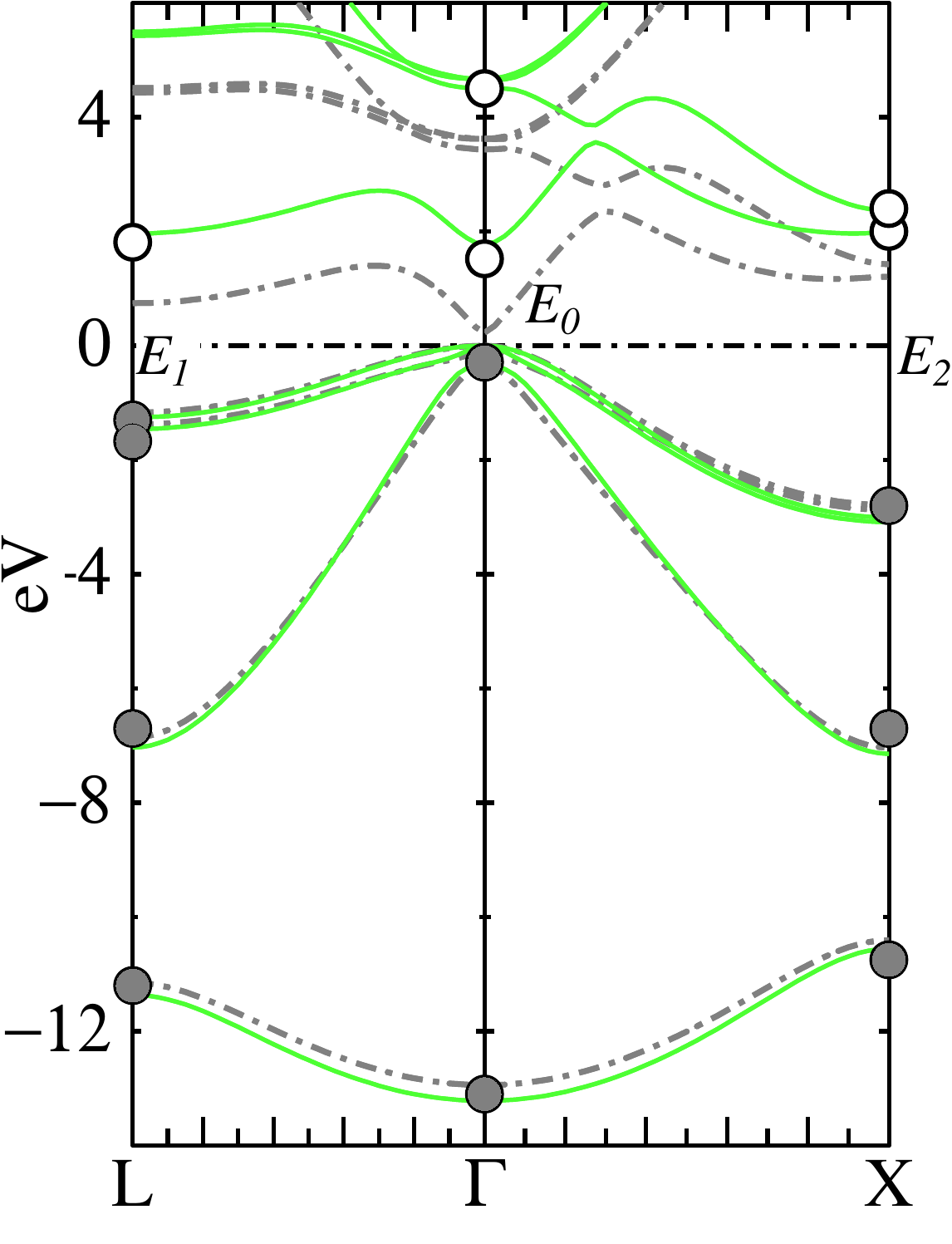}
\end{center}
\caption{QS\emph{GW} energy band structure of GaAs compared with ellipsometry data and photoemission data (open and
closed circles).  Dashed Grey lines are LDA bands.}
\label{fig:qsgw-gaas-bands}
\end{figure}

It was noticed early on~\cite{mark06qsgw} that QS\emph{GW} has a systematic tendency to overestimate bandgaps in
semiconductors and insulators.  To what extent dispersions are well described is much less discussed, in part because
there is a limited amount of experimental data reliable to the precision needed.  Perhaps the best materials family to
serve as a benchmark are the zincblende semiconductors, where critical-point analysis of the dielectric function has
been used to accurately measure in most zincblende semiconductors, not only the $\Gamma$-$\Gamma$ ($E_{0}$) transition
but also the L-L ($E_{1}$) and X-X ($E_{2}$) transitions, as indicated in Fig.~\ref{fig:qsgw-gaas-bands}, which presents
the QS\emph{GW} and LDA bands compared with ellipsometry and photoemission data.

\begin{figure*}[ht]
\begin{center}
\includegraphics[height=5.3cm]{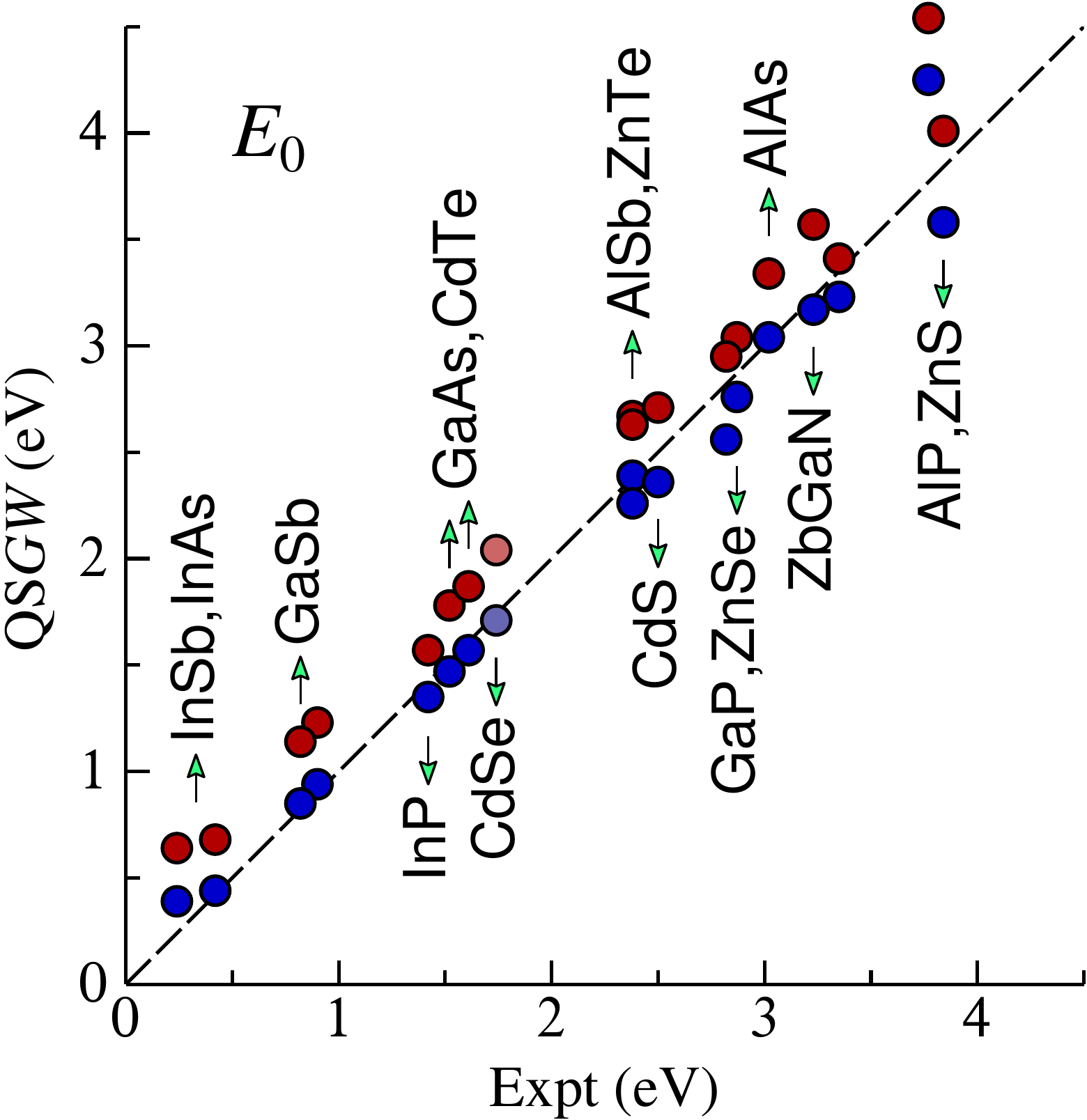}
\hspace{1em}
\includegraphics[height=5.3cm]{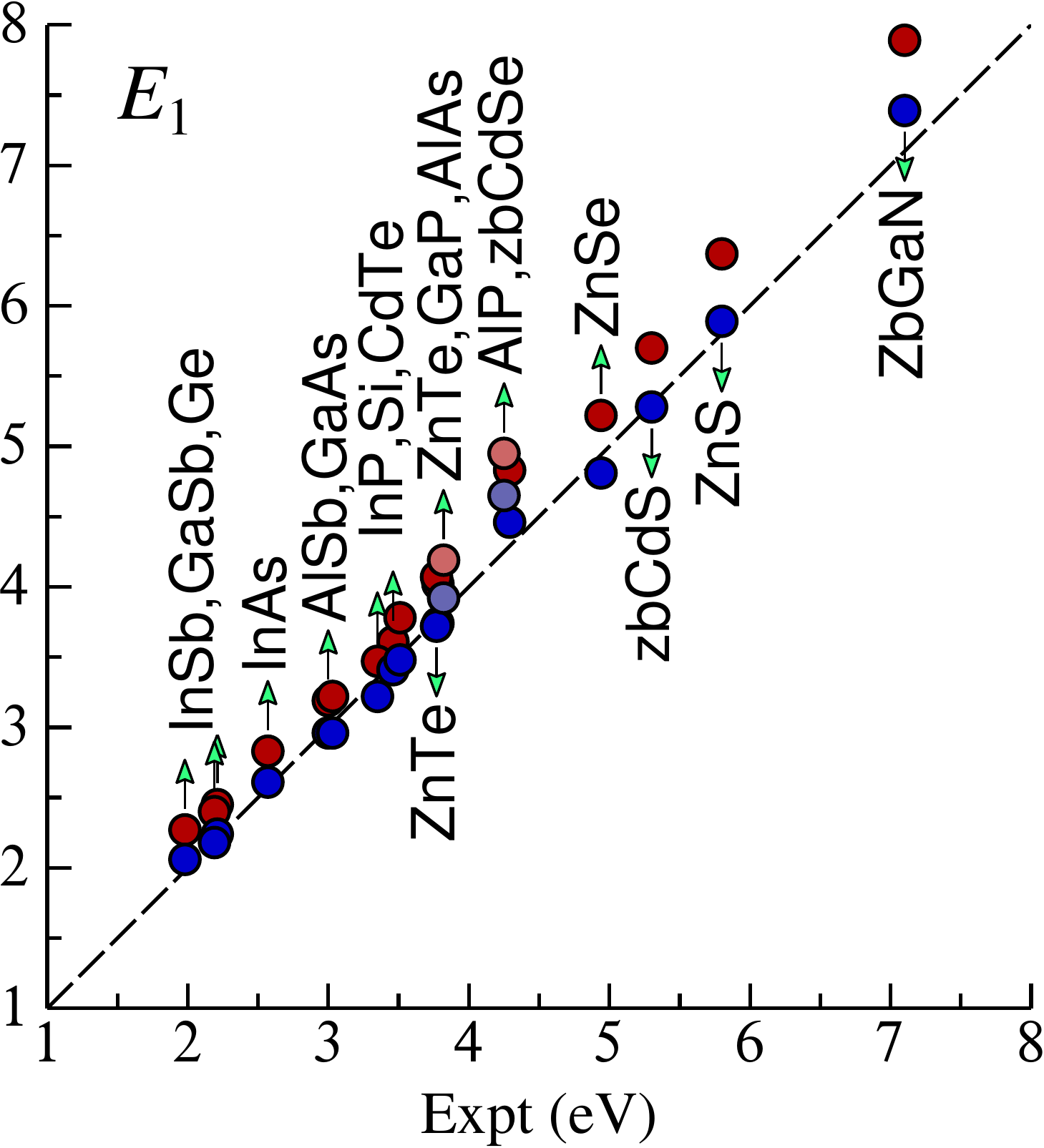}
\hspace{1em}
\includegraphics[height=5.3cm]{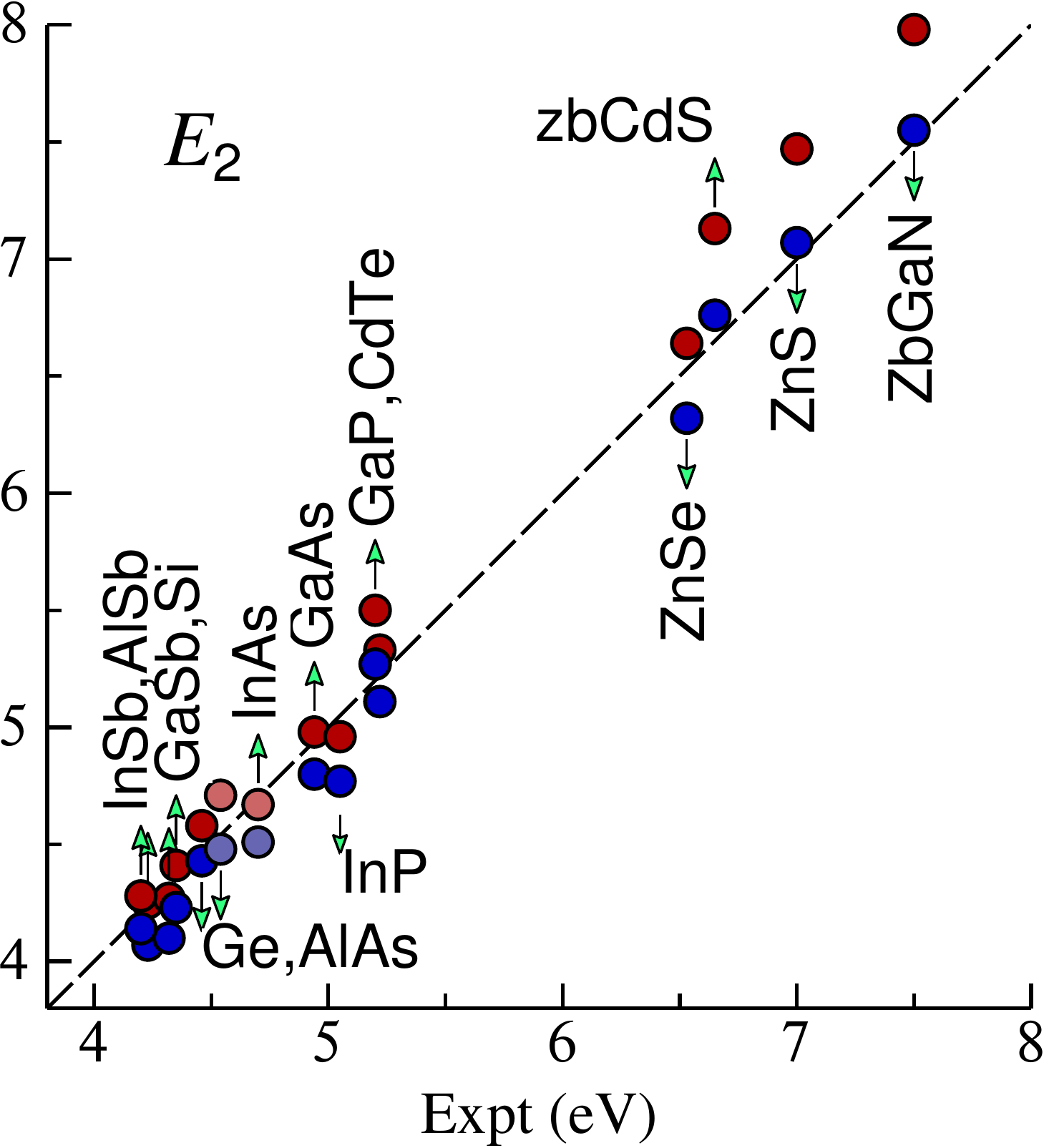}
\end{center}
\caption{Transitions $E_{0}$, $E_{1}$ and $E_{2}$ transitions in a zincblende semiconductor compared with critical point
measurements.  Red and blue circles are QS\emph{GW} results, and QS\emph{GW} with $\Sigma_0^\text{scaled}$
(Eq.~(\ref{eq:sigmascaled})), respectively.}
\label{fig:directsemi}
\end{figure*}

Fig.~\ref{fig:directsemi} shows that tendency to overestimate $E_{0}$ and $E_{1}$ is systematic and uniform, but there
is less uniformity in $E_{2}$.  Usually $E_{0}$ can be precisely measured; similarly for $E_{1}$ because there is a
sizeable volume of $k$ near L where valence and conduction bands disperse in a parallel manner.  $E_{2}$ is more
difficult, and values are less certain.  That being said, the data suggest $E_{0}{-}E_{2}$ is, on average, about 0.1\,eV
below experiment.  We have found that this error is, at least in part, due to the necessity of truncating high-lying
parts of the off-diagonal $\Sigma_0$; see $E_\text{cut}$, Sec.~\ref{sec:qsgw}.  It has recently become possible to
re-evaluate this because a short-range basis was recently developed (Sec.~\ref{sec:tblmf}) appears to interpolate
$\Sigma_0$ allowing $E_\text{cut}{\to}\infty$.

\begin{figure}[ht]
\begin{center}
\includegraphics[height=5.3cm]{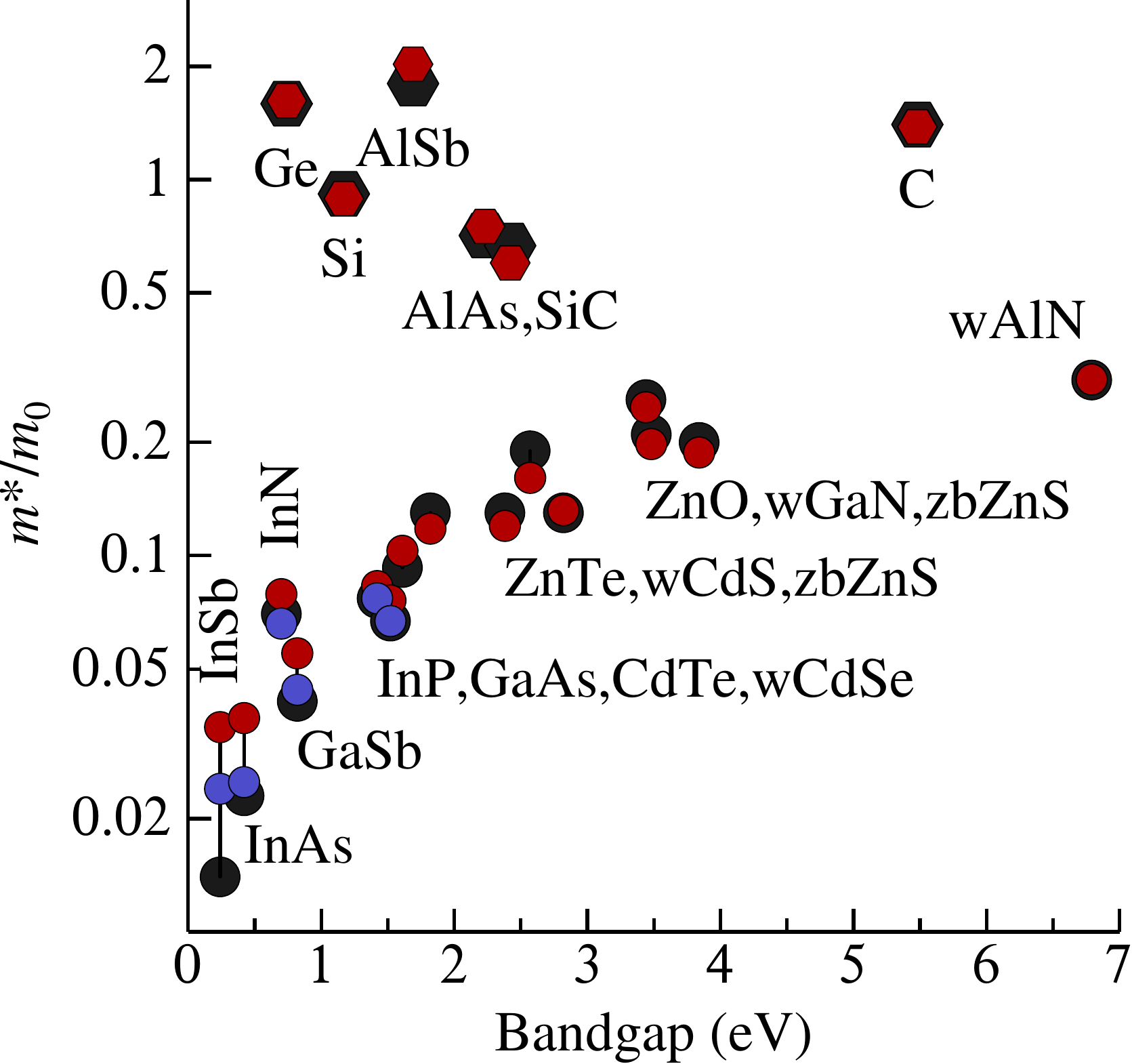}
\end{center}
\caption{Conduction band effective masses of selected zincblende compounds compared against experiment.
Red and blue circles are QS\emph{GW} results, and QS\emph{GW} with $\Sigma_0^\text{scaled}$; black
circles are experimental values.}
\label{fig:mstar}
\end{figure}

Also reliably measured are the electron effective mass $m_e^*$ (hole masses are much less well known), and for some
systems, the nonparabolicity parameter $\alpha$, characterising deviations from parabolicity near $\Gamma$, and defined
as $k^2/m^*=\varepsilon(k)[1-\alpha\varepsilon(k)]$ where $\varepsilon$ is the band energy relative to the conduction
band minimum.  $\alpha$ is an important quantity in several contexts, particularly hot-electron semiconductor
electronics.  Unfortunately $\alpha$ is difficult to measure, and values reported are usually some scalar average of the
second rank tensor.  Where it has been measured QS\emph{GW} (or scaled QS\emph{GW}, Eq.~(\ref{eq:sigmascaled})) seems to
predict $\alpha$ to within the experimental resolution; see e.g. Table~\ref{tab:gaas_lda_gw} for GaAs.
Fig.~\ref{fig:mstar} compares QS\emph{GW} and experimental effective masses for various semiconductors.  Conduction band
masses are well described, although there is a systematic tendency to underestimate it, connected to the tendency to
overestimate $E_{G}$.

\begin{figure}[ht]
\begin{center}
\includegraphics[height=5.3cm]{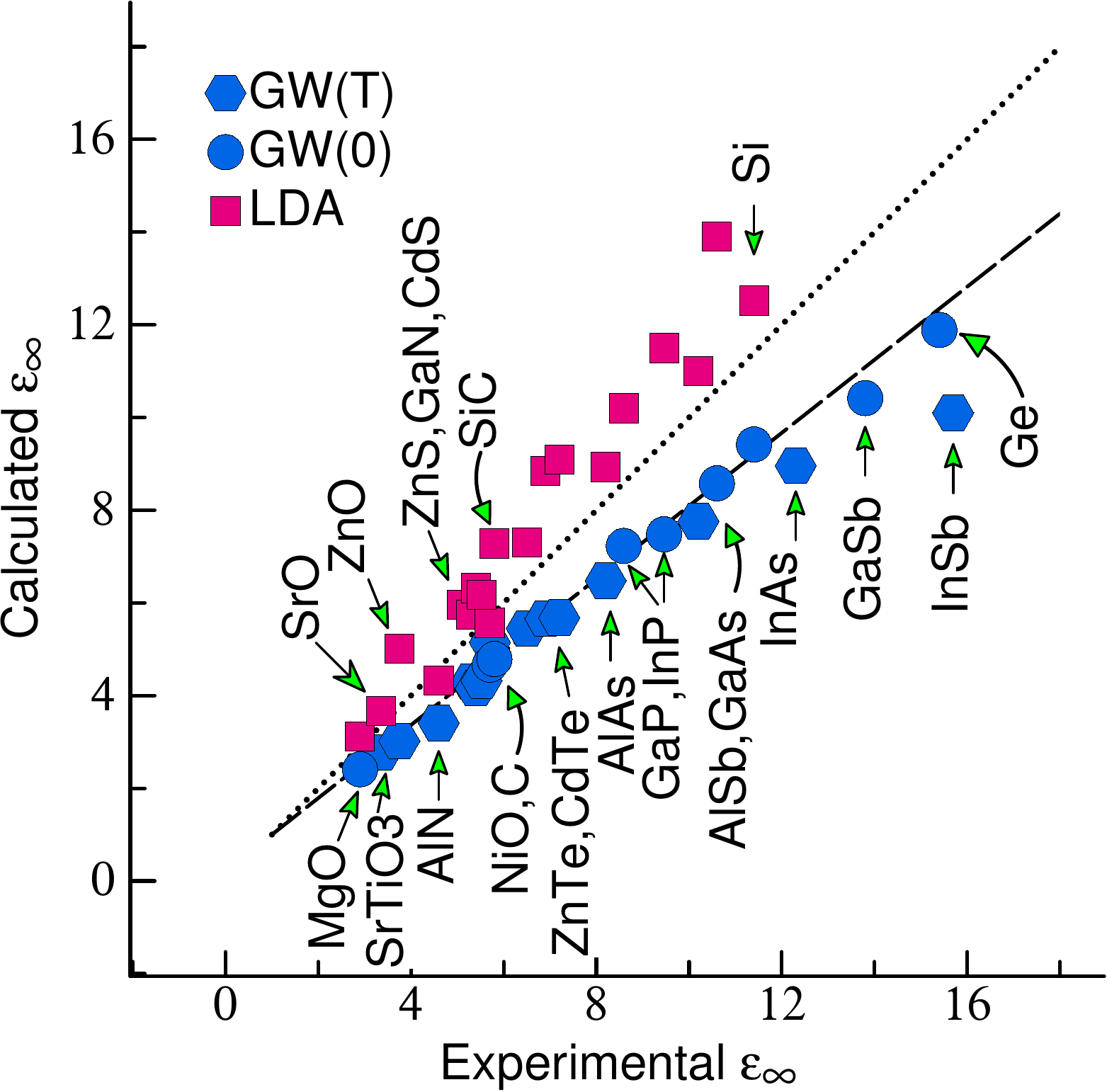}\\
\includegraphics[width=.98\columnwidth]{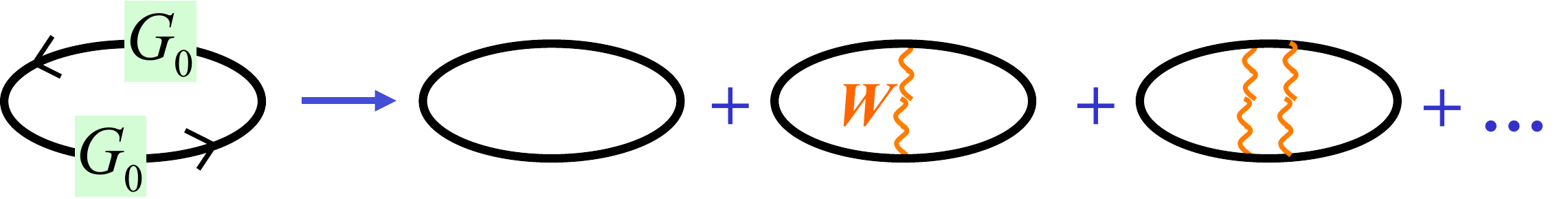}
\end{center}
\caption{Dielectric constant $\epsilon_\infty$ compared to experiment.  Blue are QS\emph{GW} results; red are LDA
results.  Also shown are the ladder diagrams that modify bubbles in the polarisability, left out in the RPA.}
\label{fig:eps}
\end{figure}

Most of this error can be traced to the RPA approximation to the dielectric function.  The optical dielectric constant
$\epsilon_{\infty}$ (Eq.~(\ref{eq:epsilonm})) in the $\omega{\to}0$ limit is uniformly too small by a factor ${\sim}0.8$
for many kinds of semiconductors and insulators (Fig.~\ref{fig:eps}).  Plasmon peaks in $\text{Im}\epsilon(\omega)$ are
almost universally blue shifted~\cite{Albrecht98,Rohlfing98} because the RPA omits the attraction between electron-hole
pairs in their virtual excitations.  A simple Kramers-Kronig analysis shows that as a consequence,
$\epsilon_{\infty}{=}\mathrm{Re\,}\epsilon(\omega{\to}0)$ should be underestimated.  Indeed we find that to be the case,
but remarkably $\epsilon_{\infty}$ is consistently underestimated, by a nearly universal factor of 0.8.  This is what
motivated a hybrid of LDA and QS\emph{GW} functionals~\cite{Chantis06a}
\begin{align}
\Sigma_0^\text{scaled} = 0.8\,\Sigma_0 + 0.2\,V^\text{LDA}_\text{xc}
\label{eq:sigmascaled}
\end{align}
The reasoning is since $W$ is dominated by the $q{\to}0,\omega{\to}0$ limit, scaling $W$ by 0.8 justifies
Eq.~(\ref{eq:sigmascaled}).  In practice, it seems that the tendency to overestimated bandgaps is almost completely
ameliorated in these systems (compare red and blue circles, Fig.~\ref{fig:directsemi}, as is the tendency to
underestimate $m_{c}^*$.  Exceptions are the small-gap InAs in InSb, where relative errors in bandgaps are still not
small even with the 0.8 scaling.

The hybrid scheme does a stellar job in many contexts: it improves on bandgaps in semiconductors
(Fig.~\ref{fig:directsemi}) and predicting the Dresselhaus splitting in \emph{sp} semiconductors~\cite{Chantis06a}.  But
the scheme is empirical, and it is limited.  It does not properly correct the blue shifts in $\epsilon(\omega)$ or take
into account other excitonic effects and fails to bring bandgaps in close agreement with experiment in systems with
strong spin fluctuations such as CoO or La$_{2}$CuO$_{4}$, and deep states such as the Ga $3d$ shift farther from
experiment (see Table below).  Recent work~\cite{Kutepov17,Cunningham20} shows that the great majority of the error in
charge susceptibility can be accounted for by ladder diagrams.  (In the case of polar semiconductors there can also be a
significant renormalisation of the gap from the Fr\"olich interaction~\cite{Frohlich54,Lambrecht17}; the electron-phonon
interaction has a modest effect in other semiconductors, especially compounds with second row elements~\cite{Cardona05},
diamond being the largest.) Without ladders the low-frequency dielectric constant is --- almost universally --- about
80\% of experiment (~\cite{Bhandari18}).  As we will show elsewhere~\cite{Cunningham20} the addition of ladders in
\emph{W} seem to dramatically improve on the charge channel even in strongly correlated cases such as CoO.  VO$_{2}$ is
an excellent test bed for optics: it has strong correlations in the monoclinic phase; yet, the conductivity is well
described by nonmagnetic QS\emph{GW}, provided ladders are taken into account~\cite{Weber19}.

Below we present more detailed properties of the band structure obtained from classical QS\emph{GW} on a single system,
selecting GaAs because reliable measurements are available and it is representative of results obtained for the entire
family of zincblende semiconductors (Table~\ref{tab:gaas_lda_gw}).  We observe the following:
\begin{enumerate}
\item dispersions $\Gamma$-L and $\Gamma$-X are mostly well described, and vastly
better than the LDA (which predicts the $\Gamma$-X dispersion to be 1\,eV);
\item the valence bands match photoemission data to within experimental resolution;
\item the Ga $3d$ (see Table below) is pushed down relative to the LDA and also relative to
$\GLDA\WLDA$.  Binding of shallow core-like states on e.g. Cd, Zn, and Cu is underestimates by ${\sim}0.4$\,eV; for
deeper states such as Ga $3d$ it is larger.  It was shown in Ref.~\cite{Gruneis14} that a low-order vertex correction to
\emph{GW} accounts for most of this discrepancy, at least for shallow core-like levels;
\item conduction bands are slightly overestimated and the mass slightly too large; and
\item the nonparabolicity parameter $\alpha$ was calculated along the [001], [110], and [111]
directions.  It varies $\sim$30\% for different directions, but since only scalar quantities are reported, we take a
geometrical average.  $\alpha$ is very sensitive to $m^*$, so we can expect it to be underestimated a little; this turns
out to be the case.  Using Eq.~(\ref{eq:sigmascaled}) all three quantities align well with experiment.
\end{enumerate}

\begin{table}
\begin{tabular}{|c|c|c|c|c|}
\hline
                           & $E_G$ & $m^*/m$ & $\alpha$ & Ga 3$d$    \\
\hline
QS\emph{GW}                & 1.78  & 0.076   & $-$0.71  & 18.5, 18.0 \\
Eq.~(\ref{eq:sigmascaled}) & 1.47  & 0.067   & $-$0.82  & 17.8, 17.3 \\
LDA                        & 0.24  & 0.020   & $-$4.0   & 15.2, 14.8 \\
Expt                       & 1.52  & 0.067   & $-$0.83  & 19.3, 18.6 \\
\hline
\end{tabular}
\label{tab:gaas_lda_gw}
\caption{GaAs: fundamental gap, effective masses, nonparabolicity $\alpha$ and location of 5/2 and 3/2 Ga $3d$ core
states relative to the valence band maximum.  Units are eV.  The measurement of $\alpha$ was taken from
Ref.~\cite{Heiblum87}, which is within 1\% of a subsequent measurement~\cite{Ruf90}.  Photoemission data taken from
Ref.~\cite{Cardona72}.}
\end{table}

Other discrepancies become apparent in the rare earths: splitting between occupied and unoccupied $4f$ levels is too
large~\cite{Chantis07c}, and multiplet splittings, which are significant for $4f$, lie beyond the scope QS\emph{GW}.

As we will show elsewhere~\cite{Cunningham20} most of the systematic errors noted above are largely corrected when
ladders are included in $W$.  Kutepov~\cite{Kutepov17} showed that the bandgaps significantly improve, but the
improvement extends to a wide range of properties and very diverse kinds of materials systems.  See for example, the
excellent description of optical conductivity in VO$_{2}$~\cite{Weber19}.

Metals and local-moment magnetic systems are similarly well described; see for example the excellent description of
known properties of Fe in the Fermi liquid regime in Ref.~\cite{Sponza17}, and the electronic structure of NiO and MnO
in Ref.~\cite{Faleev04}.

In summary, absent strong spin fluctuations, and a few other mild exceptions, QS\emph{GW} with some low-order extensions
(ladders and a low-order vertex for narrow semicore states), describe a wide range of materials throughout the periodic
table with uniform accuracy and reliability that cannot be matched by any other method.

\subsection{Limitations of QSGW}
\label{sec:qsgwlimits}

\begin{figure}[ht]
\begin{center}
\includegraphics[width=.48\columnwidth]{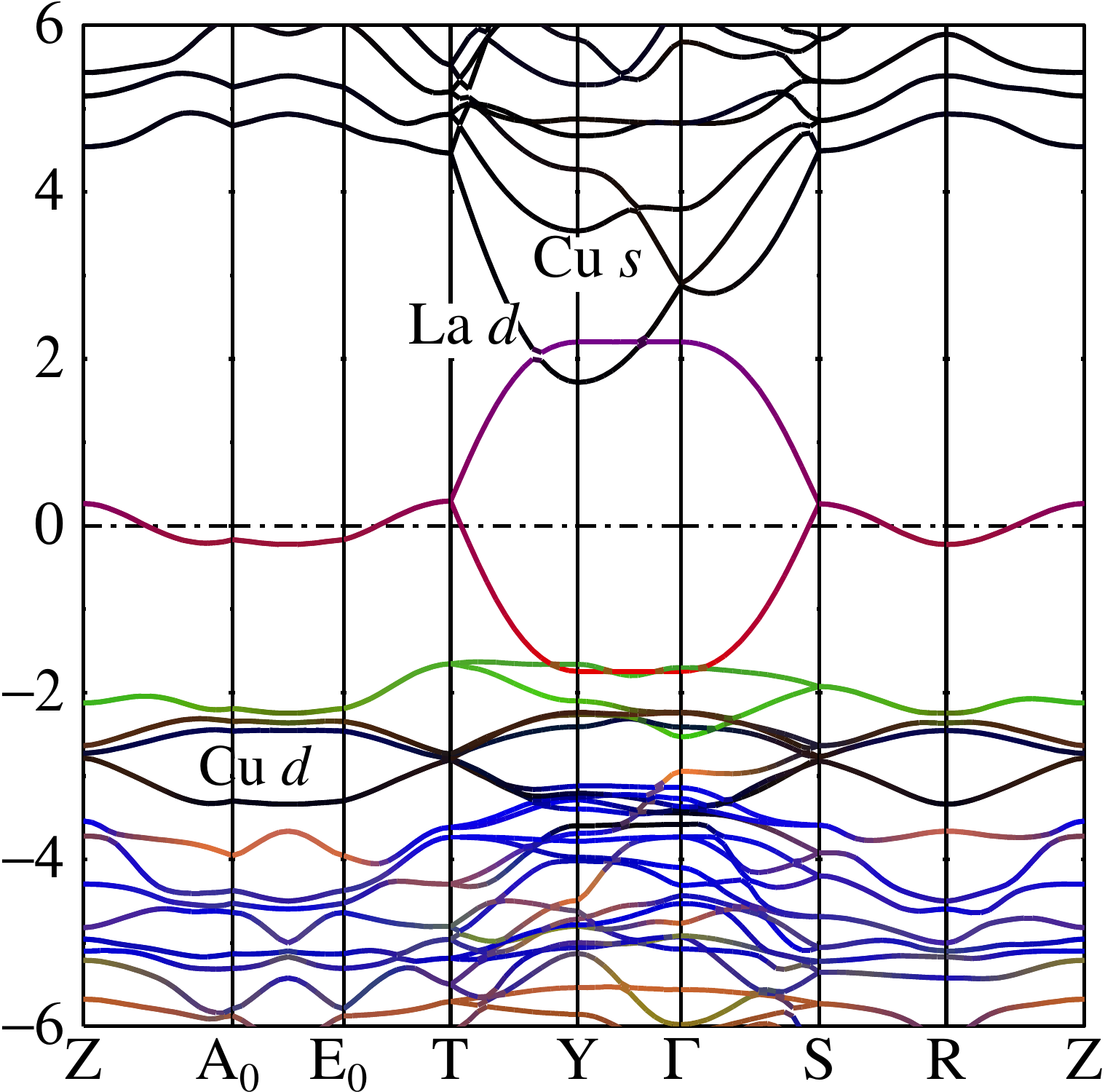}
\includegraphics[width=.48\columnwidth]{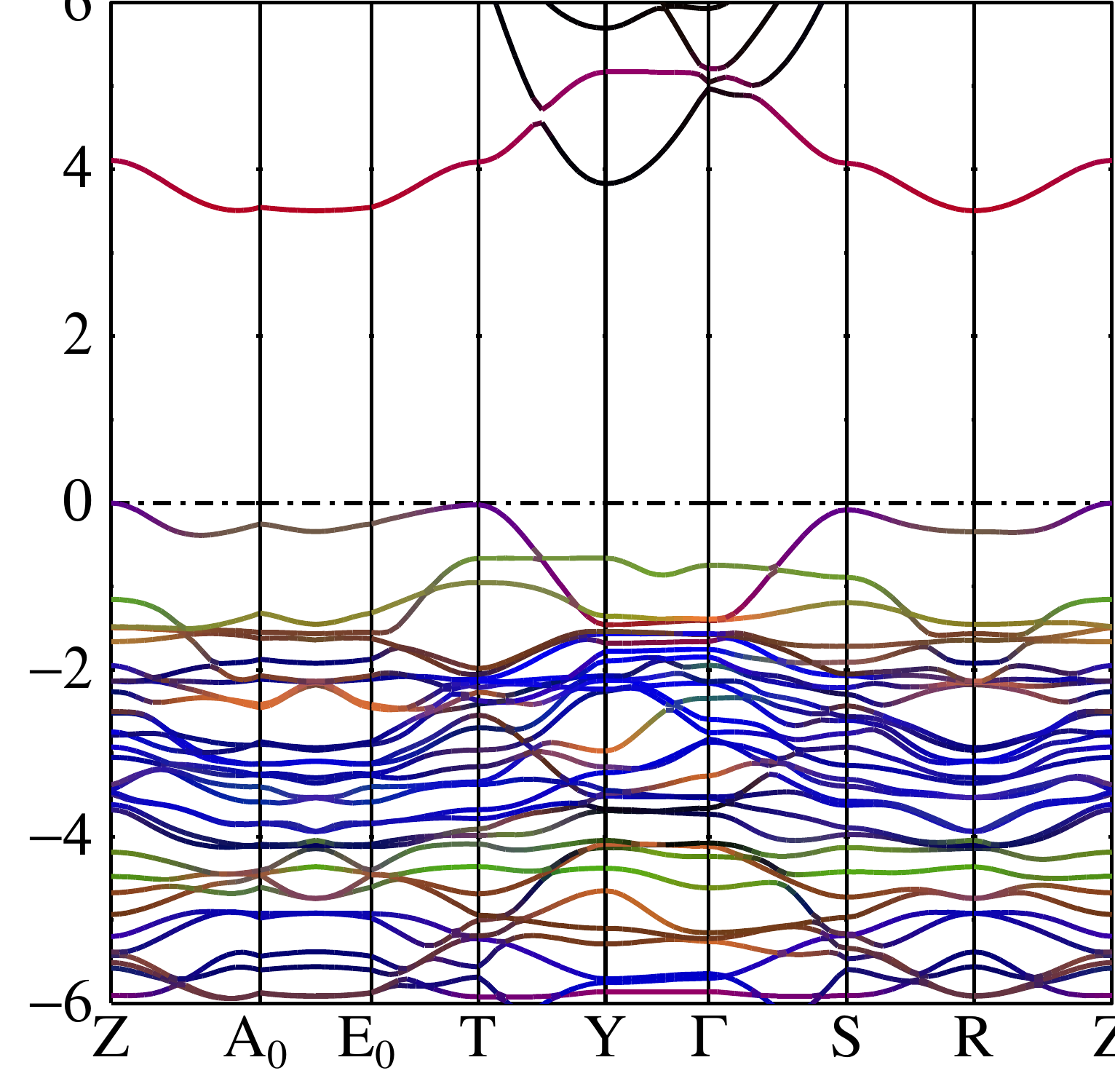}
\end{center}
\caption{
QS\emph{GW} band structure (in eV) of nonmagnetic La$_{2}$CuO$_{4}$ (left) and
antiferromagnetic LaCu$_{2}$O$_{4}$ (right).  Colours reflect the orbital character of the bands as follows:
Cu $d_{x^{2}{-}y^{2}}$ (red);
Cu $d_{3z^2-1}$ (green), near $-$2\,eV in QS\emph{GW};
O-$p_{xy}$ (blue), between $-$3\,eV  and $-$5\,eV in QS\emph{GW};
O-$p_{z}$ (orange), between $-$3\,eV  and $-$5\,eV in QS\emph{GW}.
The band near 3.3\,eV at Y has significant Cu $s$ character,
and it also admixes into the Cu $d_{x^{2}{-}y^{2}}$ near $E_{F}$.}
\label{fig:lsco}
\end{figure}

The greatest failings appear in QS\emph{GW} for systems where spin fluctuations are large.  This is not surprising since
the main many-body effect in \emph{GW} are plasmons.  \emph{GW} has no diagrams in spin beyond the Fock exchange.  When
spin fluctuations become important, the first effect is to reduce the average moment~\cite{Moriya85,Aguayo04}.  As they
increase, states begin to lose their coherence: the band picture and QS\emph{GW} both begin to break down.  A good
example of this is La$_{2}$CuO$_{4}$ (Fig.~\ref{fig:lsco}).  Undoped La$_{2}$CuO$_{4}$ undergoes a paramagnetic to an
antiferromagnetic insulator transition upon cooling below 325 K~\cite{nishihara}.  Fig.~\ref{fig:lsco} shows QS\emph{GW}
calculations in the nonmagnetic state, and the antiferromagnetic one.  Approximating the paramagnetic state by
nonmagnetic state is an inadequate approximation, though it is often done.  In this approximation La$_{2}$CuO$_{4}$ is
predicted to be metallic, with a single Cu $d_{x^2-y^2}$ band crossing $E_{F}$.  In the antiferromagnetic state
La$_{2}$CuO$_{4}$ is found to be an insulator with a 3.5\,eV bandgap.  Experimentally La$_{2}$CuO$_{4}$ is an insulator
with a gap $\sim$2\,eV.

QS\emph{GW} severely overestimates the optical gap in La$_{2}$CuO$_{4}$.  There is a similar mismatch for CoO: the gap
is overestimated by $\sim$2\,eV in these cases, but not in NiO.  Particularly telling is how for VO$_{2}$ nonmagnetic
QS\emph{GW} \emph{does} describe the M1 phase very well when both V-V pairs dimerise~\cite{Weber19}, but not the M2
phase where only one of them does.  (A gap opens up when M2 is calculated antiferromagnetically, but M2 is probably
paramagnetic at room temperature.)  M1 and M2 differ mainly through the V-V dimerisation, which suppresses spin
fluctuations.  As we will show elsewhere~\citep{Cunningham20}, including ladders considerably improves all of these
antiferromagnetic insulators, but discrepancies remain particularly for La$_{2}$CuO$_{4}$.  We believe this to be an
artefact of the interaction between spin and charge fluctuations.
\begin{table}[ht!]
\setlength{\tabcolsep}{4.8pt}
\begin{center}
\begin{tabular}{|l|rr|rr|rr|}
\hline
&          \multicolumn{2}{c|}{$\Gamma$} & \multicolumn{2}{c|}{M} & \multicolumn{2}{c|}{Z}\\
\hline
ARPES       & 9   & $-$18   & $-$22  & $-$42   & 7   &  34 \\
\hline
LDA         & 109 & 113   & $-$204 & $-$337    & 254 & 141 \\
+DMFT       & 30  & 45    & $-$110 & $-$125    & 42  &  65 \\
QSGW        & 41  & 44    & $-$107 & $-$202    & 131 &  56 \\
+DMFT       &  1  & 10    & $-$21  & $-$40     & 10  &  32 \\
SQS6        & 60  & 45    & $-$52  & $-$70     & 31  &  68 \\
\hline
\end{tabular}
\end{center}
\caption{$d_{xz}/d_{yz}$ and $d_{xy}$ QP levels (meV) near $E_{F}$ in tetragonal phase of FeSe (reference is the Fermi
energy).  LDA and QS\emph{GW} data are calculated nonmagnetically; the line marked +DMFT following each is the result
with DMFT added~\cite{fese}, as discussed in Sec.~\ref{sec:dmft}.  Line SQS is a QS\emph{GW} calculation in a
ferrimagnetic SQS structure, as described in the text.}
\label{tab:FeSe}
\end{table}

\begin{figure}[ht]
\begin{center}
\includegraphics[width=0.75\columnwidth]{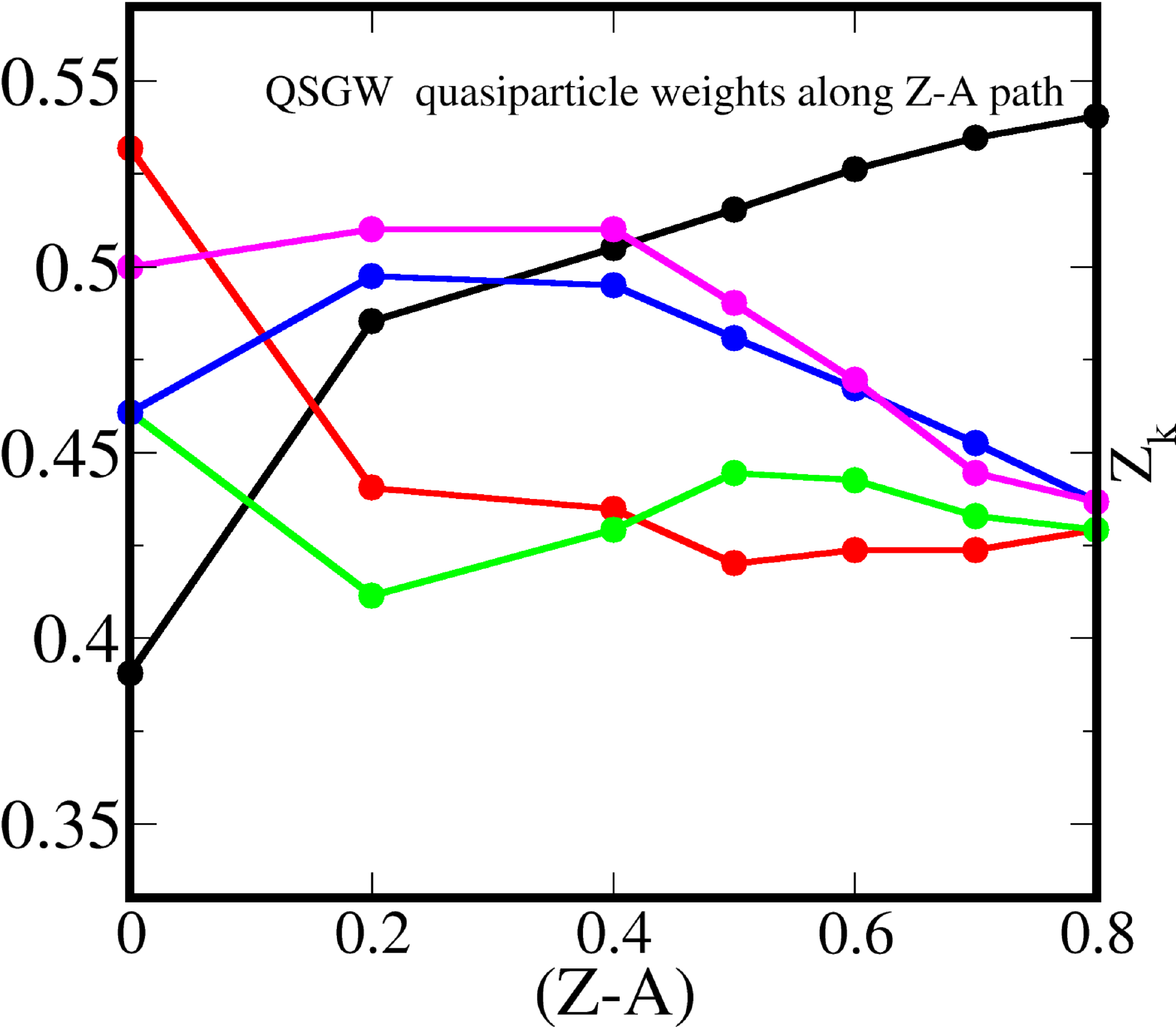}
\end{center}
\caption{The orbitally resolved quasiparticle renormalization ($Z$) factors for FeSe Fe-3d orbitals
along the high-symmetry Z-A line in tetragonal Brillouin zone; d$_{xy}$ (black), d$_{yz}$ (blue), d$_{z^2}$ (green),
d$_{xz}$ (magenta), d$_{x^2-y^2}$ (red).}
\label{fig:FeSe}
\end{figure}

FeSe is a heavily studied superconductor with large spin fluctuations: it provides an excellent testbed to compare LDA
and QS\emph{GW}.  Nonmagnetic calculations for LDA and QS\emph{GW} are shown in Table~\ref{tab:FeSe}. QS\emph{GW}
improves on the LDA, but discrepancies with ARPES are much larger than for, e.g. elemental Fe~\cite{Sponza17}.  That LDA
and QS\emph{GW} should be different is readily seen from the $k$ dependence of the $Z$ factor, shown in
Fig.~\ref{fig:FeSe}.  Local potentials such as the LDA cannot incorporate such an effect.

Spin fluctuations are important, but QS\emph{GW} does not adequately capture them.  They can be incorporated in a
mean-field manner by constructing a Special QuasiRandom Structure~\cite{zunger90} with, in this case, three spin-up and
three spin-down Fe atoms.  Two magnetic solutions can be stabilised: a low-moment and high-moment solution.
Quasiparticle spectra in the latter case are far removed from experiment, so we consider only the low-moment solution.
The moments on each Fe site are different, with a range $|m|{=}0.2{\pm}0.15\mu_B$.  The addition of magnetic terms shift
QP levels closer to ARPES measurements, but a significant discrepancy with experiment remains.  In Sec.~\ref{sec:dmft}
these results are revisited where local, high-order diagrams in spin are included.

These two systems show concretely how the shortcomings of \emph{GW} become apparent when spin fluctuations are
important.  The simplest diagram beyond \emph{GW}, the $T$ matrix, was first considered in an \emph{ab initio} framework
by Aryasetiawan and Karlsson~\cite{Aryasetiawan99}; more recently it was implemented in the \texttt{spex}
code~\cite{Sagioglu10}.  If spin fluctuations are not strong, as seems to be the case for Fe and Ni, such an approach
seems to describe spin excitations fairly well, though comparisons with experiments are too sparse to draw any strong
conclusions yet.  For strong spin fluctuations, e.g. in La$_{2}$CuO$_{4}$ or other unconventional superconductors such
as FeSe, it seems probable that many kinds of diagrams are needed, though the effective vertex entering into those
diagrams is expected to be mostly local.  Dynamical Mean Field Theory is ideally situated to address such cases; it is
discussed in Sec.~\ref{sec:dmft}.

Our view is that QS\emph{GW} with ladders in $W$ are sufficient for reliable description of susceptibilities within the
charge channel, especially when ladders are included in the self-consistency cycle~\cite{Kutepov17,Cunningham20} to make
the QS\emph{GW} $H_{0}$.  In the spin channel DMFT contributes the dominant missing diagrams to the self-energy, and the
susceptibilities can be reliably determined from local vertices, connected to $k$-dependent bubble diagrams.

\section{DFT+DMFT and \texorpdfstring{QS\emph{GW}+DMFT}{QSGW+DMFT}}
\label{sec:dmft}

Hartree Fock and DFT are band structure theories: electrons are treated as though they are independent; in solids
electron states are Bloch waves.  \emph{GW} includes correlations that go beyond the Bloch picture; still it is a
perturbation around the Bloch description.  In Landau Fermi-liquid theory, electrons are replaced by `quasi-particles'
which are adiabatically connected to the single-particle Bloch-wave representation of electrons.  This, by construction,
is a theory for excited states that adds perturbative corrections to the band picture.  In Landau adiabatic theory
electrons can have an effective mass, which is different from its free mass, and additionally, lifetime broadening
effects far from the Fermi energy.  One extreme case of this scenario is when the low-energy quasi-particle vanishes
(without magnetic ordering) and the atomic-like, high-energy excitation emerges and effective mass at the Fermi energy
tends to $\infty$.  In effect, this is an emergent, non-adiabatic, feature where single-particle spectral weight flows
to higher energies to conserve the total spectral weight and subsequently leads to suppression of low-energy
quasi-particles.  Popular rigid-band techniques~\cite{rice, GA, SB1, SB2, SB3, SB4, SB5, SB6} would catch the physics of
suppression of quasi-particles at Fermi energy, however, would fail to conserve the spectral weight since a dynamic
self-energy ($\Sigma(\omega)$) is absent from those theories.

DMFT was formulated as an effective theory that smoothly interpolates between the Bloch-wave limit and atomic limit.  It
is formulated as a local impurity embedded self-consistently in a medium or bath, and was introduced in a seminal work
by Metzner and Vollhardt~\cite{metzner} in late 80's.  Several other works from Mueller-Hartmann~\cite{hartmann}, Brandt
and Mielsh~\cite{brandt}, Jarrell \textit{et al.}~\cite{jarrell} and Georges et al.~\cite{georges-1992} helped to build
a strong foundation for DMFT.  The formulation and the implementation became popular through the seminal work by Georges
et al.~\cite{georges_rmp} and later emerged
as a way to study strong correlations in real systems, supplementing an \emph{ab initio} LDA framework
by embedding a model Anderson impurity hamiltonian inside an LDA bath~\cite{anisimov}.

An implementation of DMFT has three components: the partitioning of the Hamiltonian into impurity + bath, a means to
solve the impurity problem, and the bath-impurity self-consistency.  The bath is essential because correlated states of
interest have both atomic-like and Bloch-like properties with interesting low- and high- energy excitations.  Through
self-consistency the impurity feeds back on the bath, and ensures that all the symmetries of the entire system are
preserved.  In Questaal, QS\emph{GW} or DFT can act as the bath.  We have interfaces to two local solvers: the
hybridisation expansion flavor of continuous time Quantum Monte Carlo (CT-QMC) from \texttt{TRIQS} \cite{triqs} and the
CT-QMC written by Kristjan Haule~\cite{KH_ctqmc}.

For the partitioning, some correlated subspace must be identified and this is done using projectors.  Usually the
subspace is derived from atomic-like $d$- or $f$-states of a transition metal or $f$ shell element.  For projectors
maximally-localised Wannier orbitals have been widely used ~\cite{marzari,lechermann}.  We have so far adopted the
projection into partial waves of a particular $\ell$ in an augmentation sphere~\cite{Haule10}.  There are strong reasons
to prefer this scheme: the basis set is atom-centred and very localised with a definite $\ell$, which is the Hilbert
space where the correlations are strong.  The less localised the basis, the more it contains weakly correlated
components (e.g. O-p states in transition metal oxides).  Too much population of the
correlated subspace with uncorrelated parts conflicts with the physical interpretation of separation of bath and
interacting subspace and can reduce the reliability of the method~\cite{haule1}.  Moreover, very localised basis
functions have a much weaker energy-dependence.  As a consequence, with a very local projector it is possible to
construct an effective interaction over a wide energy window, reducing the frequency-dependence of the effective
interaction, and essentially recovering the partial waves in augmentation spheres~\cite{Choi16}.  Partial waves closely
resemble atomic orbitals, so solving a purely locally correlated Hamiltonian with the approximation of single-site DMFT
(local self energy) is a good approximation.

The thorniest issue is how to construct the effective local Hamiltonian.  It requires three related quantities:
definition of the local Green's function, an effective coulomb interaction $U$, which is partially screened by the bath,
and the double counting (DC) correction.
\begin{figure}[ht!]
\begin{center}
\includegraphics[width=\columnwidth]{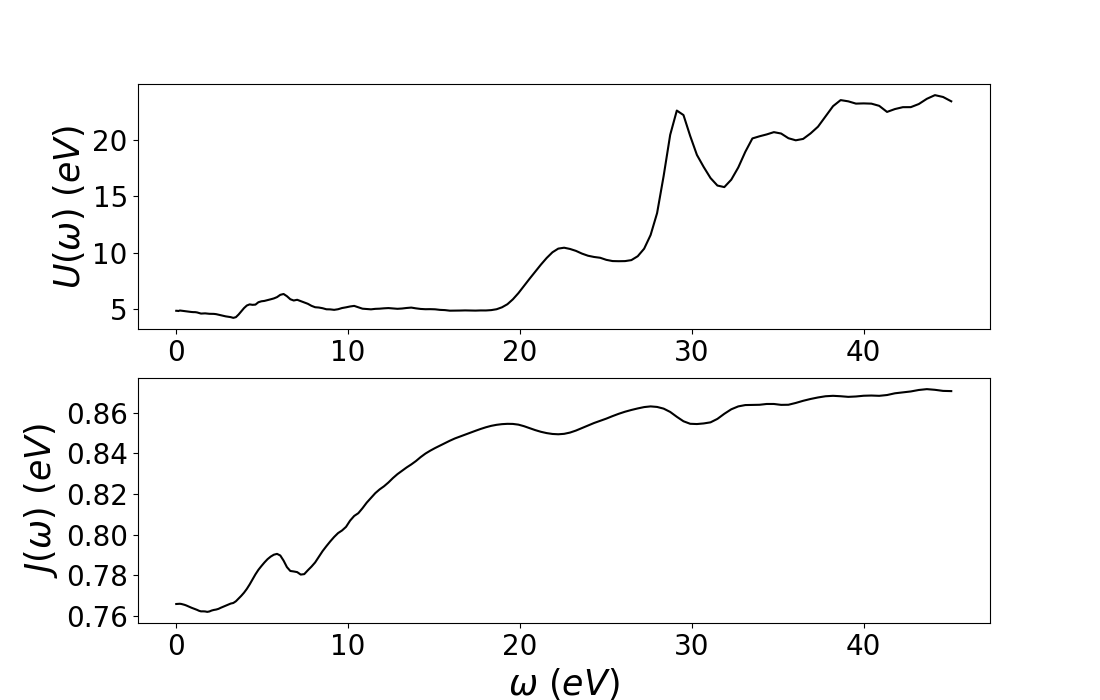}
\caption{$U(\omega)$ and $J(\omega)$ computed using our c-RPA scheme for Ni 3d orbitals starting from a QS\emph{GW}
bath.}
\label{fig:UJ_ni}
\end{center}
\end{figure}

In a fully \emph{ab initio} QS\emph{GW} + DMFT theory, these quantities must be computed from the theory itself, but how
to accomplish this is far from a settled issue.  Questaal can estimate the Hubbard \emph{U} and \emph{J} using the
constrained random-phase approximation~\cite{sasha1} where the internal transitions between states included in the DMFT
subspace are removed.  A fully internally consistent theory requires a frequency-dependent $U(\omega)$, but available
solvers can only solve the local impurity problem with such a \emph{U} in the density-density channel.  Haule's
prescription for excluding states in a large energy window partially mitigates this issue, since $U$ is closer to bare
and nearly energy-independent~\cite{Choi16}.  (It makes another approximation, namely that not all the states excluded
from screening are included in the impurity problem; these states are effectively treated only at the bath level.)

\subsection{Questaal's Implementation of DMFT}

Questaal does not have its own DMFT solver.  It has an interface to Haule's CTQMC solver~\citep{KH_ctqmc} and to the
\texttt{TRIQS} library~\citep{triqs}.  Haule's CTQMC solver uses singular value decomposition (SVD) and \texttt{TRIQS}
uses Legendre polynomial basis as compact representations for single-particle propagators.

A fully \emph{ab initio} implementation is a work in progress, that will be reported on elsewhere.  Within Questaal $U,
J$ can be calculated within constrained RPA, following Ersoy \textit{et al.}~\cite{christoph}.  It has been tested on
several materials; for instance, the Hubbard $U$ and Hund's $J$ for Ni are shown in Fig. \ref{fig:UJ_ni} which agrees
well with Ref.~\cite{christoph}.  The Hubbard \emph{U} for FeSe using QS\emph{GW} gives $U(0)=3.4$\,eV, also close to
what has been found in the literature~\cite{sasha1}.  In practical calculations today, we use a static $U$ and adopt the
fully localised limit (FLL) for double counting (Sec.\ref{sec:ldau}).
\begin{figure}
\begin{center}
\includegraphics[width=0.49\columnwidth]{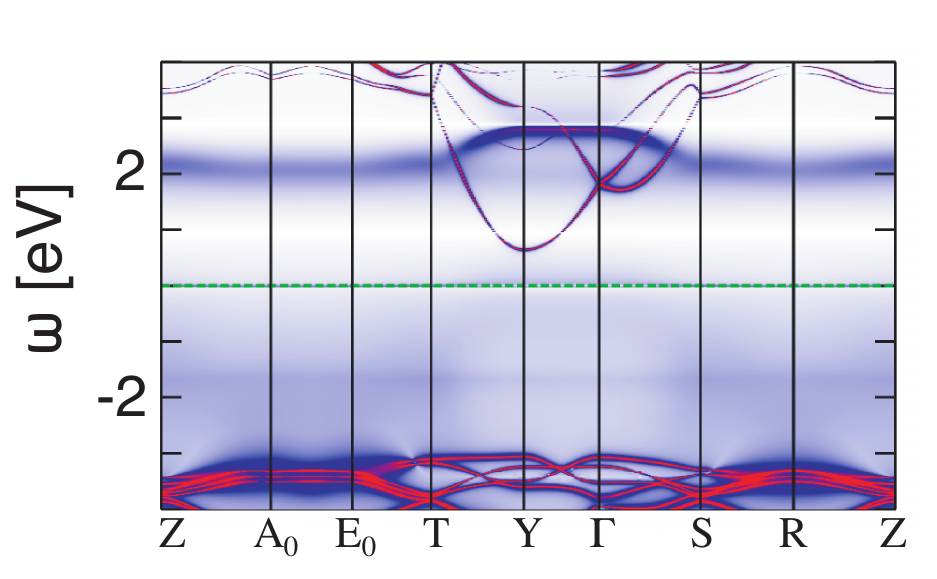}
\includegraphics[width=0.49\columnwidth]{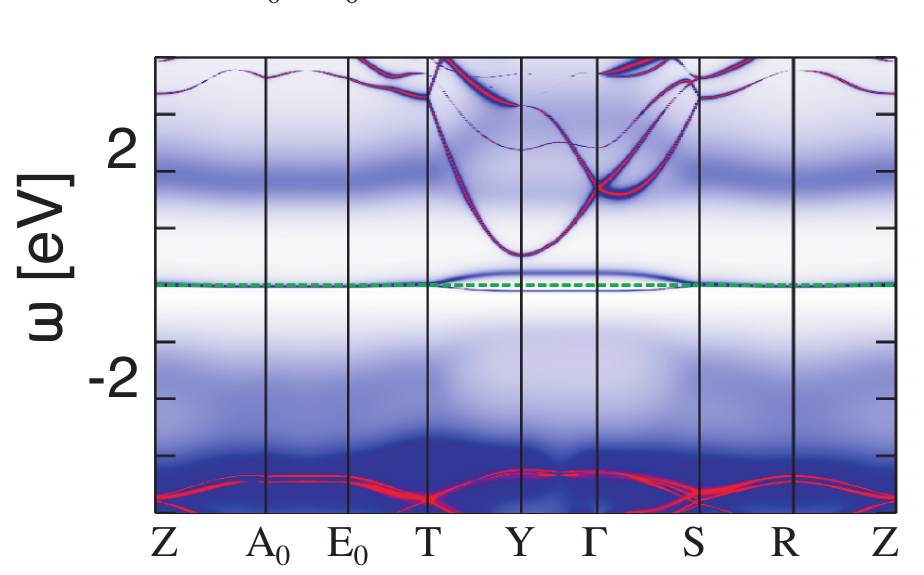}
\caption{QS\emph{GW} + DMFT spectral functions for La$_{2}$CuO$_{4}$ (left) and metallic QS\emph{GW} +
DMFT spectral functions for doped ($x$=0.12) La$_{2-x}$Sr$_{x}$CuO$_{4}$ (right) which shows the
classic three-peaked structure with upper and lower Hubbard bands at around $\pm$2\,eV, and an almost dispersionless
band at $E_{F}$.}
\label{fig:dmftsigma}
\end{center}
\end{figure}

We have used QS\emph{GW}+DMFT to study single- and two- particle properties of the hole doped cuprate,
La$_{2-x}$Sr$_{x}$CuO$_{4}$.  The parent compound at $x$=0 is a Mott insulator; while QS\emph{GW} calculated
non-magnetically leads to a metallic band structure, and an insulator with a wide gap when done antiferromagnetically
(Fig.~\ref{fig:lsco}).  The spins on the Cu in La$_{2}$CuO$_{4}$ fluctuate strongly and corresponding spin fluctuation
diagrams are missing from QS\emph{GW}.  DMFT incorporates the local spin fluctuation diagrams with an exact solver
(typically CTQMC), for the prescribed local Hamiltonian, and it can in principle, and indeed does, open a Mott
gap~\citep{Acharya18,Choi16} (charge blocking without magnetic ordering).  The five Cu-$3d$ partial waves in the
augmentation spheres comprise the correlated subspace, and the $3d_{x^2-y^2}$ splits off to form the gap
(Fig.~\ref{fig:dmftsigma}), which is the paramamagnetic analogue to the antiferromagnetic QS\emph{GW} band structure
(Fig.~\ref{fig:lsco}).  We solve the correlated impurity Hamiltonian in the Cu-$3d$ orbitals using the single-site
CT-QMC solver within paramagnetic DMFT.  Embedding local Green's functions into the QS\emph{GW} bath, we can compute the
non-local Green's functions dressed by the local DMFT self energy.  They can be used to make spectral functions and the
bubble diagrams entering into response functions (Sec.~\ref{sec:dmftrespponse}).

On doping, La$_{2-x}$Sr$_{x}$CuO$_{4}$ undergoes a metal-insulator transition when $x$ is sufficiently large.  At high
temperatures, the $x$=0.12 system is known to be metallic.  We recover a metallic solution in our QS\emph{GW}+DMFT for
$x$=0.12, using a virtual crystal approximation for $x$.  We find the classical three peak spectral feature with
atomic-like upper and lower Hubbard bands and a low energy incoherent quasi-particle peak at $E_{F}$
(Fig.~\ref{fig:dmftsigma}).  This structure is typical to strongly correlated metals in proximity of a
localisation-delocalisation transition~\citep{georges_rmp,georges_book}.  Comparing with the insulating case at $x$=0,
the spectral features for $x$=.12 has a natural explanation in the fact that spectral weight flows to low energies under
doping.

Single-site DMFT incorporates the local spin fluctuation diagrams through a local self energy.  It can not explain
Fermi-arc features of weakly hole-doped cuprates and momentum-dependent suppression of electron pockets in de-twinned
FeSe.  In weakly hole-doped cuprates Fermi arc emerges, most likely, due to spin fluctuations which are longer ranged in
nature and the corresponding diagrams are absent from single-site DMFT.  Similarly, in FeSe, there are nematic
fluctuations (which are inherently nonlocal), and the corresponding diagrams are absent from our theory.  To incorporate
such diagrams one needs to go beyond the single-site DMFT; cluster DMFT~\cite{maier} or diagrammatic techniques like
Dual Fermions~\cite{gull}.

\section{Susceptibilities} \label{sec:susceptibility}

Susceptibilities play a central role in many kinds of material properties.  In the static limit they reduce to the
second derivative of the total energy with respect to the corresponding perturbation.  This section presents some
general properties of linear susceptibilities, focusing on the transverse spin susceptibility $\chi^{+-}$, which is the
magnetic analogue of the better known charge susceptibility.  The formalism of the latter is very similar, but it is
simpler as it concerns the response to a scalar potential.  Also in the spin case, there is no analogue to the lowest
order diagram (time-dependent Hartree approximation) in the charge case.  To characterise the magnetic susceptibility
beyond the noninteracting case requires a vertex such as the $T$ matrix~\cite{Aryasetiawan99}.

Here we outline some general features; other sections explain how they are implemented in Questaal.  The spin
susceptibility relates the induced moment $\delta\myvec{m}$ to a perturbation $\delta\myvec{B}$:
\begin{align}
\delta \myvec{m}(\br,t) = -\int dt' d^3r' \chi(\br,\br',t-t') \myvec{B}(\br',t')
\label{eq:mrelatesB}
\end{align}
$\chi$ is a three-component tensor, $\chi^{\alpha\beta}$ with $\alpha$ and $\beta$ Cartesian coordinates.  The charge
susceptibility has the same form, only the perturbing potential is the scalar electric potential, which induces a
(scalar) charge density.  In general, there can be cross-coupling between spin and charge, so that the full $\chi$ is a
$4{\times}4$ matrix connecting spin and magnetisation.

We restrict ourselves to linear perturbations around an equilibrium point, where the unperturbed system is time
invariant.  $\chi$ is a function only of the difference $t-t'$, so its Fourier transform depends on a single frequency
$\omega$.  If in addition the reference system is in a periodic lattice, its Fourier transform space depends only on the
difference in translation vectors, whose Fourier transform has a single $\bq$ within the Brillouin zone with $\br$ and
$\br'$ limited to a unit cell.  Denoting $\hat{\chi}(\bq,\br,\br',\omega)$ as $\chi$ Fourier transformed in both space
and time, $\hat{\chi}$ is typically what is measured by spectroscopy.

\subsection{Spin Susceptibility}

A perturbation causes the system to generate an internal field $\myvec{B}_\text{xc}$, which adds to the total field,
$\myvec{B}_\text{tot}=\myvec{B}_\text{ext}+\myvec{B}_\text{xc}$
\begin{align}
\delta B_{\rm tot}^{\alpha} = \delta B_{\rm ext}^{\alpha} +\sum\nolimits_{\beta}
        \frac{\delta B_{\rm xc}^{\alpha}}{\delta m^\beta}\, \delta m^\beta
\label{eq:pert}
\end{align}
In collinear spin structures (all spins parallel to $z$), transverse and longitudinal components are decoupled.
Moreover, by rotating from $(x,y)$ to $x^{+-} = x \pm i y$, the three-component $\chi$ becomes diagonal with transverse
elements $\chi^{+-}$, $\chi^{-+}$, and longitudinal element $\chi^{zz}$.  Eq.~(\ref{eq:mrelatesB}) also applies to the
charge channel with the substitution $\myvec{B}{\to}V$ and $\myvec{m}{\to}n$, so that the full $\chi$ becomes a
$4\times4$ matrix, as noted above.  In general $\chi^{zz}$ is coupled to the charge channel, while $\chi^{+-}$ and
$\chi^{-+}$ are not.  Also if the equilibrium system is nonmagnetic, $\chi^{+-}{=}\chi^{zz}$.  In this work we are
concerned only with $\chi^{+-}$ and the charge susceptibility $\chi^{Q}$, or alternatively the dielectric function.

The ``magnetic'' kernel $I_\text{xc}=\partial B_{xc}/{\partial m}$ is in general a function of $(\br,\br',\omega)$,
though it must satisfy a sum rule~\cite{KotaniSW08}.  In many-body perturbation theory, the analogue of $I$ is the most
challenging quantity to obtain, but in the adiabatic time-dependent LDA it is static and local, and is written in the
transverse case as
\begin{align}
I_{\rm xc}^{+-}(\br)=\frac{B_{xc}(\br)}{m(\br')}\delta(\br-\br')
\label{eq:tdlakernel}
\end{align}
Eq.~(\ref{eq:mrelatesB}) relates $\myvec{m}$ to either $\myvec{B}_\text{ext}$ or $\myvec{B}_\text{tot}$
\begin{align}
\delta \myvec{m} = \chi_{0}\, \delta \myvec{B}_\text{tot} = \chi\, \delta \myvec{B}_\text{ext}
\label{eq:eqmxb}
\end{align}
when $\chi_{0}$ and $\chi$ are the bare (noninteracting) and full susceptibilities, respectively.  The preceding
equations imply the relation
\begin{align}
[\chi(\br,\br',\omega)]^{-1} = [\chi_{0}(\br,\br',\omega)]^{-1} - I_\text{xc}(\br,\br',\omega)
\label{eq:chirpa}
\end{align}
The noninteracting susceptibility can be computed by linearising Dyson's equation for the perturbation
$\delta\myvec{B}_\text{ext}$
\begin{align}
-\chi_0^{+-}(\br_1,\br_2,\omega) = G_\uparrow(\br_1,\br_2,\omega) \otimes G_\downarrow(\br_1,\br_2,-\omega)
\end{align}
$\otimes$ indicates frequency convolution:  $f(\omega) \otimes g(\omega) = i/{2\pi} \int_{-\infty}^{\infty} {d \omega}
f(\omega-\omega') g(\omega')$.  We use $\downarrow$ for spin 1 and $\uparrow$ for spin 2.  Similar equations with
permutations of $\uparrow$ and $\downarrow$ may be written for the other components of $\chi_0$.  In the Lehmann
representation $\chi_0^{+-}$ involves the product of four eigenfunctions
\begin{multline} 
\chi_0^{+-}(\bq,\br_1,\br_2,\omega) = \\
\phantom{+}\sum^{\rm occ}_{\bk n \downarrow} \sum^{\rm unocc}_{\bk' n'\uparrow}
\frac{
\Psi_{\bk n\downarrow}^*(\br_1)      \Psi_{\bk' n'\uparrow}(\br_1)
\Psi_{\bk' n'\uparrow}^*(\br_2) \Psi_{\bk n\downarrow}(\br_2)
}{\omega-(\epsilon_{\bk' n'\uparrow}-\epsilon_{\bk n\downarrow})+i \delta} \\
+\sum^{\rm unocc}_{\bk n \downarrow} \sum^{\rm occ}_{\bk' n'\uparrow}
\frac{
\Psi_{\bk n\downarrow}^*(\br_1)      \Psi_{\bk' n'\uparrow}(\br_1)
\Psi_{\bk' n'\uparrow}^*(\br_2) \Psi_{\bk n\downarrow}(\br_2)
}{-\omega-(\epsilon_{\bk n\downarrow}-\epsilon_{\bk' n'\uparrow})+i \delta} \\
\label{generalchi01q}
\end{multline}
where $\bk' = \bq + \bk$.  We note in passing, that in a proper MBPT formulation of susceptibility, e.g. the
T-matrix~\cite{Aryasetiawan99,Sagioglu10}, is a four-point quantity: $\chi$ is solved by a Bethe-Salpeter equation
involving $W$ and a four-point analogue of Eq.~(\ref{generalchi01q}), which at the end is contracted to two coordinates.

The Questaal code at present does not yet have a perturbative (T matrix) approach for the spin susceptibility, although
one was recently reported in a QS\emph{GW} framework~\cite{Okumura19}.  It does have the ability to include the charge
analogue of ladder diagrams to solve a Bethe-Salpeter equation for the dielectric function~\cite{Cunningham18}.  DMFT
can also be used to build a two-particle Green's function (or susceptibility), including all local diagrams, from which
a local four-point vertex can be extracted that replaces $I$.

The first QS\emph{GW} implementation of magnetic response functions~\cite{KotaniSW08} was designed for local moment
systems, by which is meant systems with rigid spins (RSA): when perturbed by a transverse
$\delta\myvec{B}^\perp_\text{ext}$ $\myvec{m}$ rotates rigidly without changing shape.  (Archetypal examples are Fe,
NiO, and MnAs~\cite{KotaniSW08}.)  In such a case the representation of $\chi^{+-}(\br,\br',\omega)$ simplifies to
\begin{align}
m(\br)\,\chi^{+-}(\omega)\,m(\br')
\end{align}
and $I$ can be completely determined by the sum rule~\cite{KotaniSW08}, thus avoiding a diagrammatic calculation for it.
This is Questaal's present perturbative approach to computing $\chi$.

In the RSA $\chi^{+-}(\br,\br',\omega)$ is discretised to a lattice model and can be written as
$\chi^{+-}_{\bR,\bR'}(\omega)$.  This makes it convenient to construct a Heisenberg model
\begin{align}
H = \sum_{RR'} J_{RR'} \delta \myvec{m}_R \delta \myvec{m}_R'
\label{eq:heisenberg}
\end{align}
and extract $J_{RR'}$ from $\chi$ as
\begin{align}
J_{RR'} = \frac{\delta^2E}{\delta {\myvec{m}_R}\delta {\myvec{m}_{R'}}} = [\chi^{+-}_{\bR,\bR'}(\omega{=}0)]^{-1}
\end{align}
The $\delta\myvec{m}$ are understood to be transverse to $\myvec{m}_R$.  Thus QS\emph{GW} can be used to determine
coefficients $J_{RR'}$ entering into the Heisenberg model.  This basic idea, with some enhancements, is what is
described in Ref.~\cite{KotaniSW08}.

\subsubsection{Magnetic Exchange Interactions in the ASA}
\label{sec:jij}

Sec.~\ref{sec:susceptibility} presents general formulations of the spin and charge susceptibility.  In the ASA, the
static transverse susceptibility is implemented in the \texttt{lmgf} code in a formulation essentially similar to the
rigid-spin approximation described there, with an additional ``long-wave'' approximation~\cite{Antropov03}
\begin{align}
\chi(\bq) \approx I\, \chi^{-1}(\bq)\, I
\label{eq:lw}
\end{align}
Using the sum rule, $I$ need not be calculated but inferred from $\chi_0$.

It is possible to compute $\chi_0$ from the full Green's function, Eq.~(\ref{eq:Gfromg}).  But \texttt{lmgf} implements
the classic Lichtenstein formula~\cite{Lichtenstein87}, in which exchange parameters $J_{RR'}$ are derived in terms of
the auxiliary $g$ and the ``magnetic force'' theorem.  This latter says that the change in total energy upon spin
rotation simplifies to the change in eigenvalue sum; it is a special instance of the Hellmanm-Feynman theorem.  It was
realised sometime later~\cite{Antropov03} that the Lichtenstein formula is correct only in the $q{\to}0$ limit (the
long-wave approximation), beginning to deviate at around $k{=}0.25{\cdot}2\pi/a$ in elemental 3$d$ magnets.  In the
notation of this paper, Lichtenstein's formula reads
\begin{align}
&{J^\ell}^{\bot}_{RR'} = \frac{1}{{2\pi }}\int^{\varepsilon _F }{d\varepsilon } \,{\mathop{\rm Im}\nolimits} {\rm{Tr}}_L
{\dots}\nonumber\\
&{\delta P_{R\ell} \left( {g_{RLR'L'}^\uparrow\,g_{R'L'RL}^\downarrow + g_{RLR'L'}^\downarrow\,g_{R'L'RL}^\uparrow}
        \right)\delta P_{R'\ell'}}
\label{eq:jlicht}
\end{align}
Questaal implements it for crystals, and for alloys within the CPA.  See Refs.~\cite{Antropov96,Antropov99,jijjap99} for
a detailed description.

\subsubsection{Comparing QSGW and LDA Spin Response Functions}

The \emph{GW} code has an implementation of spin response functions within the rigid spin approximation as described
Sec.\ref{sec:susceptibility}.  This avoids direct calculation of the vertex, and is very simple, but is restricted to
local-moment systems. Ref.~\cite{KotaniSW08} applied this approach to NiO, MnO and MnAs, with excellent results.

Questaal has the ability to compute spin waves (SWs) directly, or to extract parameters $J$ entering into the Heisenberg
model, Eq.~(\ref{eq:heisenberg}).  One application --- Mn$_{x}$Ga$_{1-x}$As alloys --- are particularly instructive
because they continue to attract interest as a spintronics material.  Here we will compare the (ASA) Lichtenstein
formula against full-potential, rigid-spin result using both LDA and QS\emph{GW} Hamiltonians.  Computing properties for
random alloys of any $x$ is feasible within the ASA, via either the Lichtenstein formula or relaxing large structures
using spin statics, Sec.~\ref{sec:noncollinear}.  But a similar study is not feasible today in QS\emph{GW}, so instead
we consider four-atom Special QuasiRandom Structures~\cite{zunger90} to simulate Mn$_{0.25}$Ga$_{0.75}$As and
Mn$_{0.75}$Ga$_{0.25}$As, and also the $x{=}1$ case.  Mn's large local moment of around 4$\mu_B$ makes the
approximations in Questaal's QS\emph{GW} magnetic response good ones.

\begin{figure}
\begin{center}
\includegraphics[width=0.49\columnwidth]{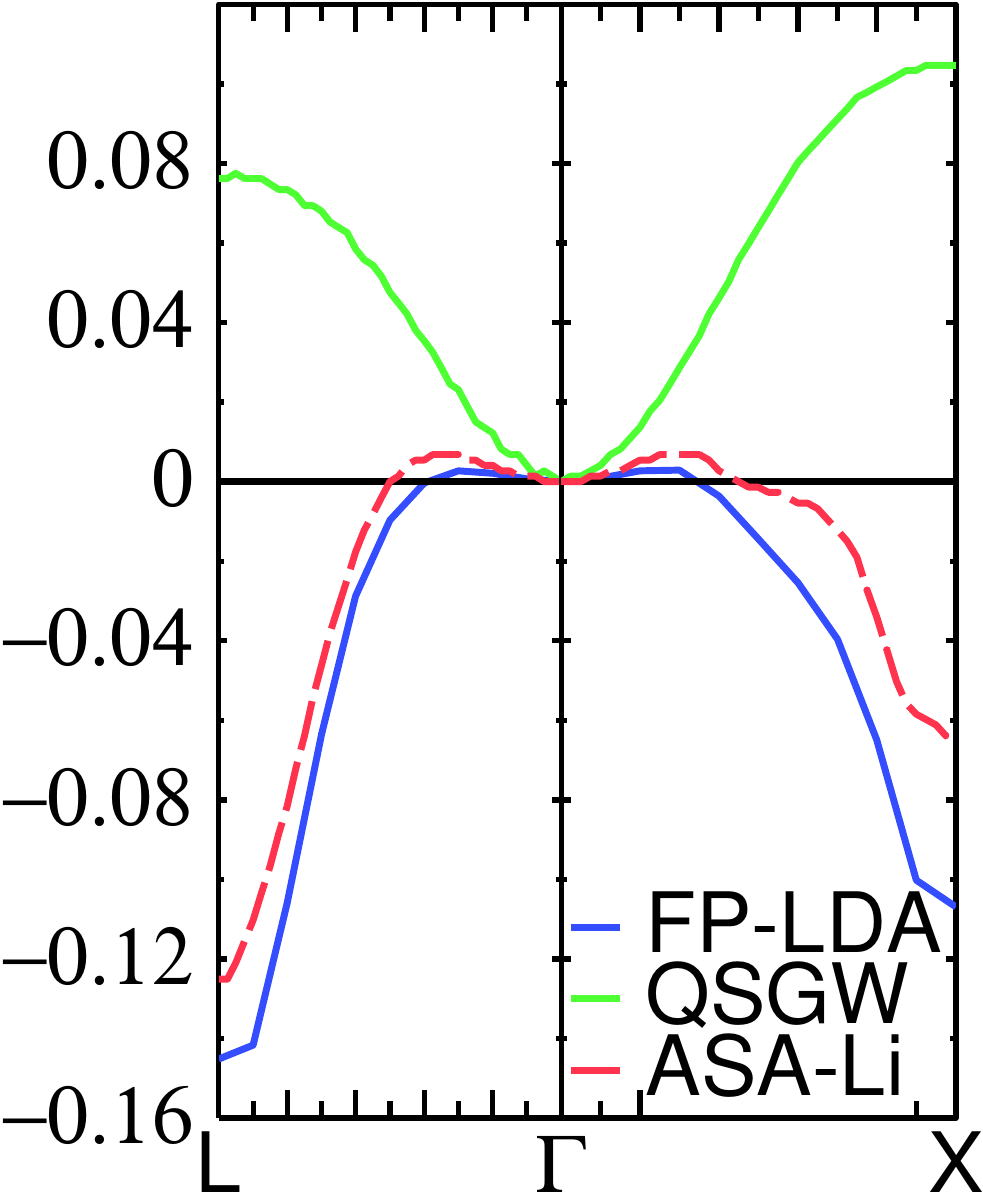}
\raisebox{-.2 em}{\includegraphics[width=0.49\columnwidth]{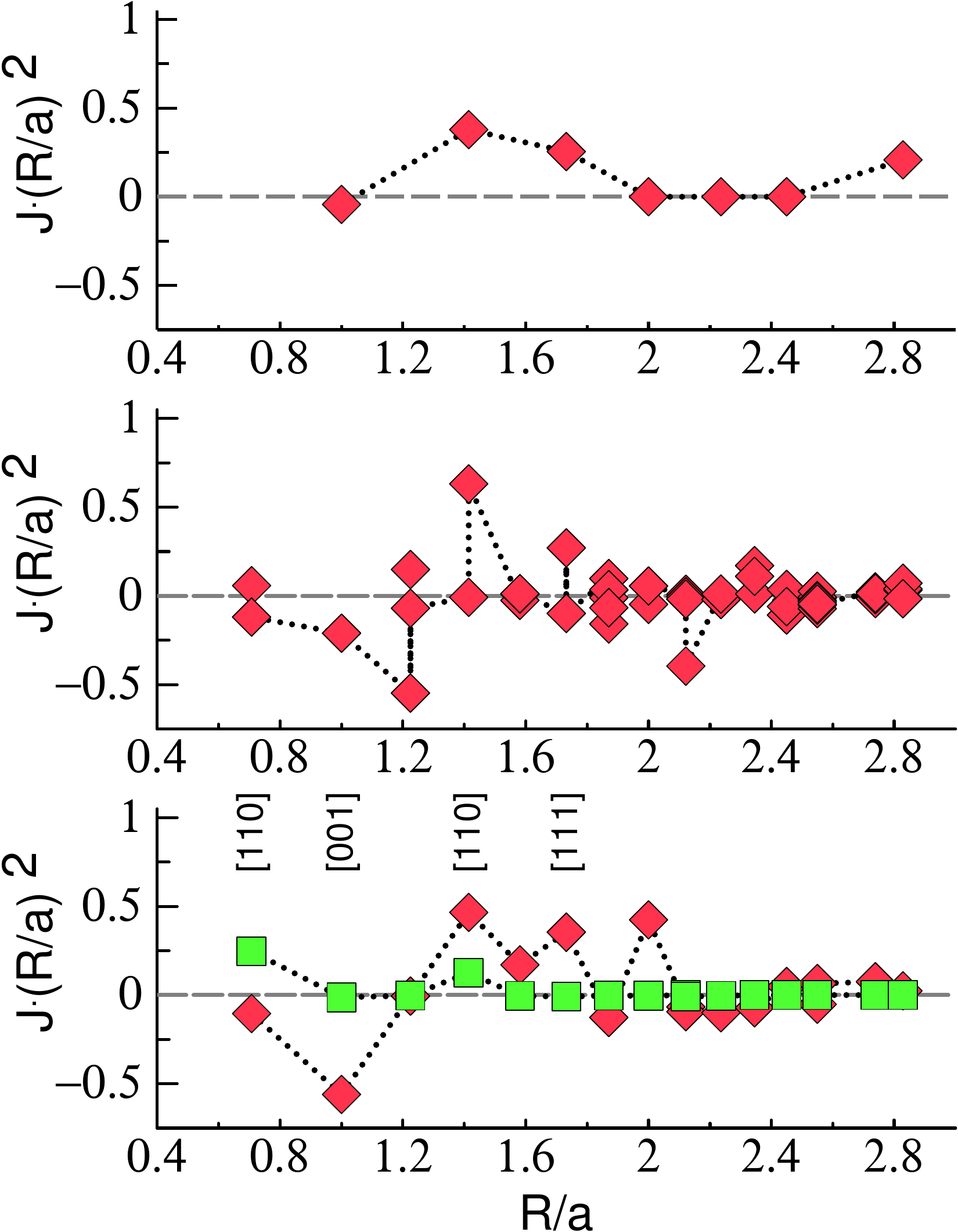}}
\caption{Left: spin waves in Zb-MnAs (eV) calculated in the \emph{GW} code with QS\emph{GW} potential (green) and LDA
potential (blue), and also in the ASA-LDA with the Lichtenstein formula, Eq.~(\ref{eq:jlicht}).  Right: Red diamonds
show Heisenberg parameters $J$ Eq.~(\ref{eq:heisenberg}) calculated in the ASA with Lichtenstein formula for a
MnGa$_{3}$As$_{4}$ SQS structure (top), MnGa$_{2}$As$_{2}$ SQS structure (middle), and Zb-MnAs (bottom), as a function
of neighbour distance.  In the SQS structures, there can be inequivalent neighbours with the same connecting vector.
LDA results calculated from the \emph{GW} machinery look similar.  Green squares: same parameters in ZB-MnAs using as
QS\emph{GW} potential. $J$ is in mRy.}
\label{fig:swmnas}
\end{center}
\end{figure}

DFT predicts Mn$_{x}$Ga$_{1-x}$As to become a spin glass when $x\,{\gtrsim}\,0.35$, though $T_{c}$ depends on what
fraction of Mn go into interstitial sites, and how the Mn are ordered on the Ga sublattice.  One application of the
Questaal code was to show how Mn in alloys that favour ordering on the (201) orientation can optimise
$T_{c}$~\citep{Franceschetti06}.  In the pure Zb-MnAs case, $x$=1, LDA predicts to be strongly antiferromagnetic, with
negative spin wave frequencies everywhere in the Brillouin zone (left panel, Fig.~\ref{fig:swmnas}).  The figures show
SWs calculated in the ASA using the Lichtenstein formula, Eq.~(\ref{eq:jlicht}).  Its long-wave approximation should be
accurate for small $q$, but it underestimates the strength of $J$ for large $q$~\citep{Antropov03}.  This is apparent in
Fig.~\ref{fig:swmnas}, comparing the LDA and Lichtenstein formulas.

This finding seems to contradict a measurement on a quantum dot of Zb-MnAs (Ref.~\cite{Okabayashi04}), which predicts it
to be ferromagnetic.  Indeed, QS\emph{GW} calculations of the same system show that spin wave frequencies are everywhere
positive (left panel, Fig.~\ref{fig:swmnas}).  $T_{c}$ based on Heisenberg parameters extracted magnon peaks in the SW
spectrum is positive, of order 600K (somewhat larger than what Okabayashi observed).

Exchange parameters $J$ were extracted for three compositions, in LDA and for the pure MnAs case, in QS\emph{GW} (right
panel, Fig.~\ref{fig:swmnas}).  At 25\% composition $J$ is still positive, but the nearest-neighbour $J$ is already very
small.  At 50\%, $J$ is negative on average, and a spin glass is predicted.  In pure Zb-MnAs, $J$ oscillate in sign with
neighbour distance, but the nearest neighbours (particularly along [001]) are strongly antiferromagnetic, leading to
negative frequencies in the SW spectrum.  QS\emph{GW} shows a marked contrast.  Only the NN is important, with $J>0$.
QS\emph{GW} and LDA are so different because the LDA underestimates the exchange splitting between spin-up and spin-down
$3d$ states, by $\sim$1\,eV.  This pushes the minority $d$ too close to $E_{F}$, and gives rise to long-range,
antiferromagnetic interactions.

\subsection{Optical Response Functions in Questaal}

The linear optical response is described by the imaginary part of the dielectric response function $\epsilon_2(\omega)$
or equivalently the real part of the frequency dependent conductivity $\sigma_1(\omega)=-i\omega
\epsilon_2(\omega)/4\pi$ at frequencies in the ultraviolet-visible (UV-VIS) range.  From it one can obtain the real part
$\epsilon_1(\omega)$ by the Kramers-Kronig relation and hence other relevant functions such as the complex index of
refraction $\tilde n(\omega)=\sqrt{\epsilon_1(\omega)+i\epsilon_2(\omega)}=n(\omega)+i\kappa(\omega)$ with
$\kappa(\omega)$ the extinction coefficient and $n(\omega)$ the real part of the index of refraction.  The reflectivity
$R(\omega)$ is then obtained via the Fresnel equations.  For instance, the normal incidence reflectivity
\begin{align}
R(\omega)=\left | \frac{\tilde n(\omega)-1}{\tilde n(\omega)+1}\right|^2
\end{align}
is a directly measurable quantity.  The absorption coefficient
\begin{align}
\alpha(\omega)=\frac{2\omega \kappa(\omega)}{c}=\frac{\omega\epsilon_2(\omega)}{n(\omega)c}
\end{align}
is measurable in transmission and spectroscopic ellipsometry allows one to measure directly $\epsilon_1(\omega)$ and
$\epsilon_2(\omega)$.

There are several levels of theory at which one can obtain the optical dielectric function supported in the Questaal
package.  Typically we need the macroscopic dielectric function:
\begin{align}
{\epsilon _M}(\omega ) = {[\epsilon_{G{=}G'{=}0}^{-1}({\mathbf{q}} \to 0,\omega )]^{ - 1}}
\label{eq:epsilonm}
\end{align}
This is essentially obtained as a byproduct of the \emph{GW} method (see Sec.~\ref{sec:gw}).  One way to obtain this
limit is to take a small but finite $\bq$ and use \texttt{lmgw} to obtain the plane wave matrix elements of the inverse
dielectric function.  The direction of the finite $\bq$ then defines the specific component of
$[\epsilon_2(\omega)]_{\alpha\alpha}$ which is actually a tensor.  If one simply calculates instead
\begin{align}
\epsilon_{G{=}G'{=}0}({\mathbf{q}} \to 0,\omega )
\end{align}
one obtains the value without local-field corrections.

Alternatively, however, one can take the limit $\bq\rightarrow0$ analytically and arrive at the Adler-Wiser formula in
the RPA,
\begin{multline}
[\epsilon_2(\omega)]_{\alpha\beta}=\frac{8\pi^2e^2}{\Omega\omega^2}\sum_n
        \sum_{n^\prime} \sum_{{\bf k}\in\text{BZ}}f_{n{\bf k}}(1-f_{n^\prime{\bf k}}) \\
\times\langle\psi_{n{\bf k}}|v_\alpha|\psi_{n^\prime{\bf k}}\rangle
        \langle\psi_{n^\prime{\bf k}}|v_\beta|\psi_{n{\bf k}}\rangle
        \delta(\omega-\varepsilon_{n^\prime{\bf k}}+\varepsilon_{n{\bf k}})
\end{multline}
Here $\Omega$ is the unit cell volume and $f_{n{\bf k}}$ are the occupation numbers, which in principle are given by the
Fermi function at finite temperature but are in practice taken to be 0 or 1 for empty and occupied states respectively.
The $v_\alpha$ are the components of the velocity operator ${\bf v}=\dot{\bf r}=(i/\hbar)[H,{\bf r}]$.  For systems with
at least orthorhombic symmetry the $\epsilon_2(\omega)$ tensor is diagonal in the Cartesian components.  This well-known
equation in the independent particle and long-wavelength form gives the $\epsilon_2(\omega)$ in terms of matrix-elements
of the velocity operator and vertical interband transitions and can also be obtained from the Kubo formula for the
optical conductivity.  Within the all-electron methods (no non-local pseudopotentials) and if \emph{GW}-self-energies
are not included, the velocity matrix elements are equivalent to the momentum matrix elements ${\bf v}={\bf p}/m$ and
can be obtained straightforwardly from the $\nabla$ operator matrix elements within the muffin-tin spheres and using a
Fourier transform for the smooth part of the wave functions in the full-potential implementation.  The matrix elements
of the $\nabla$ operator between spherical harmonics times radial functions inside the muffin-tin spheres is obtained
using the well-known gradient formula~\cite{Rose57}; more generally, see Sec.~\ref{sec:gradh} and Appendix D.

When the non-local self-energy operator or its Hermitian version are included, however, $(i m/\hbar)[H,{\bf r}]$ is no
longer equal to the momentum operator.  To correct for this, the approximation proposed by Levine and
Allan~\cite{LevineAllan89,delSole93} can be used, which consists in rescaling the matrix elements by a factor
$(\varepsilon_{n'{\bf k}}-\varepsilon_{n{\bf k}})/ (\varepsilon^{LDA}_{n'{\bf k}}-\varepsilon^{LDA}_{n{\bf k}})$.  This
is exact when the LDA and \emph{GW} eigenfunctions are the same, and it works well in weakly correlated systems.
However, it breaks down when correlations become strong, as in NiO.  Questaal also has an approximate form for the
proper nonlocal contribution to the velocity operator.  As of this writing, it takes into account only intercell
contributions.  This approximation apparently does not work well in strongly correlated systems, and results with full
matrix element will be reported in due course ~\citep{Cunningham20}.

The long-wave length approximation without local field (or excitonic) effects of $\varepsilon_2(\omega)$ can thus be
obtained both in the ASA \texttt{lm} and the full-potential \texttt{lmf} codes and, when reading in the QS\emph{GW}
self-energy, can use the corrected quasiparticle energies rather than the Kohn-Sham eigenvalues.  While approximate, it
has the advantage that the $\mathbf{q}\rightarrow0$ limit is taken analytically and that one can decompose the optical
response in contributions from each occupied-empty band pair.  By furthermore plotting vertical interband transition
energies for any given pair, one can analyse where the bands are parallel and give their largest contribution to the
joint density of states.  This provides a way to relate the critical points or van-Hove singularities in the optical
response to particular interband transitions.  \texttt{lmf} and \texttt{lm} use this to enable decomposition into
band-pair contributions if desired; it can also resolve contributions to $\epsilon$ by $k$.

The code also allows one to calculate dipole optical matrix elements between core states and valence or conduction band
states and this can be used to model X-ray absorption (XAS) and emission (XES) spectroscopies.  Furthermore, one can
calculate Resonant X-ray Emission Spectroscopy (RXES) also known as Resonant Inelastic X-ray Scattering (RIXS) using the
Kramers-Heisenberg equations~\cite{Preston11}. The interesting point about this spectroscopy is that it allows to probe
transitions between valence and conduction band states of the same angular momentum on a given site.  In fact, let's say
one considers as X-ray edge the K-edge or $1s$ core level, then the RXES probes transitions between valence bands and
conduction bands of $p$ angular momentum on the site of core hole that are in resonance with the difference of the
absorbed and subsequently coherently emitted X-ray photon transitions from these conduction and valence states to the
same core-hole.  In optical measurements in the VIS-UV region, on the other hand, one probes only $\ell$ to $\ell\pm1$
dipole allowed transitions.  Using $p$-levels as core hole RIXS thus probes $d-d$ transitions, which often may be
strongly influenced by many-body effects.  The Kramers-Heisenberg formula assumes the wavevector of the X-ray can be
neglected so that only direct transitions at the same {\bf k}-point are allowed and within this approximation it allows
one in principle to extract band-dispersions in a manner complementary to ARPES.

It was established 20 years ago that RPA approximation to the dielectric function can be dramatically improved by adding
ladder diagrams via a Bethe-Salpeter equation~\cite{Albrecht98,Rohlfing98} (BSE) in simple $sp$ semiconductors.
Traditionally the method is used by substituting \emph{GW} eigenvalues for the DFT ones obtained from one-shot \emph{GW}
(Sec.~\ref{sec:gw}).  For simple $sp$ semiconductors where LDA and \emph{GW} eigenfunctions are similar, this works
well.  It does not, however, when correlations become strong, e.g. in Cu$_{2}$O, where it was shown that
$\epsilon(\omega)$ calculated from the BSE starting from a QS\emph{GW} reference described the measured spectrum quite
well~\cite{Bruneval06b}.  Recently Cunningham \textit{et al.} studied optics from BSE using QS\emph{GW} as reference for
a number of systems, and found very good agreement with experiment even in correlated systems such as
NiO~\cite{Cunningham18} and the monoclinic phase of VO$_{2}$~\cite{Weber19}.  However, when spin fluctuations are large
and QS\emph{GW} does not describe the underlying band structure well, as in La$_{2}$CuO$_{4}$ (see
Sec.~\ref{sec:qsgwlimits}), the description begins to break down.  Recently Cunningham has added ladders to $W$ in the
QS\emph{GW} self-consistency cycle.  This seems to dramatically improve on the description of $\epsilon(\omega)$, even
in an extreme case such as La$_{2}$CuO$_{4}$.  These capabilities are embedded in the standard Questaal distribution.

In the ASA code \texttt{lm} second harmonic generation coefficients can also be calculated but this portion of the code
has not recently been updated to become applicable in the full-potential \texttt{lmf} implementation.  The calculation
of NLO coefficients is a bit more complex even within the long-wavelength independent particle approximation because of
the need to include both intra- and inter-band transitions and disentangle them in a careful manner.  The way to obtain
divergence free expression was described in a series of papers by John Sipe and
coworkers~\cite{AversaSipe,SipeGhahramani,Sipe2000}.  The version implemented in the code~\cite{Rashkeev98} uses
Aversa's length gauge formalism. The intra-band matrix elements ${\bf r}_i$ and inter-band ${\bf r}_e$ in this formalism
are
\begin{align}
\langle n{\bf k}|{\bf r}_i|m{\bf k}^\prime\rangle&=\delta_{nm}[\delta({\bf k}-{\bf k}^\prime)\xi_{nn}+i\nabla_{\bf k}
        \delta({\bf k}-{\bf k}^\prime)]\nonumber\\
\langle n{\bf k}|{\bf r}_e|m{\bf k}^\prime\rangle&= (1-\delta_{nm})\delta({\bf k}-{\bf k}^\prime)\xi_{nm}
\end{align}
with
\begin{equation}
\xi_{nm}\equiv\frac{(2\pi)^3i}{\Omega}\int_\Omega d^3r\, u_{n{\bf k}}^*({\bf r})\nabla_{\bf k}u_{m{\bf k}}({\bf r})
\end{equation}
in which $\Omega$ is the volume of the unit cell and $u_{n{\bf k}}({\bf r})$ is the periodic part of the Bloch function.
The $\xi_{nn}$ can be recognised to be a Berry connection.  The manipulation of the matrix elements of ${\bf r}_i$ in
commutators is non-trivial but described in detail in Aversa and Sipe~\cite{AversaSipe}.  An update of these parts of
the code to incorporate them in \texttt{lmf} and to be made compatible with QS\emph{GW} band structures as input is
intended.

See also Sec.~\ref{sec:dmftrespponse}.

\subsection{Response Functions within DMFT}
\label{sec:dmftrespponse}

\begin{figure}[ht]
\begin{center}
\includegraphics[width=0.33\columnwidth]{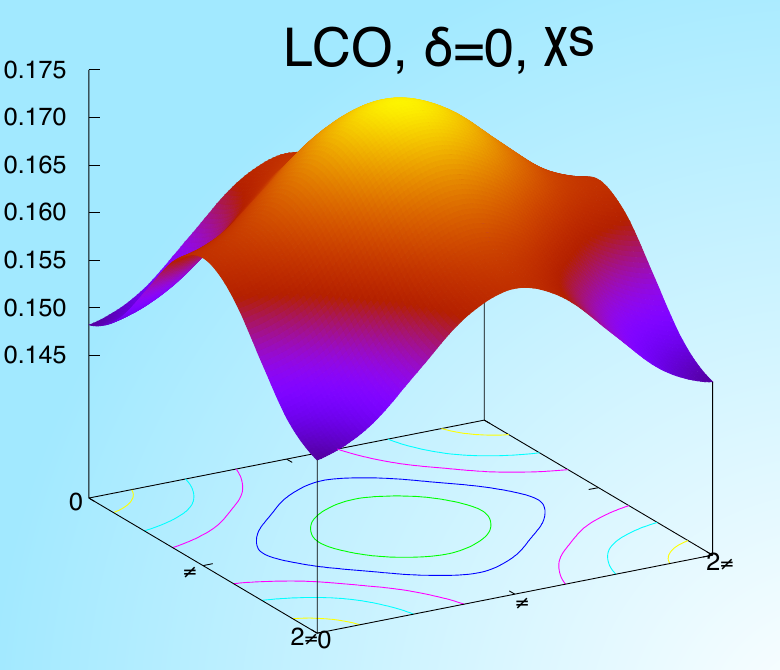}
\includegraphics[width=0.30\columnwidth]{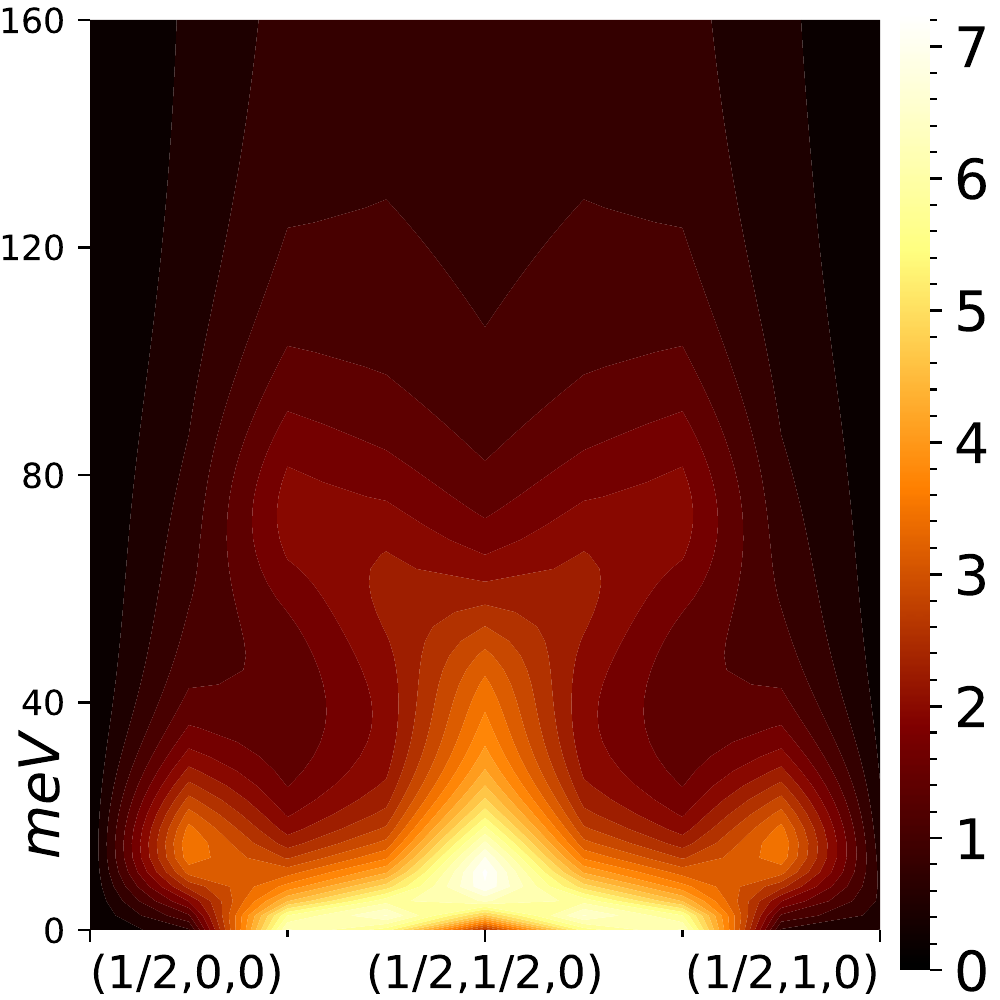}
\includegraphics[width=0.33\columnwidth]{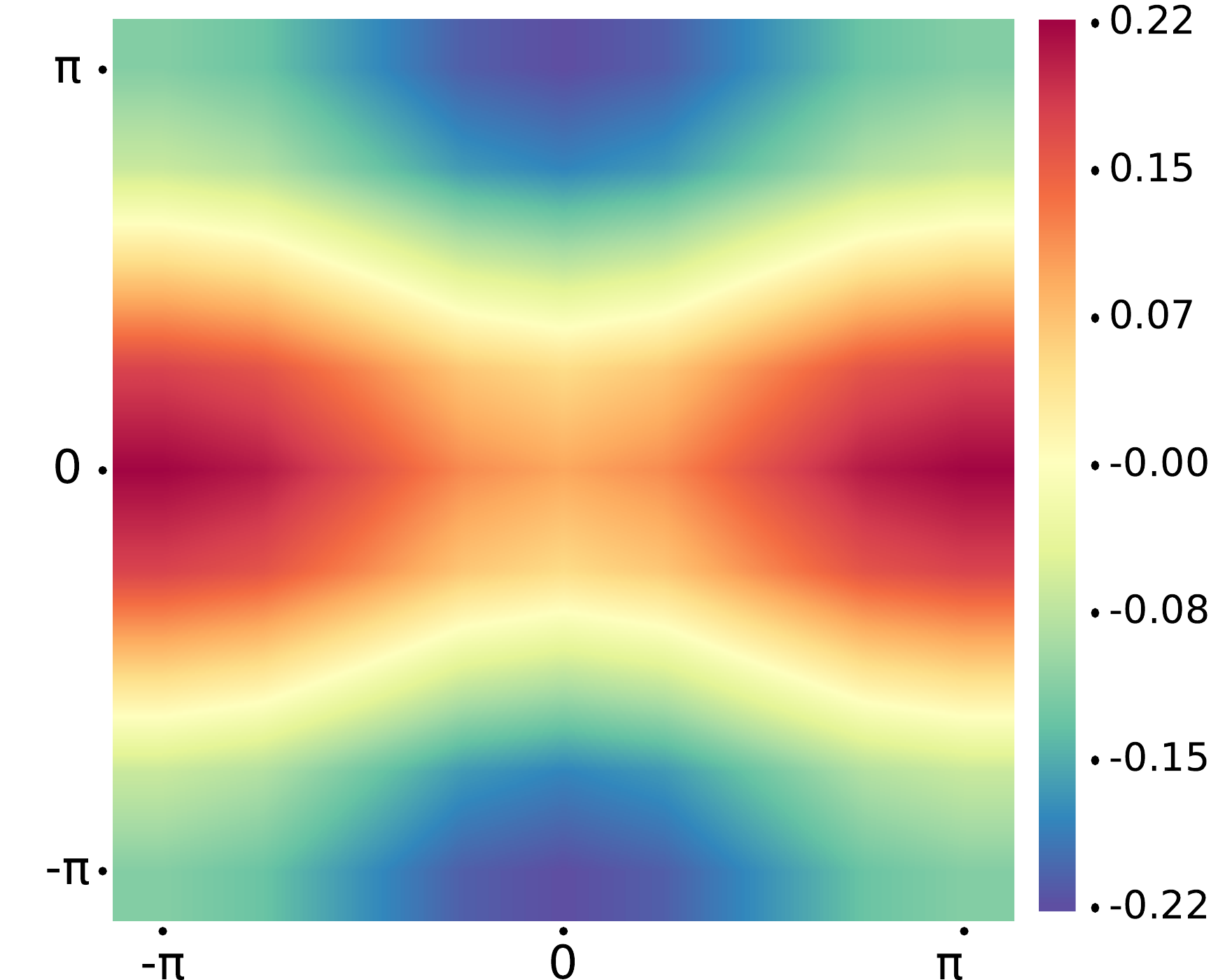}
\caption{(left) Real part of $\chi(q,\omega{=}0)$ for La$_{2}$CuO$_{4}$ showing a dominant peak at
\textbf{q}=($\pi,\pi$).  (middle) Imaginary part of $\chi(q,\omega)$ (arbitrary units) for
La$_{1.88}$Sr$_{0.12}$CuO$_{4}$ showing that spin fluctuations get gapped at ($\pi,\pi$) below 145 K.  (right) Nodal
superconducting gap structure for the hole-doped La$_{2}$CuO$_{4}$ showing d$_{x^2-y^2}$ gap symmetry. The color bar
shows the negative (in blue) and positive (in red) superconducting gap magnitudes in arbitrary units passing through
node where the gap closes (in yellow).}
\label{fig:dmftchi}
\end{center}
\end{figure}

With a converged self-energy, the CTQMC can sample the two-particle Green's function~\cite{hyowon_thesis} to obtain
local spin and charge vertices.  Finally, the non-local Bethe-Salpeter equations (BSE) in spin, charge and
superconducting channels~\cite{hyowon_thesis} can be solved from a local vertex and a non-local polarisation bubble in
the respective channels.  This allows us to compute the corresponding real~\citep{Acharya18} and imaginary part of
susceptibilities~\cite{Acharya18,sro} in those channels.  In Fig.~(\ref{fig:dmftchi}) we show the real part of the spin
susceptibilities in $x$=0 La$_{2-x}$Sr$_{x}$CuO$_{4}$, adapted from Ref.~\citep{Acharya18}. There is a peak at
\textbf{q}=($\pi,\pi$) suggesting dominant N{\'e}el anti-ferromagnetic spin fluctuations (Fig.~\ref{fig:dmftchi}).

We compute the imaginary part of the spin susceptibilities along the line ($h,k,l$) = (0,1/2,0)$-$(1,1/2,0) and find
that the peak at (1/2,1/2,0) gets spin gapped (vanishing weight at $\omega$=0) when $x$=0.12 below a certain
temperature, consistent with experimental observations~\cite{suzuki}.  Additionally, solving the BSE in the p-p
superconducting channel we recover a superconducting gap function with d$_{x^2-y^2}$ symmetry (right panel,
Fig.~\ref{fig:dmftchi}).  Our QS\emph{GW}+DMFT implementation can be successfully applied to a wide range of systems;
weakly correlated metals~\cite{Sponza17}, strongly correlated metals~\cite{sro} and correlated Mott
insulators~\cite{Acharya18}.

\section{Towards a High-Fidelity Solution of the Many-Body Problem} \label{sec:high-fidelity}

Solving the many-electron problem with high-fidelity is a formidably difficult task.  Our strategy to accomplish this
relies on the following premises:

\setcounter{Alist}{0}
\begin{list}{({\rm\roman{Alist}})\,}{\leftmargin 18pt \itemindent 0pt \usecounter{Alist}\addtocounter{Alist}{0}}
\item many-body solutions are best framed around a non-interacting starting point, and we believe that QS\emph{GW} is
the best choice among them, by construction;
\item charge fluctuations governed by long-range interactions, but they can be treated accurately with low-order
perturbation theory; and
\item spin fluctuations are governed mostly by single-site effective Hamiltonian (or action).  The short-range
interactions can be too strong to handle perturbatively, as can be seen by the non-analytic behaviour of one- and
two-particle quantities, and their high degree of sensitivity to small perturbations, e.g. change in temperature.
Nonlocal contributions are much weaker and can be treated in perturbation theory.
\end{list}
\begin{figure}[ht]
\begin{center}
\includegraphics[width=\columnwidth]{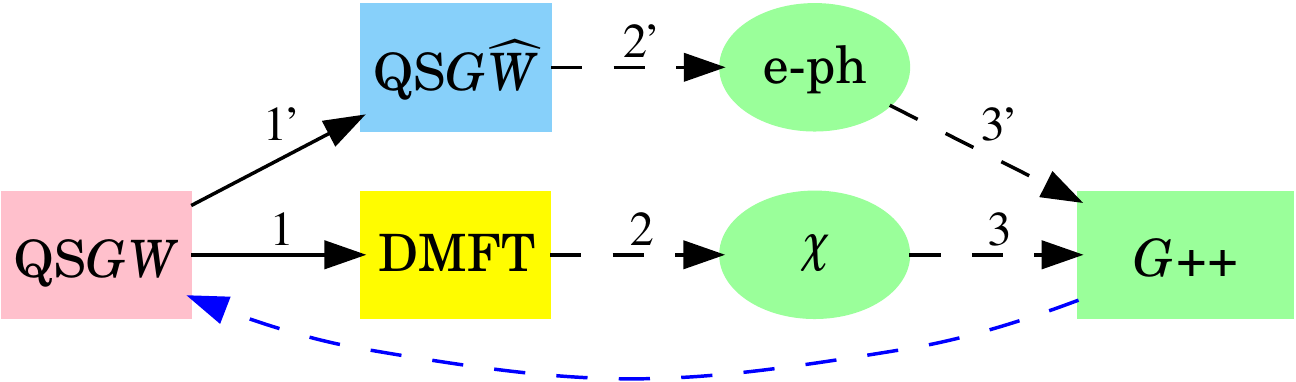}
\caption{Questaal strategy for solving the many-electron problem.}
\label{fig:questaal_strategy}
\end{center}
\end{figure}

Charge and spin fluctuations have very different characteristic energy scales: plasmon frequencies are typically of
order 5\,eV while typical magnon frequencies are of order 100\,meV or less.  As a conseqence they mutually interact
weakly so that the self-energy is predominated by independent contributions from spin and charge channels.  This is the
premise of the Bose Factor Ansatz which has has been found to be effective in a DMFT framework~\cite{casula12}.

We have framed a hierarchical strategy around these premises (Fig.~\ref{fig:questaal_strategy}).  At the lowest level is
QS\emph{GW}, which depends only minimally on DFT.  We think this is of central importance to frame a consistently
reliable theory, for reasons we have pointed out in Sec.~\ref{sec:qsgw}.

Referring to the figure, QS\emph{GW} is adequate for many purposes.  When not, there are two routes: a perturbative,
nonlocal path $(1',2',3')$ : low-order diagrams are added to $W$ to make $(W\to\widehat{W})$.  To date we have added
ladders to the charge channel; in progress is a project to add the electron-phonon interaction perturbatively
\cite{apstalksavio}, effectively adding another bosonic contribution to $W$.  Also possible, but not yet accomplished in
Questaal, is to add low order spin fluctuation diagrams such as the T matrix.  All of them make a better $G$
($G^{{+}{+}}$ in the Figure).  The second path begins as nonperturbative, local path $(1)$.  From a local interaction
nonlocal susceptibilities can be constructed $(2)$.  We have shown a few instances of this in
Sec.~\ref{sec:dmftrespponse}, and the results are remarkably good in cases we have studied so far, e.g.
Ref.~\cite{sro}.  Finally, the susceptibilities can be used to add a new diagrammatic contribution $(3)$ to $G$
($G{{+}{+}}$).  The simplest addition is Dual Fermions~\cite{gull}.

Perhaps more satisfactory would be to have a single, unified approach, for example the Diagrammatic Monte Carlo (DiagMC)
method \citep{DiagMC}.  But it is formidably difficult to make this method work practically in real systems, because of
the enormous time and memory costs that realistic (as opposed to model) Hamiltonians require.  We think that the JPO
basis alluded to in Sec.~\ref{sec:tblmf} is perhaps the most promising framework to realise DiagMC for realistic
Hamiltonians.  Even in such a case an optimal approach would likely closely resemble Fig.~\ref{fig:questaal_strategy},
but with some cross-linking between paths $(1)$ and $(1')$, using a diagrammatic Monte Carlo solver only to include
nonlocal diagrams beyond the RPA.

\section{Software Aspects}
\label{sec:software}

The software package has a long history and has featured a high level of modularisation from its early days.  The
different parts interoperate through shared interfaces and file data formats.

The implementation is mainly in procedural Fortran 77.  Fixes and improvements as well as new feature code uses modern
Fortran features (f2008+) for convenience as well as improved reliability and interoperability.  Despite f77's
shortcomings and questionable reputation, sound software engineering practices have been observed, there is (1)
extensive code documentation in an uniform format, (2) almost no shared mutable state (i.e. common blocks or module
variables); (3) action through side effect is avoided and data is passed through arguments, (4) unnecessary temporary
data copies are avoided, (5) data structures are organised with performance in mind and most heavy number crunching is
outsourced to performance libraries implementing the BLAS, LAPACK, FFTW3 APIs, (6) the code is written in an uniform
style with consistent flow patterns.  This practice has been extended through the use of modern version tracking and
continuous integration pipelines incorporating regression, coverage and performance testing, and minimal style policy
enforcement.

\subsection{Release Policy}

We use the popular distributed version tracking system \texttt{git}.  The public online hosted repository is hooked to
continuous integration pipelines executing the steps above on each push event.  New features are developed in separate
feature branches, if pushed to the public online repository, these are visible to all users.  When judged safe and
reasonably usable, feature branches are merged to the main development branch (``lm'').  After extensive use, the
development branch is merged to the release branch (``master'') and a new release is tagged with a numerical version in
the format vmajor.minor.patchlevel.  If/when issues are discovered and fixed the patchlevel is incremented in a new tag
and the new commits are merged back to the development branch and from there to the feature branches.  This approach
reduces the maintenance burden significantly, however it does mean that once the minor version is incremented there are
no more patches offered to the older versions and users are strongly encouraged to update to the latest version
available.  Since new developments are effectively done by interested users, and there is as of yet no contractual
support offered, we feel this is a justified arrangement and in the long term offers users new features and improved
performance with effectively close to no risk.  The major version is only incremented when a very significant, possibly
incompatible rewrite of core components or the input system has been performed.

New code and changes to existing code require: (1) coding standards pass; (2) regression tests suite pass (approximately
350 tests as of this writing); and (3) test coverage of at least 90\% of code.

\subsection{Parallelisation}

The programmes of the package support multilevel parallelisation multiprocessing through the message passing interface
(MPI) and simultaneous multithreading through performance libraries and (when necessary) manual OpenMP directives.  The
full-potential program uses multiprocessing mostly during the Brillouin zone sampling, outside of the $q$-loop the MPI
processes cooperate on the local potential generation, only multithreading is available within the processing of each
$q$-point.  The empirical polarisable ion tight binding program offers more flexibility in this regard by allowing
simultaneous assignment of groups of processes to each $q$-point as well as spreading different blocks of $q$-points
over groups of processes~\cite{tbe-report}.  A somewhat similar approach has been recently developed by Martin L\"uders
for the \emph{GW} code.  While it is available in a feature branch it is not yet considered production ready and has not
been merged to the main branch.  The GW code is fairly multithreaded and for systems of at least 15-20 atoms performs
well on many-core wide vector architectures like Intel's Xeon Phi family of processors (tested on x200, Knight's
Landing).  A small but performance-critical part was also ported to Nvidia's CUDA parallel platform and shows promising
performance on hardware with good double precision floating point capabilities (for example Titan, P100, V100 compute
cards).  The polarisable ion tight binding program makes more extensive use of CUDA; it is described
in~\cite{tbe-report}.

To ease parallel IO, classic fortran binary files are being moved to HDF5 based data files.  This also improves
interoperability with various postprocessing environments because the HDF5 libraries offer bindings to a variety of
popular programming languages.

\subsection{Usability}

\begin{table}[ht]
\begin{center}
\setlength{\tabcolsep}{4.8pt}
\begin{tabular}{|p{4.2em}|p{16em}|}
\hline
\multicolumn{2}{|c|}{\emph{Full-potential}}\\
\hline
\texttt{blm}         & automatic input file generator\\
\texttt{lmf}         & main band code\\
\texttt{lmfa}        & free atom solver\\
\texttt{lmfgwd}      & interface to \emph{GW}\\
\texttt{lmfdmft}     & interface to DMFT\\
\hline
\multicolumn{2}{|c|}{\emph{Atomic sphere approximation}}\\
\hline
\texttt{lm}          & ASA band code\\
\texttt{lmgf}        & ASA crystal Green's function code\\
\texttt{lmpg}        & ASA principal layer Green's function \\
\texttt{lmstr}       & ASA structure constants\\
\texttt{tbe}         & empirical tight-binding \\
\hline
\multicolumn{2}{|c|}{\emph{Post-processing and utilities}}\\
\hline
\texttt{fplot}       & a graphics package\\
\texttt{pldos}       & a DOS postprocessor\\
\texttt{plbnds}      & a bands postprocessor\\
\texttt{lmchk}       & checks structure-related quantities\\
\texttt{lmdos}       & assembles partial DOS\\
\texttt{lmfgws}      & dynamical self-energy postprocessing\\
\texttt{lmscell}     & supercell maker\\
\texttt{mcx}         & a matrix calculator\\
\hline
\end{tabular} 
\begin{tabular}{|p{9em}|p{11.2em}|}
\multicolumn{2}{|c|}{\emph{Editors}}\\
\hline
\texttt{lmf${\vert}$lm --rsedit} & restart file\\
\texttt{lm --popted}             & optics\\
\texttt{lmf --wsig$\sim$edit}    & static $\Sigma_0$\\
\texttt{lmfgws --sfuned}         & dynamic $\Sigma$\\
\texttt{lmf --chimedit}          & magnetic susceptibility\\
\texttt{lmscell --stack}         & superlattice\\
\hline
\end{tabular}
\end{center}
\caption{Main executables in the Questaal suite.}
\label{tab:executables}
\end{table}

\begin{figure}[ht]
\begin{center}
\fbox{\includegraphics[width=18em]{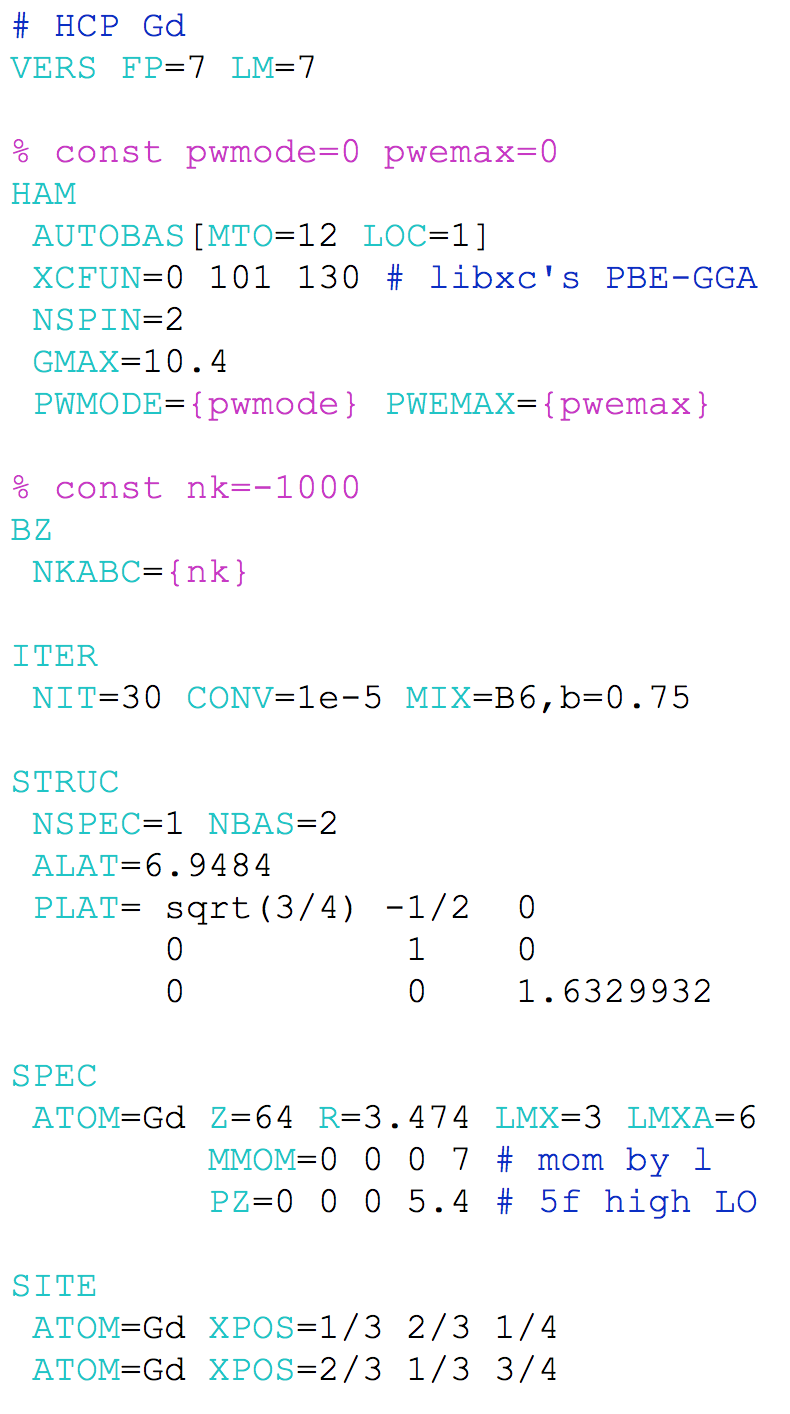}}
\end{center}
\caption{Basic input file: a full-potential GGA calculation for HCP Gd with 7$\mu_B$ $l=3$ starting moment.  A
preprocessor directive (\texttt{nk}) is used to specify the $k$-sampling, requesting 1000 points in the full Brillouin
zone: the generated grid (12,12,7) has components corresponding to the relative BZ vector lengths.  \texttt{nk}, like
\texttt{pwemax}---which flags the inclusion of LAPW basis functions---can be changed from the command line.  The LMTO
basis is to be setup automatically, caused by the \texttt{AUTOBAS} token in the \texttt{HAM} (Hamiltonian) section.}
\label{fig:ctrl}
\end{figure}

An ordinary text command line is all one needs to be able to use the programmes from the package.  The main executables
(see Table~\ref{tab:executables}) support a number of flags, notable among which is ``\texttt{--input}'', it causes the
full input understood by the executable to be printed together with default values and short documentation for each
token.  Another very valuable and possibly unique feature is the ability to override almost any input file value through
the command line through the small embedded preprocessor which renders the input transparent internally
(Fig.~\ref{fig:ctrl}).  A third is the suite of special-purpose editors (see Table).  There is extensive documentation
online together with many tutorials and ready made example script snippets.  Users are encouraged to report issues and
offer ideas through the online ticketing system and contribute improvements or fixes through pull requests or inline
patches.

\subsection{Selected Publications from Questaal}

In this section we point to original papers where new capabilities were developed within Questaal.  Some the concepts
are not original with Questaal, though many are:
\setcounter{Alist}{0}
\begin{list}{$\bullet$\,}{\leftmargin 12pt \itemindent 0pt \usecounter{Alist}\addtocounter{Alist}{0}}
\item an early DFT calculation of the Schottky barrier height at a metal/semiconductor contact, showing the importance
of screening at the interface~\citep{GaAsSB};
\item an early calculation of alloy phase diagrams combining statistical theory and
DFT~\citep{Asta92};
\item a systematic technique for deriving force theorems within
DFT~\citep{Methfessel93};
\item the first formulation of adiabatic spin dynamics within DFT~\citep{Antropov95} and its application to explain the
Invar effect in permalloy~\citep{InvarNature};
\item a formulation of self-consistent empirical tight-binding theory~\citep{Finnis98};
\item a formulation of Electron Energy Loss Spectroscopy in DFT~\citep{Paxton00};
\item the original description of Questaal's full-potential method, Sec.~\ref{sec:fp}~\citep{lmfchap};
\item the original description of Questaal's all-electron \emph{GW} approximation~\citep{GWintro}, and first calculation
of RPA total energy in a solid~\citep{Miyake02};
\item solution of the Boltzmann transport equation combined with DFT~\citep{Krishnamurthy03};
\item the original formulation of Quasiparticle Self-Consistent \emph{GW}~\citep{Faleev04}, demonstration of its wide
applicability~\citep{mark06qsgw}, and a detailed description~\citep{Kotani07};
\item Questaal's implementation of nonequilibrium electron transport in nanosystems~\citep{Faleev05}, and significant
applications to magnetic transport~\cite{Velev05,Chantis07,Wysocki09,spinlossinterface,Belashchenko19};
\item a fusion of genetic algorithms and exchange interactions to predict and optimise critical temperatures in
multi-component systems~\citep{Franceschetti06};
\item Dresselhaus terms calculated \emph{ab initio} with high fidelity in zincblende
semiconductors~\citep{Chantis06a,Krich07,Luo10};
\item first \emph{GW} description of $4f$ systems, showing the tendency to overestimate splitting between occupied and
unoccupied $f$ states ~\citep{Chantis07c};
\item spin wave theory within QS\emph{GW}~\citep{KotaniSW08};
\item prediction of Impact Ionisation rates with QS\emph{GW}~\citep{KotaniII};
\item the PMT method of Sec.~\ref{sec:pmt}~\citep{KotaniPMT10};
\item a formulation of tunnelling transport within QS\emph{GW}~\citep{Faleev12};
\item effect of spin-orbit coupling on the QS\emph{GW} self-energy~\citep{Brivio14};
\item approximate description of transient absorption spectroscopy within QS\emph{GW}~\citep{Leguy16};
\item Questaal's first application of QS\emph{GW}+DMFT, showing the need to go beyond \emph{GW} when spin fluctuations
are important~\citep{Sponza17};
\item addition of ladder diagrams to dielectric function~\citep{Cunningham18};
\item Fr{\"o}hlich contribution to renormalisation of QS\emph{GW} energy bands~\citep{Lambrecht17}; and
\item first application of QS\emph{GW}+DMFT+BSE for susceptibilities, Sec.~\ref{sec:dmftrespponse}~\citep{Acharya18}.
\end{list}

\subsection{Distribution and Licensing}

Historically the code has been distributed as a set of source tarballs, in addition registered users can git-clone our
online repository.  The code is distributed under the terms of the General Public license version 3. Contributions under
a compatible license are welcome.

\section{Conclusions}

Questaal is a descendant of one of the early all-electron methods developed in Stuttgart.  It has gradually evolved to
its present form, but as no paper has been written for any of its stages the present work is intended to summarise much
of this evolution.  Our aim was to present the many expressions from classic works, combined with previously unpublished
recent developments, in a unified way to show the connections between the parts.

We made an attempt to show Questaal's strengths as well as its limits, within varying levels of theory.  We think
Questaal provides a very promising path to efficiently solve the electronic structure problem with high fidelity.  It is
unique in its potential to span such a wide range of properties and materials, and with varying levels of approximation.

\section{Acknowledgements}

Many people have contributed to the Questaal project.  Much of the mathematics and basic algorithms in the \texttt{lmf}
implementation were largely the work of Michael Methfessel.  \emph{GW} was formulated and implemented largely by Ferdi
Aryasetiawan and Takao Kotani, and QS\emph{GW} by Sergey Faleev.  Martin L\"uders wrote a parallel version of this code
and was involved in some validation aspects.  Derek Stewart wrote the original code for Landauer-Buttiker transport, and
Faleev extended it to the nonequilibrium case.  Pooya Azarhoosh improved on the optics and performed TAS studies.  Alena
Vishina redesigned the fully relativistic Dirac code, making it self-consistent.  Herve Ness has redesigned significant
parts of the layer Green's function technique.

Brian Cunningham and Myrta Gr{\"u}ning added ladder diagrams to $W$, and Savio Laricchia is close to completing a
QS\emph{GW}-based approach to the electron-phonon interaction and related quantities.  Both of these additions will have
significant impact in the future.  Also, many thanks to Kristjan Haule and Gabi Kotliar for helping us get started in
DMFT, and to Ole Andersen for being the founder and inspiration for so many of the ideas behind Questaal.

Finally, many thanks to Andy Millis and the Simons Foundation, who enriched Questaal in fundamental ways, and to EPRSC
which made it possible to become a community code (CCP9 flagship project: EP/M011631/1).

We gratefully acknowledge PRACE for awarding us access to SuperMUC at GCS@LRZ, Germany, the UK Materials and Molecular
Modelling Hub for computational resources, which is partially funded by EPSRC (EP/P020194/1) and STFC Scientific
Computing Department's SCARF cluster (\url{www.scarf.rl.ac.uk}).  J.J.  acknowledges support under the CCP9 project
``Computational Electronic Structure of Condensed Matter'', part of the Computational Science Centre for Research
Communities (CoSeC).

\section*{Appendices}
\renewcommand{\thesubsection}{\Alph{subsection}}
\renewcommand{\theequation}{\Alph{subsection}.\arabic{equation}}\setcounter{equation}{0}\setcounter{subsection}{0}

An augmentation radius is labelled by symbol $s$.  Subscript $R$ is used to denote a site-dependent quantity, such as
the augmentation radius $s_R$.  When $R$ appears in a subscript of a function of $\br$ such as $H_{RL}(\br)$, it implies
that $\br$ is relative to $\bR$, i.e. $\mathbf{r-R}$.  When a symbol refers to angular momentum, an upper case letter
such as $L$, refers to both angular quantum number and magnetic quantum number parts, $\ell$ and $m$.

\subsection{Definition of Real and Spherical Harmonics}
\label{app:sharm}

Where spherical harmonics are used, Questaal uses the same definitions as Jackson~\cite{Jacksonbook}:
\begin{align}
Y_{\ell\,m}(\theta ,\phi ) &= \left[ \frac{(2\ell+1)(\ell - m)!}{4\pi (\ell+m)!} \right]^{\frac{1}{2}}
        P_\ell^m(\cos\theta){e^{im\phi }}\\
P_\ell^m(x) &= {(-1)^m} \frac{(1 - {x^2})^{m/2}}{2^ll!} \frac{d^{\ell+m}}{d{x^{\ell+m}}}{({x^2} - 1)^\ell}
\end{align}
The $-m$ and $+m$ functions are related by [see Jackson (3.51) and (3.53)]
\begin{align}
P_\ell^{-m}(x) &= (-1)^m \frac{(\ell - m)!}{(\ell+m)!} P_\ell^m(x) \\
Y_{\ell,-m}(\hat {\br}) &= (-1)^m Y^{*}_{\ell\,m}(\hat {\br})
\end{align}

Questaal mostly uses real harmonics $\text{Y}_{lm}$, which are related to the spherical harmonics as
\begin{align}
{\text{Y}_{\ell\,0}}(\hat{\br}) &= \ Y_{\ell\,0}(\hat{\br}) \nonumber\\
\text{Y}_{\ell\,m}(\hat{\br}) &= \frac{1}{\sqrt 2}[(-1)^m{Y}_{\ell\,m}(\hat{\br}) + {Y}_{\ell,-m}(\hat{\br})]\nonumber\\
\text{Y}_{\ell,-m}(\hat{\br}) &= \frac{1}{\sqrt 2 i}[{( - 1)}^m{Y_{\ell\,m}}(\hat{\br}) - Y_{\ell,-m}(\hat{\br})]
\end{align}
It also uses \emph{real harmonic polynomials}, [note: substituted $l\to\ell$]
\begin{align}
\mathcal{Y}_{{\ell}m}(\br)=r^\ell\text{Y}_{{\ell}m}(\hat{\br}),
\label{eq:defylpoly}
\end{align}
which are real polynomials in $x$, $y$, and $z$.

The product of two spherical harmonic polynomials can be expanded as a linear combination of these functions in the same
manner as ordinary ones
\begin{align}
\text{Y}_K(\hat{\br})\text{Y}_L(\hat{\br}) &= \sum_M C_{KLM} \text{Y}_M({\hat{\br}}) \\
\mathcal{Y}_K({\br})\mathcal{Y}_L({\br}) &= \sum_M C_{KLM}\, r^{k+\ell-m} \, \mathcal{Y}_M({{\br}})
\label{eq:expandsharm}
\end{align}
The Gaunt coefficients $C_{KLM}$ are nonzero only when ${k+\ell-m}$ is an even integer, so the r.h.s. is also a
polynomial in $(x,y,z)$.

Note also, from Eq.~(\ref{eq:defylpoly}) it follows that if $\alpha$ is purely imaginary
\begin{align}
\mathcal{Y}_{{\ell}m}(\alpha\br) = (-1)^l\, \mathcal{Y}^*_{{\ell}m}(\alpha\mathbf{r})
\label{eq:cconjgyl}
\end{align}
It also follows from Eq.~(\ref{eq:defylpoly}) and the discussion around Eq.~(\ref{eq:skgrad}) that
\begin{align}
\nabla_p\mathcal{Y}_{L}({\br}) &= (2\ell+1)\,\sum_{K} {{C}^{-}_{K{L};p}}\,\mathcal{Y}_{K}({{\br}})\nonumber\\
-\nabla_p(r^{-2l-1} \mathcal{Y}_{L}({\br})) &= (2\ell+1)\,\sum_{K} {{C}^{+}_{K{L};p}}\,\mathcal{Y}_{K}({{\br}})
\label{eq:gradrlyl}
\end{align}
for $p=1,2,3$. $r^{-2l-1} \mathcal{Y}_{L}({\br})$ and $\mathcal{Y}_{L}({\br})$ are ordinary Hankel and
Bessel functions of energy zero.  In the top (bottom) line, $k{=}\ell{+}1$ ($k{=}\ell{-}1$).  Eqs.~(\ref{eq:gradrlyl})
may be taken as a definition of ${{C}^{(\pm)}_{K{L};p}}$.  Similarly
\begin{align}
\br_p \mathcal{Y}_L(\br) = \sum_{K} \big\lbrace {{C}^{(+)}_{K{L};p}} + {{C}^{(-)}_{K{L};p}}r^2 \big\rbrace
        \mathcal{Y}_{K}(\br)
\label{eq:ryl}
\end{align}
$C^{(-)}$ is the transpose of $C^{(+)}$ :
\begin{align}
{{C}^{(+)}_{K{L};p}} = {{C}^{(-)}_{L{K};p}}
\end{align}
but we keep them separate to distinguish terms that take $Y_{\ell,m}$ into $Y_{\ell+1,m'}$ and those
that map into $Y_{\ell-1,m'}$.

\subsection{Definition of Hankel and Bessel Functions}
\label{app:hankel-bessel-defn}

To distinguish radial parts of spherical functions from solid versions, we denote spherical parts with subscript $\ell$,
and solid functions with subscript $L$, which refers to both the $\ell$ and $m$ parts.

The Questaal codes generally follow Methfessel's definitions for Hankel and Bessel functions.  Writing $k{=}\sqrt{E}$
with $\text{Im}\,k{\ge}0$, they are related to standard spherical Neumann functions $n_\ell$ and Bessel functions
$j_\ell$ as follows:
\begin{align}
H_L  &= \mathcal{Y}_L(-\nabla)\, h_0(r) \nonumber\\
     &=      k^{\ell+1} n_\ell (kr) \text{Y}_L,\quad   &E > 0 \nonumber\\
     &=  (i k)^{\ell+1} n_\ell (kr) \text{Y}_L,\quad   &E < 0 \nonumber\\
J_L  &=      k^{-\ell}  j_\ell (kr) \text{Y}_L,\quad   &E > 0 \nonumber\\
     &=  (i k)^{-\ell}  j_\ell (kr) \text{Y}_L,\quad   &E < 0 \nonumber\\
     & \text{where} \nonumber\\
h_0  &= \text{Re}\,e^{ikr}/r \quad\text{and}\quad j_0  = \sin(kr)/kr
\label{eq:defmsmHJ}
\end{align}
$H_L$ and $J_L$ are real for any real energy, and structure constants Eq.~(\ref{eq:defs0}) have the property
$S_{KL}(\bR)=S^*_{LK}(-\bR)$. They can generated from the $\ell{=}0$ functions using the operator
$\mathcal{Y}_L(-\nabla)$; see, e.g. Eq.~(\ref{eq:defhl}).  The latter are useful for derivations and also practically,
for instance, to express the gradient of $\nabla H_L$ as a linear combination of other $H_L$.  Eq.~(\ref{eq:defh0l})
shows the explicit relation to the usual spherical Hankel function of the first kind, $h_\ell^{(1)}(z)$.

Spherical harmonics $Y_L$ can be substituted for the $\text{Y}_L$ in Eq.~(\ref{eq:defmsmHJ}).

For historical reasons, Andersen made a different set of definitions, which are convenient for the ASA when $E$=0.  In
this paper we distinguish them from Questaal's standard definitions Eq.~(\ref{eq:defmsmHJ}) with a caret.  For
$E{\le}0$ and $\kappa^2{=}-E$,
\begin{align}
\hat{H}_\ell (\kappa,r) &=  - [(2\ell-1)!!(\kappa w)^{ - \ell} ]^{-1} (i\kappa w)h_\ell (\kappa r) \nonumber\\
& \xrightarrow{\kappa{\to}0} (w/r)^{\ell+1} \nonumber\\
\hat{J}_\ell (\kappa,r) &= ( - 1/2)[(2\ell-1)!!(\kappa w)^{ - \ell} ]j_\ell (\kappa r) \nonumber \\
& \xrightarrow{\kappa{\to}0} [2(2\ell+1)]^{-1} (r/w)^\ell
\label{eq:defokaKJ}
\end{align}
$w$ is a length scale which can be chosen arbitrarily (usually taken to be some average of the MT radii).

\onecolumn

\subsection{Calculations in Real and Reciprocal Space}
\label{app:smHankel-bloch-sum}

All of the formalism developed so far (with the exception of APWs) can apply to either real or reciprocal space.  At
present the Questaal codes work in reciprocal space, with the exception of \texttt{lmpg} (Sec.~\ref{sec:lmpg}), which
uses real space for the principal layer direction.  The $k$ mesh is always a uniform mesh.   Along any axis the
user can specify whether the mesh has a point coinciding with $k$=0 or points that symmetrically straddle it.

When screened MTO's are used, Bloch sums are formed directly from the short ranged orbitals.
The standard  \texttt{lmf} basis is not sufficiently short ranged, and Bloch sums are constructed
with an Ewald technique.
A Bloch sum of $\mathcal{H}_{L}$ is written as a sum over lattice vectors $\mathbf{R}$:
\begin{align*}
{\mathcal{H}^{\mathbf{k}}_{L}(\varepsilon,\gamma;\mathbf{r})}
        = \sum\limits_{\mathbf{R}} {e^{i\mathbf{k}\cdot{\mathbf{R}}}\mathcal{H}_{L}(\varepsilon,\gamma;\mathbf{r-R})}
        = {e^{i{\mathbf{k}} \cdot {\mathbf{r}}}}
        \sum\limits_{\mathbf{R}} {{e^{ - i{\mathbf{k}} \cdot ({\mathbf{r}} - {\mathbf{R}})}}
        \mathcal{H}_{L}(\varepsilon,\gamma;\mathbf{r-R})}
\end{align*}
The latter sum is periodic in $\mathbf{R}$ and can be written as a Fourier series in the reciprocal lattice vectors
$\mathbf{G}$
\begin{align*}
\sum\limits_{\mathbf{R}} {{e^{ - i{\mathbf{k}} \cdot ({\mathbf{r}} - {\mathbf{R}})}}
        \mathcal{H}_{L}(\varepsilon,\gamma;\mathbf{r-R})} =
        \frac{1}{V}\sum\limits_{\mathbf{G}} {{e^{i{\mathbf{G}} \cdot \mathbf{r}}}
        {\ftrns{\mathcal{H}}_{L}}(\varepsilon,\gamma;{\mathbf{k+G}})}
\end{align*}
so that the Bloch-summed $\mathcal{H}_{L}$ is
\begin{align}
{\mathcal{H}^{\mathbf{k}}_{L}(\varepsilon,\gamma;\mathbf{r})}
=\frac{1}{V}\sum\limits_{\mathbf{G}}
{
{\ftrns{\mathcal{H}}_{L}}(\varepsilon,\gamma;{\mathbf{k+G}})
{e^{i({\mathbf{k+G}}) \cdot \mathbf{r}}}
}
\label{eq:blochsmh}
\end{align}
$V$ is the volume of the unit cell.

\subsection{Matrix elements of Smooth Hankel Envelope Functions}
\label{sec:appendixintegrals}

This Appendix derives analytic expressions for matrix elements of the overlap, laplacian, gradient and position
operators for a pair of functions $\mathcal{H}_{k_2L_2}(\varepsilon_2,r_{s_2};{\br-\mathbf{R}_2})$ and
$\mathcal{H}_{k_1L_1}(\varepsilon_1,r_{s_1};{\br-\mathbf{R}_1})$, of the form Eq.~(\ref{eq:defhkl}).  The strategy will
be to use Parseval's identity, Eq.~(\ref{eq:powertheorem}), and evaluate the matrix elements of functions in reciprocal
space.  Analytic solutions can be obtained because products of functions $\ftrns{\mathcal{H}}^*_2\ftrns{\mathcal{H}}_1$
can be contracted to sums of functions of the same class, whose Fourier transform is known.

To begin with, consider matrix elements of the overlap and Laplacian operators.  Since $\nabla^2\mathcal{H}_{kL} =
\mathcal{H}_{k+1L}$, matrix elements of the Laplacian and overlap are different members of the same class.  Using
Eq.~(\ref{eq:defhlq}), (\ref{eq:defhkl}) and (\ref{eq:expandsharm}), products of two
$\ftrns{\mathcal{H}}_{kL}(\mathbf{q})$ read as follows.  To shorten the formulas the smoothing radius will be expressed
in terms of $\gamma{=}r^{2}_{s}/4$.
\begin{align}
\ftrns{\mathcal{H}}^*_1\,\ftrns{\mathcal{H}}_2 =
        \left[\ftrns{h}_{0}(\varepsilon_1,\gamma_1,q)\ftrns{h}_{0}(\varepsilon_2,\gamma_2,q)\right]
        \Big\lbrace
        (-q^2)^{k_1+k_2} \, (-1)^{\ell_1}\,\mathcal{Y}_{L_1}(-i{\bq}) \,\mathcal{Y}_{L_2}(-i{\bq}) \Big\rbrace
        \, e^{i\bq\cdot\mathbf{R}_1} e^{-i\bq\cdot\mathbf{R}_2}
\label{eq:fth2h1}
\end{align}
The quantity in curly braces can be expanded into a linear combination of $\mathcal{Y}_{L}$ using
Eq.~(\ref{eq:expandsharm}),
\begin{align}
(-q^2)^{k_1+k_2}(-1)^{\ell_1} \,\mathcal{Y}_{L_1}(-i{\bq})\,\mathcal{Y}_{L_2}(-i{\bq})
        &= (-1)^{\ell_1} \sum_M C_{L_1L_2M}\, (-q^2)^{k_1+k_2+(\ell_1+\ell_2-m)/2} \, \mathcal{Y}_M(-i{\bq})
\end{align}
and the radial part in square brackets is similarly expanded
\begin{align}
\ftrns{h}_{0}(\varepsilon_1,\gamma_1,q)\ftrns{h}_{0}(\varepsilon_2,\gamma_2,q)
\label{eq:h1h2}
&=\frac{{{{(4\pi)}^2}{e^{{\gamma_1}({\varepsilon_1}-{q^2})}e^{{\gamma_2}({\varepsilon_2}-{q^2})}}}}
        {{({\varepsilon_1}-{q^2})({\varepsilon_2}-{q^2})}}\nonumber\\
&=\frac{{{{(4\pi)}^2}}}{{{\varepsilon_1}-{\varepsilon_2}}}\left[{
        \frac{{{e^{{\gamma_1}({\varepsilon_1}-{\varepsilon_2})}}{e^{({\gamma_1}+{\gamma_2})({\varepsilon_2}-{q^2})}}}}
        {{{\varepsilon_2}-{q^2}}}-\frac{{{e^{{\gamma_2}({\varepsilon_2}-{\varepsilon_1})}}
        {e^{({\gamma_2}+{\gamma_1})({\varepsilon_1}-{q^2})}}}}{{{\varepsilon_1}-{q^2}}}}\right]\nonumber\\
&=\frac{{{{4\pi}}}}{{{\varepsilon_1}-{\varepsilon_2}}}\left[
        {{e^{{\gamma_2}({\varepsilon_2}-{\varepsilon_1})}}{\ftrns{h}_0}({\varepsilon_1},\gamma,q)}
        -{e^{{\gamma_1}({\varepsilon_1}-{\varepsilon_2})}}{\ftrns{h}_0}({\varepsilon_2},\gamma,q)
        \right]\\
        \label{eq:h1h2equale}
&\xrightarrow{\varepsilon_2,\varepsilon_1\to\varepsilon}
        \frac{{(4\pi)}^2}{({\varepsilon_2}-{q^2})^2}
        {e^{({\gamma_1}+{\gamma_2})({\varepsilon}-{q^2})}}\\
        \gamma &= {\gamma_1}+{\gamma_2}.
        \label{eq:avgrs}
\end{align}
When the factors in Eq.~(\ref{eq:fth2h1}) are combined, the product is seen to be the Fourier transform of a linear
combination of $\mathcal{H}_{kM}$ with smoothing radius given by $\gamma_1+\gamma_2$, at the connecting vector
$\mathbf{R}_1-\mathbf{R}_2$.  The overlap matrix can then be written
\begin{align}
\begin{split}
\left< \mathcal{H}_{k_1L_1} {\mathcal{H}}_{k_2L_2} \right>
&=\int
{{\mathcal{H}}^*_{k_1L_1}}(\varepsilon_1,\gamma_1;{\br}-{\bR_1}){{\mathcal{H}}_{k_2L_2}}(\varepsilon_2,\gamma_2;{\br}
        -{\bR_2})\,d^3r\\
&={4\pi} (-1)^{\ell_1}\sum_M C_{L_1L_2M}
        \Big\lbrace
        {\frac{e^{-\gamma_1(\varepsilon_2-\varepsilon_1)}}{\varepsilon_2-\varepsilon_1}
        {{\mathcal{H}}_{k'M}}(\varepsilon_2,\gamma;{\bR})}
        - {\frac{e^{\gamma_2(\varepsilon_2-\varepsilon_1)}}{\varepsilon_2-\varepsilon_1}
        {{\mathcal{H}}_{k'M}}(\varepsilon_1,\gamma;{\bR})}
        \Big\rbrace\\
k' &= k_1+k_2+{(\ell_1+\ell_2-m)/2}, \quad \gamma=\gamma_1+\gamma_2, \quad \bR = {\bR}_1-{\bR}_2
\end{split}
\label{eq:hhint}
\end{align}

The case $\varepsilon_1=\varepsilon_2=\varepsilon$ must be handled specially.  In that limit, the radial part simplifies
to Eq.~(\ref{eq:h1h2equale}).  This function is not contained in the ${\mathcal{H}}_{kL}$ pantheon, and a new function
must be added
\begin{align}
\begin{gathered}
\ftrns{\mathcal{W}}_{kL}(\bq) =
        {\ftrns{w}_{0}(\varepsilon,r_s;{q})}{(-iq)^{2k}}{\mathcal{Y}_L(-i{\bq})}\\
\ftrns{w}_{0}(\varepsilon,\gamma;{q}) = \frac{4\pi}{(\varepsilon-q^2)^2}e^{\gamma(\varepsilon-q^2)}
        = \frac{-1}{(\varepsilon-q^2)} \ftrns{h}_{0}(\varepsilon,\gamma;{q})
\end{gathered}
\label{eq:defw0}
\end{align}
$\ftrns{\mathcal{W}}_{kL}$ is closely related to the energy derivative of $\ftrns{{\mathcal{H}}}_{kL}$. Differentiate
Eq.~(\ref{eq:defhkl}), with respect to energy:
\begin{align}
\begin{gathered}
\ftrns{\dot{\mathcal{H}}}_{kL} \equiv \frac{\partial{\ftrns{\mathcal{H}}}_{kL}}{\partial\varepsilon}
        = {\ftrns{\mathcal{W}}}_{kL} + \gamma {\ftrns{\mathcal{H}}}_{kL}\\
\text{which are generated from radial functions}\\
\ftrns{\dot{h}}_{0}(\varepsilon,\gamma;{q}) =
        \ftrns{w}_{0}(\varepsilon,\gamma;{q}) + \gamma\, \ftrns{h}_{0}(\varepsilon,\gamma;{q})
\end{gathered}
\label{eq:defhkldot}
\end{align}
This provides a practical scheme to obtain ${\mathcal{W}}_{0L}$ in real space since analytic forms for
${\mathcal{H}}_{0L}$ and ${\dot{\mathcal{H}}}_{0L}$ are known (Eqs.~(\ref{eq:defh0}) and (\ref{eq:defh0dot})).  To make
$\mathcal{W}_{kL}$ for higher $k$ it is readily seen that
\begin{align*}
\mathcal{W}_{k+lL}(\br) = -\varepsilon\mathcal{W}_{kL}(\br) - \mathcal{H}_{kL}(\br)
\end{align*}

Using Eq.~(\ref{eq:h1h2equale}) for the quantity in square brackets, Eq.~(\ref{eq:fth2h1}), the resulting integral is
\begin{align}
\begin{split}
\int {{\mathcal{H}}^*_{k_1L_1}}(\varepsilon,\gamma_1;{\br}-{\bR_1}){{\mathcal{H}}_{k_2L_2}}
        (\varepsilon,\gamma_2;{\br}-{\bR_2})\,d^3r = &\\
{4\pi} (-1)^{\ell_1} \sum_M C_{L_1L_2M} {\ {{\mathcal{W}}_{k'M}}(\varepsilon,\gamma;{\bR})}
\end{split}
\label{eq:hhinteqe}
\end{align}
where $k'$, $\gamma$ and $\bR$ the same as in Eq.~(\ref{eq:hhint}).  Note that Eq.~(\ref{eq:hhinteqe}) encompasses
matrix elements between two generalized Gaussians, by virtue of Eq.~(\ref{eq:diffhkl}),
$H_{k+1L}(\varepsilon{=}0)=-{4\pi}G_{kL}$.

\subsubsection{The momentum and position operators acting on a smooth Hankel function}
\label{app:smHankel-rp}

Following the usual rules of Fourier transforms, the gradient operator $\nabla_\mathbf{r}$ and position operator
$\mathbf{r}$ act in reciprocal space as follows.  Use the definition Eqs.~(\ref{eq:defhlq}) and (\ref{eq:defhkl}) with
Eq.~(\ref{eq:ryl}) to obtain
\begin{align}
\begin{split}
\nabla_p\,\mathcal{H}_{kL}(\varepsilon,\gamma;\br) &=
        \frac{1}{(2\pi)^3} \int
        {\ftrns{\mathcal{H}}_{kL}}(\varepsilon,\gamma;{\bq})\,\nabla{{e^{i{\mathbf{q}} \cdot {\br}}}}\,{d^3}q\\
&=-\frac{1}{(2\pi)^3} \int
        {{e^{i{\mathbf{q}} \cdot {\br}}}}\,(-i{\mathbf{q}})
        \,\ftrns{h}_0(\varepsilon,\gamma;q)\mathcal{Y}_{L}(-i\bq)\,{d^3}q\\
&=-\frac{1}{(2\pi)^3} \int {d^3}q\,{e^{i{\mathbf{q}} \cdot {\br}}}
        \sum_{M} \big\lbrace {{C}^{(+)}_{M{L};p}}\ftrns{{\mathcal{H}}}_{kM}(\bq)
        + {{C}^{(-)}_{M{L};p}}{\mathcal{H}}_{k+1M}(\bq) \big\rbrace\\
&=-\sum_{M} \big\lbrace {{C}^{(+)}_{M{L};p}}{{\mathcal{H}}}_{kM}(\br) + {{C}^{(-)}_{M{L};p}}
        {\mathcal{H}}_{k+1M}(\br) \big\rbrace
\label{eq:gradhsm}
\end{split}
\end{align}
${C}^{(\pm)}_{M{L};p}$ has two nonzero elements for fixed $L;p$.  The position operator is more complicated, and we
restrict the derivation to $k{=}0$ case.
\begin{align*}
\br\,\mathcal{H}_{0L}(\varepsilon,\gamma;\br) &= \frac{1}{(2\pi)^3} \int
        {\ftrns{\mathcal{H}}_{0L}}(\varepsilon,\gamma;{\bq})\,\frac{\partial}{i\partial{\mathbf{q}}}
        {{e^{i{\mathbf{q}} \cdot {\br}}}}\,{d^3}q \nonumber\\
&=\frac{1}{(2\pi)^3} \int
        {{e^{i{\mathbf{q}} \cdot {\br}}}}\,
        [i{\nabla_{\mathbf{q}}}\ftrns{h}_0(\varepsilon,\gamma;q)]\mathcal{Y}_{L}(-i\bq) +
        \ftrns{h}_0(\varepsilon,\gamma;q)[i{\nabla_{\mathbf{q}}}\mathcal{Y}_{L}(-i\bq)]
        \,{d^3}q
\end{align*}
The first term contains
\begin{align*}
[i{\nabla_{\mathbf{q}}}\ftrns{h}_0(\varepsilon,\gamma;q)] &=
         2i{\mathbf{q}}\,(\ftrns{w}_0 - \gamma\ftrns{h}_0) = 2(-i{\mathbf{q}})\,\ftrns{\dot{h}}_0\\
i{\nabla_{\mathbf{q}}}\ftrns{h}_0(\varepsilon,\gamma;q)]{\mathcal{Y}_{L}(-i\bq)} &=
        2\sum_{M} \big\lbrace {{C}^{(+)}_{M{L};p}}{\ftrns{\dot{\mathcal{H}}}_{0M}(\bq)} +
        {{C}^{(-)}_{M{L};p}} {\ftrns{\dot{\mathcal{H}}}_{1M}(\bq)} \big\rbrace
\end{align*}
using Eq.~(\ref{eq:ryl}).  The second term follows directly from Eq.~(\ref{eq:gradrlyl}) to yield
\begin{align}
\br_p\,\mathcal{H}_{0L}(\varepsilon,\gamma;\br) &=
        \sum_{M} \big\lbrace {{C}^{(+)}_{M{L};p}}(2{\dot{\mathcal{H}}}_{0M}(\br)) +
        {{C}^{(-)}_{M{L};p}} ({2{\dot{\mathcal{H}}}_{1M}(\br)}+(2\ell+1){{{\mathcal{H}}}_{0M}(\br)}) \big\rbrace
\label{eq:rsm}
\end{align}

\subsubsection{Matrix elements of position and gradient operators}
\label{app:rp-integrals}

Eq.~(\ref{eq:gradhsm}) shows that $\nabla {{\mathcal{H}}_{kL}(\br)}$ generates a linear combination of the same family
of ${{\mathcal{H}}_{kL}(\br)}$, evaluated at the vector connecting atom centres.  Thus, matrix elements
of the gradient operator are linear combinations of the same functions.  A similar situation applies to matrix elements
of $\br$, though they entail matrix elements of the overlap between ${{\mathcal{H}}_{kL}}$ and
${\dot{\mathcal{H}}}_{kL}$.  These integrals can be evaluated in reciprocal space using Parseval's identity, expanding
$\ftrns{\mathcal{H}}^*_1\, \ftrns{\dot{\mathcal{H}}}_2$ analogously to Eq.~(\ref{eq:fth2h1}).  The radial part of this
product can be expanded as:
\begin{align}
\begin{gathered}
\ftrns{h}_{0}(\varepsilon_1,\gamma_1,q)\ftrns{\dot{h}}_{0}(\varepsilon_2,\gamma_2,q)
        =\frac{{{{-(4\pi)}^2}{e^{{\gamma_1}({\varepsilon_1}-{q^2})}e^{{\gamma_2}({\varepsilon_2}-{q^2})}}}}
        {{({\varepsilon_1}-{q^2})({\varepsilon_2}-{q^2})^2}}
        + \gamma_2\,\ftrns{h}_{0}(\varepsilon_1,\gamma_1,q)\ftrns{{h}}_{0}(\varepsilon_2,\gamma_2,q)\\
=\frac{4\pi}{{\varepsilon_1-\varepsilon_2}}
        \bigg\lbrace
        {{e^{\left(\varepsilon_1-\varepsilon_2\right)\,\gamma_1}}{\ftrns{w}_0}({\varepsilon_2},\gamma,q)}
        +[{\gamma_2}+{{({\varepsilon_1}-{\varepsilon_2})}^{-1}}]\,
        {e^{-{\varepsilon_2}\,{\gamma_1}-{\varepsilon_1}\,{\gamma_2}}}
        \left(e^{\varepsilon_1\,\gamma}\,\ftrns{h}_0(\varepsilon_2,\gamma,q)
        -e^{\varepsilon_2 \,\gamma}\,\ftrns{h}_0(\varepsilon_1,\gamma,q)\right)
        \bigg\rbrace
\end{gathered}
\label{eq:h2h1dot}
\end{align}

Substituting this radial function into the square brackets in Eq.~(\ref{eq:h1h2equale}), we obtain
\begin{align}
\begin{split}
\left< \mathcal{H}_{k_1L_1} \dot{\mathcal{H}}_{k_2L_2} \right> =&
        \int
        {{\mathcal{H}}^*_{k_1L_1}}(\varepsilon_1,\gamma_1;{\br}-{\bR_1})
        {{\mathcal{\dot{H}}}_{k_2L_2}}(\varepsilon_2,\gamma_2;{\br}-{\bR_2})\,d^3r\\
        =&\frac{{4\pi} (-1)^{\ell_1}}{\varepsilon_1-\varepsilon_2}\sum_M C_{L_1L_2M} \times
        \bigg\lbrace
        {{e^{\left(\varepsilon_1-\varepsilon_2\right)\,\gamma_1}}
        {{\mathcal{W}}_{k'M}}(\varepsilon_2)
        } \;+\\
        &[{\gamma_2}+{{({\varepsilon_1}-{\varepsilon_2})}^{-1}}]\,
        {e^{-{\varepsilon_2}\,{\gamma_1}-{\varepsilon_1}\,{\gamma_2}}}
        \Big(e^{\varepsilon_1\,\gamma}\,{{\mathcal{H}}_{k'M}}(\varepsilon_2)
        - e^{\varepsilon_2 \,\gamma}\,{{\mathcal{H}}_{k'M}}(\varepsilon_1)
        \Big)
        \bigg\rbrace
\label{eq:hhdotint}
\end{split}
\end{align}
where ${\mathcal{W}_{k'M}}$ and ${\mathcal{H}_{k'M}}$ are evaluated with the same arguments $k'$, $\gamma$ and $\bR$ as
in Eq.~(\ref{eq:hhint}).

Combining Eqs.~\ref{eq:gradhsm}, \ref{eq:rsm}, \ref{eq:hhint} and \ref{eq:hhdotint}
\begin{align}
\begin{split}
\left< \mathcal{H}_{k_1L_1} \nabla_p \mathcal{H}_{k_2L_2} \right>&=\int
        {{\mathcal{H}}^*_{k_1L_1}}(\varepsilon_1,\gamma_1;{\br}-{\bR_1})
        \nabla_p {{\mathcal{H}}_{k_2L_2}}(\varepsilon_2,\gamma_2;{\br}-{\bR_2})\,d^3r\\
&=-\sum_{M} \big\lbrace {{C}^{(+)}_{M{L_2};p}} \left< \mathcal{H}_{k_1L_1} \mathcal{H}_{k_2M} \right>
        + {{C}^{(-)}_{M{L};p}} \left< \mathcal{H}_{k_1L_1} \mathcal{H}_{k_2+1M} \right> \big\rbrace
\end{split}
\label{eq:hgradhint}
\end{align}
\begin{align}
\begin{split}
\left< \mathcal{H}_{0L_1} ({\br}-{\bR_2})_p \mathcal{H}_{0L_2} \right>=\int
        {{\mathcal{H}}^*_{0L_1}}(\varepsilon_1,\gamma_1;{\br}-{\bR_1}) \left({\br}-{\bR_2}\right)_p
        {{\mathcal{H}}_{0L_2}}(\varepsilon_2,\gamma_2;{\br}-{\bR_2})\,d^3r\\
=\sum_{M} \Big\lbrace {{C}^{(+)}_{M{L_2};p}} 2 \left< \mathcal{H}_{0L_1} \dot{\mathcal{H}}_{0M} \right>
        + {{C}^{(-)}_{M{L};p}} \left(2\left< \mathcal{H}_{0L_1} \dot{\mathcal{H}}_{1M} \right>
        +(2\ell_2+1)\left< \mathcal{H}_{0L_1} \mathcal{H}_{0M} \right> \right)
        \Big\rbrace
\end{split}
\label{eq:hrint}
\end{align}

\twocolumn

\begin{thebibliography}{100}
\bibitem{ldauerr}
{Note that Eq.(2) in Ref.~\cite{Larson07} incorrectly uses $50/429$ (taken from
  Ref.~\cite{Anisimov97}) instead of $1/858$ as the contribution of $F^6(ff)$
  to the average $J$, thus the actual $J$ used were about 25\,\% smaller than
  the ones listed. However, since this did not correspond to the correct
  average $J$, it makes an error in the double counting term which shifts the
  $f$-bands somewhat relative to the other bands. Also the calculation used the
  Hartree-Fock unscreened atomic values of the Slater integrals $F^k$ for
  $k\ge2$.}
\end{thebibliography}

\end{document}